\documentclass[12pt]{article}

\usepackage{amssymb}
\usepackage{amsmath}
\usepackage{amsthm}
\usepackage{cancel}
\usepackage{slashed}
\usepackage{fullpage}
\usepackage{color}
\usepackage{braket}
\usepackage{graphicx}
\usepackage{caption}
\usepackage{subcaption}
\usepackage{multirow}
\usepackage{verbatim}
\usepackage{cite}
\usepackage{bigints}
\usepackage{simplewick}
\usepackage{authblk}
\usepackage{hyperref}
\usepackage{empheq}
\usepackage{tikz}
\usetikzlibrary{calc}
\usepackage{bbm}
\usepackage{float}

\hypersetup{
    linktocpage,
     colorlinks,
     citecolor=darkgreen,
     linkcolor= darkgreen,
     urlcolor=darkgreen
}

\definecolor{darkred}{rgb}{0.5,0.0,0.0}
\definecolor{darkblue}{rgb}{0.0,0.0,0.9}
\definecolor{darkerblue}{rgb}{0.0,0.0,0.5}
\definecolor{purple}{rgb}{0.5,0.0,0.5}
\definecolor{darkgreen}{rgb}{0.0,0.5,0.0}
\definecolor{black}{rgb}{0.0,0.0,0.0}
\definecolor{brown}{rgb}{0.6,0.4,0.2}
\definecolor{newpurple}{rgb}{0.65, 0.38, 0.61}
\definecolor{newyellow}{rgb}{0.9718, 0.6093, 0.0759}
\definecolor{amber}{rgb}{1.0, 0.75, 0.0}
\definecolor{newblue}{rgb}{0.4, 0.52, 0.85}
\definecolor{newred}{rgb}{0.8524, 0.2595, 0.3294}
\definecolor{newgreen}{rgb}{0.2, 0.8, 0.2}
\definecolor{SMgreen}{rgb}{0.56, 0.69, 0.19}
\definecolor{neworange}{rgb}{0.94, 0.462, 0.162}
\newcommand{\red}{\color{darkred}}
\newcommand{\purple}{\color{purple}}
\newcommand{\orange}{\color{orange}}
\newcommand{\lightred}{\color{red}}
\newcommand{\blue}{\color{darkblue}}
\newcommand{\green}{\color{darkgreen}}

\def\be{\begin{equation}}
\def\ee{\end{equation}}
\def\cL{\mathcal{L}}
\def\cO{\mathcal{O}}
\def\cD{\mathcal{D}}

\def\cE{\mathcal{E}}
\def\cI{\mathcal{I}}
\def\cJ{\mathcal{J}}
\def\cR{\mathcal{R}}

\newcommand{\lameff}{\lambda_{\text{eff}}}

\def\msbar{\overline{\text{MS}}}

\newcommand{\barr}{\text{B}}

\newcommand{\B}{B}

\DeclareMathOperator\disc{disc}

\def\muX{\mu_X}

\renewcommand\Im{\text{Im}}
\newcommand{\Vmin}{V_{\text{min}}}

\newcommand{\abs}[1]{\left| #1 \right|}
\DeclareMathOperator{\Res}{Res}

\newcommand{\fd}[2]{\parbox{#1}{\includegraphics[width=#1]{#2}}}

\newcommand{\eqna}[1]{\begin{align}\begin{aligned}#1\end{aligned}\end{align}}
\newcommand{\eqn}[1]{\begin{align}#1\end{align}}

\numberwithin{equation}{section}
\newcommand{\mc}[1]{\ensuremath{\mathcal{#1}}}

\newcommand{\nonlintime}{T_{\text{NL}}}
\newcommand{\sloshtime}{T_{\text{slosh}}}

\def\cL{\mathcal{L}}
\def\cO{\mathcal{O}}
\def\cD{\mathcal{D}}

\def\cE{\mathcal{E}}

\def\FV{\text{FV}}
\def\TV{\text{TV}}
\def\DV{\text{DV}}
\def\R{\text{R}}
\def\omegaa{\omega_a}

\newcommand{\Veff}{V_{\text{eff}}}
\newcommand{\VSM}{V_{\text{SM}}}

\newcommand{\GeV}{\text{GeV}}
\newcommand{\Vmax}{V_\text{max}}
\newcommand{\Seff}{S_\text{eff}}
\newcommand{\meff}{m_\text{eff}}
\newcommand{\leff}{\lambda_\text{eff}}
\newcommand{\Mpl}{M_{\text{Pl}}}

\newcommand{\mpl}{\Mpl}
\newcommand{\Rmin}{R_M}

\newcommand{\cN}{\mathcal{N}}
\newcommand{\vnew}{V_{\text{new}}}
\newcommand{\bphi}{ {\bar\phi} }

\newcommand{\Det}{\text{Det}}
\newcommand{\Tr}{\text{Tr}}

\newcommand{\Du}{ \overline{D}_F}
\def\bigtau{\mathcal{T}}
\def\Dx{\cD x~}
\def\Dxspace{\hspace{-2 em}\Dx}

\title{Precision decay rate calculations in quantum field theory}

\author{Anders Andreassen\thanks{anders@physics.harvard.edu}  }
\author{David Farhi\thanks{farhi@physics.harvard.edu}  }
\author{William Frost\thanks{wfrost@physics.harvard.edu}  }
\author{Matthew D. Schwartz\thanks{schwartz@physics.harvard.edu} }

\affil{\emph{Department of Physics,
Harvard University, Cambridge, MA 02138, USA}}

\begin{document} 
\maketitle

\begin{abstract}
Tunneling in quantum field theory is worth understanding properly, not least because it controls
the long term fate of our universe. There are however, a number of features of tunneling rate
calculations which lack a desirable transparency, such as the necessity of analytic continuation,
the appropriateness of using an effective instead of classical potential, and the sensitivity to
short-distance physics. This paper attempts to review in pedagogical detail the physical origin of
tunneling and its connection to the path integral. Both the traditional potential-deformation
method and a recent more direct propagator-based method are discussed. Some new insights from
using approximate semi-classical solutions are presented. In addition, we explore the
sensitivity of the lifetime of our universe to short distance physics, such as quantum gravity,
emphasizing a number of important subtleties. 
\end{abstract}

\newpage

\tableofcontents

\newpage

\section{Introduction}
Whether the long-term future of the universe is controlled by the slow freezing and rarefaction of
cosmic acceleration
 or the sudden formation and growth of negative-energy bubbles
 is a question of visceral appeal even to non-scientists.
 There are also practical
aspects (from a particle-physics point of view) to vacuum stability, such as eliminating models of new physics or motivating new colliders to measure the top mass. Thus, it would
be good to know whether the question of stability can even be answered confidently assuming no new physics.
This paper provides a survey of some impediments to establishing that confidence. We provide a new perspective on some old
methods, such as the connection between the path integral and tunneling, and bring clarity to some recent debates, such as the UV-sensitivity of the universe's lifetime.  

One challenge to computing the rate for tunneling out of our metastable vacuum is establishing
a systematically-improvable framework for computing this rate in the first place. 
In quantum mechanics, in the absence of any approximate methods, the decay of a given initial wavefunction
can always be calculated by numerically solving Schr\"odinger's equation. In quantum field theory,
one does not have this crutch: not only is Schr\"odinger's equation infinite dimensional,
but the wave-functional (the field-theory analog of the wave-function) inspires little physical intuition.

The first few sections of this paper are devoted to reviewing how decay rates are defined, the relevant time scales,
and the derivation of various formulas used to compute them. The traditional method,
pioneered by Coleman and Callan~\cite{Callan:1977pt} (see also~\cite{Kobzarev:1974cp}) focuses on computing the imaginary part of the matrix element
$\langle a | e^{-HT} |a \rangle$, where in quantum mechanics, $|a\rangle$ is a position eigenstate.
Because this matrix element is real, one cannot simply take its imaginary part. Rather, one must analytically continue the potential
so that the false vacuum is stable, compute the matrix element, then analytically continue back. 
There are some excellent reviews of this {\it potential-deformation method}~\cite{Kleinert,Muller-Kirsten:2012wla,ZinnJustin:2002ru,Marino,Weinberg:2012pjx}.
The method seems to give the
right answer, in cases where it can be checked. 
Nevertheless,  some elements of its derivation seem to us in need of further clarification. 
For example, for physical potentials, which are bounded from below, analytic continuation gives the wrong answer.
Instead, the steepest descent contour passing through the saddle point associated with the false vacuum plays 
an essential role. 
We provide our own perspective on this method, which we hope the reader
will find illuminating.

Having digested the Callan-Coleman potential-deformation approach, one suspects that there should somehow be a more direct way to connect
tunneling rates to the path integral. Such a connection was presented in~\cite{Andreassen:2016cff} and is expounded on in Section~\ref{sec:dirmethod}. The method introduced in~\cite{Andreassen:2016cff}
is based on a direct computation of the probability for a particle to propagate through a barrier. It
has the advantage of maintaining a closer connection between the underlying physical assumptions, such as the hierarchy of timescales
required for the tunneling rate to be well-defined, than the potential-deformation method. A summary comparison of the potential-deformation
and the direct methods is given in Section~\ref{sec:compare}.

To compute a decay rate in the saddle-point approximation, one must find bounces: solutions to the Euclidean equations of motion with some particular
boundary conditions. These bounces are functions $\phi(\vec{x},\tau)$ where Euclidean time $\tau$ parametrizes a path through field space. This path
is analogous to the most-probable path that a particle passes though a barrier in the WKB approximation~\cite{Banks:1973ps}. In Section~\ref{sec:energy}, we discuss how
to think about the functional $U[\phi]$ so that it provides a close analogy with the potential energy barrier in quantum mechanics. Finding exact bounces
can be challenging, even numerically. Fortuitously, many features of the exact bounces are well described by approximate bounces which can be studied analytically, as
we discuss in Section~\ref{sec:approx}. 

In quantum field theory, it can happen that metastability arises from radiative corrections. The famous example of this is the Coleman-Weinberg model, where
a stable potential $V=\lambda \phi^4$ turns over and runs negative due to photon loops in scalar QED~\cite{Coleman:1973jx}. It is natural to anticipate
that one should therefore use the effective potential, which demonstrates the instability, to compute the tunneling rate. We argue that this is not correct. 
First of all, corrections to the effective action which vanish for constant fields may contribute equally to a rate as potential terms. Second, using
the effective potential double counts the radiative corrections: particle loops contribute both to $\Veff$ and again to the rate. Although these observations
are not deep, it is not uncommon to see $\Veff$ used as a classical potential to find bounce solutions. The appropriate use of effective actions is discussed in Section~\ref{sec:NLO}.

Finally, we reduce to the Standard Model. The lifetime of our universe has been intensively studied since the tunneling calculations in quantum field theory were first understood~\cite{Coleman:1978ae,Frampton:1976pb,Sher:1988mj,Espinosa:1995se,Isidori:2001bm,Espinosa:2007qp,Ellis:2009tp}.
Even if we assume the Standard Model is valid  
up to the Planck scale, one must be sure that quantum gravity cannot invalidate the perturbative decay rate calculation.
Current precision measurements and calculations imply that, in the 
absence of new physics, our universe will decay through the formation of ultra-tiny bubbles, with radii $R\sim (10^{17}\, \GeV)^{-1} \sim 10^{-31}\, \text{cm}$.
This bubble size
is essentially determined by the scale where the $\beta$-function for the Higgs quartic vanishes (we provide a new derivation of this result in Section~\ref{sec:approx}). 
 Although $10^{17}\, \GeV$ is close to the Planck
scale, it has been argued that it is far enough below $\mpl$ that quantum gravitational effects on $\Veff$ can be ignored~\cite{Isidori:2001bm, Espinosa:2007qp, Ellis:2009tp,DiLuzio:2015iua}.
It has also been argued that these effects cannot be ignored since the bubble takes transplanckian field values at its center~\cite{Branchina:2014rva,Branchina:2013jra,Branchina:2014efa,Branchina:2014usa}. 
The latter conclusion has been verified by other groups and we agree that the gravitational contributions to the decay rate can have important effects.
However, as we explain in Section~\ref{sec:uv}, the problem of UV sensitivity is not
just the coincidence between $\mpl$ and the flat point of Higgs quartic: the Standard Model would be Planck sensitive even if $\mpl$ were $10^{100}\, \GeV$.
Moreover, it is not correct to just use the effective potential to determine the bubble size, one really needs the full effective action, as we emphasize in Section~\ref{sec:NLO}.

A summary of some of the new perspectives provided in this paper is given in our conclusions, Section~\ref{sec:conc}.

\section{Tunneling in quantum mechanics \label{sec:QM}}
Much of our intuition for tunneling comes from one dimensional quantum mechanics. Indeed, Gamow's 1928 calculation of the relation between half-life and the energy $E$ of emitted $\alpha$-particles was seminal in establishing the validity of quantum mechanics~\cite{Gamow:1928}. 
So it is natural to start our discussion with this case. 
Gamow modeled the nuclear potential $V(x)$
as having a $\frac{1}{x}$ Coulomb tail and some kind of well for $x<a$ where the $\alpha$ particle is trapped. In 1D, the wavefunction of a state
with energy $E$  falls off exponentially between $a$ and $b$ by an amount given approximately by the WKB formula:
\be
T(E) \equiv\frac{\psi_E(b)}{\psi_E(a)} \equiv e^{-W} \approx \exp\left[ - \int_a^b d x\sqrt{2m(V(x) - E)} \right]  \label{TofE}
\ee
Here, $a$ and $b$ are the turning points where $V(a)=V(b)=E$. 
It is of course quite logical that the decay rate should be proportional to how much of the wavefunction gets through the barrier,
$\Gamma\sim |T(E)|^2$. However, if the particle is in an energy eigenstate, there is no time-dependence, so it cannot decay. To go from $T(E)$ to $\Gamma$, a step often skipped,
requires considerably more thought. 

\begin{figure}[t]\begin{center}
\begin{tikzpicture}
\node at (0,0) {\includegraphics[width=0.4\columnwidth]{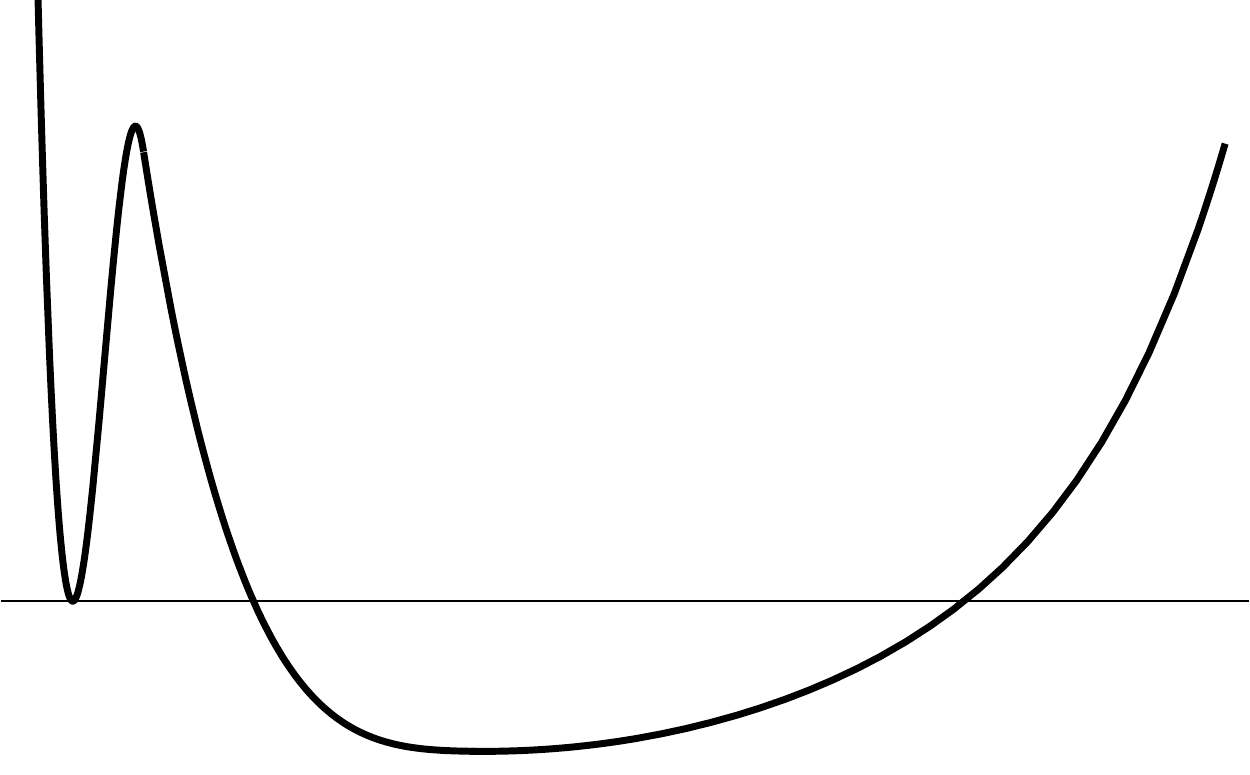}};
\draw [<->] (-3.,1.3) -- (-2.7,1.3);
\node [above] at (-2.8,1.3) {\footnotesize \FV}; 
\draw [<->] (-2.0,1.3) -- (2.6,1.3);
\node [above] at (0.3,1.3) {\footnotesize \R};
\node [above] at (3.3,-1.1) {x};
\node at (3.5,1.6) {V(x)};
\node at (-2.9,-1.39) {a};
\node at (-2.1,-1.35) {b};
\end{tikzpicture}
\begin{tikzpicture}
\node at (0,0) {\includegraphics[width=0.4\columnwidth]{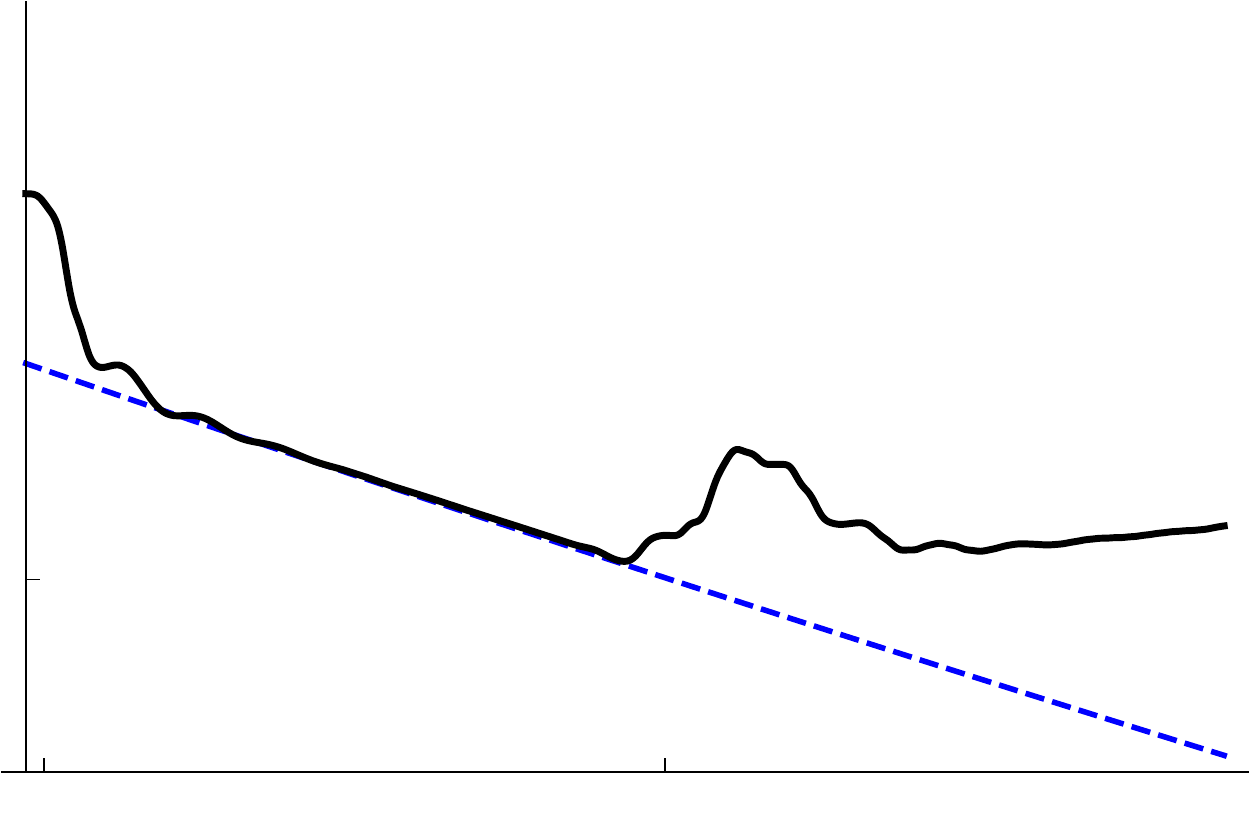}};
\node at (2.8,-0.3) {$P_{FV}$}; 
\node at (3.,-1.3) [blue] {$P_0 e^{-\Gamma T}$};
\node at (3.5,-1.95) {$T$}; 
\draw [<->] (-2.8,-0.3) -- (-2.5,-0.3);
\node [below] at (-2.6,-0.31) {$\sloshtime$}; 
\draw [<->] (-3.05,-1.3) -- (0.2,-1.3);
\node [below] at (-1.7,-1.3) {$\nonlintime$}; 
\node [above] at (-3.1,2.1) {$P$}; 
\node [left] at (-3.1,1.15) {$1$}; 
\node [above left] at (-3.1,-1.15) {$0.9$}; 
\end{tikzpicture}
\caption{
On the left, an example of a physical potential with an metastable region $\FV$, a destination region $\R$, and a barrier. We label the local minimum inside the $\FV$ region by $a$ and the turning point by $b$ (defined by $V(b)=V(a)$). On the right, the probability $P_\FV(T)$ (see Eq.~(\ref{eqn:defineP})) for this system (beginning in a Gaussian wavepacket centered at $a$) computed by numerically solving Schr\"odinger's equation. We see that the probability to find the particle in the false vacuum decays exponentially for intermediate times between the short timescale of sloshing inside the false vacuum and the long timescale on which the wavefunction begins to flow back into the false vacuum.
\label{fig:numericpotentialprobability}
}
\end{center}
\end{figure}
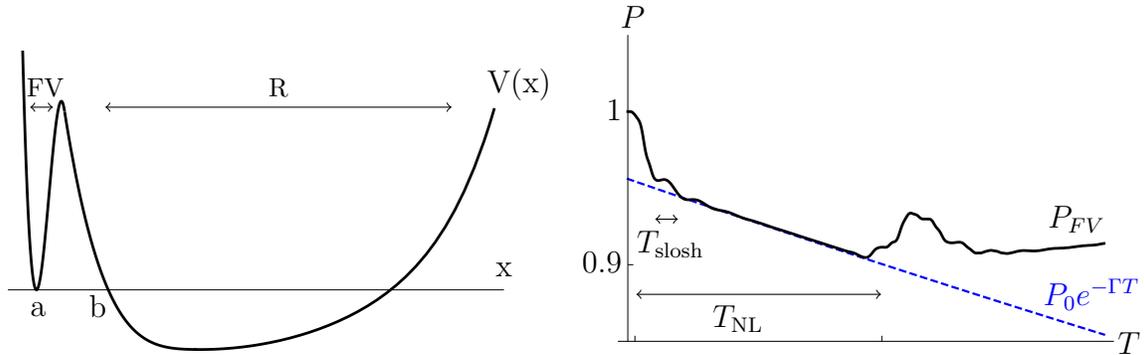

A simple picture often used to convert $T(E)$ to a decay rate depicts a particle with momentum $p = \sqrt{2mE}$,
and velocity $v = \frac{p}{m}$ in the well hitting the barrier with a rate $\frac{v}{2a}$, and each time tunneling through with probability given by the transmission coefficient, $|T(E)|^2$ (see e.g.~\cite{griffiths2005introduction}).
 With this logic,  the decay rate is
\be
\Gamma \sim \frac{p}{2am} \left|\frac{\psi_E(b)}{\psi_E(a)}\right|^2 \approx \frac{p}{2a m} e^{-2W} \label{GofT}
\ee
Indeed, if one solves the Schr\"odinger equation numerically, one can see the 
wavefunction oscillate back and forth in the well; the largest flux leaks out during the times when the wavefunction is closest
to the barrier. Fig.~\ref{fig:numericpotentialprobability} shows this exponential decay with time and the small oscillations. Snapshots of the wavefunction
oscillating in the well are shown in Fig.~\ref{fig:sloshinganimationsteps}.

\newcommand{\labelplottime}[1]{\node at (0.3,1) {t=#1};}
\begin{figure}[ht]\begin{center}
\begin{tikzpicture}
\node at (0,0) {\includegraphics[width=0.2\columnwidth]{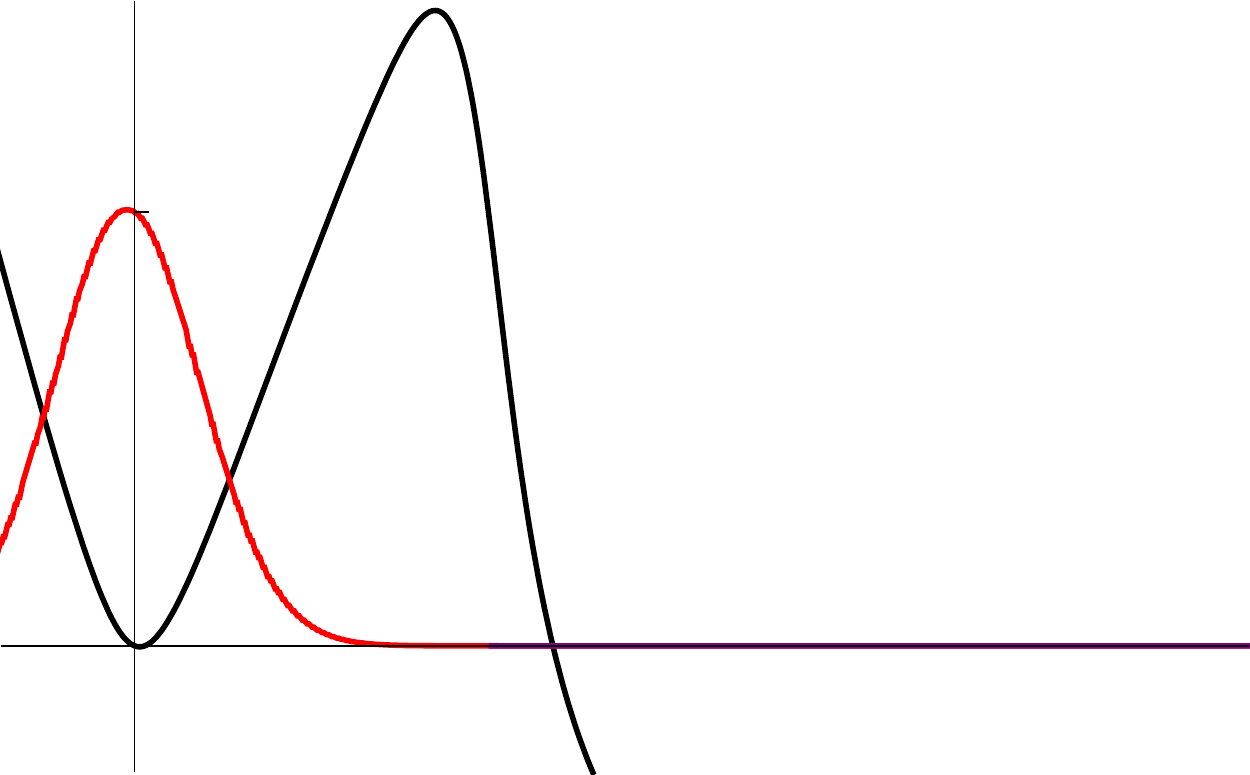}};
\labelplottime{0};
\node at (-2,-0.5) {\lightred $\abs{\psi}^2$};
\node at (-1.8,0.4) {$V$};
\end{tikzpicture}
\begin{tikzpicture}
\node at (0,0) {\includegraphics[width=0.2\columnwidth]{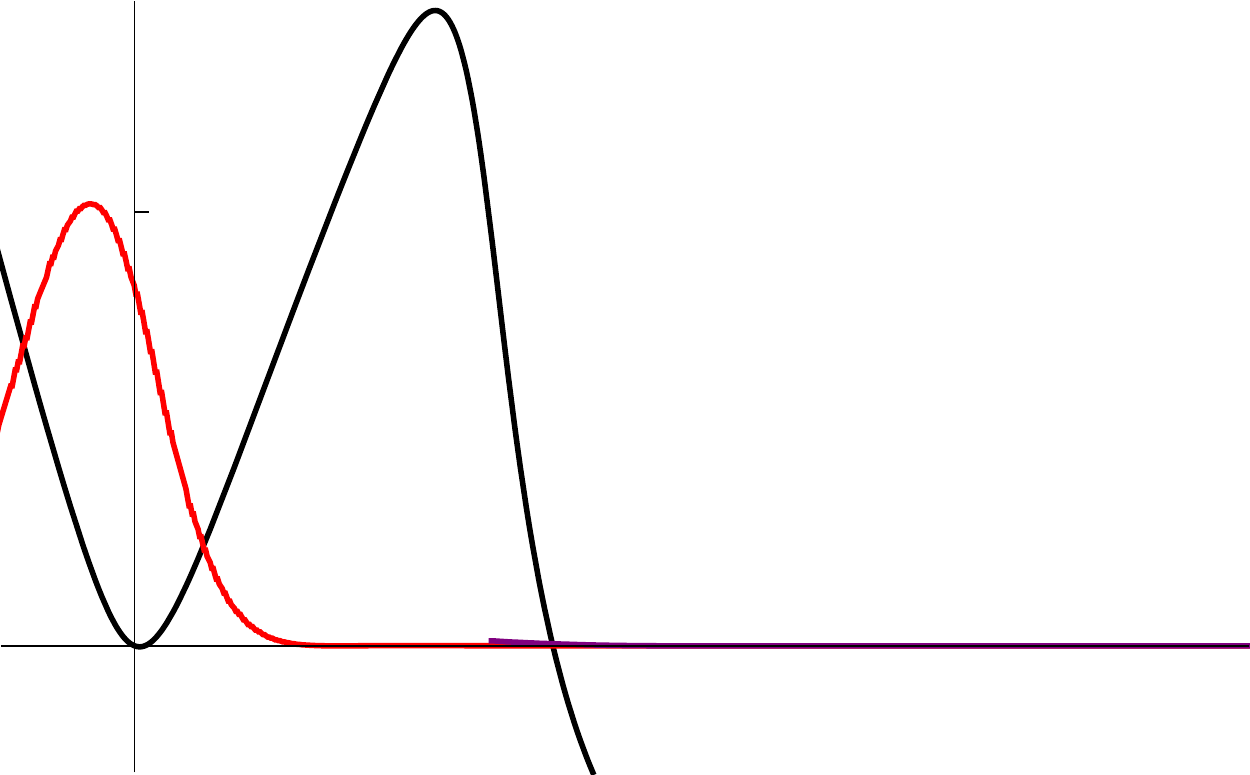}};
\labelplottime{1};
\end{tikzpicture}
\begin{tikzpicture}
\node at (0,0) {\includegraphics[width=0.2\columnwidth]{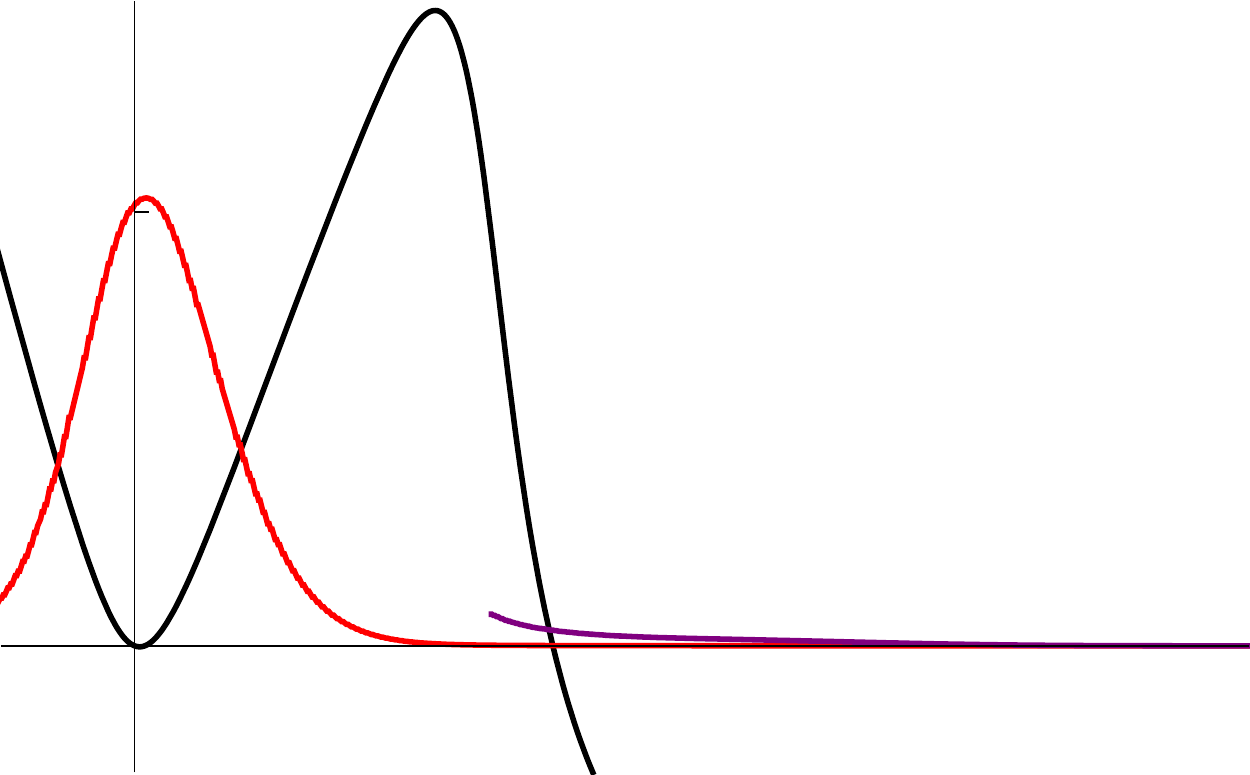}};
\labelplottime{2};
\end{tikzpicture}
\begin{tikzpicture}
\node at (0,0) {\includegraphics[width=0.2\columnwidth]{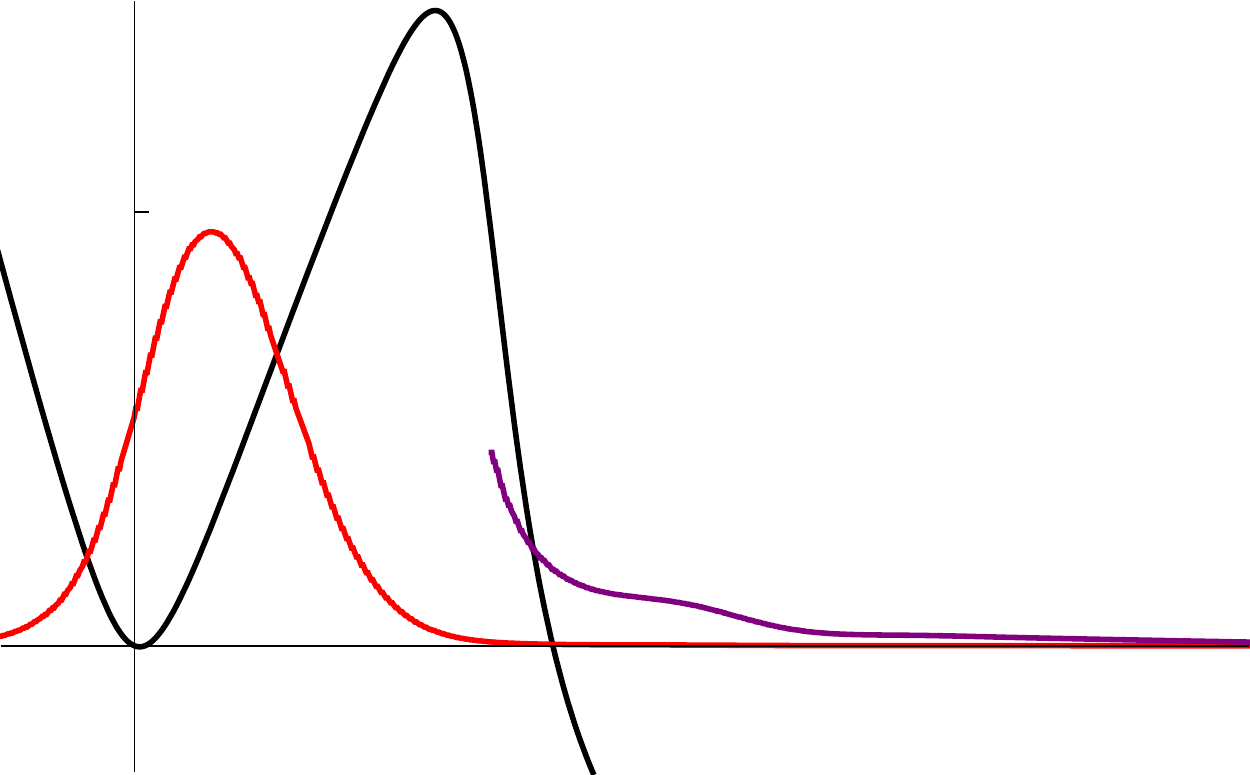}};
\labelplottime{3};
\node at (0.7,-0.2) {\color{purple} $50\abs{\psi}^2$};
\end{tikzpicture}
\begin{tikzpicture}
\node at (0,0) {\includegraphics[width=0.2\columnwidth]{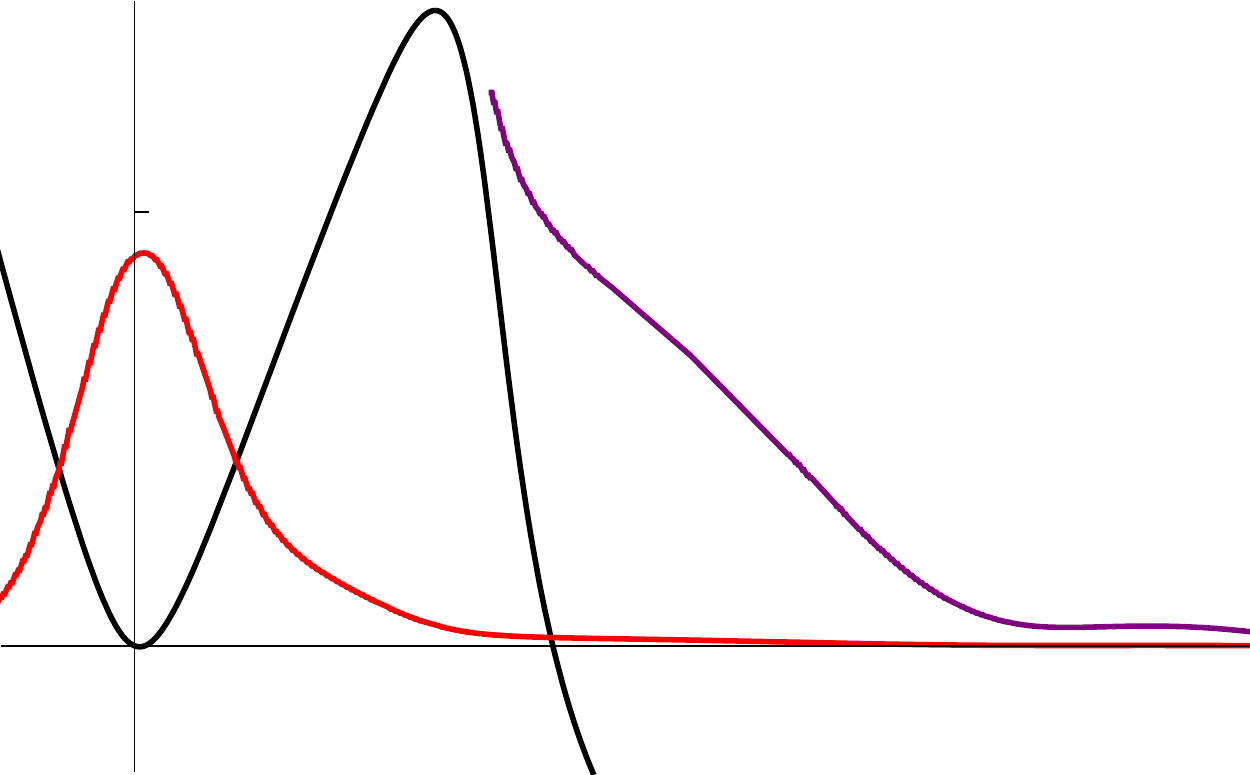}};
\labelplottime{4};
\end{tikzpicture}
\begin{tikzpicture}
\node at (0,0) {\includegraphics[width=0.2\columnwidth]{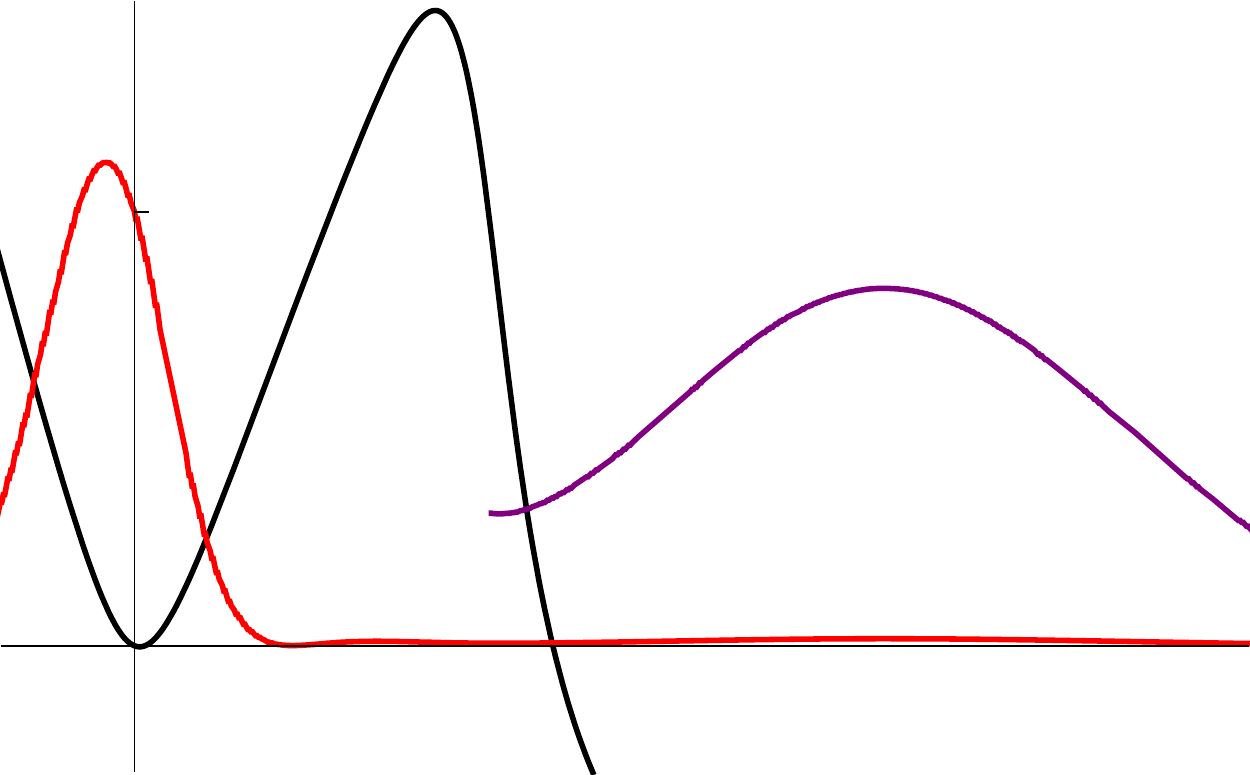}};
\labelplottime{5};
\end{tikzpicture}
\begin{tikzpicture}
\node at (0,0) {\includegraphics[width=0.2\columnwidth]{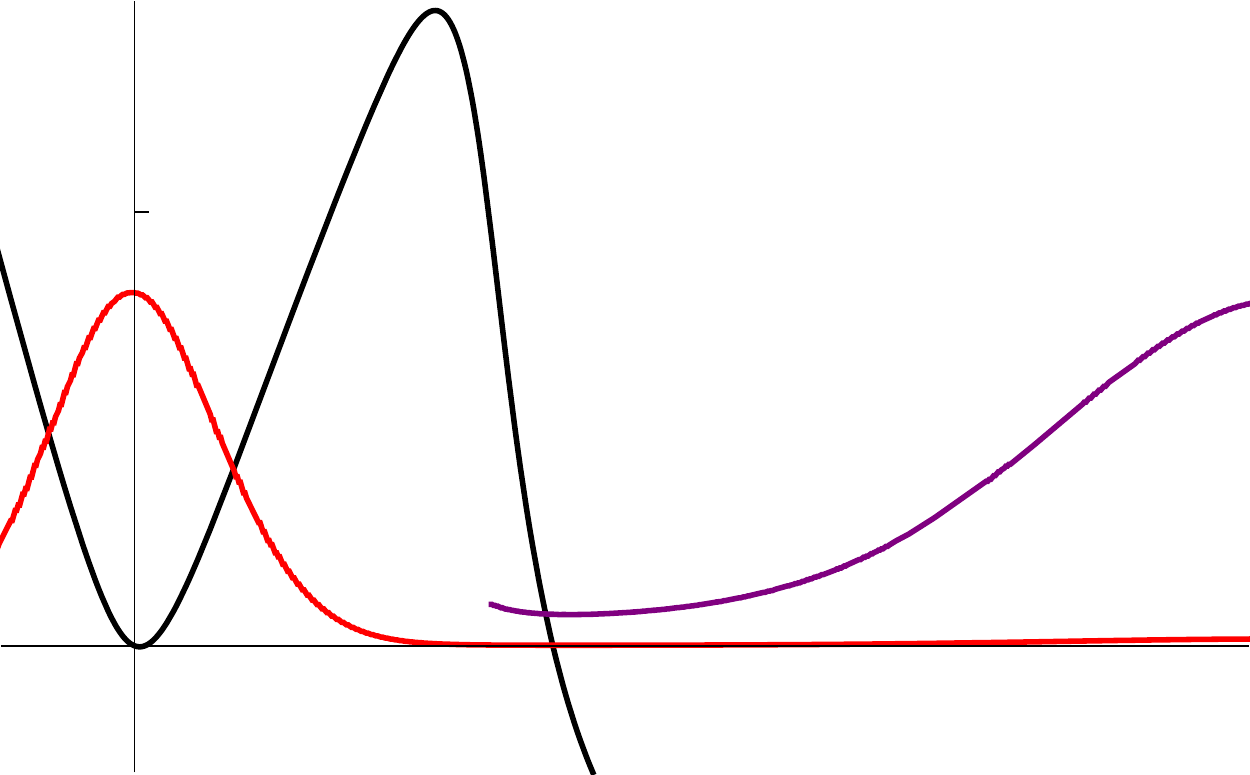}};
\labelplottime{6};
\end{tikzpicture}
\begin{tikzpicture}
\node at (0,0) {\includegraphics[width=0.2\columnwidth]{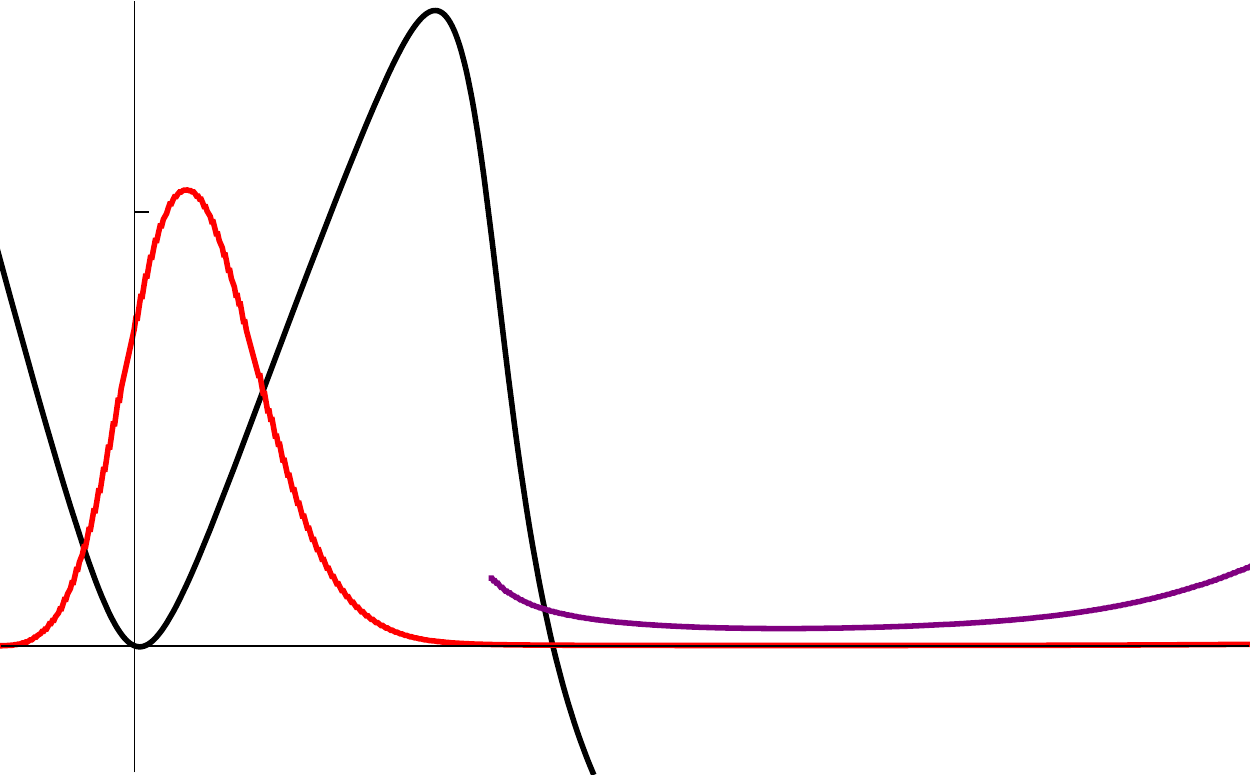}};
\labelplottime{7};
\end{tikzpicture}
\begin{tikzpicture}
\node at (0,0) {\includegraphics[width=0.2\columnwidth]{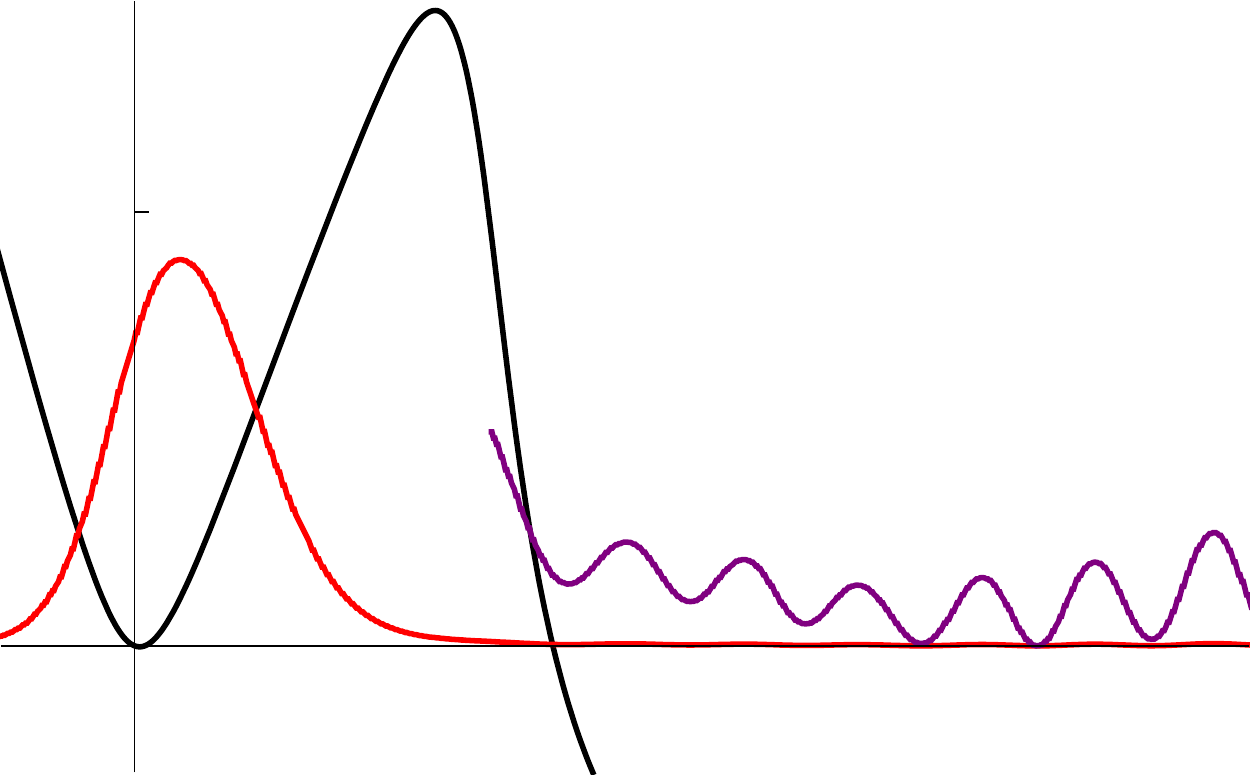}};
\labelplottime{20};
\end{tikzpicture}
\begin{tikzpicture}
\node at (0,0) {\includegraphics[width=0.2\columnwidth]{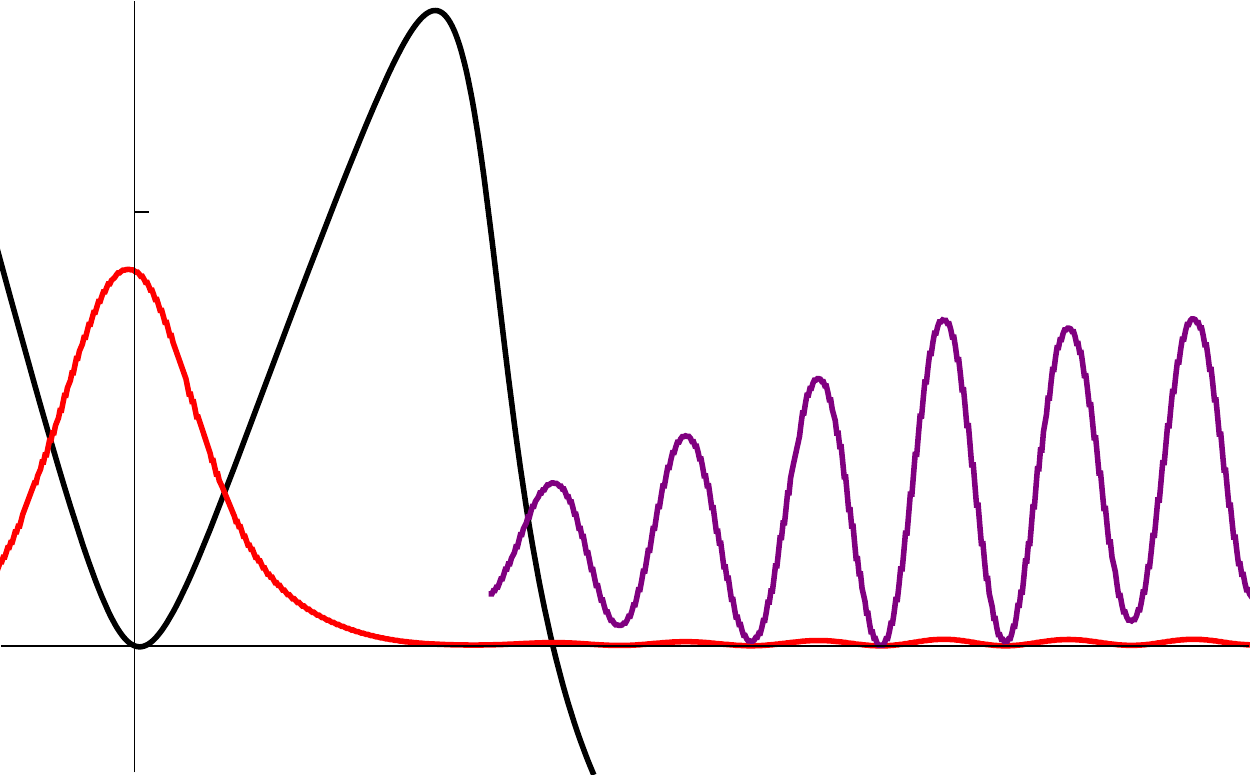}};
\labelplottime{21};
\end{tikzpicture}
\begin{tikzpicture}
\node at (0,0) {\includegraphics[width=0.2\columnwidth]{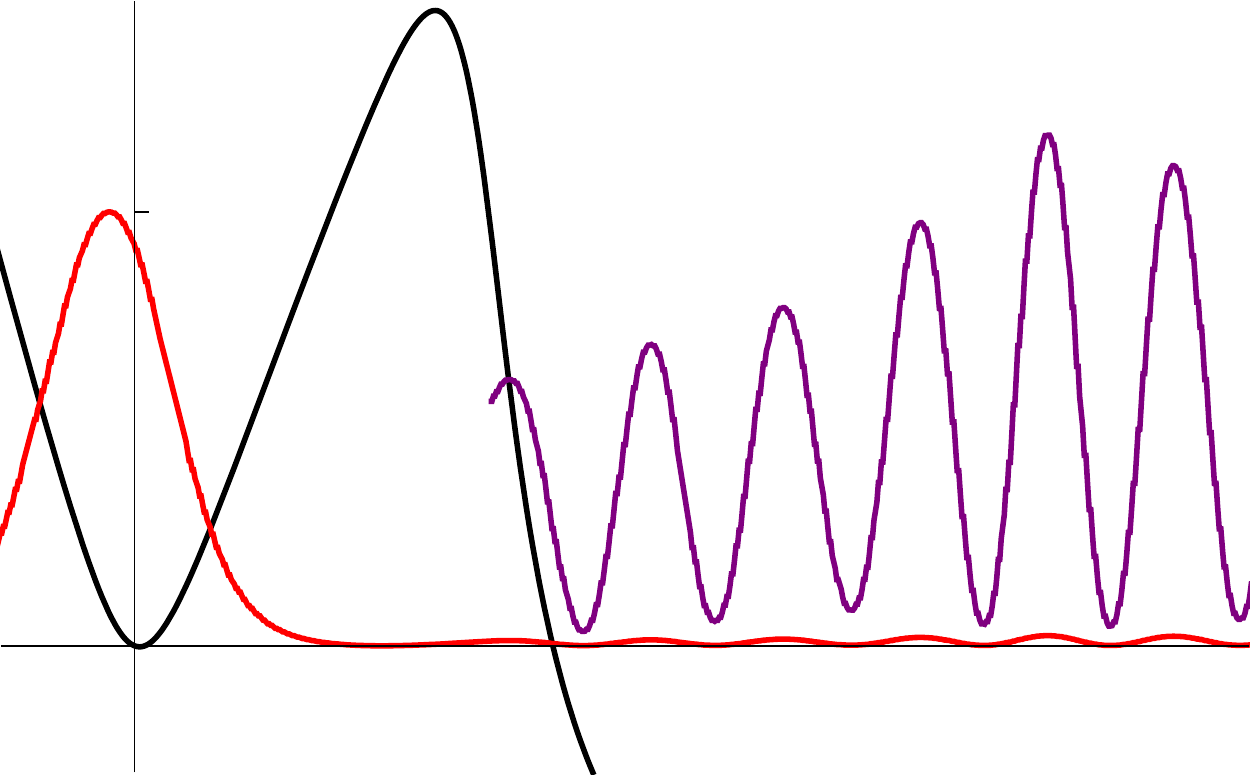}};
\labelplottime{22};
\end{tikzpicture}
\begin{tikzpicture}
\node at (0,0) {\includegraphics[width=0.2\columnwidth]{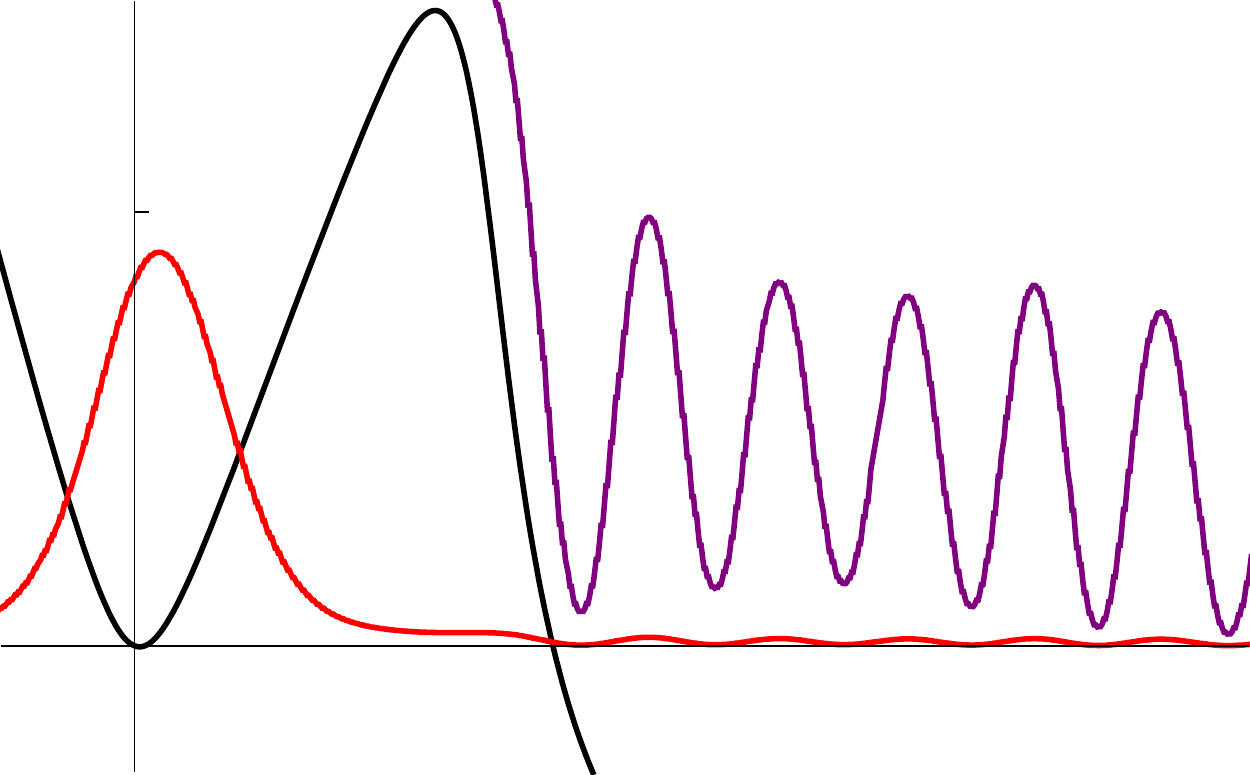}};
\labelplottime{23};
\end{tikzpicture}
\caption{
The numerical evolution of a particle initially localized in the false vacuum. At each time step, the potential is shown (black), along with the probability $\abs{\psi(x,t)}^2$ (red), and we also show the probability magnified by 50$\times$ (purple) so that we can see the small amount leaking through the barrier).
By looking at the evolution of the wavefunction we see the sloshing behavior near the false vacuum, associated with the initial Gaussian state not being an exact resonance. In the first two rows the central value of the wavefunction can be seen moving back and forth within the false vacuum well. When it hits the right wall around times 3-4, the most wavefunction amplitude escapes through the barrier. In the third row we have jumped ahead to see the nonlinear behavior when there is enough wavefunction density in the outside region that it is no longer simply flowing out.
\label{fig:sloshinganimationsteps}
}
\end{center}
\end{figure}

\subsection{Precise definition of the decay rate \label{sec:def}}

To make the above formula more precise, we need an exact definition of the decay rate to which we can then look for approximations.
A reasonable, physical, definition of the decay rate of a system comes from $P_{\FV}(t)$, the probability of finding a state $\psi$ initially confined to a false vacuum region ($\FV$) in that same region after a time $T$:
\be
\label{eqn:defineP}
P_{\FV}(T)\equiv\int_{\FV} dx\abs{\psi(x,T)}^2
\ee
We expect that for a decaying system the probability should fall exponentially:
\be
\label{eqn:Pexp}
P_{\FV}(T) \sim e^{-\Gamma T}
\ee
And so we might define:
\be
\label{eqn:GdefNaive}
\Gamma = - \frac{1}{P_\FV}\frac{d}{dT}P_\FV
\ee

Equation (\ref{eqn:Pexp}) is of course not strictly true for all times $T$, and hence $\Gamma$ defined in Eq.~\eqref{eqn:GdefNaive} is time-dependent. For a 1D system, we can calculate $P_\FV(T)$ numerically, as shown in Fig. \ref{fig:numericpotentialprobability}. The plot of $P_\FV$ makes it clear what we mean physically by the ``decay rate"; the probability falls exponentially for some particular time regime, and it is in this regime that $\Gamma$ is meaningful.

To get a time-independent rate, we can average over the oscillations which occur with frequency $\sim\frac{p}{am}$. For this to make sense, $T$ cannot be too short:
$T\gg \sloshtime = \omegaa^{-1} $ where $\omegaa$ characterizes the frequency of oscillation within the false vacuum.\footnote{
For a parabolic well, the sloshing time is just the inverse of the classical oscillation frequency $\omega_a = \sqrt{\frac{1}{m}V''(a)}$. This is also the energy difference between
excitations in the quantum system. 
For a square well, there is no classical oscillation, but $\sloshtime$ does not go to infinity. Because of the hard walls in the square well, there is still a finite $\sloshtime$, related again to the energy difference of the two lowest modes.}
 Moreover, $T$ should also not be
too long, for then the exponential decay will have significantly depleted the wavefunction and non-linearities set in. One source of non-linearities is
from the decaying wave bouncing off the potential in the true-vacuum region and returning to the false vacuum. Pooling these effects into a characteristic
scale $\nonlintime$, we also require $T \ll \nonlintime$. 

Thus, the physical decay rate is a phenomenon that happens on timescales $\sloshtime\ll T\ll\nonlintime$; for these timescales we expect the exponential fall of $P_\FV(T)$. Hence $\Gamma$ extracted from Eq.~\eqref{eqn:GdefNaive} is independent of $T$ to the extent that $\sloshtime\ll T\ll\nonlintime$ is satisfied.
These two time limits are built into what is meant by a time-independent decay rate $\Gamma$; they are not approximations we make to calculate $\Gamma$ but rather limits under which $\Gamma$ is even worth talking about. 
An ``all-orders" formula for $\Gamma$ {\it must} use these limits. Thus, a precise definition of the decay rate is:
\begin{equation}
\label{defineG}
\Gamma \equiv 
-\hspace{-1em}\lim_{\substack{T/\nonlintime\to0\\T/\sloshtime\to\infty}} \frac{1}{P_{\FV}(T)}\frac{d}{dT}P_{\FV}(T)
\end{equation}

Finally, for systems that can decay in multiple different directions (e.g. a 1-dimensional particle that can escape to the left or the right, or in multiple dimensions), we might want to know the decay rate to a particular region $\R$ (for instance, the region to the right of the barrier). Then we should define the partial decay width from the linear growth of the probability to find the particle in the region $\R$, $P_\R(T)$:
\be
\label{eqn:defineGamma}
\boxed{
\Gamma_\R =
\hspace{-1em}\lim_{\substack{T/\nonlintime\to0\\T/\sloshtime\to\infty}}
\frac{1}{P_{\FV}(T)}\frac{dP_\R(T)}{dT}
}
\ee

Another way we could intuitively have derived the decay rate to any region $\R$ would be as the probability flux in through the boundary of $\R$. (If $\R$ is everything outside the false vacuum, then this would be the total flux into $\R$, i.e. out through the boundary of the false vacuum region.) The quantum mechanical flux is defined by:
\begin{equation}
J_i(x,t) = \frac{1}{2im}\left(\psi^\star(x,t)\partial_i\psi(x,t) - \psi(x,t)\partial_i\psi^\star(x,t)\right)
\end{equation}
Then we could define the decay rate as the fraction of probability flowing through the outward-pointing boundary $\partial \R$, in the same time limits as above:
\begin{equation}\label{eqn:defineGammaFlux}
\Gamma_\R \equiv 
-\lim_{\substack{T/\nonlintime\to0\\T/\sloshtime\to\infty}}
\frac{1}{P_{FV}(T)}
\int_{\partial \R} dx_i J_i(x,T)
\end{equation}
Because of the conservation equation ($\partial_iJ_i=-\partial_t\abs{\psi}^2$), this is exactly equivalent to Eq.~\eqref{eqn:defineGamma}.

Next, we need to be able to compute $\Gamma$ in Eq.~\eqref{eqn:defineGamma}, either using the WKB approximation or with some other method.

\subsection{Real energy eigenstates and complex energy poles \label{sec:complex}}

\begin{figure}[t]\begin{center}
\begin{tikzpicture}
  \draw[->] (0,0) -- (8,0) node[right] {$x$};
  \draw[->] (1.02,0) -- (1.02,4.5) node[right] {$V(x)$};
  \node at (1,2) {$\FV$};
  \node at (2,2) {B};
    \node at (3.5,2) {$\R$};
\draw [thick,red] plot [smooth, tension=0.8] coordinates {(0,4) (1,0.5) (2,3.6) (3,0.7) (4.2,0)};
\node[above] at (1,-0.5) {$a$};
\draw [thick,red, -] (4.2,0) -- (7.8,0);
\draw[-,dashed] (1.02,0.5)--(3.13,0.5);
\draw[-,dashed] (3.13,0)--(3.13,0.5);
\node[above] at (3.13,-0.5) {$b$};
\end{tikzpicture}
\caption{Example of a potential that has a well region labeled $\FV$, a barrier region B, and is constant in the region $\R$ which extends to indefinitely to the right.}
\label{fig:potentialunboundedright}
\end{center}
\end{figure}
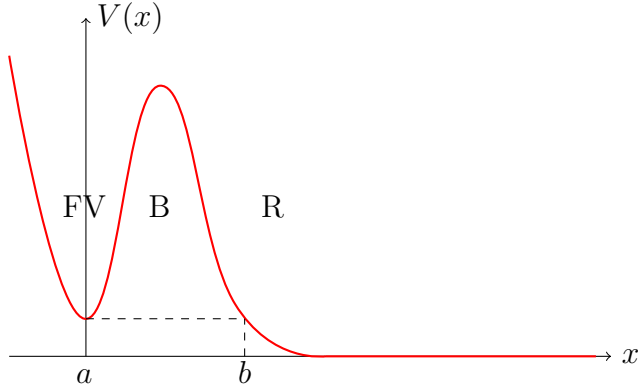

The type of potentials under consideration, such as the one in Fig.~\ref{fig:numericpotentialprobability}
comprise a well region labeled $\FV$, where the particle is initially, a barrier region $\barr$, between points $a$ and $b$ (to be specified precisely later), and an approximately free destination region $\R$. 
For now, let us assume that the potential is constant in $\R$ and extends infinitely to the right, as $V(x)$ in Fig.~\ref{fig:potentialunboundedright}.

A concrete example illustrating the points of this section is given in Appendix~\ref{app:Gammasimple}. More details and alternative derivations can be found in~\cite{Kleinert,vanDijk:1999zz,delaMadrid:2002cz,flugge1959nuclear,HolsteinSwift, Holstein,Goldberger:1964,Newton:1982qc}.

Since the system extends infinitely to the right, there will be energy eigenstates $\phi_E(x)$ for any $E$. Most of these are approximately free (plane waves) confined to region $\R$, with little support in the $\FV$ region.
Some, however, do have large support in the $\FV$ region. These are the resonances.
To be specific, we can define the resonant energies $E$ 
as those whose probability in the $\FV$ region has a local maximum:
$\partial_E P_\FV[\phi_E] = 0$ (now the probability $P_{\FV}$ defined in Eq.~\eqref{eqn:defineP} is viewed as depending on $\phi_E$ instead of on $T$, since energy eigenstates have time-independent probabilities).
In general, there will be a finite number of such resonance energies $E_1<E_2<\cdots < E_n< \Vmax$. Up to exponential corrections, these are the bound state energies for a modified potential where $\FV$ is made absolutely stable by deforming the potential $V(x)$ (e.g. by setting $V(x)=V(c)$ for $x\ge c$ where $x=c$ is the location of the maximum height of the barrier).

Now, our initial state $\psi$ cannot be an energy eigenstate, or else there would be no time dependence and $\Gamma=0$. 
However, since the initial state is, by assumption, localized in the $\FV$ region, it can be written as a linear combination
of bands of energy eigenstates close to each of the resonant energies $E_i$.
Each of these bands will have a different characteristic decay rate $\Gamma_i$. 
We must assume $\Gamma_i \ll E_i$ so that the widths are narrow and the decay is exponential.
As we will confirm, the higher energy states will decay much much faster than the lower energy states, since they have
less barrier to penetrate: $\Gamma_1 \ll \Gamma_2 \ll \cdots \ll \Gamma_n$.

As we said, we want to average over the sloshing times, $T \sim \sloshtime^i =\omega_a^{-1}$.
By the time  $T\sim \Gamma_n^{-1} \gg \sloshtime^n$, the highest energy components (around $E_n$) will have significantly leaked out of the well. 
Then at each $\Gamma_i$ threshold, another band of the original probability will have leaked out. 
Depending on the structure of the region $\R$, these leaked components may even leak back in on the time scale $T \sim \nonlintime$.
We can treat each band separately, so let us assume for simplicity that our original wavefunction only had support
from the modes close to energy $E_0$ and write $\Gamma$ for the width of this band.

The modes in the band near $E_0$ have basically the same form. They have the shape of the $E=E_0$ (approximate) bound state in the well region, decay exponentially through the barrier, and become free (plane waves) in region $\R$.

In the \FV~region, they take the form
\be
\phi_E(x)=\frac{1}{N(E)}f(E,x)
\ee
where $f(E,x)$ is a well-behaved function in $E$ and $x$, and $N(E)$ has zeros exponentially close to the resonant energies, $E_n$. We are focusing on the lowest resonant energy, which has corresponding zeros $E_0 \pm \frac{i}{2}\Gamma_0$.

To connect $\Gamma$ in Eq.~\eqref{defineG} to this pole, let us decompose our decaying wavefunction $\psi(x,t)$, initially localized in the well, into energy eigenstates. Defining
\be
 \rho(E)  = \int_0^\infty dy~\phi_E^\star(y) \psi(y) \,,
\ee
where $\psi(y)\equiv \psi(y,t=0)$. 
The  time dependence of our wavefunction $\psi(x,t)$ is given by
\be
\psi(x,t) = \int dE~\rho(E) \phi_E(x) e^{-i E t}  =  \int dy~\psi(y) \int d E \frac{f(E,x)f^\star(E,y)}{|N(E)|^2} e^{-i E t}
\ee
Closing the $E$ contour in the fourth quadrant (for convergence), and looking at times $E_0^{-1} \ll t \ll \Gamma_0^{-1}$, we have
\eqna{
\label{eqn:Npoleint}
\psi(x,t) &\approx \int dy~\psi(y)~2\pi i\Res\left[\frac{f(E,x)f^\star(E,y)}{|N(E)|^2} e^{-i E t};E=E_0-\frac{i}{2}\Gamma_0\right]\\
&= g\left(E_0-\frac{i}{2}\Gamma_0,x\right)e^{-iE_0 t}e^{-\frac{\Gamma_0}{2}t}
}
for some well-behaved function $g(E,x)$. Corrections to this formula are all suppressed by $\Gamma_0 t \ll 1$. 
Applying the definition of the decay rate, Eq.~\eqref{defineG} gives
\be
-\frac{d}{dT}\log P_\FV(T) = -\frac{d}{dT}\log \int_\FV dx\abs{\psi}^2 = -\frac{d}{dT}\log\left[ \text{const}\times e^{-\Gamma_0 T}\right] = \Gamma_0
\ee
Thus $\Gamma_0$, defined by the first pole of $N(E)$, $E=E_0-\frac{i}{2}\Gamma_0$, is indeed the decay rate, $\Gamma$.
That is, the decay rate, defined physically in Eq.~\eqref{defineG}, is given by twice the imaginary part of the complex pole. For potentials of this form, this leads us to a shortcut; we can simply calculate the poles of $\frac{1}{N(E)}$ directly, without ever explicitly time-evolving any states.

To be clear, real energy eigenstates never blow up. However,
nothing stops us from finding solutions to the Schr\"odinger equation with complex energies. Doing so, we will find that
for certain complex energies, the wavefunction does blow up. 
These poles should be close to real energy $E_0$, with a small
excursion of size $\Gamma \ll E_0$ into the complex plane. 
Actually, there will be two poles for each $E_0$, one for positive and one for
negative $\Gamma$. 
Since the Hamiltonian is Hermitian on bound states, these states cannot
have the same boundary conditions as for real energy eigenstates. 
Normally, energy eigenstates with real energies
have a probability around each point which does not change with time so there can be no outgoing or incoming flux. Thus 
nonzero flux corresponds to complex energies. 
The $\Gamma < 0$ case corresponds to incoming boundary conditions:  flux goes from region $\R$ into the $\FV$ region, and
correspondingly $P_\FV(T)$ will grow with time. 
For $\Gamma > 0$,  the flux goes from $\FV$  into $R$. These are outgoing radiating (Gamow-Siegert)
boundary conditions, the situation we are interested in. The example in Appendix~\ref{app:Gammasimple} shows more directly
the connection between radiating boundary conditions and the complex zeros of the normalization.

Note that the outgoing-only wave approximation is equivalent to
removing the back-reaction, or equivalently taking $T/\nonlintime \to 0$. This is the same limit required in Eq.~\eqref{eqn:defineGamma} to make the decay rate
well-defined (time-independent). It is reassuring that the relevant time scale is playing a role in the analysis.

With outgoing boundary conditions, the energy is $E=E_0 - \frac{i}{2}\Gamma$ with $\Gamma>0$ and the momentum in the region where $V=0$
is 
$p=p_0 - \frac{i}{2}\gamma$,
with $\gamma = \frac{m\Gamma}{p_0}$ and $p_0=\sqrt{2mE_0}[1 + \cO(\Gamma/E_0)]$.
Writing $\phi_E(x,t) = \frac{1}{N} e^{-i E t + i p x}$ in the  region $\R$ (where we are assuming $V=0$), the rate can be computed by flux conservation,
\be
\partial_t (\psi^\star \psi) = \frac{i}{2m}\partial_x\Big(\psi^\star \partial_x \psi - \psi \partial_x \psi^\star\Big)
\ee
which holds for any solution $\psi$ to the Schr\"odinger equation. 
 Integrating this from $0$ to an arbitrary point $b$ for the energy eigenstate $\phi_E$ with outgoing boundary conditions, we then get 
\be
 \label{Gofpsi}
\boxed{
\Gamma = \frac{p_b}{m} \frac{|\phi_E(b)|^2}{\int_0^b d x|\phi_E(x)|^2 }
}
\ee
where $p_b = -i \frac{\partial_x\phi_E(b)}{\phi_E(b)}$ for a plane wave, and more generally $p_b = -\frac{i}{2}\frac{\phi_E^\star \partial_x \phi_E - \phi_E \partial_x \phi_E^\star}{\phi_E^\star \phi_E}\Big|_{x=b}$. 

The expression for the decay rate in Eq.~\eqref{Gofpsi} is accurate up to exponentially small corrections. In it, $b$ 
can be any point \emph{at all}. Indeed, Eq.~\eqref{Gofpsi} is independent of
the choice of $b$, as can be seen by taking the derivative and applying Schr\"odinger's equation, and independent of the choice of $t$ because it is written in terms of spatial wavefunctions alone.

For Eq.~\eqref{Gofpsi} to be useful, we would like to be able to use it for solutions with real energies for normalizable resonance modes rather than complex energy eigenstates with outgoing
boundary conditions. After all, if we already know the complex energy, then we know the rate. 
Since the complex energy solution is exponentially close to the real energy solution in the $\FV$ region and for most of the barrier region (except near $x\approx b$), the integral
$\int_0^b |\phi_E(x)|^2$ in the denominator should be about the same if $\phi_E$ is the real or complex energy solution.
The numerator, on the other hand, involves the value of the wavefunction at $b$. Its value 
for the real and complex-energy solutions may differ by a factor of order one. 
This is because the complex energy solution has an exponentially growing component in the barrier, which can become of the same order as the exponentially decaying one at $x=b$.
Thus, while one can certainly use Eq.~\eqref{Gofpsi} with a real eigenfunction to approximate the exact answer, some precision will unfortunately be lost in doing so. 
We discuss using the WKB approximation to the energy eigenstates next and defer an explicit
example to Appendix~\ref{app:Gammasimple}.

\subsection{WKB approximation}
Once we have a formula like Eq.~\eqref{Gofpsi} which depends on the values of a wavefunction, we need to solve Schr\"odinger's equation. 
If an analytic solution is not available, we may want to approximate Eq.~\eqref{Gofpsi} with the WKB expansion. The WKB approximation tells us that
\be
\label{eqn:WKB}
\phi_E(x) = A \frac{1}{\sqrt{\abs{p(x)}}}  \exp\left[\frac{i}{\hbar}\int^x_a p(y) dy\right]\left(1+\mathcal{O}(\hbar)\right)
\ee
where $p(x)=\sqrt{2m(E_0-V(x))}$ and $A$ is an $x$-independent normalization constant that will drop out of Eq.~\eqref{Gofpsi}.
The lower limit of integration is chosen to be $a$ for convenience; changing it to something else will only change the normalization which can then be absorbed into $A$.

At leading order we ignore the $p$ prefactor and keep only the exponential. Then Eq.~\eqref{Gofpsi} gives
\be\label{eqn:SELO}
\Gamma^{\text{LO}} =\text{const} \times e^{-2\int \abs{p(x)} dx }
\ee
Where the integral is taken over the region where $V(x)>E_0$ (so that $p(x)$ is imaginary). At NLO we keep the prefactor also, giving:
\be
\Gamma^{\text{NLO}} = \frac{1}{m}\frac{ \exp\left[-\frac{2}{\hbar}\int_a^b \abs{p(x)} dx \right]}{\int_0^a\frac{dx}{p(x)}+\int_a^b \frac{dx}{p(x)}\exp\left[-\frac{2}{\hbar}\int_a^x \abs{p(y)} dy \right]}
\ee
Where $a$ and $b$ are the classical turning points.

The first factor in the denominator is exactly the classical period of oscillation around the false vacuum: $T_{\FV}=\int \frac{m}{p}dx$. 
For a roughly constant potential in the region from $0$ to $a$, this is just $T_{\FV} = \frac{a}{v}$ where $v= \frac{p}{m}$ is the velocity.
The second factor in the
denominator is exponentially smaller in the limit $\hbar\to 0$, so we can drop it even at NLO. Thus, 
\be\label{eqn:SENLO}
\Gamma^{\text{NLO}} =\frac{p}{am}\exp\left[-\frac{2}{\hbar}\int_a^b \abs{p(x)} dx \right]
\ee
This is close to Eq.~\eqref{GofT}, but differs by a factor of 2. 

Of course we had no right to expect the two formulas to agree exactly: firstly, Eq.~\eqref{GofT} was based on a imprecise
semi-classical argument; and secondly, Eq.~\eqref{eqn:SENLO} uses the WKB approximation, neglecting a careful treatment of turning points, and approximates Eq.~\eqref{Gofpsi} with real-energy
eigenstates, which is also not a controlled approximation. 
In~\cite{Muller-Kirsten:2012wla}, the WKB approximation is used in a formula like Eq.~\eqref{eqn:SENLO} for a cubic potential. They find that at NLO the prefactor differs from a presumably
more accurate result using the potential-deformation method (see Section~\ref{sec:potmethod} below), by a factor of $\frac{e}{2} \approx 1.4$.

This is not to say that WKB cannot be used to compute tunneling rates precisely. It can. For example, in \cite{Jentschura:2010zza} the WKB method was used to compute the rate from the
complex resonant energies for the quartic potential to N${}^4$LO. 
The same rate was computed to NLO using the potential-deformation method in~\cite{Jentschura:2011zza}. The two results agree exactly to the order at which they can be compared (NLO).

\subsection{WKB in multiple dimensions}\label{sec:WKBmultidim}
One might wonder what happens to the above Schr\"odinger equation,  especially Eqs.~\eqref{Gofpsi} and \eqref{eqn:SENLO}, in multiple dimensions. 
The all-orders formula Eq.~\eqref{Gofpsi} generalizes naturally enough to:
\begin{equation}
\label{eqn:WKBpathsum}
\Gamma = \frac{1}{m}\frac{\int_\Sigma db\cdot p_b |\phi_E(b)|^2}{\int_\FV d x|\phi_E(x)|^2 }
\end{equation}

The WKB approximation in multiple dimensions is more complicated, unfortunately. At an intuitive level, one would very much like to simply integrate over all paths through configuration space, and for each path apply the 1D WKB. In words, this simply says that the system can decay along any path through the barrier; for each path we apply the 1D WKB formula and then we integrate over all the paths. 

This intuitive picture is unfortunately difficult to prove precisely (an extended discussion is provided
by Banks, Bender and Wu~\cite{Banks:1973ps}). The problem is that WKB is attempting to approximate the wavefunction, which takes only a single value at each position; there is no sense in summing over all possible values it would take if we follow all possible paths to each point. All together, when we want to be precise, it is easier to apply the semiclassical approximation (which is the approximation WKB performs) in the path integral rather than trying to use multidimensional WKB.

However, the intuition from WKB is not useless. Because the trouble has to do with the sum over paths, one might expect the leading exponential behavior predicted by WKB --- that which is determined solely by the dominant path through the barrier and knows nothing of the other paths --- to be correct. 
Keeping only the leading exponentials, this says:
\begin{equation}
\Gamma\sim \int \widetilde\cD x ~e^{-2\int ds \sqrt{2V(x(s))}\label{eqn:gammarestrited}}
\end{equation}
Where the path integral here integrates over paths but not  over parametrizations of those paths; in other words we only include in $\widetilde\cD x$ paths $x(s)$ with a path-length normalization $\abs{\frac{dx_i}{ds}}^2=1$.

For any path $x(s)$, the WKB exponential is exactly the same as the minimal classical Euclidean action over all parametrizations $s(t)$:
\begin{equation}
\label{eqn:WKBEqualsAction}
\int ds \sqrt{2V(x(s))} = \min_{s(t)}\int dt\left[ \frac{1}{2}\dot x^2-V(x)\right]
\end{equation}
This can be seen because the minimum action path conserves energy; $E=\frac{1}{2}\dot x^2+V(x)$ is constant. Assuming $V=0$ at the endpoint, this means $\dot x=\sqrt{2V}$ and the minimum-action is equal to $\int dt 2V(x)$. Changing variables from $dt$ to $ds$ gives $\int ds \sqrt{2V}$. Note that both sides of Eq.~\eqref{eqn:WKBEqualsAction} only integrate over the path from the false-vacuum to the barrier; to include the return journey one adds a factor of 2.

This means that (again keeping only the dominant exponentials) integrating over the parametrizations of the Euclidean action along a single path gives the WKB factor along that path:
\begin{equation}
\int \cD s(t) e^{-S_E[x(s(t))]} \sim e^{-2\int ds \sqrt{2V(x(s))}}
\label{eqn:WKBparametrizedpath}
\end{equation}
where the left side is integrated over all paths from false-vacuum back to false vacuum which cross the barrier, and the right side integrates from the false vacuum to the turning point. Now we can remove the awkward restriction on the path measure in Eq.~\eqref{eqn:gammarestrited}:
\begin{equation}
\label{eqn:WKBPI}
\Gamma \sim \int \Dx e^{-S_E[x]}
\end{equation}
which is indeed the correct equation at leading exponential order according to the more precise path integral derivations (see Sections~\ref{sec:potmethod} and~\ref{sec:dirmethod} below), as long as one allows that the $\sim$ suppresses some sort of restriction to the bounce saddle point only.

Thus the picture of using WKB and simply integrating over paths through the barrier
(or simply taking the least-resistance path) does indeed give the correct leading order decay rate.
This ``through the barrier" restriction causes the integral in  Eq. \eqref{eqn:WKBPI} to be dominated by the bounce and not the constant false-vacuum solution (see section \ref{sec:potmethod}). 

This intuition is useful, since it says that for a given multi-dimensional potential, one can get a physical intuition for the size of the barrier by studying the potential along the least-resistance path $V_{\text{1D}}(s)\equiv V(x(s))$, where $x(s)$ is the least-resistance path which is the same as the Euclidean bounce. The leading exponential decay rate for the multidimensional problem will then be the same as it would for the 1D problem $V(s)$. 

As a side note, it is important that the integration variable $s$ in WKB be a path-length parametrization. 
In field theory (cf. Section~\ref{sec:QFT} below), we usually parametrize the path through field
space with the Euclidean time $\tau$. But if we use $\tau$ as-is then the WKB factor $\int d\tau\sqrt{2V}$ has the wrong measure. So we must first convert to a path-length parametrization:
\begin{equation}
\frac{ds}{d\tau} = \sqrt{\sum_i\left(\frac{dx_i}{d\tau}\right)^2}
\end{equation}
This is discussed further in Section~\ref{sec:energy}.

\subsection{Summary}
The first goal was to give a precise definition of the decay rate and to isolate
the conditions under which $\Gamma$ is well-defined. We did this in Section~\ref{sec:def}. The next goal was to show
how the transmission coefficient, which is what the WKB can be used to compute, is related to $\Gamma$. To do that,
we needed to discuss the time-evolution of a wavefunction $\psi$. We found, from decomposing $\psi$ into energy
eigenstates, that the rate is encoded in the zeros of the normalization of modes near resonant energies $E_0$. This normalization
has a pole at complex energies. The pole whose imaginary part gives the rate is associated with outgoing-radiation boundary
conditions. Using flux conservation, this imaginary part can be related to the energy eigenstate wavefunction which is then
approximated with WKB. Although all these steps are presumably well-known, and included in various forms in various
treatments, we nevertheless thought it could be helpful to have this whole story in one place.

Appendix~\ref{app:Gammasimple} explains a concrete  example, where the (real) energy eigenstates are solved for explicitly, and the connection between complex energies, poles in the normalization of the wavefunctions, and outgoing boundary conditions can be seen explicitly. 

\section{Potential-deformation method}\label{sec:potmethod}
The methods of the previous section rely on solving Schr\"odinger's equation, which is not practical in a many dimensional case (such as field theory). 
An alternative approach to calculating tunneling rates, which generalizes more easily to higher dimensions, works with the path integral directly~\cite{Callan:1977pt}. 
We first review this approach and point out some of its more curious aspects. Then in Sections~\ref{sec:saddle} to~\ref{sec:stablepotentials} we provide more details of particularly subtle points. 
More details of the mathematics of this method can be found in~\cite{Bender:1990pd,Collins:1977dw, ZinnJustin:2002ru,Witten:2010cx,Marino}.

\subsection{Overview}
The starting point of the calculation in~\cite{Callan:1977pt} is the relation
\be
\label{eqn:defineZ}
Z \equiv \langle x_f | e^{-H \bigtau} | x_i \rangle  = \int_{x(0)=x_i}^{x(\bigtau)=x_f} \cD x e^{-S_E[x]}
\ee
where the right-hand side is the path integral using the Euclidean action $S_E[x]$. 
By inserting a complete set of energy eigenstates, the matrix element can be written as
\be
Z = \sum_{E} e^{-E\bigtau} \phi_E(x_i) \phi_E^\star(x_f) \label{eqn:Zwf}
\ee
Then we see that the lowest energy can be deduced from
\be
E_0 = - \lim_{\bigtau\to \infty} \frac{1}{\bigtau} \ln Z \label{E0aslim}
\ee

Roughly speaking, we expect that when there is a decay, $E_0$ will have an imaginary part corresponding to the decay rate, and so\footnote{Note that there will be a sign ambiguity in the evaluation of Eq.~\eqref{eqn:Coleman}, as we will see later in this section. The calculation should always be done so that $\Gamma>0$, which corresponds to the physical decay rate.}:
\begin{equation}
\label{eqn:Coleman}
\frac{\Gamma}{2} = \Im \lim_{\bigtau\to \infty} \frac{1}{\bigtau} \ln Z
\end{equation}
There are many ways to connect the imaginary part of an energy to a decay rate, but the connection
is not automatic. For example,
in Section~\ref{sec:complex} we found the decay rate for a metastable system to be the imaginary part of a 
eigenstate of the Hamiltonian with (unphysical) Gamow-Siegert radiative boundary conditions. For normalizable modes of a Hermitian Hamiltonian, all the energies including $E_0$ are real.
 For physical potentials, which are bounded from below, the energies and Euclidean action
 are bounded from below as well. Correspondingly $Z$ is manifestly real. Hence Eq.~\eqref{eqn:Coleman} must be defined in a much more careful manner.

  \begin{figure}[t]
\begin{center}
\begin{tikzpicture}
	\coordinate (c1) at (0,0);
	\coordinate (c2) at (8,0);
	\draw[->] ($(c1)+(-3,0)$) -- ($(c1)+(3,0)$) node[right] {$x$};
	\draw[->] ($(c1)+(0,-3)$) -- ($(c1)+(0,3)$) node[above] {$V(x)$};
  \draw[domain=-2:2.9,thick,smooth,variable=\x,darkblue] plot ({\x},{-\x*\x-0.3*pow(\x,3)+0.25*pow(\x,4)+0.4517});
	\draw[->] ($(c2)+(-3,0)$) -- ($(c2)+(3,0)$) node[right] {$x$};
	\draw[->] ($(c2)+(0,-3)$) -- ($(c2)+(0,3)$) node[above] {$-V(x)$};
  \draw[domain=-2:2.9,thick,smooth,variable=\x,darkblue] plot ({\x+8},{-(-\x*\x-0.3*pow(\x,3)+0.25*pow(\x,4))-0.4517});
	\draw [red, ultra thick] ($(c2)+(0.55,0.61)+(0,-0.4517)$) circle [radius=0.125cm];;
	\node[label=above:{}] at ($(c2)+(0.55,0.61)$) {};
	\node[circle,red,fill,inner sep=3pt,label=above:{}] at ($(c2)+(-1.04,0.61)+(0,-0.4517)$) {};
	\draw[<->,thick,dashed] ($(c2)+(0.45,0.45)+(0,-0.4517)$) .. controls ($(c2)+(0.07,0)+(0,-0.4517)$) .. ($(c2)+(-0.85,0.55)+(0,-0.4517)$);
	\node[above] at ($(c1)+(-1,-0.5)$) {$a$};
	\node[above] at ($(c1)+(0.55,-0.5)$) {$~b$};
	\node[above] at ($(c1)+(1.93,-0.5)$) {$c$};
	\node[above] at ($(c2)+(-1,-0.5)$) {$a$};
	\node[above] at ($(c2)+(0.6,-0.5)$) {$~b$};
	\node[above] at ($(c2)+(1.93,-0.5)$) {$c$};
\end{tikzpicture}
\caption{Left: Generic potential with a false and true vacuum. Right: The inverted potential. The stationary path $\bar{x}(\tau)$ is the solution to the equations of motion of a ball rolling down the inverted potential with boundary conditions $x(0)=x_i$ and $x(\bigtau)=x_f$. }
\label{fig:invertedpotential}
\end{center}
\end{figure}
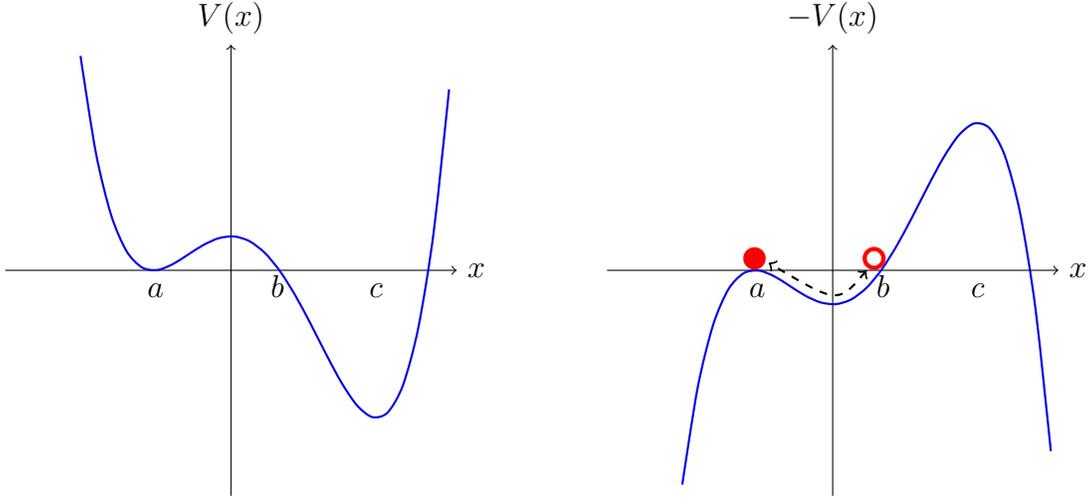
 
Consider the asymmetric double-well potential in Fig.~\ref{fig:invertedpotential}. One might hope that by taking particular boundary conditions (choice of $x_i$ and $x_f$) we can extract a metastable/resonance energy whose imaginary part gives the decay rate. However, the points $x_i$ and $x_f$ only contribute through the wavefunction factors $\phi_E(x_i)$ and $\phi_E^\star(x_f)$ in Eq.~\eqref{eqn:Zwf}, which do not contribute to $E_0$.
In order to get an imaginary part, then, we must do something more tortuous. 

Since the path integral is complicated, let us simplify things by first approximating it using the saddle-point approximation. The path integral can be approximated by summing over stationary points of the Euclidean action. For each stationary point, that is, for
each solution $\bar{x}(\tau)$ to the Euclidean equations of motion, the saddle-point approximation of the path integral around $\bar x$ evaluates to
\be
\cI_{\bar x} \equiv 
\frac{N}{\sqrt{\det[-\partial_t^2 + V''(\bar{x}) ]}}
 e^{-\frac{1}{\hbar}S_E(\bar{x})}(1+\cO(\hbar)) \label{detform}
\ee
where we have put the $\hbar$ back in for clarity, and $N$ is some constant related to the normalization of the path integral.
The stationary paths $\bar x(\tau)$ are solutions to the equations of motions for a ball rolling down a hill described by the inverted potential $-V(x)$
with the boundary conditions from the path integral: $x(0)=x_i$ and $x(\bigtau)=x_f$
 (see Fig.~\ref{fig:invertedpotential}).

\begin{figure}[t]
\begin{center}
\begin{tikzpicture}[scale=0.85]
  \draw[->,thick] (0,0) -- (10,0) node[right] {$t$};
  \draw[->,thick] (0,0) -- (0,5) node[above] {$x$};
	\node [left] at (0,1) {a};
	\node [left] at (0,2.3) {b};
	\node [left] at (0,3.5) {c};
	\node [below,darkgreen] at (1.1,0.95) {FV static};
	\node [darkblue] at (2,3.1) {the shot};
	\node [darkred] at (6.7,2) {the bounce};
	\node [orange] at (5,3.8) {TV static};
	\node [purple] at (8.5,4) {generic shot};
	\draw[thick,domain=0:10,smooth,variable=\x,orange] plot ({\x},{3.55});
	\draw[thick,domain=0:10,smooth,variable=\x,darkgreen] plot ({\x},{0.97});
        \draw[thick,domain=0.04:5,samples=500,variable=\y,darkblue]  plot ({\y},{3.35*(1-pow(1+\y,-10))});
	\draw[thick,domain=5:9.963,samples=1000,variable=\y,darkblue]  plot ({\y},{3.35*(1-pow(\y-11,-10))});
	\draw[thick,domain=0.025:5,samples=500,variable=\z,purple]  plot ({\z},{3.45*(1-pow(1.07+\z,-16))});
	\draw[thick,domain=5:9.94,samples=500,variable=\z,purple]  plot ({\z},{3.45*(1+pow(\z-11,-16))});
	\draw[thick,domain=0:3.5,smooth,variable=\w,darkred] plot ({\w},{1.1});
	\draw[thick,domain=6.5:10,smooth,variable=\w,darkred] plot ({\w},{1.1});
	\draw[thick,domain=3.5:6.5,samples=1000,variable=\w,darkred]  plot ({\w},{1.1+1.4*exp(-10*pow(\w-5,2))});
\end{tikzpicture}
\end{center}
\caption{
Different solutions to the Euclidean equations of motion for they asymmetric double well.}\label{fig:paths}
\end{figure}
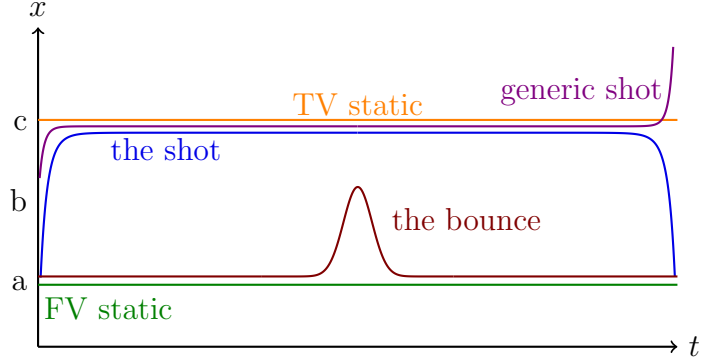

There are a range of stationary paths for this system dependent on what we choose the boundary conditions to be. For $x_i = x_f = c$ there is a solution to the Euclidean equations of motion with $\bar{x}(\tau) = c$. This is the path labeled ``{\orange TV static}'' in Fig.~\ref{fig:paths}. This has action given by $S_E[\bar x] = V(c)\bigtau$. The integral over Gaussian fluctuations around this solution produces $\cI \sim \exp(-E_c \bigtau)$, as explained in~\cite{Callan:1977pt},
where $E_c= V(c)+\frac{1}{2}\sqrt{ V''(c)}$ is the ground state energy of a harmonic oscillator using the quadratic approximation to the potential near $x=c$.
Thus the $\bigtau\to\infty$ limit produces the correct approximate ground state energy $E_0 = E_c$ for $x_i=x_f=c$, as expected.

Now say we take $x_i$ and $x_f$ arbitrary (not at $c$). 
The Euclidean equations of motion with boundary conditions $x(0)=x_i$ and $x(\bigtau)=x_f$ can always be solved by a solution starting at $x_i$ with exactly enough initial velocity
to get to the top of the hill and stay there for nearly time $\bigtau$, and then roll to $x_f$. This path is shown as the path labeled ``{\purple generic shot}'' in
Fig.~\ref{fig:paths}. This path has nearly the same Euclidean action as the  {\orange TV static} path, and matches it exactly as $\bigtau\to \infty$.
Thus we can indeed choose any points $x_i$ and $x_f$ and the true ground state energy $E_0$ results from the $\bigtau\to\infty$ limit. 

Now consider $x_i=x_f=a$. There is still a solution to the Euclidean equations of motion which stays at $x=c$ for most of the time (the path labeled the ``{\blue shot}'' in Fig.~\ref{fig:paths}) with  $\cI \sim \exp (-E_c\bigtau)$. 
With  $x_i=x_f=a$ there are actually more solutions. One, labeled ``{\green FV static}'' stays at $x=a$ for all time. It has Euclidean action $V(a) \bigtau$,
and saddle point approximation $\cI \sim \exp (-E_a\bigtau)$, where just like for $E_c$, $E_a$ has corrects due to oscillations around the $x=a$ minima.
In addition, when $x_i=x_f=a$ there is also an exact instanton solution. This solution starts very slowly from $a$. Since the potential is flat at $a$, it stays near $a$ for a long time,
then rather quickly rolls up to $b$ and back, then stays near $a$ again for a long time. This is the ``{\red bounce}'' in Fig.~\ref{fig:paths}.
 Because it spends most of the time near $a$, its saddle point differs from the FV saddle by a finite ($\bigtau$-independent) amount: $\cI = K \exp(-E_a\bigtau-S_\text{bounce})$, for some imaginary coefficient $K$.
The finite action is $S_\text{bounce}= \int_a^b dx \sqrt{2m V(x)}$ where the integral is between $a$ and the turning point $b$ on the other side of the barrier. This action is
positive, and therefore larger than the action for the {\blue shot}, which also starts and ends at $a$ in time $\bigtau$. The bounce only moves for a finite time, which is why it is called an instanton. 

So when $x_i=x_f=a$, there are contributions to the path integral from the {\blue shot}, the {\green FV static}, and unique {\red bounce} which is an exact solution to the Euclidean equations of motion.
There are also approximate solutions where the bounce is translated or multiple bounces are sewn together. 
These have actions which are exponentially close to the bounce action
and therefore contribute a large amount to the path integral even if they are not exact stationary points.
Summing all the saddle points and approximate saddle points, the result is:
\be
Z = \langle A| e^{-H \bigtau} | A\rangle \sim
\exp(- E_c \bigtau) 
+\exp(- E_a \bigtau)+ \exp(-E_a \bigtau + K e^{-S_\text{bounce}} \bigtau)
+   \cdots \label{cAaa}
\ee
where $K$ is an NLO constant arising from Gaussian integrations around the bounce \cite{Callan:1977pt}. 

One might then argue (see Coleman's discussion in \cite{Coleman:1978ae}) that the bounces are the only thing with an imaginary part and so we can keep them when computing the imaginary part in Eq.~\eqref{eqn:Coleman}, giving us
\eqn{\frac{\Gamma}{2} = \Im K e^{-S_\text{bounce}}\label{eq:naivedecay}}
But this argument is very precarious; we know for a fact that when computed exactly, $Z$ is real and the imaginary part is exactly $0$. 
The imaginary bounce contribution is exactly canceled by subdominant corrections to the true-vacuum saddle point. This cancellation can be seen in the toy examples discussed in Section~\ref{sec:saddle}.

In the following sections we will discuss some subtle points about saddle point approximations, analytic continuation and deformation of the contour of integration that will lead us to an expression like Eq.~\eqref{eq:naivedecay}. Briefly, the decay rate is actually calculated by modifying the path integral to be along a different contour of integration, the contour of steepest descent through the FV. Integrating along this contour misses the shot solution, allowing the FV 
path to dominate. The imaginary part along his contour is the same as $\frac{1}{2}$ of the imaginary part along
the steepest-descent contour passing through the bounce saddle point.
 This tells us that the decay rate associated with the false vacuum of the potential in Fig.~\ref{fig:invertedpotential} is given by
\eqn{\frac{\Gamma}{2} = \frac{1}{2}\Im K e^{-S_\text{bounce}}\label{eqn:decaywithonehalf}}
which is $\frac{1}{2}$ of the naive result given in Eq. (\ref{eq:naivedecay}).

We now turn to a careful explanation of how the contour deformation and saddle point approximation is done, and to
what extent the result is the same as one given by the analytic continuation of the path integral associated with deforming
the potential. 


\subsection{Analytic continuation, steepest descent contours, saddle points and imaginary parts}
\label{sec:saddle}
The main goal of this section is to explain a mathematically consistent procedure
for getting an imaginary part, presumably connected to the decay rate of a metastable state, out of a real path integral. This section is based to a large part on~\cite{Witten:2010cx} with insights from~\cite{Bender:1990pd,Collins:1977dw, ZinnJustin:2002ru,Marino}. In contrast to these references, we also consider
tunneling in physical bounded potentials which leads to a more nuanced picture of the origin of the imaginary part.

\begin{figure}[t]
\begin{center}
\includegraphics[width=0.41\columnwidth]{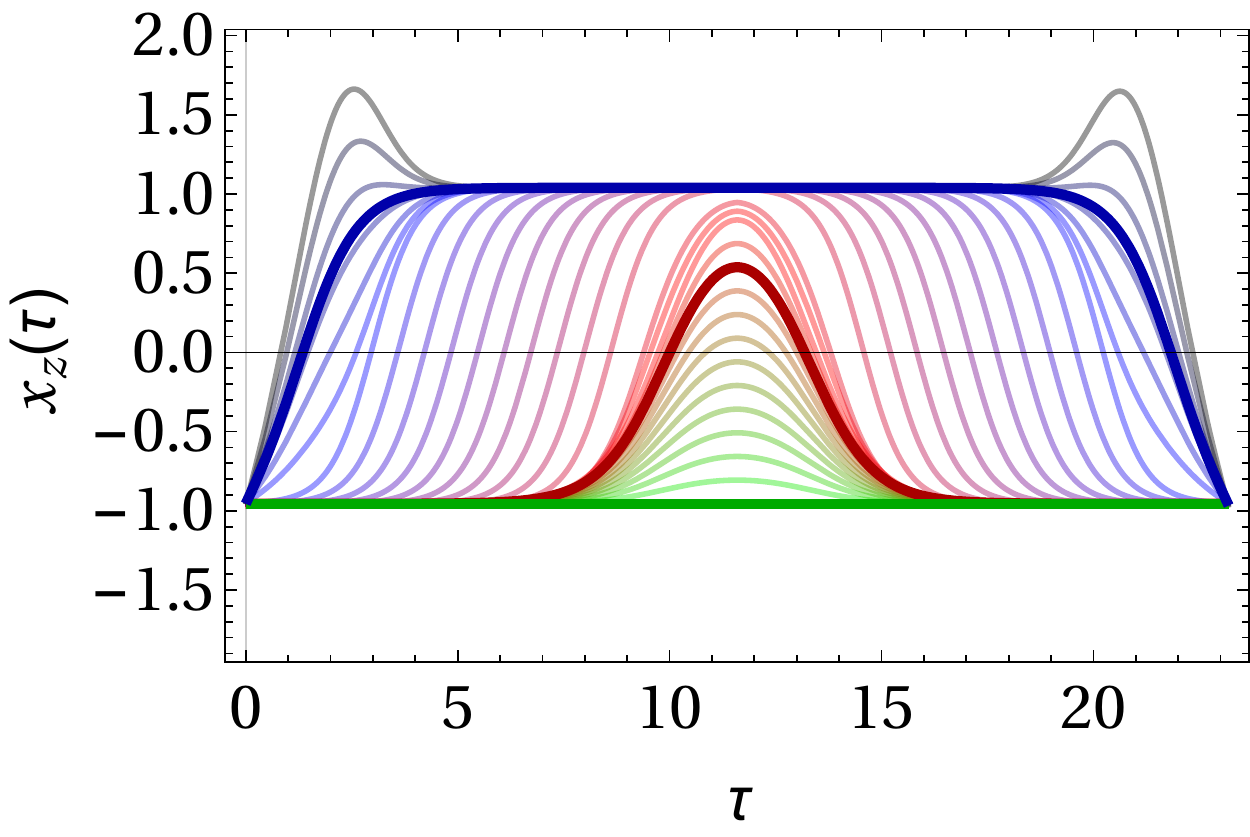}
\hspace{5mm}
\includegraphics[width=0.4\columnwidth]{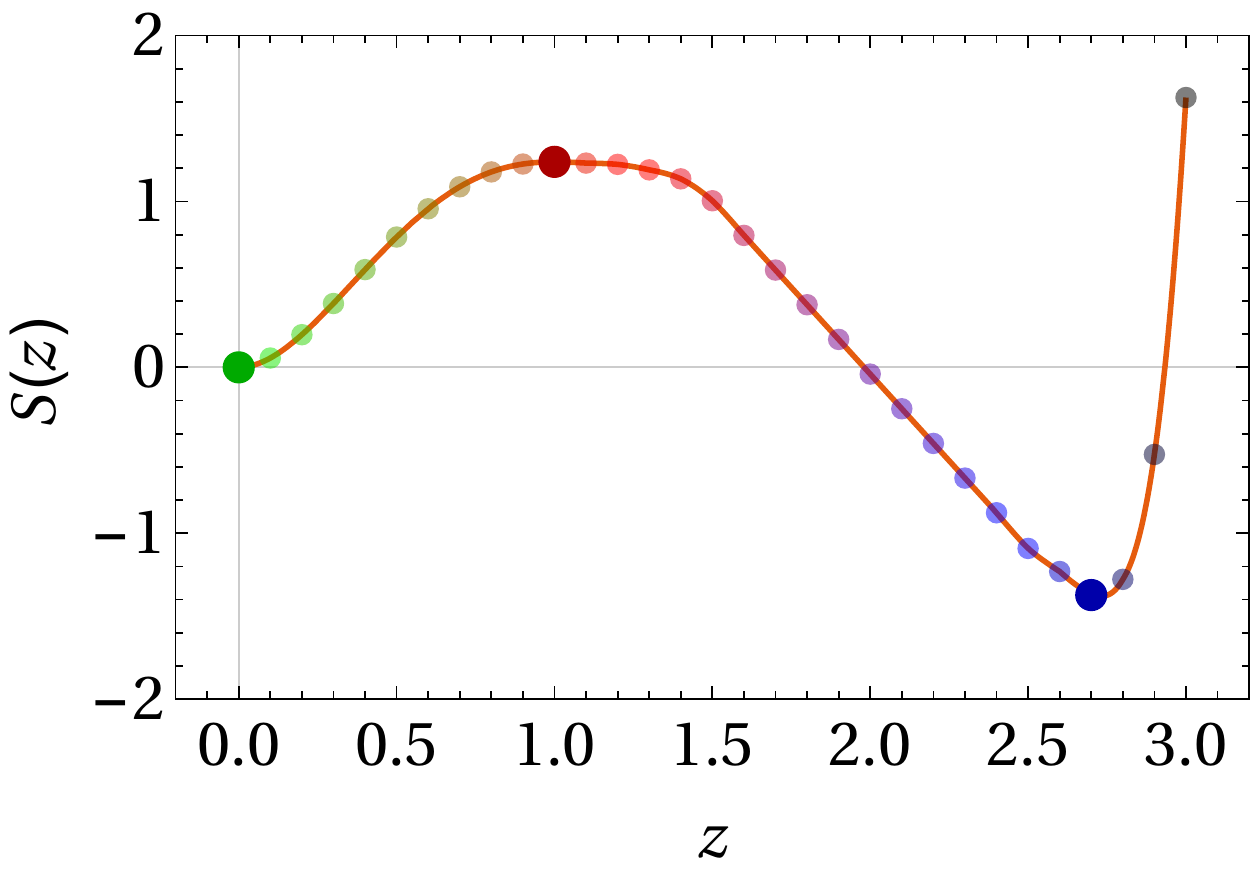}
\caption{
It is helpful to study the path integral along a one-parameter family of paths passing from the {\green static FV} through the {\red bounce} and to the {\blue shot}. These paths $x_z(\tau)$ are illustrated on the left
(found numerically using the method in Appendix~\ref{app:bounce}) for the potential $V(x) = -\frac{x}{12} - \frac{x^2}{2} + \frac{x^4}{4}$ and their actions $S(z)$ computed numerically and shown on the right. 
For the examples we consider, along families of paths like this the action $S[z]= \int d \tau \left[\frac{1}{2}(\partial_\tau x_z)^2 + V(x_z)\right]$ looks qualitatively like $V(z)$.
\label{fig:pathing}
}
\end{center}
\end{figure}

The final result, that we are trying to justify, is that the imaginary part we want comes from applying the method of steepest descent
to the Euclidean action along a family of paths passing through the FV saddle point. 
Much of the relevant mathematics can be understood most clearly by reducing the calculation to a one-dimensional integral along only this family of paths. Parametrizing the family by a parameter $z$, with $z=0$ corresponding
to the false vacuum, $z=1$ the bounce, and $z>1$ going towards the shot, we can compute directly $S(z)$ for a given potential (see Fig.~\ref{fig:pathing}). The part of the path integral of interest is then a 1D integral 
\begin{equation}
\label{eqn:Zint}
Z_C = \int_C dz e^{-\frac{1}{\hbar}S(z)}
\end{equation}
with $C$ some integration contour, in this case simply the real line. Eventually,
we will deform $C$ to some other contour in order to calculate the decay rate.

First, in Section~\ref{sec:dom}, we will discuss the saddle point approximation in general, and explain why only a subset of the stationary points of the action contribute to the decay rate. Then we will explain in a concrete example how the integral as a whole can be real even though a subdominant saddle point is imaginary
(this is what happened in the path integral in Eq.~\eqref{cAaa}).

We will unravel the origin of the imaginary parts by considering different types of potentials. 
One can have potentials in which the destination region of tunneling is 
unbounded from below, like $V(x) = \frac{x^2}{2} - \frac{x^4}{4}$. These examples are discussed in most textbook treatments~\cite{ZinnJustin:2002ru,Marino,Weinberg:2012pjx}. For such cases, the energy spectrum is unbounded from below and
the path integral is formally infinite. One can produce the tunneling rate by deforming the potential
though a parameter $g$ to the potential where $g=1$ is the original case of interest and $g<0$ makes the path integral convergent. As we will see, analytically continuing back to $g=1$ corresponds to changing the integration contour into the complex $z$ plane giving $Z$ a well defined imaginary part.

Potentials in actual physical systems are necessarily bounded from below.
One would naively expect that the same logic from the unbounded potentials should apply.
However, as we will see, analytic continuation of the potential cannot produce an imaginary part in the path integral,
because $Z$ is convergent along the real axis for any $g$. In particular, for physical potentials, 
 applying Eq.~\eqref{E0aslim} necessarily gives the \emph{real} ground state energy of the system. 

The correct procedure, which applies for all types of potentials exhibiting tunneling, is to 
compute the path integral along the steepest descent contour through the FV saddle. Along this contour, $Z$ is complex and its  imaginary part is equal to $\frac{1}{2}$ the sum over the bounces, just as in the standard formula. However, 
this understanding gives little explanation of 
why this procedure should always give the decay rate and how to calculate the rate outside of the saddle-point approximation. Those questions are answered by the alternative method presented in Section~\ref{sec:dirmethod}.

\subsubsection{Dominant and subdominant saddle points}
\label{sec:dom}

The saddle point approximation is used for arbitrary complex exponential integrals, like that in Eq.~\eqref{eqn:Zint}.
To apply the approximation, we first ignore the actual contour $C$, and instead focus on the complex saddle points of $S$, which we label $s_1,s_2,\dots,s_n$. Through each of these points we draw the steepest descent contour, $C_i$, which is defined intuitively by simply moving away from $s_i$ in the direction which increases the real part of $S$ as quickly as possible. These contours are called ``steepest descent" because the magnitude of the integrand is rapidly diminishing along the contour away from the saddle point, and thus the value of the integral can be approximated by its behavior near the maximum at $s_i$.

Along a given steepest descent contour we can approximate the integral by expanding $S$ around $s_i$:
\begin{align}
\cJ_i\equiv\int_{C_i} dz e^{-\frac{1}{\hbar}S(z)} 
& \sim\int_{C_i} dz e^{-\frac{1}{\hbar}S(s_i)-\frac{1}{2\hbar}S''(s_i)(z-s_i)^2+\cdots}\\
&\sim\sqrt{\frac{2\pi \hbar }{S''(s_i)}}e^{-\frac{1}{\hbar}S(s_i)}\left(1+\mc{O}(\hbar)\right)\equiv\cI_i
\label{Iis}
\end{align}
Here, $\cI_i$ is an approximation to $\cJ_i$, approaching it exactly in the $\hbar \to 0$ limit. 
The subleading corrections in the ``$\mc{O}(\hbar)$'' can be calculated in an asymptotic series in $\hbar$. Because the series is asymptotic, summing
the series will not reproduce $\cJ_i$ exactly. It will only produce $\cJ_i$ up to terms exponentially suppressed in $\frac{1}{\hbar}$. 

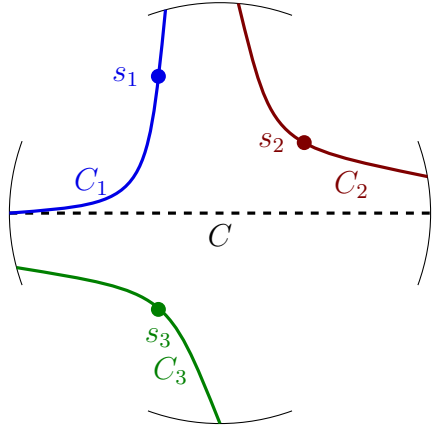
\begin{figure}[t]\begin{center}
\begin{tikzpicture}[scale=0.7]
\coordinate (c1) at (0,0);
\draw ($(c1) + (-20:4cm)$) arc (-20:20:4cm);
\draw ($(c1) + (160:4cm)$) arc (160:200:4cm);
\draw ($(c1) + (70:4cm)$) arc (70:110:4cm);
\draw ($(c1) + (250:4cm)$) arc (250:290:4cm);
\draw[dashed,very thick] (-4,0) -- (4,0);
\draw[very thick,darkblue] (-4,0) .. controls (-1.4,0.2) .. ($(c1) + (105:4cm)$);
\draw[very thick,darkred] ($(c1) + (10:4cm)$) .. controls (1,1.3) .. ($(c1) + (85:4cm)$);
\draw[very thick,darkgreen] ($(c1) + (270:4cm)$) .. controls (-1,-1.5) .. ($(c1) + (195:4cm)$);
\node [below] at (0,0) {$C$};
\node [left] at (-1.9,0.6) {\color{darkblue}$C_1$};
\node [below] at (2.5,1) {\color{darkred}$C_2$};
\node [left] at (-0.4,-3) {\color{darkgreen}$C_3$};
\node[circle,darkblue,fill,inner sep=2pt,label=left:{\color{darkblue}$s_1$}] at (-1.17,2.6) {};
\node[circle,darkred,fill,inner sep=2pt,label=left:{\color{darkred}$s_2$}] at (1.6,1.34) {};
\node[circle,darkgreen,fill,inner sep=2pt,label=below:{\color{darkgreen}$s_3$}] at (-1.17,-1.83) {};
\end{tikzpicture}
\caption{
An example of the complex $z$-plane with the critical points of $S$ marked as dots. Each saddle point $s_i$ has a steepest descent contour $C_i$ passing through it. Black arcs around the edges note the directions in the $x$-plane in which the exponent $-S$ goes to $-\infty$; the integral only converges along contours which start and end in these directions.
\label{fig:saddlepoint}
}
\end{center}
\end{figure}

We can then make a plot of the complex $z$-plane, marking the saddle points of $S$ as well as their steepest descent contours, as shown in Fig.~\ref{fig:saddlepoint}. Suppose the integral we want to compute is along the original contour $C$. We can deform the contour $C$ into a sum of steepest descent contours $C_i$; in the example of Fig.~\ref{fig:saddlepoint} this would be:
\begin{equation}
C = C_1+C_2
\end{equation}
 There are no poles, so the deformation is allowed as long as the endpoints remain the convergent regions
indicated by the arcs in Fig.~\ref{fig:saddlepoint}.
Note that although there are three saddle points in this example, only two contribute because of the contour.

The integral along $C$ is the sum of the integrals along these steepest descent contours, yielding:
\begin{equation}
\label{eqn:ZexpandJ}
Z = \cJ_1+\cJ_2
\end{equation}
Note that the fact that some or all of the $s_i$ are complex does not matter; one still includes those $s_i$ whose contours are involved in the sum, regardless of whether the saddle point itself is real or not\cite{Behtash:2015zha}.

Now we can use the saddle point approximation along each contour. We write the saddle point approximation to $\cJ_i$ as $\cI_i$, so that
\begin{equation}
Z \sim \cI_1+\cI_2
\end{equation}
where $\sim$ indicates that corrections are exponentially small. 

However, note that each $\cI_i$ approximates an exact contour integral $\cJ_i$ up to exponentially suppressed terms, and the $\cI_i$ can be exponentially different. In other words, one of the terms in Eq.~\eqref{eqn:ZexpandJ} (say $\cJ_1$) is exponentially larger than the rest. So if we are going to perform an expansion which is accurate up to exponentially small corrections, we cannot keep the subdominant terms.  At the level of Eq.~\eqref{eqn:ZexpandJ}, we write this as:
\begin{equation}
\label{eqn:boxedJsum}
Z = \boxed{\cJ_1}+\cJ_2
\end{equation}
to indicate that while the equation is exact at this level, the boxed term is exponentially dominant so, when approximating, the second term is meaningless. Unfortunately, it may
be $\cJ_2$ that has the imaginary part.

\begin{figure}[t]
\begin{center}
\begin{tikzpicture}
\coordinate (c2) at (0,0);
\draw ($(c2) + (-22.5:2.5cm)$) arc (-22.5:22.5:2.5cm);
\draw ($(c2) + (157.5:2.5cm)$) arc (157.5:202.5:2.5cm);
\draw ($(c2) + (67.5:2.5cm)$) arc (67.5:112.5:2.5cm);
\draw ($(c2) + (247.5:2.5cm)$) arc (247.5:292.5:2.5cm);
\draw[thick, darkgreen] ($(c2) + (0,2.5)$) -- ($(c2) + (0,0.8)$) .. controls ($(c2) + (0.0,0.1)$) .. ($(c2) + (0.8,0.1)$) -- ($(c2) + (2.5,0.1)$);
\draw[thick, darkred] ($(c2) + (-0.1,2.5)$) -- ($(c2) + (-0.1,0)$) -- ($(c2) + (0.1,0)$)  -- ($(c2) + (0.1,-2.5)$);
\draw[thick, newgreen] ($(c2) + (-2.5,-0.1)$) -- ($(c2) + (-0.8,-0.1)$) .. controls ($(c2) + (0,-0.1)$) .. ($(c2) + (0,-0.8)$) -- ($(c2) + (0,-2.5)$);
\node at ($(c2)+(3.5,3)$) { $ \boxed{S(z)=-\frac{z^2}{2}+\frac{z^4}{4}}$};
\draw[dashed,very thick] ($(c2)+(-2.5,0)$) -- ($(c2)+(2.5,0)$);
\node at (-0.35,0.35) {\color{darkred}$s_0$};
\node[circle,newgreen,fill,inner sep=4pt,label=below:{\color{newgreen}$s_{-1}$}] at (-1.6,0) {};
\node[circle,darkgreen,fill,inner sep=4pt,label=above:{\color{darkgreen}$s_{1}$}] at (1.6,0) {};
\node[circle,darkred,fill,inner sep=4pt] at (c2) {};
\node [below] at (1.6,0) {$C=\mathbb{R}$};
\node [right] at (-0.1,1.5) {\color{darkgreen}$C_1$};
\node [left] at (0.1,-1.5) {\color{newgreen}$C_{-1}$};
\node [left] at (0,1.5) {\color{darkred}$C_0$};
\coordinate (c3) at ($(c2)+(7,0)$);
\node at ($(c3)$) {\includegraphics[width=0.35\columnwidth]{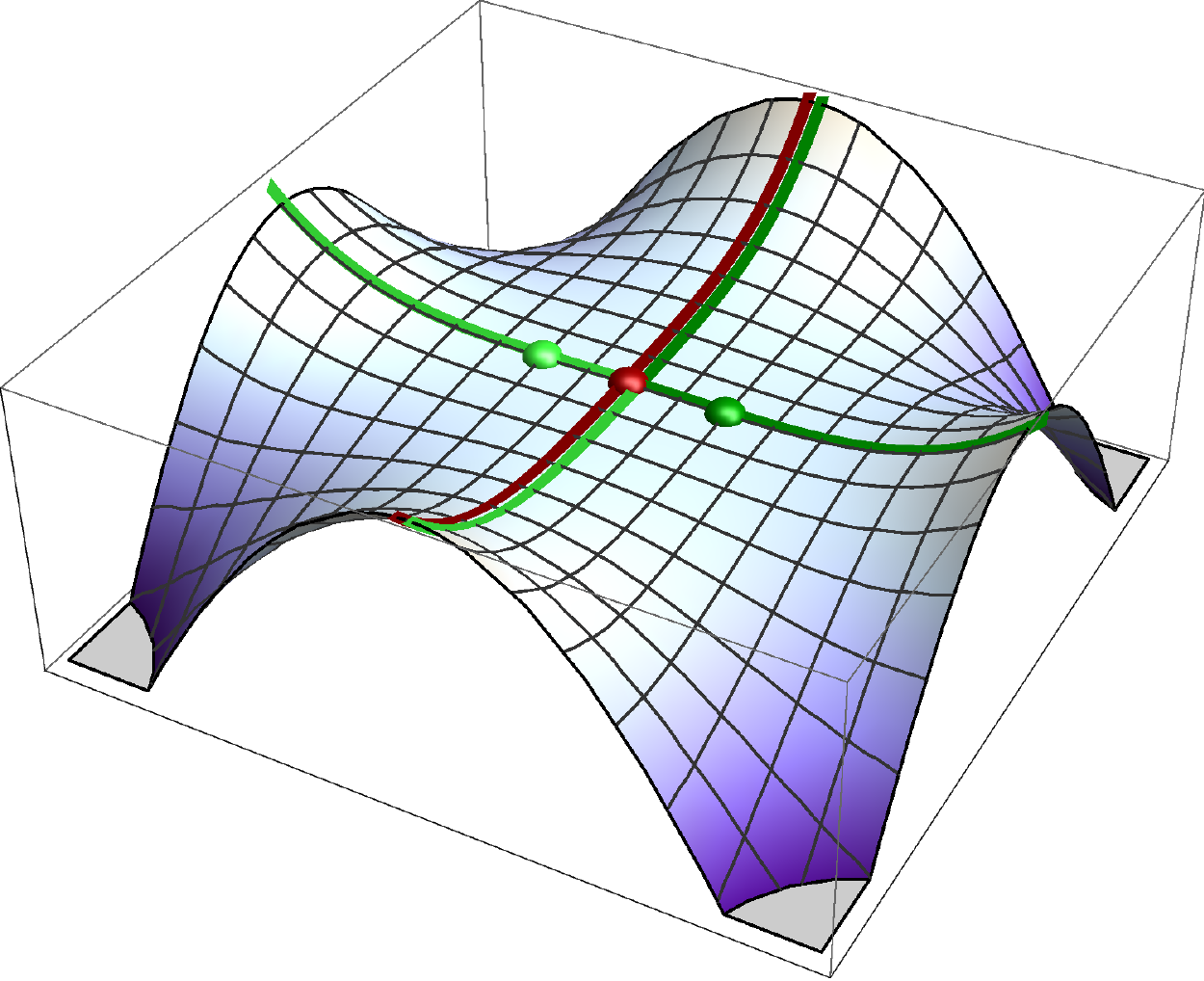}};
\draw[->] ($(c3)+(-1.5,-1.48)$)--($(c3) + (-1,-1.68)$);
\node [below] at ($(c3)+(-1.3,-1.55)$) {$\footnotesize{\text{Re}[z]}$};
\draw[->] ($(c3)+(1.935,-1.4)$)--($(c3) + (2.2,-0.98)$);
\node [below] at ($(c3)+(2.5,-1)$) {$\footnotesize{\Im[z]}$};
\end{tikzpicture}
\caption{
The saddle points $\{-1,0,1\}$ and associated steepest descent contours for 
$S(z)=- \frac{z^2}{2}+ \frac{z^4}{4}$.
Left: the real line (dotted) can be written as a sum of all three contours $\mathbb{R}={\color{newgreen} C_{-1}}+{\color{darkred} C_0}+{\color{darkgreen} C_1}$. Actually, there is an ambiguity for this action (it is on a Stokes line);
the complex conjugate contours are equally valid. Right: a plot of $\text{Re}\left(S[z]\right)$ in the complex plane where we can clearly see the lines of steepest descent through each saddle point.
\label{fig:saddleimaginaryparts}
}
\end{center}
\end{figure}

The problem of subdominant imaginary contributions is perhaps easiest to appreciate through an example. Suppose we have the following function $S$:
\begin{equation}
\label{eqn:quarticS}
S(z) =-\frac{z^2}{2}+ \frac{z^4}{4}
\end{equation}
and we want to integrate as in Eq.~\eqref{eqn:Zint} along the real line. This function has saddle points at $z=-1,0,1$, with approximations:
\begin{equation}
\cI_{-1} = \sqrt{\pi \hbar}\exp\left(\frac{1}{4\hbar}\right)
\hspace{1cm}
\cI_0 = \sqrt{-2\pi \hbar}
\hspace{1cm}
\cI_{1} = \sqrt{\pi \hbar}\exp\left(\frac{1}{4\hbar}\right)
\end{equation}

Around $z=0$, the quadratic action $S(z)=-\frac{z^2}{2}$ has increasing real part along the imaginary axis. Around $z=1$, $S(z) = -\frac{1}{4} + (z-1)^2$ which has increasing real part along the real axis. Thus going from $z=1$, the steepest descent contour moves along the real axis, until it hits $z=0$, where it must turn 
to either the positive or negative imaginary direction. This ambiguity (this action is said to be on a Stokes line) is easily resolved by giving the action a small imaginary part. For one choice,  the steepest descent contours are sketched in Fig.~\ref{fig:saddleimaginaryparts}. The steepest descent contours are sometimes called Lefshetz thimbles. For the other choice, the thimbles would be the complex conjugates of those in Fig~\ref{fig:saddleimaginaryparts} (see Fig.~\ref{fig:factorofhalfcontours} below for another example which shows the conjugate contours).

We see from Fig~\ref{fig:saddleimaginaryparts} that in this case, the original contour deforms to a sum of all three
contours ($\mathbb{R}={\color{newgreen} C_{-1}}+{\color{darkred} C_0}+{\color{darkgreen} C_1}$) and all three saddle points contribute. So we have:
\begin{equation}
\label{eqn:quarticZJ}
Z = \boxed{\cJ_{-1}}+\cJ_0+\boxed{\cJ_{1}}
\end{equation}
and we have boxed both $\cJ_{-1}$ and $\cJ_1$ because they are exactly degenerate and exponentially larger than $\cJ_0$. If we perform the naive saddle point approximation, we would obtain:
\begin{equation}
\label{eqn:quarticZapproxI}
Z \sim \cI_{-1}+\cI_0+\cI_1
\end{equation}
Now we see a confusion; since $\cI_0$ is imaginary, it seems that $Z$ might have an imaginary part in the approximation, even though it is a convergent real integral and thus is clearly actually real. What is going on?

As one can see in Fig.~\ref{fig:saddleimaginaryparts}, if we really compute $\cJ_1$ exactly, integrating along $C_1$, it will have a real part from the real-line part of the contour, and an imaginary part from the imaginary-line part. The sum of the imaginary parts of $\cJ_1$ and $\cJ_{-1}$ is exactly the negative of the imaginary part of $\cJ_0$ ---  they are simply integrating along the same contour in opposite directions.
However, when we use the saddle point approximation, $\cI_1$ and $\cI_{-1}$ are both real; their imaginary parts are (rightfully) discarded as they are exponentially small.
As we discussed before, in the saddle point approximation, like Eq.~\eqref{eqn:quarticZapproxI}, exponentially subdominant terms are meaningless. Thus to be consistent, we should also drop $\cI_0$,
since $\cJ_0$ is exponentially suppressed, as indicated in Eq.~\eqref{eqn:quarticZJ}. The integral is in fact real at any order in any expansion, if all the pieces of the same order are consistently kept.

Thus, although the saddle point approximation can seem to produce an imaginary part in a real quantity, this is an illusion. Within a consistent expansion, real integrals are real. To get an imaginary part,
we really do need to change the original contour of integration, as we explain next.


\subsubsection{Unstable potentials and analytic continuation}
\label{sec:unstablecase} 
To understand the imaginary part and factor of $\frac{1}{2}$, a standard example~\cite{ZinnJustin:2002ru,Marino}
 is the following action function
\eqn{S_g(z) = \frac{z^2}{2}-g\frac{z^4}{4} \label{eqn:quarticfactorofhalfS}}
Although this action is unbounded from below and unphysical, we will see in Section~\ref{sec:stablepotentials} how stabilizing it by adding a term $\frac{z^6}{60}$ to the action will lead to a similar result.

As above, we would like to study the imaginary part of
\eqn{Z_g \equiv \int^\infty_{-\infty} dz~ e^{-\frac{1}{\hbar}S_g(z)}}
for $g =1$. 

For any $g\neq 0$ the action has the three saddle points: $s_j = \left\{-\sqrt{\frac{1}{g}},0,\sqrt{\frac{1}{g}}\right\}$ for $j = -1,0,1$ respectively. The $z=s_0=0$ saddle point plays the role of the false vacuum solution and the $z=s_{\pm1}=\pm \sqrt{\frac{1}{g}}$ saddle points play the role of the bounce. 
Applying the saddle approximation, 
\begin{equation}
\cI_{-1} = \sqrt{-\pi \hbar} \exp\left(-\frac{1}{4g \hbar} \right)
\hspace{1cm}
\cI_0 = \sqrt{2\pi \hbar}
\hspace{1cm}
\cI_{1} = \sqrt{-\pi \hbar} \exp\left(-\frac{1}{4g \hbar} \right)
\label{cisg}
\end{equation}
we find that $\cI_0$ is real and $\cI_{\pm 1}$ are imaginary. 

As explained in the previous section, we cannot trust the saddle point approximation because exponentially small imaginary contributions from the $\cI_0$ have been dropped. Even worse in this case, $Z$ itself does not converge at $z=\pm \infty$ 
so we cannot integrate $Z$ along the real axis. However, $Z_g$ can be integrated along the real axis for $g<0$.
The steepest descent contours are shown for $g=1$ and $g=-1$ in Figure~\ref{fig:factorofhalfcontours},
with the black arcs indicated regions of convergence.
So one thing we can do is calculate $Z_g$ for $g<0$ along the real axis, and analytically
continue that back to $g=1$. We call that result $Z^\text{cont.}_g$.

\begin{figure}[H]
\begin{center}
\begin{tikzpicture}[scale=0.8]
\coordinate (c5) at (-2.9,-2.9);
\coordinate (c6) at (-2.9,2.9);
\coordinate (c1) at (2.9,-2.9);
\coordinate (c3) at (2.9,2.9);
\coordinate (c2) at (8,-2.9);
\coordinate (c4) at (8,2.9);
\node at ($(c1) + (-90:2.8cm)$) { $g=-1+i\epsilon$};
\draw ($(c1) + (-22.5:2.5cm)$) arc (-22.5:22.5:2.5cm);
\draw ($(c1) + (157.5:2.5cm)$) arc (157.5:202.5:2.5cm);
\draw ($(c1) + (67.5:2.5cm)$) arc (67.5:112.5:2.5cm);
\draw ($(c1) + (247.5:2.5cm)$) arc (247.5:292.5:2.5cm);
\draw[thick, red] ($(c1) + (0,2.5)$) -- ($(c1) + (0,0.8)$) .. controls ($(c1) + (0.0,0.1)$) .. ($(c1) + (-0.8,0.1)$) -- ($(c1) + (-2.5,0.1)$);
\draw[thick, darkgreen] ($(c1) + (-2.5,-0.05)$) -- ($(c1) + (0,-0.05)$) -- ($(c1) + (0,0.05)$)  -- ($(c1) + (2.5,0.05)$);
\draw[thick, darkred] ($(c1) + (2.5,-0.1)$) -- ($(c1) + (0.8,-0.1)$) .. controls ($(c1) + (0,-0.1)$) .. ($(c1) + (0,-0.8)$) -- ($(c1) + (0,-2.5)$);
\node[circle,darkgreen,fill,inner sep=3pt] at (c1) {};
\node[circle,red,fill,inner sep=3pt] at ($(c1) + (90:1cm)$) {};
\node[circle,darkred,fill,inner sep=3pt] at ($(c1) + (-90:1cm)$) {};
\node at ($(c2) + (-90:2.8cm)$) { $g=-1- i\epsilon$};
\draw ($(c2) + (-22.5:2.5cm)$) arc (-22.5:22.5:2.5cm);
\draw ($(c2) + (157.5:2.5cm)$) arc (157.5:202.5:2.5cm);
\draw ($(c2) + (67.5:2.5cm)$) arc (67.5:112.5:2.5cm);
\draw ($(c2) + (247.5:2.5cm)$) arc (247.5:292.5:2.5cm);
\draw[thick, darkred] ($(c2) + (0,2.5)$) -- ($(c2) + (0,0.8)$) .. controls ($(c2) + (0.0,0.1)$) .. ($(c2) + (0.8,0.1)$) -- ($(c2) + (2.5,0.1)$);
\draw[thick, darkgreen] ($(c2) + (-2.5,0.05)$) -- ($(c2) + (0,0.05)$) -- ($(c2) + (0,-0.05)$)  -- ($(c2) + (2.5,-0.05)$);
\draw[thick, red] ($(c2) + (-2.5,-0.1)$) -- ($(c2) + (-0.8,-0.1)$) .. controls ($(c2) + (0,-0.1)$) .. ($(c2) + (0,-0.8)$) -- ($(c2) + (0,-2.5)$);
\node[circle,darkgreen,fill,inner sep=3pt] at (c2) {};
\node[circle,darkred,fill,inner sep=3pt] at ($(c2) + (90:1cm)$) {};
\node[circle,red,fill,inner sep=3pt] at ($(c2) + (-90:1cm)$) {};
\node at ($(c4) + (-90:2.8cm)$) { $g=1-i\epsilon$};
\node at ($(c4) + (-1.1,1.3)$) {{$\color{darkgreen} C^{-}$}};
\draw ($(c4) + (22.5:2.5cm)$) arc (22.5:67.5:2.5cm);
\draw ($(c4) + (202.5:2.5cm)$) arc (202.5:247.5:2.5cm);
\draw ($(c4) + (112.5:2.5cm)$) arc (112.5:157.5:2.5cm);
\draw ($(c4) + (292.5:2.5cm)$) arc (292.5:337.5:2.5cm);
\draw[thick, darkred] ($(c4) + (44:2.5cm)$) .. controls ($(c4) + (0.75,0)$) .. ($(c4) + (-44:2.5cm)$);
\draw[thick, red] ($(c4) + (136:2.5cm)$) .. controls ($(c4) + (-0.75,0)$) .. ($(c4) + (-136:2.5cm)$);
\draw[thick, darkgreen] ($(c4) + (0.0,0.0)$) -- ($(c4) + (0.5,0.0)$) .. controls ($(c4) + (0.9,0.0)$) and ($(c4) + (0.88,-0.12)$) .. ($(c4) + (1.05,-0.5)$) .. controls ($(c4) + (1.3,-1.0)$) .. ($(c4) + (-45.2:2.5cm)$);
\draw[thick, darkgreen] ($(c4) + (0.0,0.0)$) -- ($(c4) + (-0.5,0.0)$) .. controls ($(c4) + (-0.9,0.0)$) and ($(c4) + (-0.88,0.12)$) .. ($(c4) + (-1.05,0.5)$) .. controls ($(c4) + (-1.3,1.0)$) .. ($(c4) + (134.8:2.5cm)$);
\node[circle,darkgreen,fill,inner sep=3pt] at ($(c4)$) {};
\node[circle,red,fill,inner sep=3pt] at ($(c4) + (180:1cm)$) {};
\node[circle,darkred,fill,inner sep=3pt] at ($(c4) + (0:1cm)$) {};
\node at ($(c3) + (-90:2.8cm)$) { $g=1+i\epsilon$};
\node at ($(c3) + (1.1,1.3)$) {{$\color{darkgreen} C^{+}$}};
\draw ($(c3) + (22.5:2.5cm)$) arc (22.5:67.5:2.5cm);
\draw ($(c3) + (202.5:2.5cm)$) arc (202.5:247.5:2.5cm);
\draw ($(c3) + (112.5:2.5cm)$) arc (112.5:157.5:2.5cm);
\draw ($(c3) + (292.5:2.5cm)$) arc (292.5:337.5:2.5cm);
\draw[thick, darkred] ($(c3) + (44:2.5cm)$) .. controls ($(c3) + (0.75,0)$) .. ($(c3) + (-45:2.5cm)$);
\draw[thick, red] ($(c3) + (135:2.5cm)$) .. controls ($(c3) + (-0.75,0)$) .. ($(c3) + (-136:2.5cm)$);
\draw[thick, darkgreen] ($(c3) + (0.0,0.0)$) -- ($(c3) + (0.5,0.0)$) .. controls ($(c3) + (0.9,0.0)$) and ($(c3) + (0.88,0.12)$) .. ($(c3) + (1.05,0.5)$) .. controls ($(c3) + (1.3,1.0)$) .. ($(c3) + (45.2:2.5cm)$);
\draw[thick, darkgreen] ($(c3) + (0.0,0.0)$) -- ($(c3) + (-0.5,0.0)$) .. controls ($(c3) + (-0.9,0.0)$) and ($(c3) + (-0.88,-0.12)$) .. ($(c3) + (-1.05,-0.5)$) .. controls ($(c3) + (-1.3,-1.0)$) .. ($(c3) + (-134.8:2.5cm)$);
\node[circle,darkgreen,fill,inner sep=3pt] at ($(c3)$) {};
\node[circle,red,fill,inner sep=3pt] at ($(c3) + (180:1cm)$) {};
\node[circle,darkred,fill,inner sep=3pt] at ($(c3) + (0:1cm)$) {};
\node at ($(c3)+(90:3.3cm)$) { $ \boxed{S_g(z)=\frac{z^2}{2}-g\frac{z^4}{4}}$};
\node[above] at ($(c5)+(0,2)$) {$S_{-|g|}(z)$};
\node[right] at ($(c5)+(2,0)$) {$z$};
\draw[->] ($(c5)+(-2,0)$) --($(c5)+(2,0)$);
\draw[->] ($(c5)+(0,-2)$) --($(c5)+(0,2)$); 
\draw[domain=-1.4:1.4,thick,smooth,variable=\x,black] plot ({\x+(-2.9)},{-2.9+0.5*\x*\x+0.25*pow(\x,4)});
\node[above] at ($(c6)+(0,2)$) {$S_{|g|}(z)$};
\node[right] at ($(c6)+(2,0)$) {$z$};
\draw[->] ($(c6)+(-2,0)$) --($(c6)+(2,0)$);
\draw[->] ($(c6)+(0,-2)$) --($(c6)+(0,2)$); 
\draw[domain=-2:2,thick,smooth,variable=\x,black] plot ({\x+(-2.9)},{2.9+0.5*\x*\x-0.25*pow(\x,4)});
\end{tikzpicture}
\caption{
The saddle points and steepest descent contours for $S_g(z)=\frac{z^2}{2} - g\frac{z^4}{4}$ in Eq.~\eqref{eqn:quarticfactorofhalfS}. 
For $g=1$, integrating along the real axis is divergent (as indicated by the lack of arcs at $z=\pm \infty$). 
For $g=-1$, the  $\FV$ contour (green line) falls along the real axis. Rotating $g$ back from $-1$ to $1$, the $\FV$ contour remains convergent, but depends on whether one rotates $g$ clockwise or counterclockwise in the complex
plane.
\label{fig:factorofhalfcontours}
}
\vspace{5mm}
\begin{tikzpicture}
\coordinate (c6) at (0,0);
\coordinate (c3) at (3.5,0);
\coordinate (c4) at (7,0);
\coordinate (c1) at (-7,0);
\coordinate (c2) at (-3.5,0);
\node at ($(c1) + (-90:1.9cm)$) { $\underset{\text{\normalsize{(deforms to $C^-$)}}}{\theta = -\pi}$};
\draw ($(c1) + (22.5:1.5cm)$) arc (22.5:67.5:1.5cm);
\draw ($(c1) + (202.5:1.5cm)$) arc (202.5:247.5:1.5cm);
\draw ($(c1) + (112.5:1.5cm)$) arc (112.5:157.5:1.5cm);
\draw ($(c1) + (292.5:1.5cm)$) arc (292.5:337.5:1.5cm);
\draw[dashed,thick]  ($(c1)-(1.5,0)$)--($(c1)+(1.5,0)$);
\draw[thick,red] ($(c1)+(-1.5/1.414,1.5/1.414)$)--($(c1)+(1.5/1.414,-1.5/1.414)$);
\node at ($(c2) + (-90:1.8cm)$) { $\theta=-\frac{\pi}{2}$};
\draw ($(c2) + (0:1.5cm)$) arc (0:-45:1.5cm);
\draw ($(c2) + (180:1.5cm)$) arc (180:135:1.5cm);
\draw ($(c2) + (270:1.5cm)$) arc (270:225:1.5cm);
\draw ($(c2) + (90:1.5cm)$) arc (90:45:1.5cm);
\draw[dashed,thick]  ($(c2)-(1.5,0)$)--($(c2)+(1.5,0)$);
\draw[thick,red] ($(c2)+(-1.5*0.9238,1.5*0.3827)$)--($(c2)+(1.5*0.9238,-1.5*0.3827)$);
\node at ($(c6)+(90:2.3cm)$) { $ \boxed{~S_{g}(z)=\frac{z^2}{2}-g\frac{z^4}{4},~~~g=-e^{i\theta}~}$};
\node at ($(c6) + (-90:1.8cm)$) { $\theta = 0$};
\draw ($(c6) + (22.5:1.5cm)$) arc (22.5:-22.5:1.5cm);
\draw ($(c6) + (202.5:1.5cm)$) arc (202.5:157.5:1.5cm);
\draw ($(c6) + (112.5:1.5cm)$) arc (112.5:67.5:1.5cm);
\draw ($(c6) + (292.5:1.5cm)$) arc (292.5:247.5:1.5cm);
\draw[thick,red] ($(c6)-(1.5,0)$)--($(c6)+(1.5,0)$);
\draw[dashed,thick]  ($(c6)-(1.5,0)$)--($(c6)+(1.5,0)$);
\node at ($(c3) + (-90:1.8cm)$) { $\theta=\frac{\pi}{2}$};
\draw ($(c3) + (0:1.5cm)$) arc (0:45:1.5cm);
\draw ($(c3) + (180:1.5cm)$) arc (180:225:1.5cm);
\draw ($(c3) + (270:1.5cm)$) arc (270:315:1.5cm);
\draw ($(c3) + (90:1.5cm)$) arc (90:135.5:1.5cm);
\draw[dashed,thick]  ($(c3)-(1.5,0)$)--($(c3)+(1.5,0)$);
\draw[thick,red] ($(c3)-(1.5*0.9238,1.5*0.3827)$)--($(c3)+(1.5*0.9238,1.5*0.3827)$);
\node at ($(c4) + (-90:1.9cm)$) { $\underset{\text{\normalsize{(deforms to $C^+$)}}}{\theta = \pi}$};
\draw ($(c4) + (22.5:1.5cm)$) arc (22.5:67.5:1.5cm);
\draw ($(c4) + (202.5:1.5cm)$) arc (202.5:247.5:1.5cm);
\draw ($(c4) + (112.5:1.5cm)$) arc (112.5:157.5:1.5cm);
\draw ($(c4) + (292.5:1.5cm)$) arc (292.5:337.5:1.5cm);
\draw[dashed,thick]  ($(c4)-(1.5,0)$)--($(c4)+(1.5,0)$);
\draw[thick,red] ($(c4)-(1.5/1.414,1.5/1.414)$)--($(c4)+(1.5/1.414,1.5/1.414)$);
\end{tikzpicture}
\vspace{-5mm}
\caption{The contour (red) used for analytic continuation as a function of $\theta$ compared to the original real axis contour (dashed line). Because there are no poles in $S(z)$, we can deform the contours however we like as long as they end in the same convergent regions (black arcs are where $\text{Re}\left(g z^4\right)<0$). 
\label{fig:analyticcontinuationpic}
}
\end{center}
\end{figure}

The convergent regions in complex $z$-plane of $Z^\text{cont.}_g$ change as a function of $g$. The analytic extension of away from $g<0$ is unique if we fix the endpoints of the contour to the convergent regions as we change $g$. 
For example, we can parametrize the analytic continuation from $g=-1$ to $g=1$ by varying $g = -e^{i\theta}$ from $\theta = 0$ to $\theta = \pi$. 
For $\theta=0$ the integration contour is given by the real line. As we change $\theta$ we can have the end points of the contour following the convergent region by rotating the contour of integration by an angle $\frac{\theta}{4}$ when we rotate $g=-1$ to $g = -e^{i\theta}$ (see Fig. \ref{fig:analyticcontinuationpic}). 
Note that rotating by $\theta=-\pi$ gives the conjugate result to rotating by $\theta=\pi$, and thus there is a branch cut along the positive real axis if we restrict the Riemann surface to a single sheet. 
In Fig. \ref{fig:factorofhalfcontours} the steepest descents are given for $g$ slightly above and below the positive real axis to break the degeneracy of steepest descent curves.

Fixing the endpoints
of the contour, the rest of the contour can safely be deformed since $S_g(z)$ contains no poles. 
We can deform the $\theta = \pi$ contour of integration to the contour of steepest descent through the false vacuum saddle point, $C^+$. This can easily be seen by comparing Fig.~\ref{fig:factorofhalfcontours} and Fig.~\ref{fig:analyticcontinuationpic}. Similarly, rotating by $\theta=-\pi$ gives the contour $C^{-}$.

We therefore find
\begin{equation}
Z^{\text{cont.}\pm}_{g=1} = \int_{\color{darkgreen} C^{\pm}} dz \exp\left(-\frac{z^2}{2} + \frac{z^4}{4}\right)
\end{equation}
where $\left(Z^{\text{cont.}+}_{g}\right)^\star = Z^{\text{cont.}-}_{g}$. 

\begin{figure}[t]\begin{center}
\begin{tikzpicture}
\coordinate (c1) at (-2.9,2.9);
\node at ($(c1) + (1.1,1.3)$) {{$\color{darkgreen} C^{+}$}};
\node at ($(c1) + (-1.1,1.3)$) {{$\color{darkgreen} C^{-}$}};
\node at ($(c1) + (-1.5,0)$) {{$\color{red} C_{-1}$}};
\node at ($(c1) + (1.4,0)$) {{$\color{darkred} C_{1}$}};
\draw ($(c1) + (22.5:2.5cm)$) arc (22.5:67.5:2.5cm);
\draw ($(c1) + (202.5:2.5cm)$) arc (202.5:247.5:2.5cm);
\draw ($(c1) + (112.5:2.5cm)$) arc (112.5:157.5:2.5cm);
\draw ($(c1) + (292.5:2.5cm)$) arc (292.5:337.5:2.5cm);
\draw[thick, darkred] ($(c1) + (44:2.5cm)$) .. controls ($(c1) + (0.75,0)$) .. ($(c1) + (-44:2.5cm)$);
\draw[thick, red] ($(c1) + (136:2.5cm)$) .. controls ($(c1) + (-0.75,0)$) .. ($(c1) + (-136:2.5cm)$);
\draw[thick, darkgreen] ($(c1) + (0.0,0.0)$) -- ($(c1) + (0.5,0.0)$) .. controls ($(c1) + (0.9,0.0)$) and ($(c1) + (0.88,0.12)$) .. ($(c1) + (1.05,0.5)$) .. controls ($(c1) + (1.3,1.0)$) .. ($(c1) + (45.2:2.5cm)$);
\draw[thick, darkgreen] ($(c1) + (0.0,0.0)$) -- ($(c1) + (-0.5,0.0)$) .. controls ($(c1) + (-0.9,0.0)$) and ($(c1) + (-0.88,0.12)$) .. ($(c1) + (-1.05,0.5)$) .. controls ($(c1) + (-1.3,1.0)$) .. ($(c1) + (134.8:2.5cm)$);
\draw[thick, darkgreen] ($(c1) + (0.0,0.0)$) -- ($(c1) + (0.5,0.0)$) .. controls ($(c1) + (0.9,0.0)$) and ($(c1) + (0.88,-0.12)$) .. ($(c1) + (1.05,-0.5)$) .. controls ($(c1) + (1.3,-1.0)$) .. ($(c1) + (-45.2:2.5cm)$);
\draw[thick, darkgreen] ($(c1) + (0.0,0.0)$) -- ($(c1) + (-0.5,0.0)$) .. controls ($(c1) + (-0.9,0.0)$) and ($(c1) + (-0.88,-0.12)$) .. ($(c1) + (-1.05,-0.5)$) .. controls ($(c1) + (-1.3,-1.0)$) .. ($(c1) + (-134.8:2.5cm)$);
\node[circle,darkgreen,fill,inner sep=3pt] at ($(c1)$) {};
\node[circle,red,fill,inner sep=3pt] at ($(c1) + (180:1cm)$) {};
\node[circle,darkred,fill,inner sep=3pt] at ($(c1) + (0:1cm)$) {};
\end{tikzpicture}
\caption{
The difference between the two $\FV$ contours (green lines, $C^+$ and $C^-$) is equivalent to the sum of the two bounce contours ($C_{-1}$ and $C_1$).
(
Computing the discontinuity $\disc Z^{\text{cont.}}_{g=1}$ rather than just $Z^{\text{cont.}}_{g=1}$ involves subtracting two integration contours, yielding the combined contour which can be deformed into a sum of the red and blue steepest-descent curves, with no presence of the green curve.
\label{fig:disccontour}
}
\end{center}
\end{figure}
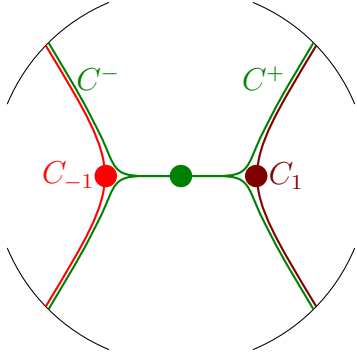

Next, we want to approximate this integral using a saddle point approximation. Unfortunately,
the saddle approximation along the contour {$\color{darkgreen} C^{+}$} (or {$\color{darkgreen} C^{-}$}) around the FV saddle point is purely real, to all orders
 in the expansion. The saddle-point approximation probes the function only in an asymptotically small region close to the
saddle-point. This is completely insensitive to the part of the contour {$\color{darkgreen} C^{+}$} (or {$\color{darkgreen} C^{-}$}) which passes into into the complex plane. However, it is precisely this part of the contour we are interested in, since that is where the imaginary part will
come from. A way around this is to use the fact that $Z^{\text{cont.} \pm}_g$ are complex conjugates of each other to write
\eqn{\Im Z^{\text{cont.} +}_{g=1} = \frac{Z^{\text{cont.} +}_{g=1}-Z^{\text{cont.} -}_{g=1}}{2 i} = \frac{\disc\left(Z^\text{cont.}_{g=1}\right)}{2i}}
As can been seen in Fig.~\ref{fig:disccontour} integrating along ${\color{darkgreen} C^{+}}- {\color{darkgreen} C^{-}}$ is equivalent to integrating along
{$\color{red} C_{-1}$} + {$\color{darkred} C_{1}$}, that is, along the  contours of steepest descent passing through the bounces. 

Now that we have directly related the imaginary part of $Z^{\text{cont.} +}_{g=1}$ 
to an integral along paths which are complex at the saddle points, we can safely take the saddle-point approximation. 
Thus,
\begin{equation}
\disc Z^\text{cont.}_{g=1} = {\color{red} \mc{J}_{-1}} + {\color{darkred}\mc{J}_{1}} \sim
{\color{red} \mc{I}_{-1}} + {\color{darkred}\mc{I}_{1}}
\end{equation}
and therefore
\begin{equation}
\Im Z^{\text{cont.} +}_{g=1} \sim \frac{1}{2}\Im\left({\color{red} \mc{I}_{-1}} + {\color{darkred}\mc{I}_{1}}\right)
\end{equation}
That is, the desired imaginary part is give by \emph{half} the sum over the relevant bounces.

\subsubsection{Bounded potentials and steepest descent contours}
\label{sec:stablepotentials}
One concern with the procedure described above is that the action we used in Section~\ref{sec:unstablecase}, $S_g[z] = \frac{1}{2}z^2 - g \frac{z^4}{4}$ was unbounded from below, and therefore unphysical. One might worry that the justification for analytic continuation and changing the integration contour from the real axis was to make the integral well-defined; for a physical potential, perhaps the imaginary part remains zero. With that motivation, consider the action function 
\be
S_g(z) =\frac{ z^2}{2} - g\frac{ z^4}{4} + \frac{z^6}{60}
\ee
This is similar to the action from Section~\ref{sec:stablepotentials}, but now the integral $Z$ over the real axis is convergent.

Now the action is 6th order, so $S'(z)=0$ has 5 solutions for 5 saddle points. For $g=1$, these are around
$z\approx \{-3.0,-1.0,~0,~1.0,~3.0\}$, all along the real axis. There are 6 convergent regions, including the region at $z=\pm \infty$ for any $g$. So in this case, the original contour of the real line is perfectly fine for any $g$. Indeed, the function $Z_g = \int^\infty_{-\infty} dz \exp(-S_g(z))$ is an analytic function of $g$; it is real for real $g$ and has no discontinuity near $g=1$. Thus, whatever we do, we are certainly not analytically continuing $Z_g$.

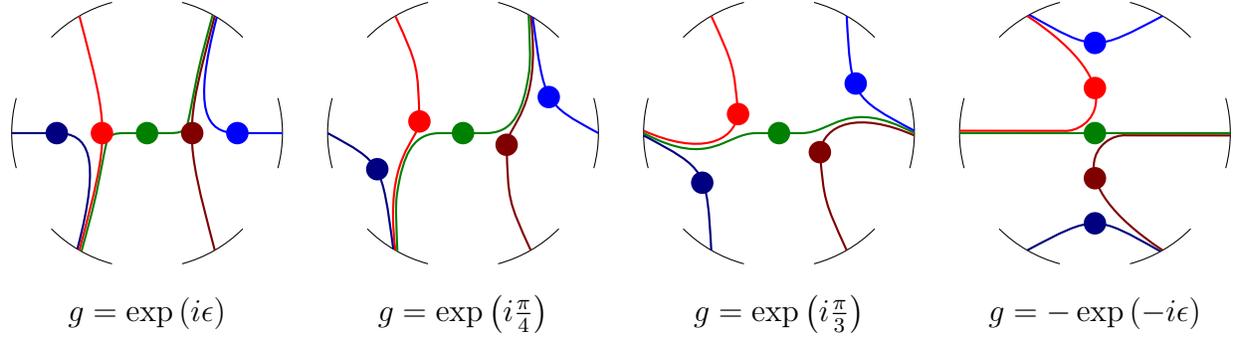
\begin{figure}[t]\begin{center}
\begin{tikzpicture}[scale=0.6]
\coordinate (ls) at (0,-4);
\coordinate (c1) at (0,0);
\node at ($(c1) + (ls)$) {$g=\exp\left(i \epsilon\right)$};
\draw ($(c1) + (15:3cm)$) arc (15:-15:3cm);
\draw ($(c1) + (45:3cm)$) arc (45:75:3cm);
\draw ($(c1) + (105:3cm)$) arc (105:135:3cm);
\draw ($(c1) + (165:3cm)$) arc (165:195:3cm);
\draw ($(c1) + (-45:3cm)$) arc (-45:-75:3cm);
\draw ($(c1) + (-105:3cm)$) arc (-105:-135:3cm);
\draw[thick, darkred] ($(c1) + (60:3cm)$) .. controls ($(c1) + (0.82,0)$) .. ($(c1) + (-60:3cm)$);
\draw[thick, red] ($(c1) + (120:3cm)$) .. controls ($(c1) + (-0.82,0)$) .. ($(c1) + (-120:3cm)$);
\draw[thick, darkgreen] ($(c1) + (0.0,0.0)$) -- ($(c1) + (0.5,0.0)$) .. controls ($(c1) + (0.9,0.0)$) and ($(c1) + (0.88,0.12)$) .. ($(c1) + (0.96,0.5)$) .. controls ($(c1) + (1.15,1.5)$) .. ($(c1) + (61:3cm)$);
\draw[thick, darkgreen] ($(c1) + (0.0,0.0)$) -- ($(c1) + (-0.5,0.0)$) .. controls ($(c1) + (-0.9,0.0)$) and ($(c1) + (-0.88,-0.12)$) .. ($(c1) + (-0.96,-0.5)$) .. controls ($(c1) + (-1.15,-1.5)$) .. ($(c1) + (-119:3cm)$);
\draw[thick, blue] ($(c1) + (0:3cm)$) -- ($(c1) + (2,0.0)$) .. controls ($(c1) + (1.3,0.0)$) and ($(c1) + (1,0.5)$) .. ($(c1) + (59:3cm)$);
\draw[thick, darkerblue] ($(c1) + (-180:3cm)$) -- ($(c1) + (-2,0.0)$) .. controls ($(c1) + (-1.3,0.0)$) and ($(c1) + (-1,-0.5)$) .. ($(c1) + (-121:3cm)$);
\node[circle,darkgreen,fill,inner sep=3pt] at ($(c1)$) {};
\node[circle,red,fill,inner sep=3pt] at ($(c1) + (180:1cm)$) {};
\node[circle,darkred,fill,inner sep=3pt] at ($(c1) + (0:1cm)$) {};
\node[circle,darkerblue,fill,inner sep=3pt] at ($(c1) + (180:2cm)$) {};
\node[circle,blue,fill,inner sep=3pt] at ($(c1) + (0:2cm)$) {};
\coordinate (c2) at (7,0);
\node at ($(c2) + (ls)$) {$g=\exp\left(i\frac{\pi}{4}\right)$};
\draw ($(c2) + (15:3cm)$) arc (15:-15:3cm);
\draw ($(c2) + (45:3cm)$) arc (45:75:3cm);
\draw ($(c2) + (105:3cm)$) arc (105:135:3cm);
\draw ($(c2) + (165:3cm)$) arc (165:195:3cm);
\draw ($(c2) + (-45:3cm)$) arc (-45:-75:3cm);
\draw ($(c2) + (-105:3cm)$) arc (-105:-135:3cm);
\draw[thick, darkred] ($(c2) + (-60:3cm)$) .. controls ($(c2) + (1,-1.5)$) .. ($(c2) + (-15:1cm)$) .. controls ($(c2) + (1,0.2)$) and ($(c2) + (1.75,0.5)$) .. ($(c2) + (60:3cm)$);
\draw[thick, red] ($(c2) + (120:3cm)$) .. controls ($(c2) + (-1,1.5)$) .. ($(c2) + (165:1cm)$) .. controls ($(c2) + (-1,-0.2)$) and ($(c2) + (-1.75,-0.5)$) .. ($(c2) + (-120:3cm)$);
\draw[thick, darkgreen] ($(c2) + (0.0,0.0)$) -- ($(c2) + (0.5,0.0)$) .. controls ($(c2) + (1,0)$) and ($(c2) + (1.25,0.3)$) .. ($(c2) + (1.4,1)$) .. controls ($(c2) + (1.5,1.5)$) .. ($(c2) + (61:3cm)$);
\draw[thick, darkgreen] ($(c2) + (0.0,0.0)$) -- ($(c2) + (-0.5,0.0)$) .. controls ($(c2) + (-1,0)$) and ($(c2) + (-1.25,-0.3)$) .. ($(c2) + (-1.4,-1)$) .. controls ($(c2) + (-1.5,-1.5)$) .. ($(c2) + (-119:3cm)$);
\draw[thick, blue] ($(c2) + (0:3cm)$) .. controls ($(c2) + (1.8,0.65)$) .. ($(c2) + (59:3cm)$);
\draw[thick, darkerblue] ($(c2) + (-180:3cm)$) .. controls ($(c2) + (-1.8,-0.65)$) .. ($(c2) + (-121:3cm)$);
\node[circle,darkgreen,fill,inner sep=3pt] at ($(c2)$) {};
\node[circle,red,fill,inner sep=3pt] at ($(c2) + (165:1cm)$) {};
\node[circle,darkred,fill,inner sep=3pt] at ($(c2) + (-15:1cm)$) {};
\node[circle,darkerblue,fill,inner sep=3pt] at ($(c2) + (-1.9,-0.8)$) {};
\node[circle,blue,fill,inner sep=3pt] at ($(c2) + (1.9,0.8)$) {};
\coordinate (c3) at (14,0);
\node at ($(c3) + (ls)$) {$g=\exp\left(i\frac{\pi}{3}\right)$};
\draw ($(c3) + (15:3cm)$) arc (15:-15:3cm);
\draw ($(c3) + (45:3cm)$) arc (45:75:3cm);
\draw ($(c3) + (105:3cm)$) arc (105:135:3cm);
\draw ($(c3) + (165:3cm)$) arc (165:195:3cm);
\draw ($(c3) + (-45:3cm)$) arc (-45:-75:3cm);
\draw ($(c3) + (-105:3cm)$) arc (-105:-135:3cm);
\draw[thick, darkred] ($(c3) + (-60:3cm)$) .. controls ($(c3) + (1,-1.5)$) .. ($(c3) + (-25:1cm)$) .. controls ($(c3) + (1,0.2)$) and ($(c3) + (1.75,0.5)$) .. ($(c3) + (-1:3cm)$);
\draw[thick, red] ($(c3) + (120:3cm)$) .. controls ($(c3) + (-1,1.5)$) .. ($(c3) + (155:1cm)$) .. controls ($(c3) + (-1,-0.2)$) and ($(c3) + (-1.75,-0.5)$) .. ($(c3) + (179:3cm)$);
\draw[thick, darkgreen] ($(c3) + (0.0,0.0)$) -- ($(c3) + (0.5,0.0)$) .. controls ($(c3) + (1.2,0)$) and ($(c3) + (1.5,0.8)$) ..  ($(c3) + (0:3cm)$);
\draw[thick, darkgreen] ($(c3) + (0.0,0.0)$) -- ($(c3) + (-0.5,0.0)$) .. controls ($(c3) + (-1.2,0)$) and ($(c3) + (-1.5,-0.8)$) ..  ($(c3) + (180:3cm)$);
\draw[thick, blue] ($(c3) + (1:3cm)$) .. controls ($(c3) + (1.6,0.8)$) .. ($(c3) + (60:3cm)$);
\draw[thick, darkerblue] ($(c3) + (-179:3cm)$) .. controls ($(c3) + (-1.6,-0.8)$) .. ($(c3) + (-120:3cm)$);
\node[circle,darkgreen,fill,inner sep=3pt] at ($(c3)$) {};
\node[circle,red,fill,inner sep=3pt] at ($(c3) + (155:1cm)$) {};
\node[circle,darkred,fill,inner sep=3pt] at ($(c3) + (-25:1cm)$) {};
\node[circle,darkerblue,fill,inner sep=3pt] at ($(c3) + (-1.7,-1.1)$) {};
\node[circle,blue,fill,inner sep=3pt] at ($(c3) + (1.7,1.1)$) {};
\coordinate (c3) at (21,0);
\node at ($(c3) + (ls)$){$g= - \exp\left( - i \epsilon\right)$};
\draw ($(c3) + (15:3cm)$) arc (15:-15:3cm);
\draw ($(c3) + (45:3cm)$) arc (45:75:3cm);
\draw ($(c3) + (105:3cm)$) arc (105:135:3cm);
\draw ($(c3) + (165:3cm)$) arc (165:195:3cm);
\draw ($(c3) + (-45:3cm)$) arc (-45:-75:3cm);
\draw ($(c3) + (-105:3cm)$) arc (-105:-135:3cm);
\draw[thick, darkred] ($(c3) + (-1:3cm)$) -- ($(c3) + (0.7,-0.05)$) .. controls ($(c3) + (0.1,0)$) and ($(c3) + (-0.9,-1)$) .. ($(c3) + (-60:3cm)$);
\draw[thick, red] ($(c3) + (179:3cm)$) -- ($(c3) + (-0.7,0.05)$) .. controls ($(c3) + (-0.1,0)$) and ($(c3) + (0.9,1)$) .. ($(c3) + (120:3cm)$);
\draw[thick, darkgreen] ($(c3) + (0:3cm)$) -- ($(c3) + (-180:3cm)$);
\draw[thick, blue] ($(c3) + (119:3cm)$) .. controls ($(c3) + (0,1.8)$) .. ($(c3) + (60:3cm)$);
\draw[thick, darkerblue] ($(c3) + (-120:3cm)$) .. controls ($(c3) + (0,-1.8)$) .. ($(c3) + (-61:3cm)$);
\node[circle,darkgreen,fill,inner sep=3pt] at ($(c3)$) {};
\node[circle,red,fill,inner sep=3pt] at ($(c3) + (90:1cm)$) {};
\node[circle,darkred,fill,inner sep=3pt] at ($(c3) + (-90:1cm)$) {};
\node[circle,darkerblue,fill,inner sep=3pt] at ($(c3) + (-90:2cm)$) {};
\node[circle,blue,fill,inner sep=3pt] at ($(c3) + (90:2cm)$) {};
\end{tikzpicture}
\caption{
For the action function $S_g[z] =\frac{ z^2}{2} - g\frac{ z^4}{4} + \frac{z^6}{60}$, the real axis is
a valid integration contour for any $g$. However, as we change $g$, the steepest descent contours
change. We see that for $g=\exp(i\epsilon)$ the steepest descent contour ends along the ray $\theta = \frac{\pi}{3}$ rather than the real axis, $\theta=0$. 
\label{fig:physicalpotentialx6}
}
\end{center}
\end{figure}

So what can we do? We saw in the previous section that for unbounded potentials, analytic continuation is equivalent to integrating along the steepest descent contour through the false vacuum saddle point. In the case of a bounded potential like we have here, analytic continuation and integrating along the contour of steepest descent are necessarily different; the steepest descent will always move off the real axis and will end in a different region of convergence. We therefore introduce the path integral along the steepest descent contour through the FV saddle point
\eqn{Z^{C^\pm} = \int_{C^\pm} dz~ e^{-S_g(z)}\label{eqn:Zsteepestdecent}}
where $C^\pm$ are the steepest descent contours for $g=1\pm i\epsilon$.  In Fig.~\ref{fig:physicalpotentialx6} we see how the contours and saddle points move about as we rotate $g$. 

By using Eq.~\eqref{eqn:Zsteepestdecent} for both bounded and unbounded potentials, we will always find that the imaginary part of $Z^{C^+}$ is given by 
\be
\Im Z^{C^+}=\frac{1}{2i}\text{disc} \left(Z^{C^+}\right)
\ee
which in the saddle point approximation is equal to $\frac{1}{2}$ times the sum over the relevant bounce saddle points.
With this method, modifying the action far away from the region relevant to the false vacuum and the bounce does not seem to affect the prediction for the tunneling rate by very much, which is reassuring.

\subsubsection{Dependence on the choice of contour} \label{sec:conts}
We have intimated that the key to finding the decay rate is to integrate not along the real axis but along a steepest descent
contour passing through the false vacuum. If the potential is deformed to be convex with the $\FV$ at the true minimum,
the real axis will coincide with this contour. To clarify the importance of the $\FV$ saddle, let us now look at how different results arise when different saddle points are stabilized.

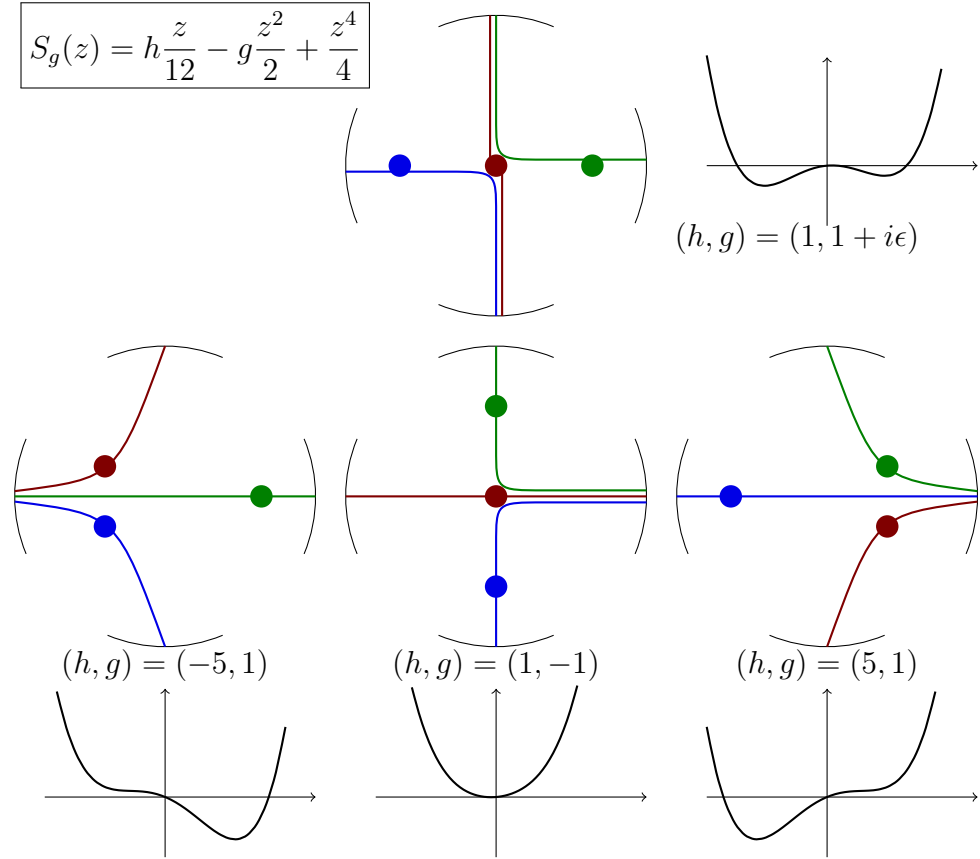
\begin{figure}
\begin{center}
\begin{tikzpicture}[scale=0.8]
\coordinate (c1) at (0,0);
\coordinate (c4) at (-5.5,-5.5);
\coordinate (c3) at (0,-5.5);
\coordinate (c2) at (5.5,-5.5);
\coordinate (c11) at (5.5,0);
\coordinate (c24) at (-5.5,-10.5);
\coordinate (c23) at (0,-10.5);
\coordinate (c22) at (5.5,-10.5);
\draw ($(c1) + (-22.5:2.5cm)$) arc (-22.5:22.5:2.5cm);
\draw ($(c1) + (157.5:2.5cm)$) arc (157.5:202.5:2.5cm);
\draw ($(c1) + (67.5:2.5cm)$) arc (67.5:112.5:2.5cm);
\draw ($(c1) + (247.5:2.5cm)$) arc (247.5:292.5:2.5cm);
\draw[thick, darkgreen] ($(c1) + (0,2.5)$) -- ($(c1) + (0,0.8)$) .. controls ($(c1) + (0.0,0.1)$) .. ($(c1) + (0.8,0.1)$) -- ($(c1) + (2.5,0.1)$);
\draw[thick, darkred] ($(c1) + (-0.1,2.5)$) -- ($(c1) + (-0.1,0)$) -- ($(c1) + (0.1,0)$)  -- ($(c1) + (0.1,-2.5)$);
\draw[thick, darkblue] ($(c1) + (-2.5,-0.1)$) -- ($(c1) + (-0.8,-0.1)$) .. controls ($(c1) + (0,-0.1)$) .. ($(c1) + (0,-0.8)$) -- ($(c1) + (0,-2.5)$);
\node[circle,darkblue,fill,inner sep=3pt] at (-1.6,0) {};
\node[circle,darkgreen,fill,inner sep=3pt] at (1.6,0) {};
\node[circle,darkred,fill,inner sep=3pt] at (c1) {};
\node at ($(c1) + (5,-1.2)$) {$(h,g)=(1,1+i\epsilon)$};
\draw ($(c2) + (-22.5:2.5cm)$) arc (-22.5:22.5:2.5cm);
\draw ($(c2) + (157.5:2.5cm)$) arc (157.5:202.5:2.5cm);
\draw ($(c2) + (67.5:2.5cm)$) arc (67.5:112.5:2.5cm);
\draw ($(c2) + (247.5:2.5cm)$) arc (247.5:292.5:2.5cm);
\draw[thick, darkgreen] ($(c2) + (90:2.5cm)$) .. controls ($(c2) + (0.8,0.3)$) .. ($(c2) + (2:2.5cm)$);
\draw[thick, darkred] ($(c2) + (-90:2.5cm)$) .. controls ($(c2) + (0.8,-0.3)$) .. ($(c2) + (-2:2.5cm)$);
\draw[thick, darkblue] ($(c2) + (-2.5,0)$) -- ($(c2) + (2.5,0)$) ;
\node[circle,darkblue,fill,inner sep=3pt] at ($(c2)+(-1.6,0)$) {};
\node[circle,darkgreen,fill,inner sep=3pt] at ($(c2)+(1,0.5)$) {};
\node[circle,darkred,fill,inner sep=3pt] at ($(c2)+(1,-0.5)$) {};
\node at ($(c2) + (-90:2.8cm)$) {$(h,g)=(5,1)$};
\draw ($(c3) + (-22.5:2.5cm)$) arc (-22.5:22.5:2.5cm);
\draw ($(c3) + (157.5:2.5cm)$) arc (157.5:202.5:2.5cm);
\draw ($(c3) + (67.5:2.5cm)$) arc (67.5:112.5:2.5cm);
\draw ($(c3) + (247.5:2.5cm)$) arc (247.5:292.5:2.5cm);
\draw[thick, darkgreen] ($(c3) + (0,2.5)$) -- ($(c3) + (0,0.8)$) .. controls ($(c3) + (0.0,0.1)$) .. ($(c3) + (0.8,0.1)$) -- ($(c3) + (2.5,0.1)$);
\draw[thick, darkred] ($(c3) + (180:2.5cm)$) -- ($(c3) + (0:2.5cm)$);
\draw[thick, darkblue] ($(c3) + (2.5,-0.1)$) -- ($(c3) + (0.8,-0.1)$) .. controls ($(c3) + (0,-0.1)$) .. ($(c3) + (0,-0.8)$) -- ($(c3) + (0,-2.5)$);
\node[circle,darkblue,fill,inner sep=3pt] at ($(c3)+(0,-1.5)$) {};
\node[circle,darkgreen,fill,inner sep=3pt] at ($(c3)+(0,1.5)$) {};
\node[circle,darkred,fill,inner sep=3pt] at ($(c3)+(0,0)$) {};
\node at ($(c3) + (-90:2.8cm)$) {$(h,g)=(1,-1)$};
\draw ($(c4) + (-22.5:2.5cm)$) arc (-22.5:22.5:2.5cm);
\draw ($(c4) + (157.5:2.5cm)$) arc (157.5:202.5:2.5cm);
\draw ($(c4) + (67.5:2.5cm)$) arc (67.5:112.5:2.5cm);
\draw ($(c4) + (247.5:2.5cm)$) arc (247.5:292.5:2.5cm);
\draw[thick, darkred] ($(c4) + (90:2.5cm)$) .. controls ($(c4) + (-0.8,0.3)$) .. ($(c4) + (178:2.5cm)$);
\draw[thick, darkblue] ($(c4) + (-90:2.5cm)$) .. controls ($(c4) + (-0.8,-0.3)$) .. ($(c4) + (-178:2.5cm)$);
\draw[thick, darkgreen] ($(c4) + (-2.5,0)$) -- ($(c4) + (2.5,0)$) ;
\node[circle,darkgreen,fill,inner sep=3pt] at ($(c4)+(1.6,0)$) {};
\node[circle,darkred,fill,inner sep=3pt] at ($(c4)+(-1,0.5)$) {};
\node[circle,darkblue,fill,inner sep=3pt] at ($(c4)+(-1,-0.5)$) {};
\node at ($(c4) + (-90:2.8cm)$) {$(h,g)=(-5,1)$};
\draw[->] ($(c11) + (-2,0)$) -- ($(c11) + (2.5,0)$);
\draw[->] ($(c11) + (0,-1)$) -- ($(c11) + (0,1.8)$); 
\draw[domain=-2:1.9,thick,smooth,variable=\x,black] plot ({\x+5.5},{+(1/12)*\x-0.5*pow(\x,2)+0.25*pow(\x,4)});
\draw[->] ($(c22) + (-2,0)$) -- ($(c22) + (2.5,0)$);
\draw[->] ($(c22) + (0,-1)$) -- ($(c22) + (0,1.8)$); 
\draw[domain=-2:1.8,thick,smooth,variable=\x,black] plot ({\x+5.5},{-10.5+(5/12)*\x-0.5*pow(\x,2)+0.25*pow(\x,4)});
\draw[->] ($(c23) + (-2,0)$) -- ($(c23) + (2.5,0)$);
\draw[->] ($(c23) + (0,-1)$) -- ($(c23) + (0,1.8)$); 
\draw[domain=-1.4:1.35,thick,smooth,variable=\x,black] plot ({\x},{-10.5+(1/12)*\x+0.5*pow(\x,2)+0.25*pow(\x,4)});
\draw[->] ($(c24) + (-2,0)$) -- ($(c24) + (2.5,0)$);
\draw[->] ($(c24) + (0,-1)$) -- ($(c24) + (0,1.8)$); 
\draw[domain=-1.8:2,thick,smooth,variable=\x,black] plot ({\x-5.5},{-10.5-(5/12)*\x-0.5*pow(\x,2)+0.25*pow(\x,4)});
\node at ($(c1)+(-5,2)$) { $ \boxed{S_g(z)=h\frac{z}{12}-g\frac{z^2}{2}+\frac{z^4}{4}}$};
\end{tikzpicture}
\caption{
The function $S_{h,g}(z) =h\frac{z}{12}-g\frac{z^2}{2}+\frac{z^4}{4}$ can be deformed different ways from the physical
case $(h,g)=(1,1)$ (top). 
For $(h,g)=(-5,1)$, the $\FV$ is stabilized, and the real axis lines up with the $\FV$ contour (green). The correct factor
of $\frac{1}{2}$ results. 
 For $(h,g)=(1,-1)$, the bounce is stabilized (red). The imaginary part computed this way is the naive one, missing the
 factor of $2$. For $(h,g)=(5,1)$, the shot is stabilized (blue). 
\label{fig:discwithouthalf}
}
\end{center}
\end{figure}

To explore how different steepest descent contours may affect the result, consider the following two-parameter family of actions
\be
\label{potgh}
S_g(z) = h \frac{z}{12} -g\frac{z^2}{2}+\frac{z^4}{4}
\ee
The case of interest is $g=h=1$. There are three saddle points: a $\FV$ at $z\approx -1.0$, a bounce at $z\approx 0.0$ and a shot at $z\approx 1.0$. The saddle points and steepest descent contours
are shown in Fig.~\ref{fig:discwithouthalf}. 
 
First, we consider keeping $h=1$ and rotating to $g=-1$. For $h=1$ and $g=-1$, the potential has one minimum at $z\approx 0.0$: the bounce has been stabilized. In this case, the integration contour
along the $z$ axis coincides with the bounce saddle point contour. When we rotate back to $g=1$, this contour lines up with the imaginary axis. Thus, integrating along the contour will give the complete
imaginary part of the bounce saddle-point integration, {\it without} the factor of $\frac{1}{2}$. Of course, this had to happen: by stabilizing the bounce, we matched the integration contour with the bounce contour.
When we rotate back, it remains lined up and therefore the full integral over the bounce contour is kept.

Next, consider keeping $g=1$ but rotating $h$ from 1 to something negative and large enough to remove the other minimum, such as $h=-5$. For example, we can rotate as $h=-2+3e^{i\theta}$ with $0\le \theta \le \pi$. 
For $h=-5$, the saddle point on the real axis is the $\FV$ saddle, and the other two have moved into the complex plane. When we rotate back to $h=1$, this $\FV$ saddle moves along the real axis and then up {\it half}
of the bounce saddle. Thus for the $h$ deformation, we do get the extra factor of $\frac{1}{2}$, as expected.

Finally, consider rotating $h$ from $1$ to $5$. This stabilizes the shot. Rotating back to $h=1$, we see that the shot contour lines up with the other hand of the bounce contour than when we stabilized the $\FV$.
Thus we do get a factor of $\frac{1}{2}$ in this case. The sign of the imaginary part in any case has to be fixed by physics. When we stabilize the shot, we can choose the sign to be negative so that $\Gamma < 0$. This makes sense physically because flux enters the true vacuum region , so the probability grows with time. This corresponds to incoming Gamow boundary conditions (as in Section~\ref{sec:complex} or Appendix~\ref{app:Gammasimple}) and one expects $\Gamma <0$.

\subsection{Summary of potential deformation method}
In this section, we discussed how to compute a decay rate from the Euclidean path integral, filling in some details and examining some peculiarities not mentioned in~\cite{Callan:1977pt,Coleman:1978ae}
or elsewhere in the literature to our knowledge. 
In this method, one starts with a Euclidean path integral or partition function
\be
Z  = \int_{x(0)=a}^{x(\bigtau)=a} \cD x e^{-S_E[x]}
\ee
which is real for all $\bigtau$. 

To get an imaginary part, we reduced the problem to integrating along a 1-parameter family of curves, passing through
the static $\FV$ path, the bounce, and the shot. 
Around the bounce saddle point $z=s_1$, the 1D integral is approximately
\begin{equation}
Z = \int d z e^{-S(z)}\approx \sqrt{\frac{2\pi}{S''(s_1)}} e^{-S(s_1)}
\end{equation}
which is imaginary because of the negative curvature around the bounce. However, one cannot just pick a single saddle point. One must use the proper integration contour, which can include multiple saddles, and the imaginary parts can (and do) cancel. Physically, the $\bigtau \to \infty$ limit always picks out the true vacuum, and the path integral
is dominated by the shot not the bounce. The path integral around the shot is real and exponentially larger than the path integral around the bounce. Instead, we want the metastable resonance state near the false vacuum to dominate. 

To isolate the resonance near the $\FV$, one approach
is to deform the potential so that the $\FV$ is the true minimum.
We write $g=1$ for the original potential, and $g=-1$ for when the $\FV$ is stable.
Unfortunately, at $g=-1$, the
path integral is still real. And moreover one cannot simply analytically continue the result back to $g=-1$, for then the true minimum
would be the true vacuum again, and $Z$ still real. 

The right way to isolate the resonance is to deform the theory at $g=-1$ in a different way: we pin the contour
with the steepest descent contour ${\color{darkgreen} C^+}$ passing through the $\FV$ saddle. Near $g=-1$ this is identical to the analytic function $Z_g$. However, as the potential deforms back, the analytic continuation remains real, while pinning the contour to ${\color{darkgreen} C^+}$ does not. This is a consequence of the fact that through the deformation, the system crosses a Stokes line, changing the region (the black arcs in Fig. \ref{fig:discwithouthalf} for example) in which the steepest descent through the $\FV$ saddle ends.

Of course, if at the end of the day all we want is to stick to the FV contour, we do not have to bother with the analytic continuation at all. The procedure is simply that
\begin{itemize}
\item  The decay rate is given by the imaginary part of the path integral along the steepest descent  contour
${\color{darkgreen} C^+}$ 
 passing through the false vacuum saddle point ({\it not} the bounce saddle point, and {\it not} the path integral over real paths).
\end{itemize}
Pinning the contour to focus on the $\FV$ is exactly what we want physically: we want the resonance, associated with the $\FV$ to dominate, even when it is not the dominant saddle point. This procedure enforces this dominance, albeit in a somewhat
artificial mathematical way. 

The integral along the $\FV$ contour will produce something complex $Z=\cR+i \cI$. Then the decay rate, in the limit $\cI \ll \cR$ is
\begin{equation}
\frac{\Gamma}{2} = \frac{1}{\bigtau}\Im \ln Z \approx \frac{1}{\bigtau} \frac{\cI}{\cR}
\end{equation}
Unfortunately, to compute $Z$ one cannot directly use the saddle point approximation, since close to the saddle point the $\FV$ contour it will give something real, $\cR$ alone. This real part $\cR$ is indeed given by the saddle-point approximation around $\FV$, $\cR\approx 
\exp\{-S[\bar{x}_{\FV}]\}$. The 
imaginary part $\cI$ comes from a region on the contour far away from the $\FV$ saddle. Thus to get the imaginary part under a saddle-point approximation, we
have to compute the discontinuity between the two degenerate steepest descent contours which is equivalent to evaluating $\frac{1}{2}$ of the steepest descent contour through the bounce saddles. That is,
\begin{itemize}
\item  The imaginary part of the path integral can be computed along the steepest descent  contour passing through the bounce saddle point,
times $\frac{1}{2}$.
\end{itemize}
Thus to leading order
\be
\frac{\Gamma}{2} \sim \frac{1}{2\bigtau} \tilde{K} \frac{\exp\{-S[\bar{x}_\text{bounce}]\}}{ \exp\{-S[\bar{x}_{\FV}]\}}
\ee
with the prefactor $\tilde{K}$ computable systematically in the saddle-point approximation.

We have performed some helpful sanity checks on this algorithm. 
First, we considered examples from the literature
that use actions unbounded from below, where one can get a nontrivial result using analytic continuation 
because the integral $Z$ along the real axis diverges.
Thus it looks like $Z$ {\it must} be defined through analytic continuation. 
But when considering bounded potentials, we saw that analytic continuation does not give the right answer. Instead 
pinning to the steepest descent contour through the false vacuum gives consistent results
in both the bounded and unbounded cases we have considered. Moreover, deforming the potential away from the $\FV$ region and the barrier has little effect on the rate, as desired. 

Second, we checked that the deformation must stabilize the false vacuum, not any of the other saddle points. We found that if the bounce is stabilized, one gets an answer that is a factor of $2$ too large. We also found that
stabilizing the shot gives an answer with the wrong sign. These observations are consist with the general physical argument that a proper derivation must include consideration that $T$ not be too large. For very large $T$,
only the true vacuum is relevant. By deforming the potential so the false vacuum is the ground state, the false vacuum bound states can be found. After deforming back, these presumably turn into the resonances, with outgoing
boundary conditions and imaginary energies.

There are two nagging questions we have been unable to answer in our explorations of this method.
\begin{enumerate}
\item How can the rate be calculated without using the saddle-point approximation, for example, non-perturbatively?
\end{enumerate}
If the procedure were just analytic continuation, one could in principle compute the path integral as a function of $g$ and analytic continue it from around $g=-1$ back to $g=1$. However, fixing one dimension of an infinite-dimensional integral
to a particular contour does not have an obvious non-perturbative analog.
\begin{enumerate}
\item[2.] Can one prove the pinning the contour to ${\color{darkgreen} C^+}$ is identical to imposing the limits 
$\sloshtime \ll T \ll \nonlintime$?
\end{enumerate}
In other words, 
how do we know the number $\Gamma$ computed through this  mathematically-consistent procedure always gives the decay rate exactly?

 The next section discusses a more physical approach, with fewer mathematical subtleties, that we can (in principle) compare to the potential-deformation approach order-by-order. 


\section{Direct method}\label{sec:dirmethod}
The potential-deformation method described in Section~\ref{sec:potmethod} connects the Euclidean action to the decay rate in a roundabout manner. 
It relies critically on an understanding of the subtleties of analytic continuation, steepest descent contours and saddle approximations of the path integral in order to obtain an imaginary number.
In this section, we describe an alternative derivation that connects the path integral directly to the decay rate.\footnote{Another approach, employing coherent states, is described in~\cite{Hammer:1978xu}.
}

Let us assume we have a potential with a false vacuum region ($\FV$), a barrier, and a true vacuum region  ($\R$), as in Fig.~\ref{fig:numericpotentialprobability}. The energy eigenstates which have support
in the $\FV$ region are in bands of width $\Gamma_i$ around resonance energies $E_i$. In Section~\ref{sec:complex}, we were led to assume, for simplicity, that our initial state only 
had support for energies near $E_0$. We would still like this to be true, but it is more convenient with path integrals to work with position eigenstates than energy eigenstates. So let
us assume now that the initial state is localized at the point $x=a$ where the minimum of the well is located. As we take $T$ large, this wavefunction will have dominant support from within the lowest energy band. It will also
have some support for the higher energy bands, but the higher energy components die off much faster than the $E_0$ components, so if we focus on time scales $T \gg E_0^{-1}$, we
should be able to ignore those components.

The decay rate to a region $\R$ is defined as in Eq.~\eqref{eqn:defineGamma}:
\be
\label{eqn:defineGammaAgain}
\Gamma_\R =
\hspace{-1em}\lim_{\substack{T/\nonlintime\to0\\T/\sloshtime\to\infty}}
\frac{1}{P_{\FV}(t)}\frac{dP_\R(T)}{dT}
\ee
where the probability $P_\R$ is defined as:
\be
P_\R(t) = \int_\R d x_f | \langle  x_f; t|a; 0\rangle |^2 = \int_\R d x_f \left| \cN  D_F(a,0;x_f,t)\right|^2
\label{Pform}
\ee
The factor of $\cN$ in Eq.~\eqref{Pform} is a normalization factor pulled out of the path-integral form of the propagator, so that
the Feynman propagator $D_F$ is simply
\be\label{eqn:defineD}
D_F(a,0;x_f,t) =  \int_{x(0) =a}^{x(t)=x_f} \Dxspace e^{i S[x]}
\ee
Normally there would be a factor of $\cN$ in Eq.~\eqref{eqn:defineD} but we have put the $\cN$ in Eq.~\eqref{Pform} instead. 

Let us denote by $b$ the point on the boundary of $\R$ where the potential is degenerate with the initial point $a$: $V(b)=V(a)$. 
By splitting every path into the part before it first hits $b$ (at time $t_0$) and the part after $t_0$ we can write
\be
D_F(a,0; x_f,t) = \int_0^{t} d t_0  \Du(a,0; b, t_0) D_F(b, 0; x_f,t-t_0)  \label{Dsplit}
\ee
where
\be
\label{eqn:defineDbar}
\Du(a,0; b,t_0) \equiv \int_{x(0)=a}^{x(t_0)=b}\Dxspace e^{i S[x] }\delta(t_b[x]-t_0)
\ee
Here, $t_b[x]$ is the functional returning the time the path $x(t)$ first crosses $b$. So $\Du$ 
is the Feynman propagator over paths on the interval $(0,t_0)$ that hit $b$ exactly once, at $t=t_0$. 

The separation in Eq.~\eqref{Dsplit} works so long as all paths in the original propagator pass through $b$ at least once. In the path integral, for each path $x(t)$ into $\R$ there is a time $t_0$ when the path exits the barrier for the first time.
Since we have taken $b$ to be the classical turning point on the boundary of $\R$, any path into $\R$ must hit $b$, so we can indeed use Eq.~\eqref{Dsplit}.
Thus we find
\be
\label{eqn:PIexpandedconjugate}
P_\R(t) =\cN \cN^\star \int d x_f \int_0^T d t_0  \int_0^T d t_0'  \Du(a,0; b, t_0) \Du^\star(a,0; b, t_0') D_F(b, 0; x_f,T-t_0) D_F^\star(b, 0; x_f,T-t_0')
\ee

Now we want to use the fact that once the particle gets to region $\R$ it stays in region $\R$; this is the limit $T/\nonlintime\to0$.
This fact lets us replace the sum of $|x_f\rangle \langle x_f |$ over points in $\R$ to a sum over all points.
Such a replacement will modify $P_\R$ only by terms which are exponentially small, for example suppressed by extra factors of the $e^{-W}$
WKB penetration factor. 
Such exponentially small corrections are an irreducible ambiguity in what is meant by a decay rate. Thus,
to the extent that $\Gamma$ is well-defined at all, we can drop them.
Then,
\begin{equation}
\label{eqn:applyTNLapprox}
\begin{aligned}
\int d x_F D_F(b, 0; x_f,T-t_0) D_F^\star(b, 0; x_f,T-t_0') &= \int d x_f \langle b| e^{-i H(T-t_0)} | x_f \rangle \langle x_f| e^{i H(T-t_0')} | b \rangle \\
& \approx  D_F(b,0; b, t_0'-t_0)
\end{aligned}
\end{equation}

So we now have
\be
\label{eqn:threepropagators}
P_\R(t) = \cN \cN^\star \int_0^T d t_0  \int_0^T d t_0'  \Du(a,0; b, t_0) \Du^\star(a,0; b, t_0') D_F(b, 0; b,t_0'-t_0)
\ee
Breaking the two integrals into the regions with $t_0' > t_0$ and $t_0 > t_0'$, we can use Eq.~\eqref{Dsplit} on the two halves along with $D_F^\star(x,0; y,t) = D_F(y,0; x,-t)$ to get
\be
P_\R(t) = \cN \cN^\star\int_0^T d t_0 \left[ \Du(a,0; b, t_0) D_F^\star(a,0; b, t_0)  + \Du^\star(a,0; b,t_0) D_F(a,0 ;b,t_0)\right]
\ee
Then expanding the definitions of $D_F$ and $\Du$ in Eqs.~\eqref{eqn:defineD} and \eqref{eqn:defineDbar} and plugging into the definition of $\Gamma_\R$ in Eq.~\eqref{eqn:defineGammaAgain}
produces
\be
\Gamma_\R = \frac{\cN\cN^\star}{P_{\FV}(T)}\left( \int_{x(0)=a}^{x(T)=b} \Dxspace  e^{i S[x] } \delta(t_b[x]-T)\right)
\left( \int_{x(0) =a}^{x(T)=b} \Dxspace e^{i S[x]}\right)^\star + c.c. \label{Gpath}
\ee

Now let us consider how to calculate these two path integrals. 
Because the path integral has an imaginary exponent, it is not convergent (when integrated along the real $x$).
For this reason the actual definition of the Minkowski path integral involves a strange imaginary integration path over $x$, or more simply, evaluating it for $\bigtau=iT$ real and then analytically continuing.

For the first path integral in  Eq.~\eqref{Gpath} we analytically continue to $\bigtau = i T$. Then the boundary conditions are $x(0) = a$ and $x(\bigtau)=b$ as before, which are equivalent to $x(-\bigtau)=a$ and $x(0)=b$.   For the second integral, which is complex conjugated, we must analytically continue to $\bigtau' = - i \tau=-\bigtau$ to ensure convergence. This leads to $x(0) = a$ and $x(-\bigtau)=b$ as boundary conditions, or equivalently, $x(0)=b$ and $x(\bigtau)=a$. Because the endpoints
have switched, the two Euclidean path integrals can then be recombined leading to
\be
\label{eqn:Gammaimaginerydeltas}
\Gamma_\R =\frac{\cN\cN^\star}{P_{FV}(T)} \int_{x(-\bigtau)=a}^{x(\bigtau)=a} \Dxspace e^{- S_E[x] }
\Big[\eta_+ \delta(-i\tau_b[x]) + \eta_- \delta(i\tau_b[x])\Big]
\ee
To get to this line, we have replaced $t_b[x]$ with $-i\tau_b[x]$ in the first $\delta$-function and $i\tau_b[x]$ in the second $\delta$-function. However, changing variables within
a $\delta$-function has to be done carefully, since the $\delta$-function is not defined for imaginary arguments. Thus we have added phase factors $\eta_\pm$ in front, which we will now fix.

 In order to analytically continue correctly, we should first remove the $\delta$-function mode, then analytically continue, and then put it back. This will lead to $\delta(\pm if[x])\equiv\eta_{\pm}\delta(f[x])$
for some $\eta_\pm$. To see what these $\eta_\pm$ are, we focus only on a single degree of freedom restricted by the $\delta$-function. We are trying to analytically continue like this:
\begin{multline}
\int dx \Big[Ae^{i\alpha S(x)}\delta(\alpha t[x])+A^\star e^{-i\alpha S(x)}\delta(\alpha t[x]) \Big]
\\
=\left(\int dx\Big[ \eta_- Ae^{-r S(x)}\delta(i r t[x])+ \eta_+ A^\star e^{-s S(x)}\delta(-ist[x])\Big]\right)_{\substack{r=-i\alpha\\s=i\alpha}}
\end{multline}
where $A$ represents the integral over all the other modes, and $\alpha$ represents a phase-factor coefficient in front of $\tau$ which we will rotate from 1 to $i$. The notation means that we evaluate the integrals for $r$ and $s$ positive (where the other path integrals converge) and then make the substitutions. Integrating over the $\delta$-functions gives
\begin{align}
\label{eqn:etaplusminus}
\frac{Ae^{i\alpha S(x_{*})}+A^\star e^{-i\alpha S(x_{*})}}{\abs{\alpha t'(x_*)}}
&= \left(\frac{\eta_-A e^{-rS(x_{*})}}{r\abs{t'(x_*)}}+\frac{\eta_+A^\star e^{-sS(x_{*})}}{s\abs{t'(x_*)}}\right)_{\substack{r=-i\alpha\\s=i\alpha}}\\
&= \frac{1}{r\abs{t'(x_*)}}\left(\eta_-Ae^{-rS(x_{*})} - \eta_+A^\star e^{rS(x_{*})}\right)_{r=-i\alpha}
\end{align}
from which we see that $\eta_+=-\eta_-=i\,\text{sign}(\alpha)$ will give us the right answer. 
Thus we see that the analytic continuation of the $\delta$-function is
\begin{multline}
\int dx\Big[ Ae^{i\alpha S(x)}\delta(\alpha t[x])+A^\star e^{-i\alpha S(x)}\delta(\alpha t[x])\Big] \\
= -\text{sign}(\alpha)2\text{Im}\left(A\int dx e^{-r S(x)}\delta(r t(x))\right)_{\substack{r>0\\r=-i\alpha}}
\end{multline}
where the notation means that we evaluate the expression for $r>0$, and then afterwards analytically continue to $r=-i\alpha$. 
The sign ambiguity is due to a branch cut: the answer depends on which way we rotate the argument,
and we will have to fix it with physical argument. 

Fixing the sign by requiring $\Gamma>0$, the precise version of Eq.~\eqref{eqn:Gammaimaginerydeltas} is then
\be
\Gamma =\lim_{\bigtau\to\infty}\abs{ \frac{\cN\cN^\star}{P_{FV}(T)} 2\text{Im}\left(\int_{x(-\bigtau)=a}^{x(\bigtau)=a} \Dxspace e^{- S_E[x] }\delta(\tau_b[x])\right)_{\substack{\bigtau>0\\\bigtau=iT}}}
\ee
The normalization $\frac{P_{FV}(T)}{\cN\cN^\star}$ can be manipulated in a similar manner in the limit $T/\nonlintime\to 0$,
which ultimately gives us:
\be
\boxed{
\Gamma =\lim_{T\to\infty}\abs{2\text{Im}\left(
\frac{\int\Dx e^{- S_E[x] }\delta(\tau_b[x])}
{\int \Dx e^{- S_E[x] }}
\right)_{\substack{\bigtau>0\\\bigtau=iT}}}
\label{GpathE}
}
\ee
where both path integrals are evaluated with boundary conditions $x(\pm\bigtau)=a$. This formula provides an exact expression 
for the decay rate defined in Eq.~\eqref{eqn:defineGamma}.  

In Section~\ref{sec:potmethod}, in the potential-deformation method, we were worried that we might be accidentally taking the imaginary part of a real quantity by making an invalid saddle point approximation. So let us now
discuss in detail where the imaginary parts are coming from in this direct method.
Eq.~\eqref{Gpath}, before the analytic continuation to imaginary time, is all-orders exact and manifestly real; one could in principle make a lattice and calculate it numerically. Then we analytically continued Eq.~\eqref{Gpath} to arrive at Eq.~\eqref{GpathE}. 
The path integral $\int \cD x e^{-S_E}$, without the $\delta$-function, would be real for real $\bigtau$, but the $\delta$-function will introduce a factor with dimensions of time, which becomes imaginary when we plug in $\bigtau\to iT$. 
Thus, we are taking the imaginary part of something purely imaginary in Eq.~\eqref{GpathE}.
We will discuss the saddle point approximation's interaction with this story in Sections~\ref{sec:sp1} and~\ref{sec:minkowskisaddle}.
 
Before showing how Eq.~\eqref{GpathE} can be evaluated, let us contrast it with the potential-deformation method discussed in Section~\ref{sec:potmethod}.
To make a precise connection to  Eq.~\eqref{eqn:Coleman}, let us first change from $\bigtau$ to $\bigtau/2$ (since the time is going to infinity, the factor of 2 has no effect). To match the other formula, we need to re-introduce the time-translation degeneracy. Isolating the term
\eqn{\mc{E} \equiv 
\frac{\int\Dx e^{- S_E[x] }\delta(\tau_b[x])}
{\int \Dx e^{- S_E[x] }}}
which is time-translation-invariant for all $\tau$, that is
\eqn{\mc{E} = \mc{E}_\tau \equiv 
\frac{\int\Dx e^{- S_E[x] }\delta(\tau_b[x]-\tau)}
{\int \Dx e^{- S_E[x] }}}
allows us to rewrite it as
\eqna{\mc{E} &= \frac{1}{\bigtau}\int^{\bigtau/2}_{-\bigtau/2} d\tau~\mc{E}_\tau\\
&= \frac{1}{\bigtau} 
\frac{\int_{\text{paths hit $b$}}\Dx e^{- S_E[x] }}
{\int \Dx e^{- S_E[x] }}
}
from which we arrive at: 
\begin{align}
\Gamma=\lim_{T\to\infty}\abs{2\text{Im}\left(\frac{1}{\bigtau}
\frac{\int_{\text{paths hit $b$}}\Dx e^{- S_E[x] }}
{\int \Dx e^{- S_E[x] }}
\right)_{\substack{\bigtau>0\\\bigtau=iT}}}
\end{align}
 The $\delta$-function in the numerator has been removed by the $\int d\tau$, except that it leaves the requirement that the path must hit $b$ at some time, so that $\tau_b$ is defined.
Thus, the path integral in the numerator will exclude the constant false-vacuum solution which dominates the denominator. In this way, the need to determine the contour of steepest descent as we did in the potential-deformation method
is sidestepped completely.

\subsection{Saddle point approximations} \label{sec:sp1}
As discussed in Section~\ref{sec:dirmethod}, $\int \Dx e^{-S_E}$, computed to all orders, is real. 
And as we saw in Section~\ref{sec:saddle}, when we approximate the path integral with a sum over saddle points, some of the saddle points might be imaginary.
The imaginary parts will cancel if all the saddle points are kept associated with the integration contour, but if some can be dropped, the result may be complex. 
In the traditional method, deformation of the contour of integration in the path integral is used to justify dropping some saddle points giving a well defined imaginary part. 
In the direct method, the imaginary part comes out with less gymnastics. 
The path integral is real for real $\bigtau$ but simply becomes imaginary for imaginary $\bigtau$.  

In this section we will show that when performing the saddle point approximation
for real $\bigtau$, the true vacuum solution (the shot) dominates, but when evaluated for imaginary $\bigtau$, the instanton solution dominates.
Thus we are justified in using only the instanton because we are looking at imaginary $\bigtau$. In particular, there is no tension with the instanton's saddle point expansion (which matters for imaginary $\bigtau$) being imaginary when the original path integral is real (for real $\bigtau$).

As in the potential deformation method, the path integrals in Eq.~\eqref{GpathE} are approximated by a sum of saddle points:
\begin{equation}
\label{eqn:allsaddles}
\frac{\exp(-S_{\text{shot'}})+\exp(-S_{\text{bounce}})}{\exp(-S_{\text{shot}})+\exp(-S_{\text{bounce}})+\exp(-S_{\FV})}
\end{equation}
Consider first the denominator. It contains contributions from exactly the same paths as in the potential-deformation method, shown in Fig.~\ref{fig:paths}: the static $\FV$ solution, the bounce, and the shot.
Because of the forms of these solutions, it is clear that the $\bigtau$ dependence, for large $\bigtau$, must have a linear dependence for the long stationary times, and a constant piece for the brief times 
when the particle is rolling fast:
\begin{align}
S_{\text{shot}} &= E_{\TV}\bigtau+S_S^0
\\
S_{\FV} &= E_{\FV}\bigtau
\\
S_{\text{bounce}} &= E_{\FV}\bigtau+S_B^0
\end{align}
Also we note that $S_S^0 > S_B^0$ since the shot must go faster than the bounce and hence has more  energy. 

Recall that in the potential-deformation method, the shot dominated for the actual path integral with the physical potential, but when we deformed to $g<0$, then the false vacuum 
dominated. 
With the direct method, rather than deforming the potential, we performing the standard $\bigtau\to iT$ Wick rotation. For real $\bigtau$, the shot dominates. 
But we are not interested in which dominates for real $\bigtau$, rather which dominates for $\bigtau\to iT$. Then,
\begin{align}
S_{\text{shot}} &= iE_{\TV}T+S_S^0
\\
S_{\FV} &= iE_{\FV}T
\\
S_{\text{bounce}} &= iE_{\FV}T+S_B^0
\end{align}
Since $S_B^0 < S_S^0$, due to the $e^{-S}$ factors in the saddle point approximation, the bounce exponentially dominates over the shot. However, both of these are dominated by the $\FV$ solution which has
no exponential suppression at all. Thus for the denominator, if we drop exponentially suppressed pieces, only the $\FV$ contribution remains. 

The numerator of Eq.~\eqref{eqn:allsaddles} is similar to the denominator, but has been modified by the $\delta(\tau_b)$. In particular, the $\FV$ solution, which never hits the point $b$, is removed entirely by the $\delta$-function. 
The shot is also removed, since it hits $b$ before $\tau=0$ (it hits the $\TV$ region at $\tau=0$), but there is a solution qualitatively similar to the shot
that we call the {\it modified shot}, or shot$^\prime$ as in Eq.~\eqref{eqn:allsaddles}.\footnote{There will nevertheless still be a lower-action solution hitting $b$ at $\tau=0$ (we know this because the bounce
still has a negative eigenvalue~\cite{Callan:1977pt,Coleman:1987rm}). The minimum action solution probably looks like the bounce up to $\tau=0$ spliced to a rescaled shot for $\tau>0$. The shot part has to be rescaled to return to the $\FV$
at $\tau=\bigtau$. The extra kick needed to splice these solutions at $\tau=0$ is allowed because the $\delta$-function can introduce discontinuities in $\partial_\tau x(\tau)$.
We call the actual minimum action solution the modified shot.}
In any case, the argument for the numerator is then exactly the same as for the denominator; for real $\bigtau$ the modified shot dominates, but when we rotate $\bigtau\to iT$, the constant part of the action now controls the size of $e^{-S}$ and so the bounce dominates. Since the false vacuum is not present in the numerator at all, the result is given by the bounce alone. 

In summary, performing the saddle point approximation  to Eq.~\eqref{GpathE} for imaginary $\bigtau$ carefully, we find the bounce dominates the numerator and the $\FV$ dominates the denominator.
For real $\bigtau$, this would \emph{not} be the correct set of saddle points to use (the correct saddle points would be the shot and modified shot). The point is that
 there is no tension between the dominant saddle points being imaginary (for imaginary $\bigtau$) and the full path integral being real (for real $\bigtau$).

\subsection{Saddle-point approximation and NLO formula}\label{sec:minkowskisaddle}

Having the all-orders formula, Eq.~\eqref{GpathE}, we want to apply the saddle-point approximation to it to get something we can actually calculate. If $\bar{x}$ is the bounce solution to the Euclidean equations of motion and $x_{\FV}$ is the static solution which stays at the false vacuum, then we see that at leading order:
\begin{equation}
\Gamma_\R^{\text{LO}} = \#\frac{e^{-S_E[\bar{x}]}}{e^{-S_E[x_{FV}]}}
\end{equation}
which is the usual leading-order formula~\cite{Coleman:1977py}. It also agrees with Eq.~\eqref{eqn:SELO},  since for $E_0=0$ in the false vacuum, $S_E=2\int\sqrt{2mV}dx$.

To go to NLO, we would like to perform a Gaussian approximation on Eq.~\eqref{GpathE}. Expanding around the bounce we write $x=\bar x+\delta x$ and
\begin{equation}
\Gamma =\frac{e^{-S_E[\bar{x}]}}{e^{-S_E[x_{FV}]}} \lim_{\tau\to\infty}\abs{ \frac{2\text{Im}\int [\cD\,  \delta x]~ e^{-\frac{1}{2} S_E''[\bar x] \delta x^2 }\delta(\tau_b[\bar x+\delta x])}{\int [\cD\, \delta x]~ e^{- \frac{1}{2}S_E''[x_{FV}] \delta x^2 }}}  \label{Gnloform}
\end{equation}
Normally, for a Gaussian integral, we expand the path in orthonormal modes
\begin{equation}
x^{\xi_0, \xi_1,\dots}(\tau) = \bar x(\tau)+\sum_{i=0}^\infty\xi_i x_i(\tau) \label{param1}
\end{equation}
where $x_i(\tau)$ are the eigenvectors of $S_E''[\bar x]$ with eigenvalues $\lambda_i$. Plugging this back into Eq.~\eqref{Gnloform} would then give Gaussian integrals $\int d \xi_i e^{-\frac{1}{2}\lambda_i \xi_i^2}$. 
However, since one of the modes has a zero eigenvalue ($\lambda_0 = 0$ for $\xi_i(\tau) \propto \partial_\tau \bar{x}$) this does not quite work. To resolve the divergent zero-mode
integral, we must replace $\xi_0$ with a collective coordinate $\tau_0$~\cite{Gervais:1974dc,Gervais:1975pa,Callan:1975yy,Jevicki:1976kd,Coleman:1978ae} (see also \cite{Kleinert, Marino, Zinn-Justin}). This means that instead of Eq.~\eqref{param1}, we 
parametrize our paths as:
\begin{equation}
\label{param2}
x^{\tau_0, \zeta_1,\dots}(\tau) = \bar x(\tau-\tau_0)+\sum_{i=1}^\infty\zeta_i x_i(\tau-\tau_0)
\end{equation}

In the potential-deformation method, the integral over $\tau_0$ gives a factor of $\bigtau$ (due to the exact translation symmetry) that resolves the $\int d\xi_0$ singularity, along with a Jacobian
factor from going between $\xi_0$ and $\tau_0$; one then divides by $\bigtau$ to find the rate. 
In the direct method, we can remove the collective coordinate with the $\delta$-function. With the parametrization in Eq.~\eqref{param2}, the functional $\tau_b$ is
\begin{equation}
\tau_b\left[x^{\tau_0, \zeta_1,\dots}\right]=\tau_0+\tau_b\left[x^{0, \zeta_1,\dots}\right]
\end{equation}
So depending on the $\{\zeta_i\}$, one of two things happens: 
\begin{enumerate}
\item If the path $x^{0,\zeta_1,\dots}$ hits $b$ at some time, then $\delta(\tau_b[x])$  simply removes the $\tau_0$ integral and fixes $\tau_0$ to some value $\tau_*(\zeta_i)$
\item If the path $x^{0,\zeta_1,\dots}$ never hits $b$, then the $\delta$-function is always 0, and this point in $\zeta$-space does not contribute at all.
\end{enumerate}
So we get:
\begin{align}
\Gamma^{\text{NLO}} &=\frac{e^{-S_E[\bar{x}]}}{e^{-S_E[x_{FV}]}} \lim_{\bigtau\to\infty}\abs{ \frac{2\text{Im}\int d^n\zeta~\Theta[\zeta_ix_i(0)] J[\tau_*(\zeta), \zeta]e^{-\frac{1}{2}\sum\lambda_i\zeta_i^2}}{\int \cD \delta x~e^{- \frac{1}{2}S_E''[x_{FV}] \delta x^2 }}}
\end{align}
where the $d^n\zeta $ indicates infinitely many integrals. We will now explain the addition of the theta function and the Jacobian factor, and we will see how an appealing feature of this method is the explanation of the factor of $\frac{1}{2}$. 
 
Since the path $\bar x$ just barely hits $b$ at its maximum, the constraint that $x=\bar x +\delta x$ must hit $b$ forces $\delta x(0)\geq0$.
Since $\delta x = \sum_i\zeta_i x_i$, we can enforce this positivity constraint with a step function $\Theta[\zeta_ix_i(0)]$. 
Now, since we are working at Gaussian order only and this is a constraint on a simple linear combination of the $\zeta$, we can use symmetry of the other terms under $\zeta \to -\zeta$ to drop the step function and divide by 2.
This factor of 2, which arises in the Euclidean approach from a subtle analytic continuation argument (cf. Section~\ref{sec:saddle}), arises naturally in the direct method from the requirement that the $\delta$-function fire. More physically, it is the requirement that the path enter the destination region $\DV$, which excludes exactly half the variations around $\bar x$.

Finally we must discuss the Jacobian $J(\tau_0, \zeta)$ arising when one goes from the orthonormal basis of fluctuations in Eq.~\eqref{param1} to the collective coordinate
parametrization in Eq.~\eqref{param2}. $J$ is non-singular after fixing $\tau_0$, and it has some expansion in $\zeta$. At NLO, we only need to keep the constant, $\zeta$-independent piece. So we can replace
\begin{equation}
J(\tau_*(\zeta), \zeta) \to J(\tau_*(0), 0) = J(0,0)
\end{equation}
This Jacobian at leading order is well-known~\cite{Callan:1977pt, Kleinert, Marino, Zinn-Justin} and discussed further in Appendix~\ref{app:jacobian}\footnote{In the existing literature (e.g. \cite{Kleinert}), authors often calculate $J(\tau_0=0)$, which is all that we need for our derivation. However, for \emph{their} derivations using the potential-deformation method, they need the stronger derivation of $J(\tau_0)$ for general $\tau_0$. For this reason in appendix~\ref{app:jacobian}, we prove that $J$ is a constant function of $\tau_0$, even though in our case we could simply ignore the $\tau$ dependence.}:
\begin{equation}
J(0,0) = \sqrt{S_E(\bar x)/m}
\end{equation}

Putting together the Jacobian factor and the factor of $\frac{1}{2}$, we get
\begin{equation}\label{eqn:PINLOconfusingunits}
\Gamma^{\text{NLO}} = \frac{e^{-S_E[\bar{x}]}}{e^{-S_E[x_{FV}]}} \sqrt{S_E[\bar x]/m}\abs{ \frac{1}{\sqrt{\pi}}\text{Im}\left(\frac{\det'\frac{1}{2}S_E''[\bar x]}{\det \frac{1}{2}S_E''[x_{FV}]}\right)^{-1/2}}
\end{equation}
where $\det'$ indicates the determinant omitting the 0-eigenvalue and the boundary conditions of the determinants' domains are $x(\pm\infty)=a$. The $\pi$ comes because the denominator path integral has one more Gaussian integral than the numerator. 

While the dimensions of Eq.\eqref{eqn:PINLOconfusingunits} are the correct dimensions of rate, they have become obscured by the combination of the $\sqrt{S_E/m}$ and the determinants. To make the units clearer, let us pull out $m/2$ from the determinant, using $\frac{\det'A}{\det A}\sim\frac{1}{A}$:
\begin{equation}
\Gamma^{\text{NLO}} = \frac{e^{-S_E[\bar{x}]}}{e^{-S_E[x_{FV}]}} \sqrt{S_E[\bar x]/2\pi}\abs{ \frac{\det'S_E''[\bar x]/m}{\det S_E''[x_{FV}]/m}}^{-1/2}
\end{equation}
Expanding $S''$ then gives
\begin{equation}\label{eqn:PINLO}
\boxed{
\Gamma^{\text{NLO}} = \frac{e^{-S_E[\bar{x}]}}{e^{-S_E[x_{FV}]}}\sqrt{\frac{S_E[\bar x]}{2\pi}}\abs{ \frac{\det'\left(-\partial_t^2+\frac{V''(\bar x(t))}{m}\right)}{\det \left(-\partial_t^2+\frac{V''(a)}{m}\right)}}^{-1/2}
}
\end{equation}
This agrees exactly with the formula surmised from the potential-deformation method~\cite{Marino,Kleinert}.

\subsection{The Direct Method in $d >1$}
In more than 1 dimension, the main change is that we must 
 extend the turning point $b$ to a surface $\Sigma$ of possible turning points, since paths can enter the destination region from any direction. The critical  Eq.~\eqref{Dsplit} becomes in multiple dimensions:
\be
\label{DsplitMdim}
D_F(a,0; x_f,t) = \int_\Sigma db \int_0^{t} d t_0  \Du(a,0; b, t_0) D_F(b, 0; x_f,t-t_0)  
\ee
for $\Sigma$ any codimension-1 surface which all paths go through. The only subtlety is that, to avoid overcounting of paths that enter and leave, the functional $t_b[x]$ in $\Du$ only returns the first time $x(t)$ hits $b$ if that is the first time the path crosses $\Sigma$ at all (and returns $\infty$ otherwise).

From there the steps go through the same as the one-dimensional case. Eqs.~\eqref{eqn:PIexpandedconjugate} through \eqref{eqn:threepropagators} will contain two integrals $\int_\Sigma db\int_\Sigma db'$.  Eq.\eqref{Gpath} will thus include an integral $\int_\Sigma db$, which stays through the end. Thus we see:
\be
\Gamma_\R =\abs{ \frac{\cN\cN^\star}{P_{FV}(\infty)} 2\text{Im}\int_\Sigma db\int_{x(-\infty)=a}^{x(\infty)=a} \Dxspace e^{- S_E[x] }\delta(\tau_b[x])}
\ee
where $\int_\Sigma db \delta(t_b[x])=\delta(t_\Sigma[x])$, where $t_\Sigma$ is the operator which returns the first time $\vec x(t)$ crosses $\Sigma$. Thus:
\be
\label{GsplitMdim}
\Gamma_\R =\abs{ \frac{2\text{Im}\int \Dx e^{- S_E[x] }\delta(\tau_\Sigma[x])}{\int \Dx e^{- S_E[x] }}}
\ee
where now $\Sigma$ is the entire surface which bounds $\R$, just like $b$ was the turning point at the boundary of $\R$. Both path integrals go from $x(-\infty)=a$ to $x(\infty)=a$.

\subsection{Comparisons of the potential-deformation and direct approaches \label{sec:compare}}
While similar in many ways, the derivation in Sections~\ref{sec:potmethod} and \ref{sec:dirmethod} have a few key differences.
\begin{enumerate}
\item The potential-deformation method starts from $\frac{\Gamma}{2}=|\text{Im}E_0|$.
While the decay rate is the imaginary part of an energy, as explained in Section~\ref{sec:complex}, it is certainly not the imaginary part of the ground state energy $E_0$.
There is an implicit assumption that deforming the potential 
(or, more honestly, integrating over complex paths intersecting the $\FV$ saddle) somehow isolates the energy of interest. 
The direct method instead starts from a physical definition, in  Eq.~\eqref{eqn:defineGamma} and there is no leap of faith required. 

\item In both approaches, the path integral only has an imaginary part after an analytic continuation/contour deformation. In the direct approach, this is the usual $\bigtau\to iT$ Wick rotation, which came naturally in the derivation. In the Euclidean approach, this contour deformation had to be put in by hand; instead of just taking the imaginary part of $E_0$, we had to deform the contour of integration in the path integral, in order to get a non-zero imaginary part.
\item Calculationally, the formulas remove the divergent Gaussian integral of the time-translation zero mode in the path integral differently. The deformation approach schematically
generates $\Gamma \bigtau = \int \Dx e^{-S_E}$. The divergence at large $\bigtau$ is removed by dividing by $\bigtau$. In the direct approach, $\Gamma = \int \Dx e^{-S_E}\delta(\tau_b)$ and the would-be divergent integral
is removed by the $\delta$-function.
\item  In both methods, the NLO rate is given by a path integral around a bounce configuration divided by a path integral around the static  $\FV$ solution. 
In the direct method, the $\FV$ solution does not contribute to the path integral in the numerator because it never fires the $\delta(\tau_b)$.
In the deformation approach, to prevent the $\FV$ dominating one needs to calculate the discontinuity between the two degenerate steepest descent contours.
\item In the deformation formula, the factor of $\frac{1}{2}$ in the NLO approximation 
comes because an integral along the $\FV$ contour has half of the imaginary part of an integral along the bounce contour. 
In the direct formula, it comes from the fact that only half of all small variations of the bounce solution enter the destination region $\R$. 
\item In the direct approach, one never has to worry about summing over approximate instantons or the validity of the dilute gas approximation. One simply systematically calculates the expansion of a path integral in $\hbar$. 
\end{enumerate}
\section{Tunneling in Quantum Field Theory\label{sec:QFT}}
Quantum field theory is just quantum mechanics, where the Hilbert space happens to be an infinite dimensional Fock space. Thus all the fundamental facts of quantum mechanics apply, including all the facts about tunneling. Understanding tunneling in QFT is exactly the same as understanding tunneling in QM. However, some confusing language can make it more obscure than it needs to be.

Let us begin with theories with a single scalar field $\phi$ with a (classical) action of the form
\be
S[\phi] =\int d^4 x\left[ \frac{1}{2}(\partial_\mu \phi)^2 - V(\phi)\right]= \int d^3 x dt\left[ \frac{1}{2} \dot\phi^2-\frac{1}{2} (\vec\nabla \phi)^2 - V(\phi)\right]
\ee
where $\dot\phi \equiv \partial_t \phi$. We assume $V(\phi)$ has a false vacuum at $\phi=\phi_F$ where the potential is $V(\phi_F)=V_F$. It is particularly convenient to shift the overall potential so that $V_F=0$
since this is the unique value for which the total potential energy of the false vacuum $\int d^3 x V_F$ is finite. 
Thus, when convenient we will follow the usual convention and assume $\phi_F=0$ and $V_F=0$.

When going from finite-dimensional QM to infinite-dimensional QFT, it is important to keep in mind that the notation and language changes as well:
\begin{enumerate}
\item The set of classical configurations is not an n-dimensional set of values $\vec x$, but the infinite-dimensional space of all field configurations, $\phi(x)$.
\item A quantum state, such as the true ground state, is not specified by a wavefunction, $\psi(x)$, but a wave-functional $\Psi[\phi]$. 
\end{enumerate}
In particular, in 1D quantum mechanics, we argued in Section~\ref{sec:complex} that the decay rate is independent of the initial wavefunction, as long as it has some support of the lowest resonance in the false-vacuum region.
Thus, for the path-integral derivation, we took it to be $\psi(x) = \delta(x-a)$. We could have equally well have taken it to be the ground state of a harmonic oscillator in the quadratic approximation to the false-vacuum well:
$\psi(x) \propto \exp(-\frac{1}{2} m\omegaa x^2)$ where $\omegaa = V''(a)$. Or we could have included admixtures of higher excited states or continuum modes if the domain of $x$ is unbounded. 

In quantum field theory, the ground state in the false vacuum is specified by a functional
\be
\Psi[\phi] = \mathcal{N}\exp\left\{-\frac{1}{2}\int \frac{d^3 k}{(2\pi)^3}\omega_k \tilde{\phi}(k)\tilde{\phi}(-k)\right\}
\ee 
where $\omega_k=\sqrt{k^2+m^2}$ and $\tilde{\phi}(k)= \int d^3 x e^{ikx}(\phi(x)-\phi_F)$ are the Fourier modes of $\phi(x)$. States with a finite number of particles are finite energy excitations on top of this vacuum. The wave-functional corresponding to the
first excited state of the harmonic oscillator has an infinite amount of energy (proportional to volume) more than the ground state. 
 For the purpose of the tunneling rate calculation, we can, analogously to QM, take $\Psi[\phi] = \delta[\phi-\phi_F]$. This is what we mean when we say the state is initially localized at $\phi_F$. 

Another important difference is:
\begin{enumerate}
\item[3.] Despite the fact that the Lagrangian has an object called $V$, the energy of a static field configuration $\phi(\vec x)$ is not given just by $V[\phi]$ but also includes
gradient energy contribution. Thus we define the potential-energy functional as
\be
U[\phi(x)] \equiv \int d^3 x\left[\frac{1}{2} (\vec\nabla \phi)^2 + V(\phi)\right] \label{Udef}
\ee
\end{enumerate}
The fact that the intuitive ``potential" object is not the full potential energy through which the tunneling occurs leads to some effects that seem counter-intuitive at first. For example, suppose $V(\phi)=V_F$ is constant,
or has a slight downward slope $V(\phi)=V_F - \varepsilon \phi$. Then one might imagine since there is no barrier in $V$, tunneling would occur infinitely fast, dominating over the classical rolling. However, there
is a potential barrier in $U[\phi]$ due to the gradient energy term, so the tunneling rate is finite even as $\varepsilon \to 0$. See~\cite{Lee:1985uv} for more details. We discuss below the case 
$V(\phi) = - \lambda \phi^4$ where one also might worry that tunneling would be infinitely fast.

In quantum field theory, the decay rate can be calculated to all orders using Eq.~\eqref{GsplitMdim} with an infinite number of degrees of freedom:
\be
\boxed{
\Gamma_\R =\lim_{\tau\to\infty}\abs{ \frac{2\text{Im}\int \cD \phi~e^{- S_E[\phi] }\delta(\tau_\Sigma[\phi])}{\int \cD \phi~e^{- S_E[\phi] }}}
\label{GammaRQFT}
}
\ee
Now $\Sigma$ is a codimension-1 surface in the enormous configuration space of possible classical field configurations $\phi(\vec x)$ which bounds the destination region $\R$ (also a region in field configuration space).
The surface $\Sigma$ is naturally taken to be the set of field configurations for which  $U[\phi]=U[\phi_{\FV}]$. Indeed, because semi-classical methods conserve energy, the endpoint of quantum tunneling must be on this surface.

Some interesting examples of the competition between tunneling rates to different regions were considered by Brown and Dahlen~\cite{Brown:2011um}. 
In particular, these authors explored examples where the rate for one of the  tunneling processes abruptly goes from a finite value to zero after a small
change in the potential. They also point out that in quantum field theory, the surface $\Sigma$ on which $U[\phi]=U[\phi_{\FV}]$ is connected: one can interpolate between one region $\R$ and another with
non-spherically symmetric configurations comprising subcritical and supercritical bubbles. Nevertheless, after tunneling occurs, subcritical bubbles will implode. 
In field theory, there is a well-defined set of quasistable local minima to which fields will evolve classically after tunneling occurs.
Possible regions $\R$ are therefore naturally taken to be bounded by the separate parts of $\Sigma$ from which fields will evolve separately to each minima.

In section \ref{sec:energy} we will look at how we can visualize the decay in terms of the potential-energy functional along the particular field directions.

\subsection{Bounces in QFT}

If we want to work in the saddle point approximation, then we need to find the stationary configurations of the Euclidean action:
\be
S_E[\phi] =\int d^4 x\left[ \frac{1}{2}(\partial_\mu \phi)^2 + V(\phi)\right]
=\int d^3 x dt \frac{1}{2} \dot\phi^2+\int dt U[\phi]
\ee
The dominant bounce will always be $O(4)$ symmetric~\cite{Coleman:1977th}, that is $\phi_B(x) = \phi_B(\rho)$,
where $\rho=\sqrt{x^2 +y^2+z^2 + \tau^2}$. 
For these solutions, the Euclidean equations of motion reduce to
\be
\partial_\rho^2 \phi_B + \frac{3}{\rho} \partial_\rho \phi_B- V'[\phi_B] =0 \label{bouncede}
\ee
This equation has the analog mechanics interpretation of a particle rolling down a potential $-V(\phi)$ with $\rho$ as a time coordinate and the $\frac{3}{\rho}\dot{\phi}$ representing a kind of time-dependent friction.
Matching on to the false vacuum with $\phi=\phi_F$ at $\tau=\pm\infty$ translates to the boundary condition $\phi(\rho=\infty)=\phi_F=0$. For the solution to be smooth at $\rho=0$, we must also have $\partial_\rho\phi|_{\rho=0}=0$. With these boundary conditions, the analog classical system
has the particle starting at a point $\phi_0$ at rest (at $\rho=0$) and ending at rest at $\phi=\phi_F=0$ at $\rho=\infty$. 

One should think of $\phi_B(\vec x,\tau)$ as providing a path through field space $\phi(\vec x)$ parametrized by $\tau$. This path goes from the false vacuum at $\tau=-\infty$ to the bubble at $\tau=0$ and then
back to the false vacuum at $\tau=\infty$.
This path is much like a path through configuration space $\vec x$ in multidimensional quantum mechanics parametrized by some path length $s$. 
The bubble $\phi_B(r) = \phi_B(\vec x, \tau=0)$ is the analog of the turning point $x=b$. 

As the potential 
is time-independent, energy is conserved along the path in time (or Euclidean time). Thus the potential energy of the bubble is the same as that of the false vacuum $U[\phi_B(r)]=U[\phi_F]=0$. Since
total energy is conserved in $\tau$, we also have
\be
U[\phi_B ]  = \int d^3x \frac{1}{2}( \partial_\tau \phi_B)^2 \label{Ufb}
\ee
And the Euclidean action on the bounce is
\be
S_E[\phi_B] = \int d^4 x\left[ \frac{1}{2}(\partial_\tau \phi_B)^2+\frac{1}{2} (\vec \nabla \phi_B)^2 + V[\phi_B]\right]= \int_{-\infty}^{\infty} d\tau\, 2U[\phi_B]
\label{SEeq}
\ee

There are a few known exact solutions to Eq.~\eqref{bouncede}. An important case is when $V = -\lambda \phi^4$ and the solution has the form
\be
\phi_C(\rho) = \sqrt{\frac{2}{\lambda}} \frac{R}{R^2+\rho^2} \label{confbubb}
\ee
for any $R$ and $\lambda>0$. We call these solutions ``quartic bounces''. 
There are a handful of other known exact solutions~\cite{FerrazdeCamargo:1982sk,Lee:1985uv,Dutta:2011rc,Aravind:2014pva}.
To explore the relation between the bounce shape and the potential, it is sometimes easier to construct exact potentials given a bounce (by integrating Eq.~\eqref{bouncede}) than to construct bounces given potentials.
Alternatively, one can build up exact solutions perturbatively if there is some small parameter in the potential.

More generally, it is straightforward to solve Eq.~\eqref{bouncede} numerically with the shooting method. The idea is that it is easy to solve differential equations with initial conditions, but not with
our boundary conditions $\phi'(0) = 0$ and $\phi(\infty) = \phi_F$. So one solves with  $\phi'(0) = 0$ and $\phi(0) = \phi_0$ and varies $\phi_0$ until the solution satisfies $\phi(\infty) = \phi_F$. If $\phi_0$ is too large, the particle will overshoot $\phi=\phi_\FV$ at
large $\rho$ and if $\phi_0$ is too small, it will undershoot. Thus one iteratively hones in on the right $\phi_0$. One hurdle to rapid convergences with this method is that due to the $\frac{3}{\rho}$ term,
one cannot actually start rolling from $\rho=0$ without a numerical singularity. However, this hurdle is easily surmounted using a trick described in Appendix~\ref{app:bounce}.

\subsection{Visualizing $U[\phi]$ \label{sec:energy}}
As mentioned after Eq.~\eqref{Udef}, the gradient energy present in $U[\phi]$ but not in $V(\phi)$ can mislead our intuition. Unfortunately, because $U[\phi]$ is a functional rather than a function,
we cannot simply plot it. More generally, QFT field configuration space, in all its infinite dimensional glory, can be challenging to visualize. One approach which we now explore
is to consider $U[\phi]$ restricted to a one or two parameter slice of field space.
We can plot $U(\alpha)\equiv U[\phi_\alpha]$ along the parameter(s) $\alpha$.
In such a plot all of the intuition from quantum mechanics applies, including intuition for the tunneling rate coming from the  WKB approximation (cf. Section~\ref{sec:WKBmultidim}).

One natural choice for $\alpha$ is the Euclidean time $\tau$ parametrizing the path through field space of the bounce. 
For any bounce $\phi_B(\rho)$, the $\tau$ path is given by:
\be
\phi_\tau(\vec x) = \phi_B\left(\sqrt{r^2+\tau^2}\right)
\ee
Now we are using $\tau$ as simply a useful coordinate on the slice of field space of interest. 
For example with a potential $V(\phi)=-\lambda \phi^4$, the bounce is given by Eq.~\eqref{confbubb}, so that
\be
\label{eqn:tauparam}
U(\tau) = U[\phi_C] = \frac{\pi^2 R^2 \tau^2}{2\lambda (R^2+\tau^2)^{5/2}}
\ee

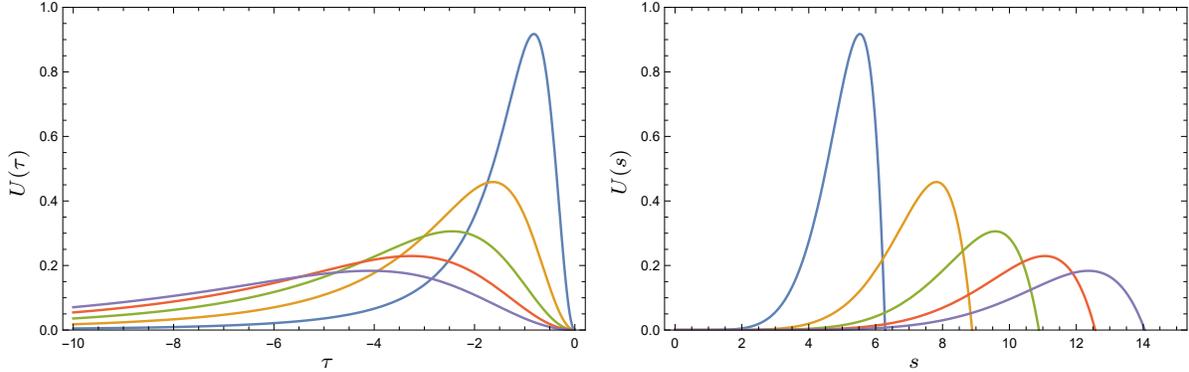
\begin{figure}[t]
\begin{center}
\begin{tikzpicture}
\node at (-3,0) {\includegraphics[width=0.44\columnwidth]{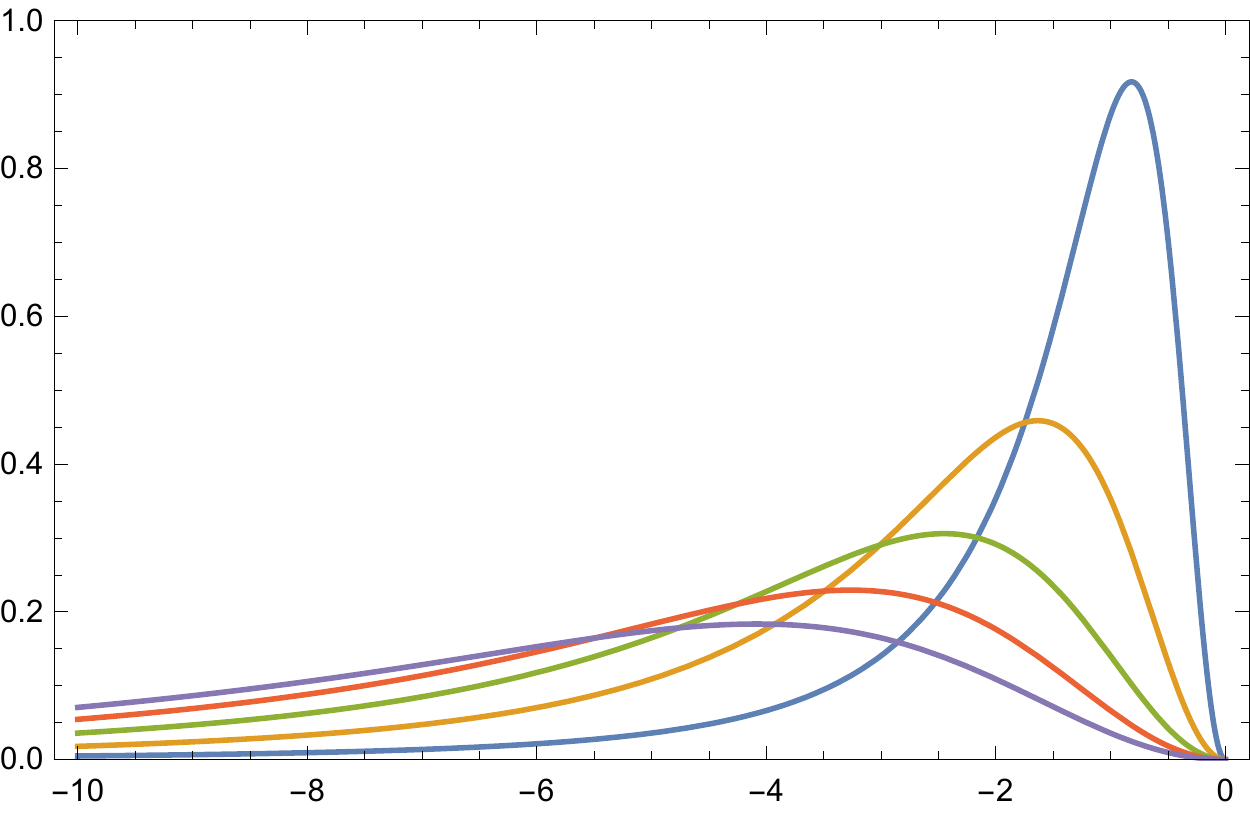}};
\node at (5,0) {\includegraphics[width=0.44\columnwidth]{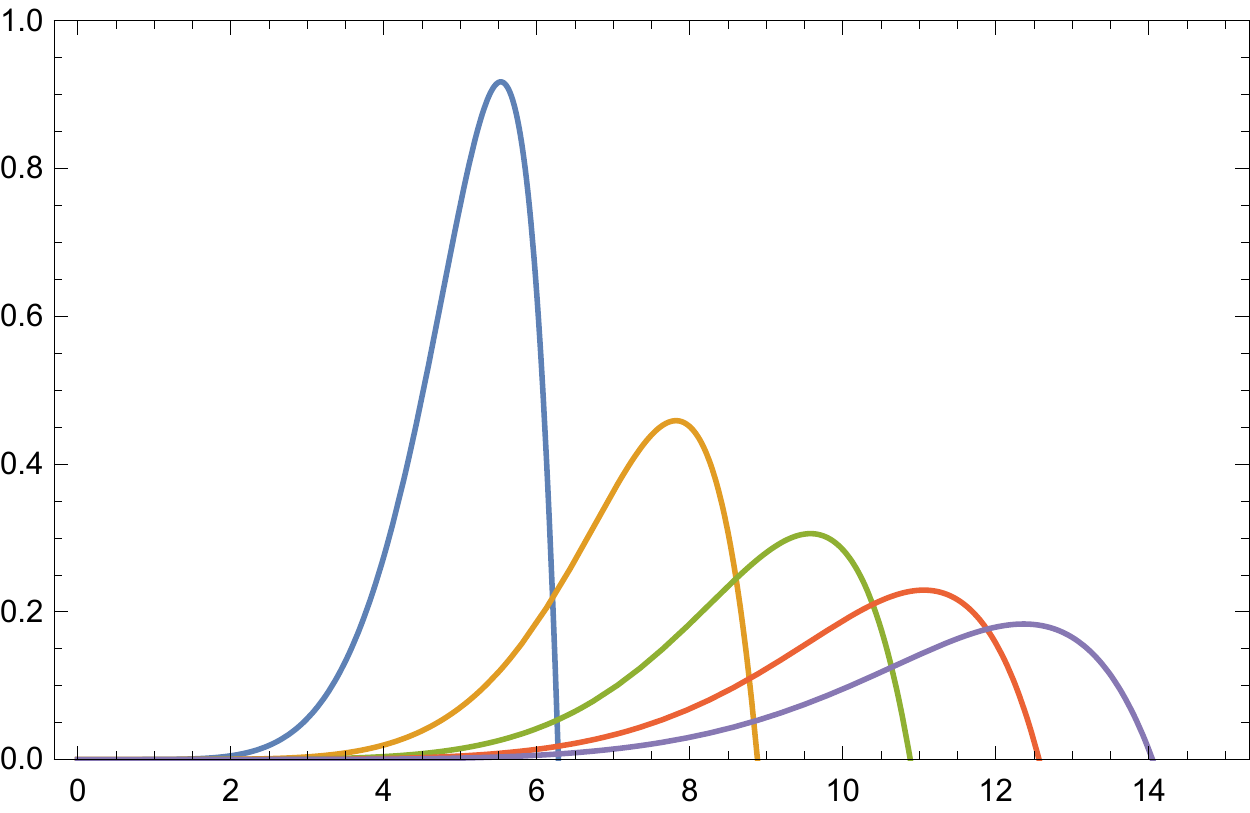}};
\node[left,rotate=90] at (-6.9,0.5) {\scriptsize $U(\tau)$};
\node[left,rotate=90] at (1.1,0.5) {\scriptsize $U(s)$};
\node at (-2.8,-2.5) {\scriptsize $\tau$};
\node at (5,-2.5) {\scriptsize $s$};
\end{tikzpicture}
\end{center}
\caption{The proper analog of the potential in QM is the effective energy functional $U$. This examples uses $V(\phi) = -\lambda \phi^4$ with $\lambda=1$. Left shows $U(\tau)$, and the integral of $4U(\tau)$ from $-\infty<\tau<0$ gives $S_E$. Right shows $U(s)$, and the integral of $\sqrt{2U(s)}$ over $0<s<\sqrt{\frac{4R}{\lambda}} \pi$ gives $S_E$. 
One can see the degeneracy in the fact that each of these barriers has the same integral.
Remember that these curves show the height of the barrier in five \emph{different} directions in field configuration space, which happen to all have the same tunneling rate at LO.
Curves from steep to shallow have $R=1,2,3,4$ and $5$. }
\label{fig:conformalUtau}
\end{figure}

This function is shown for a selection of $R$ in Fig.~\ref{fig:conformalUtau}, on the left. We see that while $V(\phi)$ has no barrier, $U[\phi]$ does. Also, 
from Eq.~\eqref{SEeq} we see that the area under the curves gives the Euclidean action for that bounce. In this case, the areas are all equal as the potential is classically scale invariant
and $S_E = \frac{2\pi^2}{3\lambda}$ is independent of $R$. 

Parametrization of paths with $\tau$ is fine, but to really think of $U(\tau)$ as a potential, the distance along the path should be
properly normalized. Recall that in going from quantum mechanics to quantum field theory, position is replaced by field values $\phi(\vec x)$.
Thus, instead of $ds^2 = dx^2 +dy^2+\ldots$, the measure along a path in field space is given by the somewhat odd-looking line element $ds^2 = \int d^3x [d \phi(x)]^2$. 
To change from $\tau$ to $s$, we use
\be
\left(\frac{ds}{d\tau}\right)^2 = \int d^3x \left(\frac{\partial \phi_B}{\partial \tau}\right)^2 = 2 U[\phi_B] \label{softau}
\ee
Thus Eq. \eqref{SEeq} becomes
\be
S_E = \int_{-\infty}^{\infty} d\tau\, 2U[\phi_B]=2\int_{-\infty}^0 d\tau\,2 U[\phi_B]=  2 \int d s \sqrt{2 U[\phi_B]} \label{twoforms}
\ee
The right-hand side of this equation is the exact field theory analog of the exponent in the WKB formula in Eq.~\eqref{eqn:WKBparametrizedpath}. 
The right side of Fig.~\ref{fig:conformalUtau} shows $U$ as a function of $s$. It is the integral of the square-root of these curves which gives the tunneling
rate, as in the finite-dimensional WKB formula.

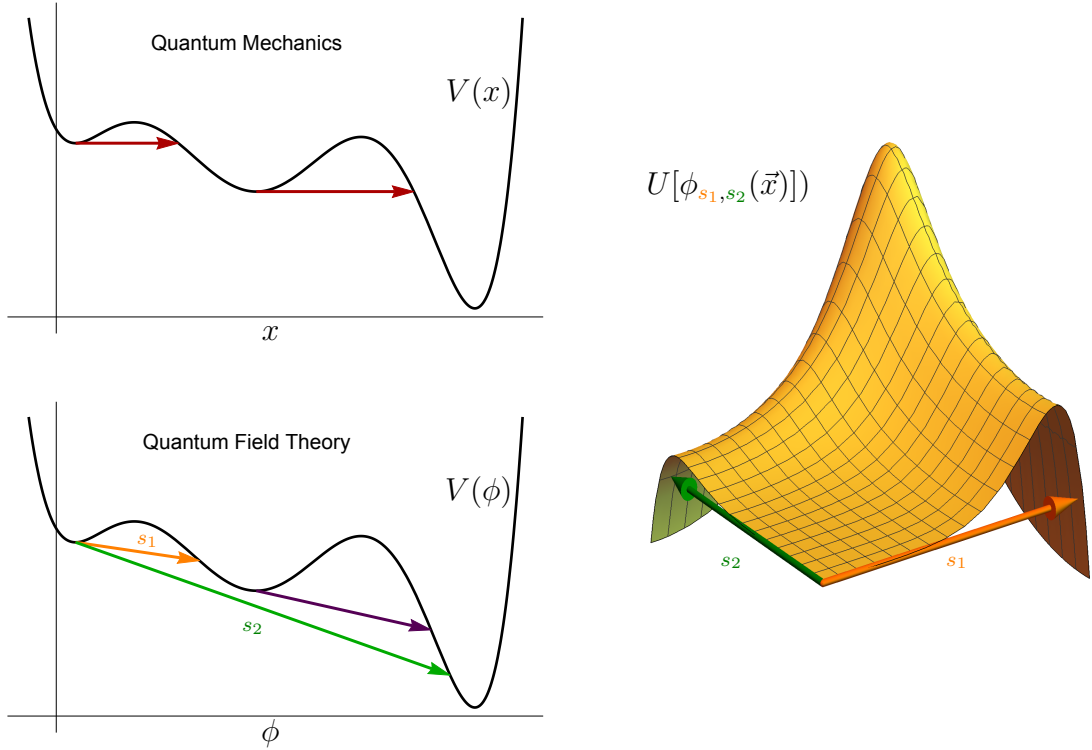
\begin{figure}[t]
\begin{center}
\begin{minipage}{0.48\columnwidth}
\begin{tikzpicture}
\node at (0,0) {\includegraphics[width=0.9\columnwidth]{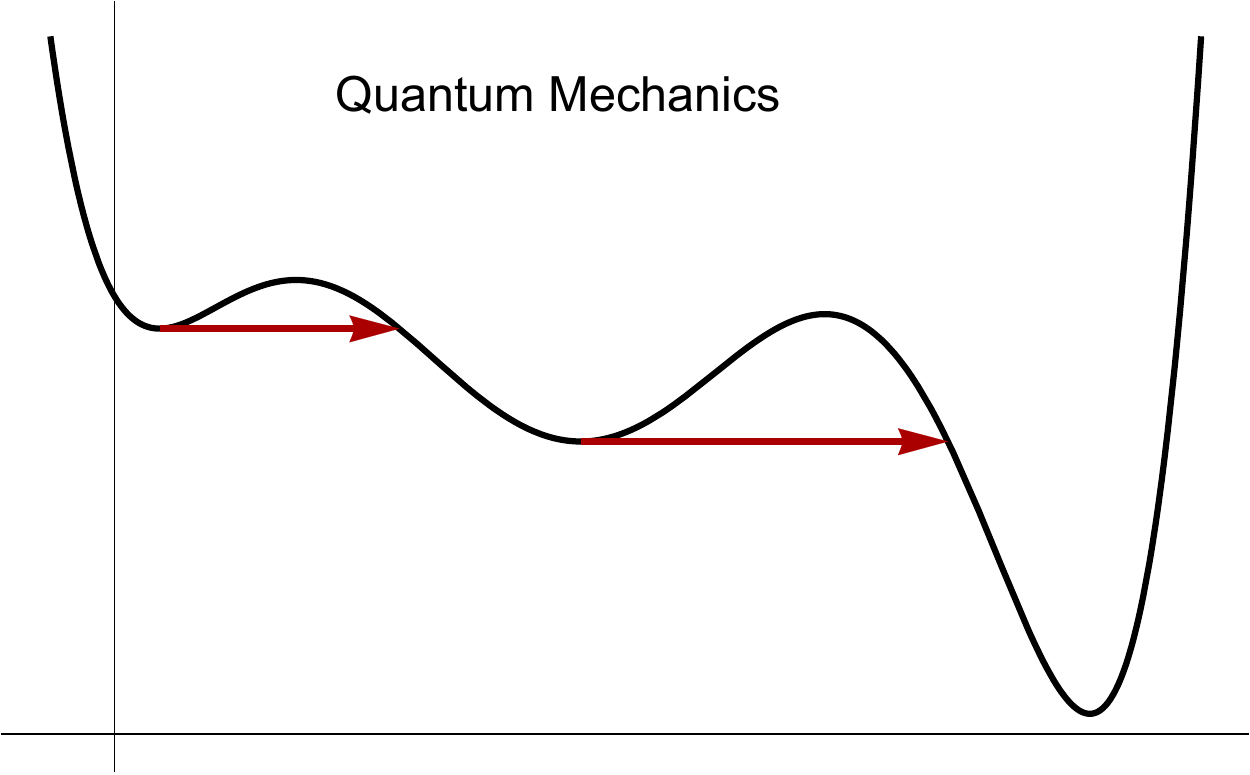}};
\node[left] at (0.2,-2.2) {$x$};
\node[left] at (3.3,1) {$V(x)$};
\end{tikzpicture} 
\\[5mm]
\begin{tikzpicture}
\node[left] at (0.2,-2.2) {$\phi$};
\node[left] at (3.3,1) {$V(\phi)$};
\node at (-1.7,0.35) {\scriptsize ${\orange s_1}$};
\node at (-0.3,-0.8) {\scriptsize ${\green s_2}$};
\node at (0,0)  {\includegraphics[width=0.9\columnwidth]{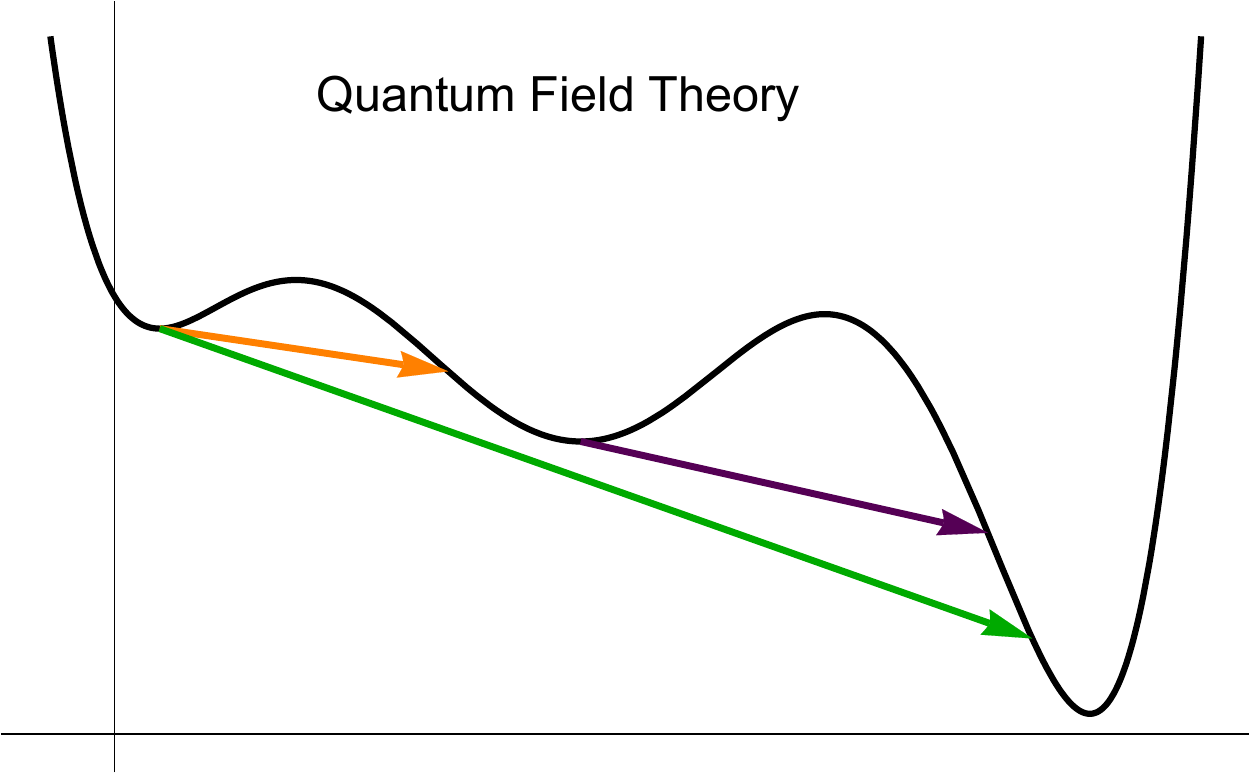}};
\end{tikzpicture} 
\end{minipage}
\begin{minipage}{0.48\columnwidth}
{
\begin{tikzpicture}
\node at (0,0)  {\includegraphics[width=0.9\columnwidth,trim={4cm 5cm 3cm 0},clip]{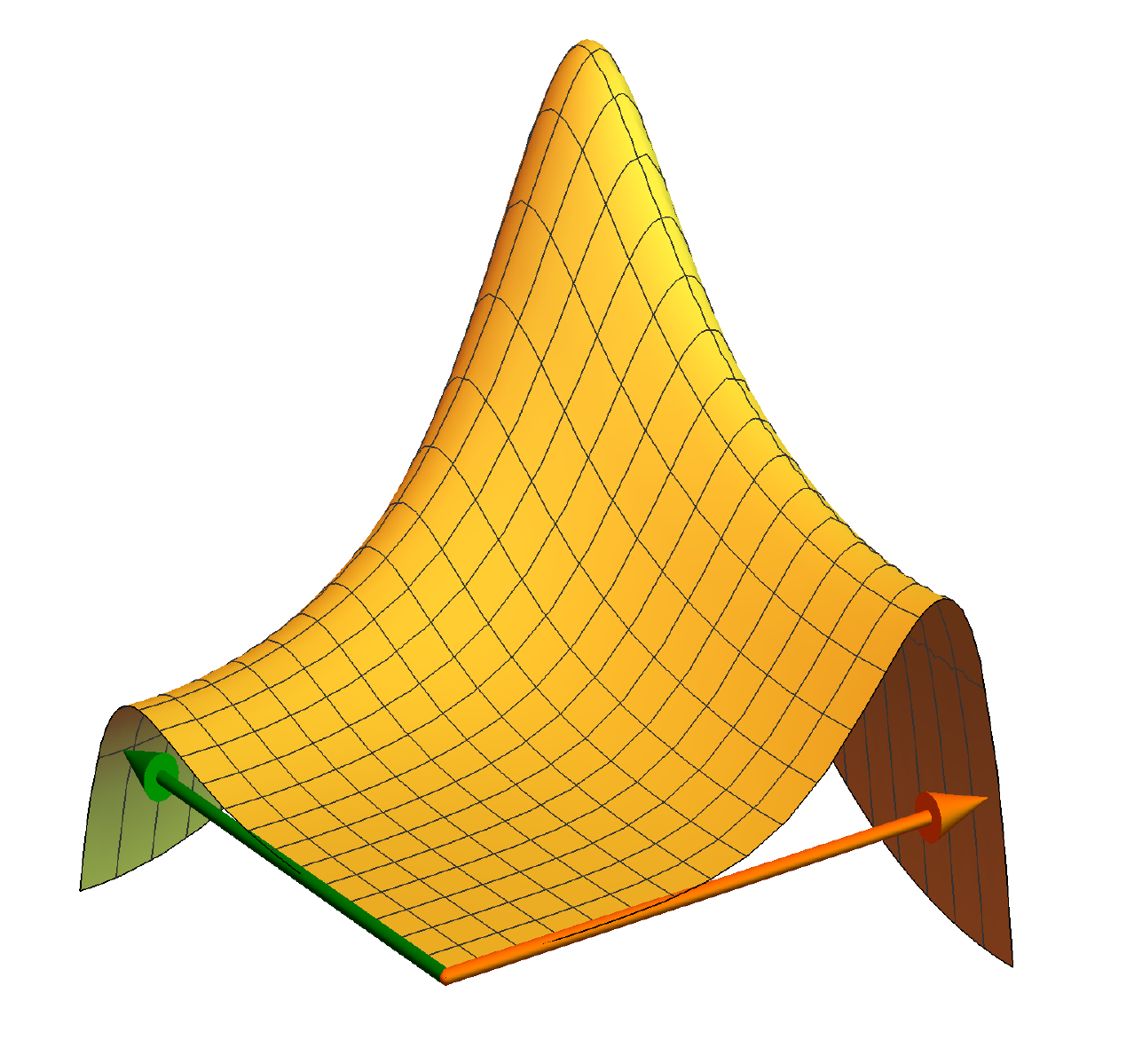}};
\node at (1,-2.5) {\scriptsize ${\orange s_1}$};
\node at (-2,-2.5) {\scriptsize ${\green s_2}$};
\node[left] at (-0.8,2.5) {$U[\phi_{{\orange s_1},{\green s_2}}(\vec{x}) ])$};
\end{tikzpicture} 
}
\end{minipage}
\end{center}
\caption{
In quantum mechanics (top left), the tunneling rate is determined by the nearest potential barrier to the closest turning point of $V(x)$.
In quantum field theory, tunneling can proceed in different directions of field space. For a potential $V(\phi)$ of the same shape as $V(x)$,
tunneling may proceed through the path labelled ${\orange s_1}$ corresponding to a family of fields $\phi_{\orange s_1}(\vec{x})$ or
through the path labelled ${\green s_2}$ corresponding to a family of fields $\phi_{\green s_2}(\vec{x})$. The energy functional on a 2-parameter family $\phi_{{\orange s_1},{\green s_2}}(\vec{x})$ is shown on the right.}
\label{fig:potentialtunneling}
\end{figure}

More generally, for any family $\phi_\alpha$ of field configurations parametrized by some coordinate $\alpha$, we can find the appropriate normalization using

\be
ds^2 = \left[\int d^3x \left(\frac{d\phi_\alpha}{d\alpha}\right)^2\right]d\alpha^2
\ee

Once normalized, the decay rate calculated from the minimal action along the subspace of field configurations will approximate the correct decay rate. It will be inexact for two reasons. First, in the fluctuations around the minimum-action path, we will entirely ignore all the modes of fluctuation that are not contained in our subspace. If all we care about is LO accuracy, then this is not a worry.
 Second, the minimum-action path through our subspace may be very different from the true minimum path, and so the dominant $e^{-S_E}$ in the subspace may exponentially overestimate the leading-order decay rate.

Thus to make our 1D slice \emph{useful}, it should contain the dominant path, as the parametrization by Euclidean time $\tau$ Eq. (\ref{eqn:tauparam}) automatically does (since $\tau$ is exactly the parameter along the dominant path).

Although we often restrict to 1D slices through field space, it is important to keep in mind that QFT has an infinite variety of field configurations and potential barriers can be subverted through excursions to large field values. Indeed, one very important distinction between quantum mechanics and quantum field theory is that in quantum field theory
the tunneling rate depends on the potential at arbitrarily large field values. In quantum mechanics, tunneling always proceeds through the closest barrier first, as illustrated in the top-left plot in Fig.~\ref{fig:potentialtunneling}. In QFT, 
there can be multiple competing tunneling directions. This is indicated
by the orange and green arrows in the lower-left plot of Fig.~\ref{fig:potentialtunneling}.  The endpoints of the arrows indicate the value of the field at the center of the bubble which forms. Thus, these two arrows correspond to 2 paths through
field space $\phi_{\orange s_1}(\vec{x})$ and $\phi_{\green s_2}(\vec{x})$, corresponding to two different types of bubbles forming. The barrier
(according to $U[\phi]$)  along the shorter direction (smaller change in $\phi(0)$) is not necessarily lower than the barrier along the longer  direction. In fact, tunneling rates to bubbles with large field values can be exponentially larger than tunneling rates to smaller field values. 
This is important for the Standard Model, as it makes the stability of our universe unavoidably sensitive to arbitrarily high-scale physics, as we discuss in Section~\ref{sec:highscale}. 

\subsection{Using approximate solutions}\label{sec:approx}
We can visualize the energy along a path through field space as a function of $\tau$ if we already know the bounce solution. But what if we do not already have a bounce in hand? 
Since $\phi_B$ is an extremum of the action, we can use a variational approach to get a handle on it by considering a family of approximate solutions.

For most potentials of interest, it is sufficient to consider a two parameter family of spherically-symmetric fields, with the two parameters representing the {\it width} of the bubble $R$ which (roughly) controls the gradient energy/surface-tension contribution and the {\it height}
of the bubble $\phi_0$ which we can define as the value at the center $\phi_0 = \phi(r=0)$. For example, a useful two-parameter family are
the ``Gaussian'' bubbles:
\be
\phi_G(\rho) = \phi_0 e^{-\rho^2/R^2} \label{gdef}
\ee
Other simple families are the quartic bubbles, $\phi = \phi_0\frac{1}{1+\rho^2/R^2}$ and thin wall bubbles for which $\phi'=0$ outside of a wall of thickness 
$t$.\footnote{This parametrization is more general that the thin-wall approximation originally used by Coleman~\cite{Coleman:1977py}. The thin-wall approximation requires the two minima to be nearly degenerate
so that the bounce stays the true vacuum long enough to neglect the damping term. For potentials which are not nearly degenerate, it has been shown that the thin-wall approximation is not in good quantitative
agreement with exact numerical results~\cite{Samuel:1991mz,Shen:1988si}.  We will not neglect the damping term. Using Gaussian bubbles seems to give results in good agreement with exact numerical solutions in
all the cases we have tried.
}

The Euclidean action for the Gaussian bubbles is
\begin{align}
S_E[\phi_G] &= 2\pi^2 \int d\rho \rho^3 \left[\frac{1}{2} (\partial_\rho \phi_G)^2 + V[\phi_G(\rho)] \right]\\
& = \frac{\pi^2}{2} R^2 \phi_0^2 + a \frac{\pi^2}{4} R^4 V(\phi_0) \label{SEV}
\end{align}
The second term has a potential-dependent number $a$ in it. For smooth potentials we expect $a\sim 1$. For example, if
$V=-\lambda \phi^4$, then $a=\frac{1}{4}$. 

To find the bounce, we must keep in mind that the bounce is not the true minimum of the action, but only a saddle point. Indeed, the negative-action fluctuation is precisely what is needed to give
the path integral the required imaginary part to compute the decay rate. 
We know a decay can occur if and only if $V(\phi_0)<0$ for some $\phi_0$, but then $S_E$ in Eq.~\eqref{SEV} is unbounded from below as $R\to\infty$.
The true minimum is the shot: a path which starts at the false vacuum with 
enough kinetic energy to make it to the true vacuum quickly, stay there for a long time, and eventually return. 
Recall that the fluctuation around the bounce with negative eigenvalue (the fluctuation towards the minima) has no nodes and therefore is nonzero at $\tau=0$. 
Thus, while the bounce hits the zero energy surface at $\tau=0$, the shot or any other spherically-symmetric lower-action configuration cannot. So
to find the bounce and not the shot, we can minimize the action over $\phi_0$ and $R$ restricting to $U[\phi_G(\tau=0)]=0$.

The energy of the Gaussian bubbles (at $\tau=0$) is
\begin{align}
U &= 4\pi\int r^2 dr \left[ \frac{1}{2} (\partial_r \phi_G(r))^2 + V(\phi_G(r))  \right] \\
&=  \frac{3 \pi^{3/2}}{4\sqrt{2}} R \phi_0^2
+ b \frac{\pi^{3/2}}{\sqrt{2}} R^3 V[\phi_0] \label{bdef}
\end{align}
The second term here has another potential-dependent order one number, $b$. For example, if
$V=-\lambda \phi^4$, $b=\frac{\sqrt{2}}{8}$. Restricting to $U=0$ means:
\be
R^2= -\frac{3}{4b}\frac{\phi_0^2}{V(\phi_0)} \label{Rofphi}
\ee
Note that $V(\phi_0)<0$, since there is a friction term that forces the field to start out slightly beyond the turning point where $V=0$, so $R^2>0$. 
Using this restriction on $R$, the Euclidean action becomes
\be
\label{eqn:gaussianactionbound}
S_E = \frac{3\pi^2}{64 b^2}\left(8b-3a\right)\frac{\phi_0^4}{-V(\phi_0)}
\ee
More suggestively, if we define the effective quartic as
\be
\lameff(\phi) \equiv \frac{V(\phi)}{\phi^4}
\label{eqn:defleff}
\ee
then the extremum is where
\be
\frac{d}{d\phi}\lameff=0 \label{dleff}
\ee
Note that this condition is independent of $a$ and $b$; it depends only on the Gaussian bubbles being a reasonable approximation to the true solution\footnote{The condition does require $8b>3a$, which is true for all the examples we have considered.
}.

The above manipulations provide us with a useful shortcut to deduce the approximate features of the bounce simply by looking at the potential: 
the field value at the center of the bounce $\phi_0$ is where $\lameff$ is flat; the size of the bubble is given by Eq.~\eqref{Rofphi}, and the action on the bounce
is no larger than Eq.~\eqref{eqn:gaussianactionbound}. For example, with $V=-\lambda \phi^4$, the exact Euclidean action on the true, quartic, bounce is
$S_E[\phi_C]=\frac{2\pi^2}{3\lambda} \approx \frac{6.5}{\lambda}$. The extremum along the paths of zero-energy Gaussian bubbles has $S_E[\phi_C] \approx \frac{9.8}{\lambda}$ (although it is something of a degenerate case because the action along the zero-energy Gaussian bubbles is exactly constant).

 This effective coupling is commonly used in the literature on the Standard Model effective potential. There it is often invoked in the context of a resummed
effective potential where the $\msbar$ scale $\mu$ is set equal to the Higgs background field value, and minimizing $\lameff$ is sometimes associated with
a renormalization group condition $\partial_\mu \lameff(\mu)=0$. Here we see that the relevance of $\lameff$ to the bounce solutions has nothing to do with renormalization:
it follows generically in classical potentials. 

Another fun exercise with these Gaussian bubbles is to determine how $R$ would depend on time in a classical field theory.
Consider a Gaussian bubble 
\be
\phi_G(r,t)=\phi_0 \exp\left[-\frac{r^2}{R(t)^2}\right]
\ee
To determine the time dependence, we can integrate over $d^3x$ in the Minkowski-space
action to get an effective 1-dimensional Lagrangian:
\begin{align}
L &= 4\pi \int dr r^2 \left[\frac{1}{2}(\partial_t\phi_G)^2- \frac{1}{2} (\partial_r \phi_G)^2 - V(\phi_G) \right]\\
&= \frac{15 \pi^{3/2}}{16\sqrt{2}} R \dot{R}^2 \phi_0^2 -\frac{3\pi^{3/2}}{4\sqrt{2}} R \phi_0^2 - b \frac{\pi^{3/2}}{\sqrt{2}} R^3 V(\phi_0)
\end{align}
where $b$ is the same order one constant as in Eq.~\eqref{bdef}. The last two terms in this expression for $L$ are just $-U$, from Eq.~\eqref{bdef}. 
Calculating the Euler-Lagrange equations from this Lagrangian gives:
\be
\ddot R = -\frac{\dot R^2}{2R} - \frac{8b V(\phi_0)R}{5 \phi_0^2} -\frac{2}{5R}
\ee
We immediately confirm that for bubbles to grow $V(\phi_0)$ must be negative (as expected). 
For bubbles at rest ($\dot R=0$) the condition for bubble growth ($\ddot R>0$) is that
\be
R^2 > -\frac{\phi_0^2}{4 b V(\phi_0)}
\ee
Comparing to Eq.~\eqref{Rofphi},
we see that these bubbles grow if $R^2 > \frac{1}{3} R_{E=0}^2$, where $R_{E=0}$ is the condition for the bubble
to have zero energy. In particular, Gaussian bubbles which form on the $U=0$ surface $\Sigma$ do indeed grow with time.  

\section{Tunneling at NLO and effective actions \label{sec:NLO}}
The previous sections have discussed tunneling in quantum mechanics and quantum field theory, both through an exact, non-perturbative, definition of the decay rate and the perturbative approximation to the rate coming
from expanding around saddle points. All of these calculations assumed that a potential $V(\phi)$ of a single scalar field $\phi$ was given. In quantum field theory, the physics of tunneling is often associated
with an {\it effective potential} $\Veff$, generated by integrating out quantum corrections.
 In some cases, the instability is even induced from radiative corrections. 
One might imagine that the effective potential can simply be used in place of the classical potential in the tunneling formulas. However this is incorrect; it will overcount the quantum corrections.

A classic example where the instability is a quantum effect is the Coleman-Weinberg potential \cite{Coleman:1973jx}. Coleman and Weinberg considered 
the theory of a complex scalar field and an Abelian gauge boson, i.e. massless scalar QED. In the Coleman-Weinberg model, the classical potential $V(\phi)=\lambda |\phi|^4$ is scale invariant and has  an absolutely stable minimum
at $\phi=0$ (for $\lambda>0$). In contrast, the effective potential has a minimum at $|\phi| >0 $, and thus $\phi=0$ is unstable. The scale for the new minimum is determined by dimensional transmutation from the renormalization group
scale of the couplings. What is the correct procedure to compute the tunneling rate in this model? One cannot use $V(\phi)$ in the tunneling rate formulas above, since there are no bounce solutions for this potential to expand around. 
Nor can one use $\Veff$, as the quantum fluctuations have already been integrated out, so there is no longer a path integral.

The same problem occurs in the Standard Model, where the classical potential for the Higgs field has an absolute minimum at the electroweak vacuum expectation value $\langle h \rangle = 246~\GeV$. 
The effective potential indicates tunneling to a very high scale. However, how can we calculate this tunneling rate accurately?

\subsection{Effective actions}
First, let us quickly review what is meant by the terms {\it effective potential} and {\it effective action}. Unfortunately, these same terms are used for a few different but related functions.

The simplest way to compute an effective action for a field $\phi$ is to integrate out (perform the path integral over) all the {\it other} fields in the theory. For example, with two fields $\phi$ and $\chi$,
one could integrate out $\chi$ to get an effective action for $\phi$:
\be\label{eqn:defineSAeff}
e^{-\Seff[\phi]} \equiv \int \cD \chi e^{-S[\phi,\chi]}
\ee
The path integral on the right is to be calculated for a fixed (but possibly position-dependent) background configuration $\phi$. In particular, no loops involving virtual $\phi$ particles are to be included 
in the calculation on the right-hand side. 
Although computing effective actions this way for general field configurations $\phi$ in any theory is essentially impossible, a momentum expansion is feasible. 
For example, the Euler-Heisenberg action is an effective action of this type, where the electron is integrated out in QED and the background $A^\mu$ is assumed constant.

Even if the classical action for $\phi$, $S[\phi,0]$ has no instability but the effective action computed this way does, one can proceed to calculate the tunneling rate using 
the effective action. There is no double counting
because the fluctuations of some fields ($\chi$ in Eq.~\eqref{eqn:defineSAeff}) are included in the calculation of the effective action while the fluctuations of other fields ($\phi$) are only included
when the rate is computed, as in Eq.~\eqref{GammaRQFT}. This approach was explored by Weinberg in~\cite{Weinberg:1992ds}. In particular, Weinberg observed that in scalar QED integrating out $A^\mu$
generates the instability. He also observed that one can additionally integrate out the imaginary part of $\phi$, leaving an effective action that depends only on the real part of $\phi$. 
A derivative expansion, justified with $\lambda \sim e^4$ scaling was critical to Weinberg's argument, as we review below.

It may turn out that the fluctuations of the field $\phi$ itself are required to generate the instability. In that case, one cannot simply integrate out $\phi$, or there is nothing left for the effective
action to depend on. For example, one might try to replace $\phi \to \phi + \bar{\phi}$ with $\bphi$ a fixed external background field and then to integrate out $\phi$. That is, we can compute
\be\label{defineSeff}
Z[0]  =\int \cD \phi e^{-S[\phi]} = \int \cD \phi e^{-S[\phi+\bphi]}
\ee
Since the path integral is invariant under field reparametrizations, $Z[0]$ computed this way is just a number. That is, it will not depend on $\bar{\phi}$ at all. Nevertheless, this is the same $Z$ used in the 
potential-deformation method for computing the tunneling rate. Thus one should expect $\Im \ln Z[0]$ to be related to the decay rate. Note that even if the classical potential has no instability, as long as the full theory
has an instability, one can still, in principle, compute the decay rate in this way. However, one cannot use the saddle point approximation,
since we have no classical bounce solution.

The usual procedure for computing an effective action that admits a bounce is to begin with the generating functional
\be
W[J] = - \ln Z[J] = - \ln \int \cD \phi e^{-S[\phi] -\int d^4x \phi(x) J(x)}
\ee
and take its Legendre transform
\be
\Seff[\bphi] = W[J_{\bphi}] - \int J_{\bphi}(x) \bphi(x)
\ee
Here, $J_\bphi$ is defined so that $\frac{\delta W}{\delta J}\big{|}_{J=J_\bphi} = \bphi$. Since $\frac{\delta W}{\delta J}\big{|}_{J=J_\bphi}$ is the expectation value of the field in the quantum theory, this says
that $J_\bphi$ is the background current required to make the expectation value of $\phi = \bphi$. 
Since $\frac{\delta \Seff}{\delta \bphi} = -J_\bphi$, we conclude that true vacuum of the theory $\bphi_{0,\TV}$ is given by an extremum of $\Seff$.

Formally, the Legendre transform requires that there is a one-to-one correspondence between $J_{\bphi}$ and $\bphi$. That is, the functionals $W[J]$ and $\Seff[\bphi]$ must be convex for the
Legendre transform to exist. However, convex actions never admit tunneling!

The incompatibility of convexity and tunneling is not as problematic as it sounds. First of all, in practice one computes the effective action not through the Legendre transform but through
the computation of 1PI diagrams with background fields. The action generated by background fields does not have to be convex (in the Standard Model, for example, it is not). 
Weinberg and Wu have argued~\cite{Weinberg:1987vp} that this 1PI effective action is more physical than the Legendre transform version since it does not include linear superpositions of vacua. Although their
argument is compelling, it does not help if we want to use the Legendre transform definition to compute the tunneling rate.

A more convincing argument is that in the potential-deformation method, we
actually {\it do} want a convex potential, whose single minimum corresponds to the false-vacuum.
We can get this by deforming the potential, or more directly by sticking to a steepest-descent contour.
Following the logic from Section~\ref{sec:potmethod} we can define the effective action $\Seff[\bphi_0]$ through the Legendre transform along the convex steepest-descent contour through the false vacuum. We then write the imaginary part as the discontinuity
of the effective action defined along the steepest-descent contour through the bounce. So
\begin{equation}
\frac{1}{2}\frac{\Gamma}{V} = \frac{1}{2}\Im \left(\frac{1}{\bigtau V}e^{-\Seff[\bphi_0]} \right)
\label{eq:decayrateZg1}
\end{equation}
where 
$\bphi_0$ is the bounce solution to the equation of motion $\Seff'[\bphi_0]=0$. In principle we should normalize this by $e^{-\Seff[0]}$, but this is equal to 1 by our choice of origin $\phi=0$.  
The requirement to only include the bounce saddle point is built into the way we compute in perturbation theory; we only include small fluctuations around the argument of $\Seff$.
The gauge-dependence of this way of computing the tunneling rate, particularly with regard to boundary conditions, is discussed in~\cite{Plascencia:2015pga}.

Let us briefly review how one evaluates the one bounce contribution to the path integral $Z$ (see Section \ref{sec:NLO2ways}) using the functional determinant and the classical action. There $Z=Ke^{-S[\phi_C]}$ at NLO, where $K\equiv \left(\frac{\Det\left[S''[\phi_C]\right]}{\Det\left[S''[0]\right]}\right)^{-\frac{1}{2}}$. Due to the zero modes from translations, $K=\bigtau V K'$ where $K'$ has the determinants evaluated with the zero modes removed, and $\bigtau V$ is the volume of Euclidean space-time. $K'$ has units of mass${}^{4}$, and it must come from some characteristic length scale $R$ from the bounce solution, so we can re-express this as $K'\sim \frac{1}{R^4}$ up to some dimensionless number we expect to be of order one \cite{Branchina:2014rva}.   

We expect the same thing to happen for the effective action using Eq.~\eqref{eq:decayrateZg1}, but to our knowledge there exists no complete proof. The bounce solution $\bphi_0$ is space-time translation invariant, so $Z$ should be proportional to $\bigtau V$. To get the right dimensions, we must compensate with some characteristic scale $R$ from the bounce solution.  We then find
\begin{equation}
\label{GammaSeff}
\frac{\Gamma}{V}=\frac{1}{R^4} \text{Im}\left[e^{-\Seff[\bphi_0]}\right]
\end{equation}
We will show that the tree level and logarithmic terms agree between the effective potential method and the functional determinant method in the next section. We will also see that if we wish to check the prefactor of Eq.~\eqref{GammaSeff} and the $\cO(1)$ terms in the exponent, we would have to know all the higher derivative terms in the effective action. 

In summary, in situations where the classical potential is stable but there is an instability in the quantum theory, the tunneling rate is given by the exponential of the 
effective action evaluated on the solution $\bphi_0$ to its equations of motion. 
Of course, the effective action can be used to calculate the decay rate even in situations where the classical potential admits tunneling. 
However, this approach is only useful to the extent that the effective action can be
computed exactly, or in some approximation consistent with a perturbative expansion of the decay rate. 
Next, we explore what the effective action looks like and whether it is useful for computing tunneling rates in an example. 

\subsection{NLO tunneling in scalar field theory\label{sec:NLO2ways}}
Consider the theory of a real scalar field with classical potential $V(\phi) = \lambda \phi^4$. For $\lambda < 0$, the vacuum at $\phi=0$ is unstable. The Euclidean equations of motion are solved by the quartic bounces in Eq.~\eqref{confbubb}: $\phi_C(\rho) = \sqrt{\frac{2}{|\lambda|}} \frac{R}{R^2+ \rho^2}$. The tree-level Euclidean action
evaluated on the quartic bounce is $S_E[\phi_C] = \frac{2 \pi^2}{3|\lambda|}$, independent of the bubble size $R$.
The decay rate at NLO can be computed either using Eq.~\eqref{defineSeff}, summing over the Gaussian fluctuations around $\phi_C$, or using Eq.~\eqref{GammaSeff} where the field fluctuations around an arbitrary background field configuration contribute to the form of the effective action. 

To compute the rate by integrating over Gaussian fluctuations around $\phi_C$ we follow the approach of~\cite{Isidori:2001bm} and~\cite{Dunne:2005rt}. We directly calculate
\be
\frac{\Gamma}{V} =\frac{S_E[\phi_C]^2}{4\pi^2} \left[ \frac{-\det'S_E''[\phi_C]}{\det S_E''[0]} \right]^{-1/2} e^{-S_E[\phi_C]}
\ee
Here $\det'$ refers to the functional determinant where the zero modes corresponding to translations have been removed. This functional determinant can be
computed by solving the eigenvalue equation $[-\partial^2 + W(\rho)]\phi = \lambda \phi$ and taking the product of the eigenvalues. 
We relegate details of this calculation to Appendix~\ref{sec:nlofunctionaldeterminants}. In $\msbar$, the result is that
\be
\frac{\Gamma}{V}  = \frac{1}{R^4}\exp\left[  -\frac{2 \pi^2}{3|\lambda(\mu)|}+  3 \ln(R\mu) +\cO(1)\right]
\label{directcalc}
\ee
where $\cO(1)$ is some order one number coming from the evaluation of the functional determinant, and $R$ takes a specific value which saturates the path integral over the associated collective coordinate.

As a quick check, we can verify that the rate is independent of $\mu$ to order $\lambda$, using $\beta_\lambda=\mu \frac{\partial}{\partial \mu} \lambda = \frac{9}{2\pi^2}\lambda^2$.
Note that the NLO rate breaks the degeneracy in $R$. In fact, for a given value of $\mu$, the rate appears to go to zero if $R\to 0$, or the rate is unbounded from above if $R\to \infty$. As the rate increases, however, the logarithm also grows to the point
where subleading orders become relevant. The logarithms are minimized at the scale $\mu$ where $\beta_\lambda = 0$. This is consistent with the prescription for finding the bubble shape for a general potential discussed in
Section~\ref{sec:approx}. 

The second method, using Eq.~\eqref{GammaSeff}, requires knowing the exact effective action, so that its equations of motion can be solved and its solution used to evaluate the rate. If we write $\bphi_0 = \phi_C + \hbar \phi_1 +\cdots$
and $\Seff = S_0 + \hbar S_1 + \cdots$, then  
\be
\Seff[\bphi_0] = S_0[\phi_C] + \hbar S_0'[\phi_C] + \hbar S_1[\phi_C]+ \cO(\hbar^2)
\ee
Since $S_0'[\phi_C]=0$, to NLO we only need to compute $S_1[\phi_C]$. That is, the corrections to the shape of the bubble come in first at next-to-next-to-leading order. In particular, we do not have to compute
or solve the equations of motion for the effective action (which could have been very difficult considering that $\Seff$ is non-local). 

Computing the full effective action $\Seff[\phi]$ is essentially impossible, even at 1-loop. We can however compute it order-by-order in a momentum expansion. The leading contribution, with no derivatives,
is the effective potential:
\be
\Veff = \lambda \phi^4 + \frac{9}{4 \pi^2} \lambda^2 \phi^4\left(\ln\frac{12 \lambda \phi^2}{\mu^2} - \frac{3}{2}\right)
\ee
This is computed in the background field method assuming the background fields are constant. To get the terms with 2 derivatives, we compute diagrams with background fields with non-zero momenta
and take two derivatives with respect to those momenta, then set the momenta to zero. More details are given in Appendix \ref{app:derivcorr}.
The result is that to 1-loop with up to 4-derivatives, the effective action in Euclidean space is 
\eqn{\Seff[\phi] &= \int d^4x \Big[\lambda\phi^4 + \frac{9}{4\pi^2}\lambda^2\phi^4\left(\ln\frac{12\lambda\phi^2}{\mu^2} - \frac{3}{2}\right)\nonumber\\
&~~~~~~~~~~~~~~~~~~~~+ \frac{1}{2}(\partial_\mu \phi)^2\left(1+\frac{1}{4\pi^2}\lambda\right)- \frac{1}{2}\left(\square \phi\right)^2 \frac{1}{480\pi^2}\frac{1}{\phi^2}\nonumber\\
&~~~~~~~~~~~~~~~~~~~~ + \frac{1}{2}(\partial_\mu \phi)^2\square\phi \frac{1}{720\pi^2}\frac{1}{\phi^3}
-\frac{1}{8}(\partial_\mu \phi)^2(\partial_\nu \phi)^2\frac{1}{360\pi^2\phi^4}+\mc{O}(\partial^6) \Big]}
We can now check whether the momentum expansion is justified for use in the calculation of the decay rate. 

The effective potential (zero-derivative terms) contributes to the action as
\be
\Seff^{V}[\phi_C] = 2\pi^2 \int d\rho \rho^3  \Veff[\phi_C(\rho)] = - \frac{2\pi^2}{3|\lambda|} -\frac{19}{4} - 3 \ln\frac{R\mu}{2\sqrt{6}}\pm i \frac{3\pi}{2}
\label{Vbounce}
\ee
where the $\pm i \frac{3\pi}{2}$ comes from $\ln\frac{\lambda}{|\lambda|}$ and $\lambda<0$. This imaginary term in the action makes $e^{-\Seff}$ imaginary since $e^{\pm i \frac{3\pi}{2}}=\mp i$. The sign is ambiguous, but we pick the sign such that $\Gamma>0$.  
The two-derivative terms 
\be
\Seff^{\text{2 der}}[\phi] = \int d^4x \left(1+ \frac{\lambda}{4 \pi^2}\right) \frac{1}{2} (\partial_\mu \phi)^2
\ee
contribute to the action on the bounce as
\be
\Seff^{\text{2 der}}[\phi_C] = \frac{4\pi^2}{3|\lambda|} -\frac{1}{3}
\label{eqn:S2derphiV}
\ee
Using Eq.~\eqref{GammaSeff}, and ignoring the $\cO(1)$ numbers, we find 
\begin{align}
\frac{\Gamma}{V}&=\frac{1}{R^4}\text{Im}\left[e^{-i\frac{3\pi}{2}}e^{-\frac{2\pi^2}{3|\lambda|}+3\ln(R\mu)+\cO(1)}\right]\\
&= \frac{1}{R^4}\exp\left[-\frac{2\pi^2}{3|\lambda|}+3\ln(R\mu)+\cO(1)\right]
\end{align}
We see that the logarithm and $\frac{1}{\lambda}$ term in Eq.~\eqref{directcalc} is reproduced exactly. This of course had to happen, since all the $\mu$ dependence at 1-loop must be compensated by the potential and kinetic term in the classical action.

Note that the NLO part of Eq.~\eqref{eqn:S2derphiV} (the $\frac{1}{3}$) has the same scaling as the NLO contributions from the potential, Eq.~\eqref{Vbounce}. Thus, adding derivatives does not seem to give
additional suppression.

To understand whether higher order terms give additional suppression, let us turn to the 4-derivative terms in the 1-loop effective action (see Appendix \ref{app:derivcorr}):
\be
\Seff^{\text{4 der}} = -\frac{1}{2} (\Box \phi)^2 \frac{1}{480\pi^2} \frac{1}{\phi^2} + \frac{1}{2}(\partial_\mu \phi)^2\Box\phi \frac{1}{720\pi^2 \phi^3}
-\frac{1}{8}(\partial_\mu \phi)^2(\partial_\nu \phi)^2\frac{1}{360\pi^2 \phi^4}
\ee
Since the bounce $\phi_C \sim \frac{1}{\sqrt{|\lambda|}}$, when we evaluate the action on the bounce, all these terms will scale like $\lambda^0$. Thus they will be of the same order as the potential and two-derivative terms.

More generally, 1-loop terms with any number of derivatives contribute to $\Seff[\phi_C]$ at the same order as the 1-loop effective potential. To see this, note that whenever a factor of momentum is pulled out
of a loop graph, it must be compensated by an effective mass, by dimensional analysis. In $\lambda \phi^4$ theory,
the effective mass is $m_\text{eff}^2 =12\lambda \phi^2$. Since $\phi_C \sim \frac{1}{\sqrt{\lambda}}$, adding two derivatives and $\frac{1}{m_\text{eff}^2}$ does not change the power counting.

We conclude that the derivative expansion is {\it not} justified for calculating decay rates. That is, to compute the decay rate at NLO in $\lambda \phi^4$ theory using the effective action, we need the complete effective action to 1-loop,
not just the leading-momentum dependent terms. Since this effective action is nearly impossible to compute, this method is not feasible for computing NLO decay rates in quantum field theory.
Similar conclusions were reached through a calculation in quantum mechanics in~\cite{Bender:1984jc}.

Keep in mind that we have only shown the derivative expansion does not work for $\lambda \phi^4$ theory. In other theories, it may be useful, but it  depends on the circumstances. 
For example, 
in the Coleman-Weinberg
model, there are two couplings, $\lambda$ and $e$. For $e$ small, the running of the couplings is perturbative and $\lambda$ goes from $-\infty$
to $+\infty$ while $e$ remains small. Thus, the theory is completely specified by a single small number $e$. 
The loop corrections
in this theory induce spontaneous symmetry breaking at a scale $\mu$ where $\lambda(\mu) \sim e(\mu)^4$. 
 Having $\lambda \sim e^4$ is 
more than just a numerical association: one must power count with this scaling so that physical quantities, such as the vector-to-scalar
mass ratio, are gauge invariant \cite{Andreassen:2014eha}.  
In the Coleman-Weinberg model, the effective masses relevant at 1-loop are of the form $\meff \sim e \phi$.
When $\lambda \sim e^4$, then  $\frac{\Box}{\meff^2} \sim \frac{\lambda}{e^2} \sim e^2$, so higher derivative terms are power suppressed.

Weinberg and Metaxas~\cite{Metaxas:1995ab} used the $\lambda\sim e^4$ power counting to show that the tunneling rate in the Coleman-Weinberg model is gauge invariant at NLO.
But although $\lambda \sim e^4$ scaling is critical for spontaneous symmetry breaking, it is not the appropriate scaling for
the calculation of tunneling rates. As discussed earlier in this section, the scale appropriate for tunneling is where $\beta_\lambda(\mu)=0$. 
Since $\beta_\lambda \sim e^4 + e^2 \lambda + \lambda^4$, $\beta_\lambda = 0$ is compatible with the normal loop power counting
where $\lambda \sim e^2$.
With this counting $\meff \sim 1$ as in $\lambda \phi^4$ theory and $\frac{\Box}{\meff}$ corrections are as important
as corrections to the effective potential.

\section{The Standard Model \label{sec:theSM}}

In the previous sections, we have discussed general features of tunneling calculations in quantum mechanics and quantum field theory. In this section, we apply some of those insights to 
the Standard Model. In particular, we consider the question of how to calculate the lifetime of the metastable Standard Model vacuum with $\langle h \rangle = 246~\GeV$ in a systematically improvable way.
We also discuss the UV sensitivity of the decay rate calculation.

\subsection{The Effective Potential and Gauge-Invariance \label{sec:SM2loop} }
A discussion of vacuum stability in the Standard Model usually begins with the Standard Model effective potential. Although, as we have seen, this is not all we need to accurately describe tunneling.
This potential $\VSM(h,\mu)$ is a function of a constant real scalar background Higgs field $h$, the
renormalization group scale $\mu$, and the various couplings in the theory.
It has been computed to 2-loop accuracy with resummation of large logarithms to the 3-loop level.

An important feature of the Standard Model effective potential is that it is nearly scale-invariant. The only mass scale in the Standard model is the single dimensionful parameter $v$ (the Higgs vev). For $h\gg v$,
we can set $v=0$, along with the Higgs mass and all other masses. Then all of the scale-dependence comes from quantum corrections and dimensional transmutation. To make this clear, we often write
\be
\VSM(h,\mu) = \frac{1}{4} h^4 \leff(h,\mu) \label{vsmlmu}
\ee
The function $\leff(h,\mu)$  is dimensionless, matches the Higgs quartic at the weak scale, and changes slowly (logarithmically) as $h$ is increased. 
Here $\mu$ is the $\msbar$ scale and all of the $\mu$ dependence is either implicit, through the gauge couplings $g_i(\mu)$, top Yukawa $\lambda_t(\mu)$, etc., or explicit in terms of the form $\ln\frac{h}{\mu}$. 
The explicit and implicit $\mu$ dependence are related by the renormalization group equation
\be
\left( \mu \frac{\partial}{\partial  \mu} - \gamma h \frac{\partial}{\partial h} + \beta_i \frac{\partial}{\partial \lambda_i}
 \right) \VSM(h,\mu) = 0
\label{vsmrge}
\ee
with $\gamma$ the Higgs field anomalous dimension.

It is commonplace to resum the effective potential by setting $\mu=h$~\cite{Buttazzo:2013uya}. Indeed, this is the natural choice, as $h$ is the only scale around.
However, setting $\mu=h$ is dangerous. For example, if one is interested in extrema of $\VSM$, then solving $\frac{\partial}{\partial h} \VSM(h,\mu)=0$ and setting $\mu=h$ afterwards does
not give the same value of $h$ as solving $\frac{\partial}{\partial h} \VSM(h,h)=0$. Or to calculate the tunneling rate one must evaluate the effective action on a solution to the equations
of motion before setting $\mu=h$, rather than after. 

Another issue to keep in mind is that $\VSM$ is not gauge invariant. In fact, explicit gauge dependence is {\it required} for Eq.~\eqref{vsmrge} to hold since the anomalous dimension
$\gamma$ is gauge-dependent. Moreover, not all the gauge-dependence can be associated with $\gamma$; there is further gauge-dependence in the non-logarithmic terms in $\VSM$ as well. The gauge dependence of
effective potentials is fairly well understood. 
Recall that the effective potential describes the energy of the system in the presence of a background current $J$. Since this current is a charged source, it depends on gauge,
and therefore the potential will be gauge dependent whenever $J\ne 0$. This explanation also implies that at the extrema, where $J=0$, the effective potential is gauge-invariant. Thus the energy of
any metastable minimum is gauge-invariant and possibly physical. This is indeed true, and has been checked explicitly. A more rigorous proof that the vacuum energy is gauge invariant
relies on the Nielsen identity~\cite{Nielsen:1975fs} (see also~\cite{Fukuda:1975di}). 

Since the tunneling rate is physical, it should be gauge-invariant. In fact, as it appears to be given by the Euclidean effective action at its extremum (the bounce), its gauge-invariance also follows from the Nielsen identity. 
Some subtleties in establishing gauge-invariance, involving boundary conditions on the path integral, were recently explored in~\cite{Plascencia:2015pga}
(see also \cite{DiLuzio:2015iua}). Unfortunately, we know of no explicit demonstrations of gauge-invariance, say
at next-to-leading order, even in scalar QED.\footnote{
In~\cite{Metaxas:1995ab}, Weinberg and Metaxas confirm the Nielsen identity at 1-loop using $\lambda \sim \hbar$ scaling, but do not explicit show gauge invariance of the 1-loop rate. }
Such demonstrations may have practical implications for tunneling rate calculations, as they had for absolute stability bounds. 

In~\cite{Andreassen:2014gha}, it was shown that for the absolute stability bound in the Standard Model to be gauge invariant, two modifications of the usual procedure were necessary.
To be clear, by ``usual procedure'' we mean using the resummed effective potential with $\mu=h$ in Landau gauge as in~\cite{Buttazzo:2013uya}.
The modifications were that:
\begin{enumerate}
\item The effective potential had to be known to fixed order in $\hbar$ assuming that $\lambda \sim \hbar$. This is the same scaling as introduced
by Coleman and Weinberg, but had not been used in the SM.
\item The RGE for the effective potential should not be solved, and we should not set $\mu=h$, but rather the couplings run to a scale $\muX$ where
the leading-order (in the modified $\hbar$ scaling) potential is minimized and the potential evaluated at fixed order. 
\end{enumerate}
Let us now consider whether these two modifications also must be applied to the calculation of tunneling rates.

First, we would like to know if we should use $\lambda \sim \hbar$ as in the absolute stability
calculation, or $\lambda \sim \hbar^0$ as in ordinary perturbation theory?  The reason the modified scaling is appropriate for absolutely stability is because an extremum of the potential requires
that the growth of the monotonic classical potential $V = \frac{1}{4} \lambda h^4$ be canceled by loop effects scaling like $\hbar$. This is only possibly if $\lambda$ is anomalously small, $\lambda \sim \hbar$. For absolute
stability, we  need the potential in only a small neighborhood of an extremum, so it is consistent for $\lambda\sim \hbar$ within the entire neighborhood. 
However, for tunneling, which involves the potential connecting different extrema,
we necessarily need a large range of scales and it is inconsistent for $\lambda \sim \hbar$ throughout that range. It is easy to see that in the Standard Model, $\lambda(\mu)$ is indeed not anomalously small away from extrema of $\VSM$.  
We conclude that for a tunneling rate, one should not take $\lambda \sim \hbar$, but instead take $\lambda \sim \hbar^0$ and use ordinary perturbation theory. 
As discussed in Section~\ref{sec:NLO2ways}, this means that all higher derivative terms in the Standard Model effective action will be as important as corrections to the effective potential. 

For the second point, we have a new problem: the effective potential describes a minimum near $h=0$ (our vacuum), a maximum where the Landau-gauge field is around $h\sim 10^{9}~\GeV$ and another minimum at around $h \sim 10^{30}~\GeV$. These scales are so far apart that there are necessarily large logarithms in the effective potential and presumably resummation is critical. Indeed, with $\mu \approx 200~\GeV$, the resummed effective
potential is not at all well described by fixed-order perturbation theory. But if we require resummation, which mixes orders in $\hbar$, checking gauge-invariance order-by-order in $\hbar$ is impossible.

To investigate further, we write the SM effective potential to 2-loop order (in Landau gauge) as
\be
\VSM^{(2)}(h) = h^4\left( A + B \ln\frac{h}{\mu} + C \ln^2 \frac{h}{\mu}\right) \label{v2}
\ee
Here, $A,B$ and $C$ are calculable functions of the SM couplings, which in turn depend on $\mu$. A lot of this $\mu$ dependence is canceled by the explicit $\mu$ dependence in Eq.~\eqref{v2}, but not all of it. 
At tree-level with $\mu=200$ GeV, $A=\frac{\lambda}{4}=0.031$ and $B=C=0$. Values for $A$, $B$, and $C$ at two-loops for different choices of $\mu$ are given in Table \ref{tab:abc}.

\begin{figure}
\begin{center}
\begin{tikzpicture}
\node at (0,0) {\includegraphics[width=0.7\textwidth]{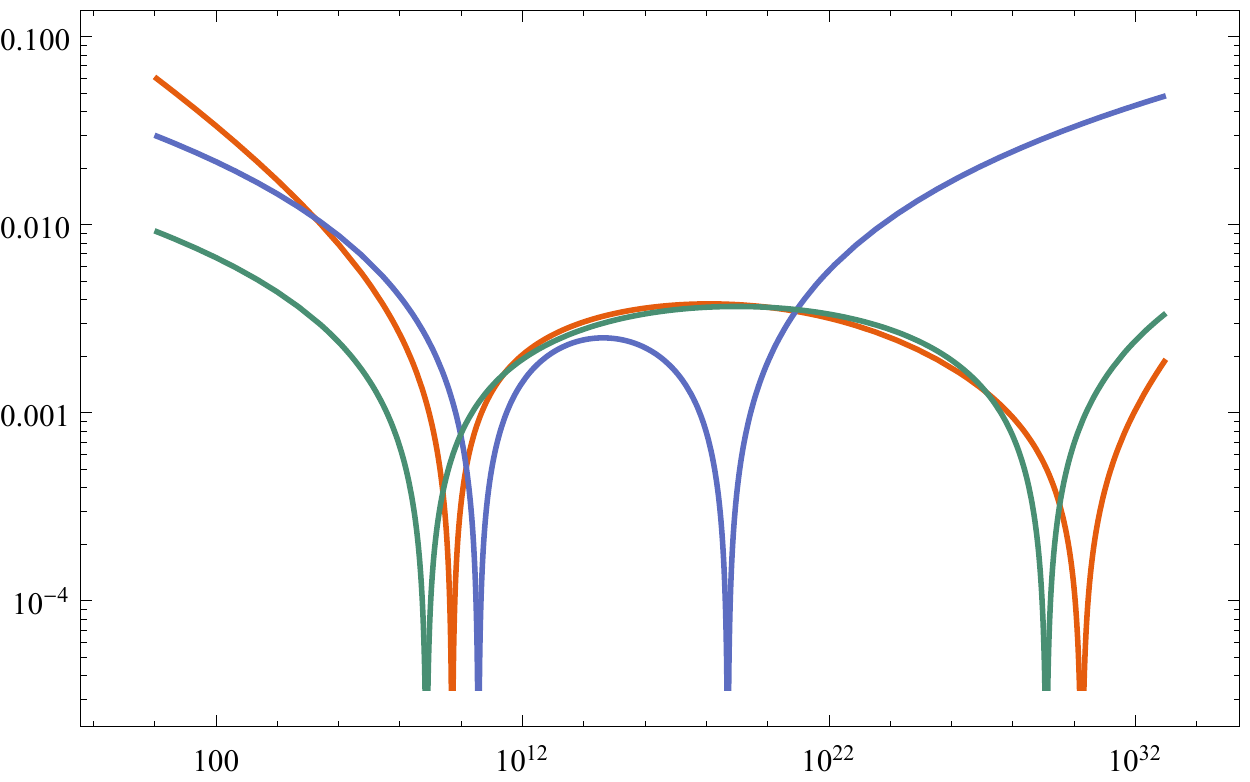}};
\node at (0.5,-4) {$h$};
\node[rotate=90] at (-6.5,0) {$\left|\log\left(\frac{\VSM^{(2)}(h)}{h^4}\right)\right|$};
\end{tikzpicture}
\end{center}
\caption{The effective potential in the Standard Model in Landau gauge resummed with $\mu=h$ (red curve) is compared
to the effective potential at 2-loops with $\mu = 10^{10}~\GeV$ (blue) and $\mu = 10^{17}~\GeV$ (green).} 
\label{fig:SMfits}
\end{figure}

Conveniently, we find that there is good agreement between the traditional resummed potential
$\VSM(h,h)$ and the fixed order potential $\VSM^{(2)}$ when $\mu=10^{17}~\GeV$ as can be seen in Fig.~\ref{fig:SMfits}. By good agreement, we mean that the extrema are fairly close. Both have maxima at around $h = 10^{9}$ GeV
and minima around $h= 10^{30}$ GeV. (In contrast, taking $\mu=10^{10}$ GeV leads to a maximum at $h = 10^{10}$ GeV and a minimum at $10^{18}$ GeV). 

In conclusion, we have shown that fixed-order perturbation theory can be used to calculate the Standard Model effective potential in good quantitative agreement with the resummed potential. 
This implies that there is no impediment in trying to establish {\it explicitly} the gauge-independence of the tunneling rate in the Standard Model using the fixed-order effective action if one can calculate all the higher derivative corrections relevant to that order.
Even in the absence of such an explicit demonstration, it seems reasonable to expect that, since resummation mixes up orders in perturbation theory, one should always use the fixed-order potential
and never the resummed one.

\begin{table}
\begin{center}
\begin{tabular}{|c|c|c|c|}
\hline
	$\mu$ (GeV) & A & B & C\\
\hline
$200$ & $0.0387$ & $-0.00777$ & $8.38 \times 10^{-4}$ \\
$10^5$ & $0.151$ & $-0.00248$ & $1.64 \times 10^{-4}$ \\ 
$10^{10}$ & $0.000751$ & $-6.08 \times 10^{-4}$ & $2.84 \times 10^{-5}$ \\
$10^{15}$ & $-0.00313$ & $-1.48 \times 10^{-4}$ &  $9.32 \times 10^{-6}$ \\
${\mathbf{10^{17}}}$ & {\bf -0.00352} &  ${\mathbf{-6.19 \times 10^{-5}}}$ &  ${\mathbf{6.76 \times 10^{-6}}}$ \\
$10^{20}$ & $-0.00347$ & $2.28 \times 10^{-5}$ & $4.86 \times 10^{-6}$ \\
\hline
\end{tabular}
\end{center}
\caption{Values of $A$, $B$, and $C$ in the fixed order SM effective potential $V_\text{SM}^{(2)}(h) = h^4\left( A + B \ln\frac{h}{\mu} + C \ln \frac{h}{\mu}\right)$
as in Eq.~\eqref{v2} for various choices of $\mu$ in Landau gauge.
We have used $m_h^{\text{pole}} = (125.14 \pm 0.24)$ GeV and $m_t^{\text{pole}} = (173.34 \pm 1.12)$ GeV. 
 For $\mu = 10^{17}$ GeV, the 2-loop potential is in good agreement with the resummed potential.
}
\label{tab:abc}
\end{table} 

\subsection{SM tunneling rate}
The Standard Model effective potential has a minimum at our electroweak vacuum ($h\sim 0$) with $\VSM(0)=0$ and a maximum value of $\Vmax \approx (10^{10}~\GeV)^4$. It then turns over and runs negative, eventually hitting
another minimum at around $\Vmin \sim -(10^{30}~\GeV)^4$~\cite{Andreassen:2014gha}. The field values where these extrema are taken
are gauge-dependent, but the energy densities at the extrema are gauge-independent. One should not take the value $10^{30}~\GeV$ for the scale of the absolute minimum very seriously: quantum gravity will obviously modify this scale, as 
we discuss below in Section~\ref{sec:uv}.

The fact that the potential grows, has a maximum, and then decreases to a minimum indicates that there is a potential barrier through which the Higgs field can tunnel. However, as we saw in Section \ref{sec:QFT}, this is not the correct picture. It is not the barrier of the effective potential of height $10^{10}~\GeV$ that we should think about tunneling through. Instead we must study the barrier in the potential energy functional $U[h]$ defined in Eq.~\eqref{Udef}. 
To see this, recall the general result from Section~\ref{sec:approx} that for a potential $V(h)$ the bubble size is determined by the condition $\frac{d}{d h} \frac{V(h)}{h^4} = 0$. For the Standard Model, this scale is $h \sim 10^{17}~\GeV$. 
As we can see from Table~\ref{tab:abc}, near the scale $\mu=10^{17}~\GeV$, the Standard Model effective potential is basically just 
\be
\VSM(h \sim 10^{17} \GeV) \approx \frac{\lambda_0}{4} h^4,\qquad \frac{\lambda_0}{4} = -0.00352  \label{VSMl0}
\ee
Here $\lambda_0$ is approximately the value of the Higgs quartic at its minimum, $\lambda_0 \approx \lambda(\mu=10^{17})$.

There are solutions to the Euclidean equations of motion following from Eq.~\eqref{VSMl0} for any bubble size $R$. These
are just the quartic bounces as in Eq.~\eqref{confbubb}:
$h_C^R(\rho) = \sqrt{\frac{8}{|\lambda_0|}} \frac{R}{R^2+\rho^2}$, 
and the energy functional $U$ for the different bubble shapes is shown in Fig.~\ref{fig:conformalUtau}.
The tree-level action on any of these bounces is
$S[h_C^R] = \frac{8\pi^2}{3|\lambda_0|}$. The decay rate per unit volume  is therefore
\be
\frac{\Gamma^{\text{LO}}}{V} = \exp{\left( - \frac{8\pi^2}{3|\lambda_0|} \right)} \approx 10^{-812}
\label{LOrate}
\ee
independent of $R$. The right-hand side has the wrong units, as the units come in at NLO. However, the rate is so small that to a first approximation, the units do not even matter. The difference between choosing a scale of the Planck mass and a scale of the size of the universe (to the fourth power) gives only a factor around $10^{200}$!

If the degeneracy in $R$ persisted to all orders, it would invalidate the method of computing the tunneling rate in the Gaussian approximation: varying $R$ would lead to a zero eigenvalue of the quadratic fluctuations, and
be unsuppressed in the path integral. We could possibly treat the scale-invariance as we treat translation symmetries: integrate over $R$ as a collective coordinate to produce a ``volume" factor. However,
the $R$ dependence is in fact broken by loop corrections.

There is a shortcut to determining the $R$ dependence of the NLO decay rate. The decay rate, being physical, satisfies an RGE of the form
\be
\left( \mu \frac{\partial}{\partial  \mu} + \beta_i \frac{\partial}{\partial \lambda_i} \right) \Gamma(R,\mu) = 0
\label{gammarge}
\ee
The $\mu$ dependence is known from the SM $\beta$-functions, and then the $R$ dependence is determined by dimensional analysis. Explicitly, we must have
\begin{align}\label{eqn:SMTunnelingRate}
\frac{1}{V} \Gamma^{\text{NLO}} = R^{-4} \exp \Bigg{[}
-\frac{8\pi^2}{3|\lambda_0|} 
-\left(-4+2\frac{\lambda_t^2}{|\lambda_0|}- \frac{2 g_2^2 + g_Z^2}{2 |\lambda_0|} +\frac{\lambda_t^4}{\lambda_0^2}- \frac{2 g_2^4 + g_Z^4}{16 \lambda_0^2}\right) \ln(R\mu) &\\
 + 
\text{$R$-independent}& \Bigg{]}\nonumber
\end{align}
in agreement with~\cite{Isidori:2001bm}.

Since the action depends on $R$, the degeneracy over $R$ is broken. At this order, however, the rate can be made arbitrary small for any choice of $\mu$, since
it is a monotonic function of $R$. At the next order there will be terms quadratic in $\ln R\mu$. If we had those terms, we could solve for an exact, well-defined minimum of the 2-loop rate, 
which could only give us $\Rmin \sim\mu^{-1}$. Thus even at NNLO, the result would depend on an arbitrary scale $\mu$. 
The full exact rate, however, is independent of the artificial scale $\mu$. Thus, there must be an actual scale $\Rmin$ for which the rate is maximal. By dimensional analysis, 
$\Rmin^{-1}$ is almost certainly near the scale $\mu_0$ where $\beta_\lambda(\mu_0)=0$~\cite{Isidori:2001bm,Branchina:2014rva}.
However, to our knowledge this has not been rigorously shown. Assuming $\mu_0$ is the correct
scale, we take $R=\Rmin=\mu_0^{-1}$ and then
$\mu=\mu_0$ to minimize the large logarithms.
The conclusion is that the decay rate is given the formula in Eq.~\eqref{LOrate} plus NLO corrections.
 The leading order rate is so small, that the NLO corrections are not even worth computing. 

The take-home lesson from the analysis of the SM decay rate is that everything hinges on a single dimensionful scale $\mu_0$. This scale sets the size of the bubble $R=\Rmin \sim \mu_0^{-1}$,
which in turn provides the dimensions of the decay rate $\Gamma \sim \frac{1}{R^4}$. The decay rate is exponential in $\lambda_0^{-1}$, where $\lambda_0=\lambda(\mu_0)$.
The scale $\mu_0$ is completely undetermined by the tree-level action, which is scale invariant. It is also not determined by the NLO corrections to the decay rate: quadratic fluctuations
in the direction of changing $R$ are not exponentially suppressed. Thus one needs to go to at least NNLO to fix $\mu_0$. Looking at the effective potential at NNLO, we see that the degeneracy is broken
an a scale $\mu_0$ where $\leff'(h=\mu_0)=0$, with $\leff \equiv \frac{4}{h^4}\VSM$. However, the full effective action (with derivative terms) could change this conclusion and is not known to NNLO.
Of course, the simplest procedure  to determine  $\mu_0$ is to maximize the leading order rate $\Gamma \sim \exp(-\frac{8\pi}{|\lambda(\mu_0)|})$, as is often done~\cite{Isidori:2001bm,Branchina:2014rva,DiLuzio:2015iua}.
The discussion here has investigated to what extent that procedure can be rigorously justified.

\subsection{Higher-dimension Operators \label{sec:uv}}
In the discussion above, we assumed that there was no physics beyond the Standard Model which could affect the lifetime of our vacuum. This assumption is not valid. At minimum, there 
will be contributions from gravity which come in at a scale $\Mpl \sim 10^{19}~\GeV$, but there is also the possibility of new physics at scales well-below $\mpl$. 
It has been argued relatively recently by Branchina et al. that even contributions at $\Mpl$ can destabilize the vacuum~\cite{Branchina:2013jra,Branchina:2014rva,Branchina:2014efa,Branchina:2014usa}.
Their argument relies on the coincidence between the field value $\phi_0$ at the center of the critical bubble (ultimately determined by where $\beta_\lambda=0$) in the Standard Model and the Planck scale.
In this section, we discuss how sensitive the SM tunneling rate may be to physics at a new scale. We confirm the Planck sensitivity, but also show how it would persist even without the coincidence between $\Mpl$
and $\phi_0$. 

To begin, we recall that the Standard Model is qualitatively very similar to a simple toy scalar field theory with potential $V=\frac{\lambda}{4} \phi^4$ with $\lambda<0$. As long as we are concerned with
energy scales well-below the scale of the new physics, we can perform the path integral over the new particles to generate a low-energy effective action which can be expanded in derivatives.
This leads us to consider a modified potential of the form
\be
\label{eqn:Vpoly}
V =\frac{\lambda}{4} \phi^4 -\frac{1}{ 6\Lambda^2} \phi^6 + \frac{1}{8M^4} \phi^8 
\ee
where $\Lambda$ and $M$ are two parameters with dimension of mass. 
For $\lambda<0$, $\Lambda>0$ and $M>0$, the potential has a local maximum at $\phi=0$ and a
minimum at $\phi_{\text{min}}\approx \frac{M^2}{\Lambda}+\mathcal{O}\left(\frac{\lambda\Lambda^3}{M^2}\right)$. 

To be clear, since we have completely integrated out the new physics, we can put this classical potential into the action
to find the tunneling rate; there is no double counting since the fluctuations of new physics are integrated out in producing the potential and only the fluctuations of $\phi$  is used to calculate the rate.
This is consistent with the discussion in Section~\ref{sec:NLO}.

While we do not have analytical solutions to the Euclidean equations
of motion for this potential, we can easily find numerical solutions (see Appendix~\ref{app:bounce}). We find, numerically, that the starting
point for the bounce (field value at the center) and value of the Euclidean action on the bounce are
\be
\phi_0 \approx 0.85 \times \frac{M^2}{\Lambda},\qquad
S_E \approx 290 \times \frac{\Lambda^4}{M^4} \label{eqn:num68}
\ee
respectively.

\subsubsection{Approximate Solutions}
It is perhaps informative to compare this exact numerical result to the approximate result coming from using approximate solutions, the Gaussian bubbles
discussed in Section~\ref{sec:approx}. These are bubbles of the form $\phi_G(\rho) = \phi_0 \exp(-\rho^2/R^2)$ with two parameters $\phi_0$
and $R$, as in Eq.~\eqref{gdef}. The Euclidean action with this potential on these bubbles is
\be
S_E[\phi_G] =\frac{1}{2}\pi^2 R^2 \phi_0^2+ \frac{\pi^2 R^4\phi_0^4}{64} \left(\lambda- \frac{8\phi_0^2}{27 \Lambda^2} + \frac{\phi_0^4}{8M^4}  \right)
\ee
The energy of the bubbles is
\be
U[\phi_G] =\frac{3 \pi ^{3/2} R \phi_0^2}{4\sqrt{2}}
   +\phi_0^4R^3\frac{\pi ^{3/2}}{16\sqrt{2}}
	\left(\frac{\lambda}{\sqrt{2}}
	-\frac{4\phi_0^2}{9 \sqrt{3} \Lambda ^2}
	+\frac{\phi_0^4}{8 M^4}\right)
\ee
We recall that before minimizing, we need to restrict to the $U=0$ surface. 
So the zero energy surface is:
\be
R^{-2} =-\lambda\frac{\phi_0^2 }{12\sqrt{2}}
+\frac{\phi_0^4}{27 \sqrt{3} \Lambda ^2}
-\frac{\phi_0^6}{96 M^4}
\ee
This gives us the action on the zero energy surface as a function of $\phi_0$ only:
\be\label{eqn:deltaSpoly}
S_E[\phi_0] =
\frac{162 \pi ^2 \Lambda ^2 M^4 \left(64 \left(4 \sqrt{3}-3\right) M^4 \phi _0^2-27 \Lambda ^2 \left(8 \left(4 \sqrt{2}-3\right) \lambda  M^4+5 \phi _0^4\right)\right)}{\left(27 \Lambda ^2 \phi _0^4+4 M^4 \left(27 \sqrt{2} \lambda  \Lambda ^2-8 \sqrt{3} \phi _0^2\right)\right){}^2}
\ee
Minimizing with respect to $\phi_0$ leads to:
\be
\phi_0 
   = 1.1\times \frac{M^2}{\Lambda},\qquad
R =8.9 \times \frac{\Lambda^3}{M^4}, \qquad
S_E[\phi_G] = 224.7 \times \frac{\Lambda^4}{M^4}
\label{extrem}
\ee
in parametric and reasonable numerical agreement with Eq.~\eqref{eqn:num68}.

Alternatively, solving $\leff'(\phi_0)=0$ gives
\be
\phi_0 = \sqrt{\frac{2}{3}} \frac{M^2}{\Lambda} = 0.82 \times \frac{M^2}{\Lambda}
\ee
also in good agreement with Eq.~\eqref{eqn:num68}.

\subsubsection{High-Scale Operators in the Standard Model \label{sec:highscale}}
Now let us consider what happens when we add such higher dimension operators to the Standard Model.
 Importantly, there are two qualitatively different types of effects. 
First, adding 
a potential $\vnew$ will cause a perturbative change to the Euclidean action on the Standard Model bounce, given by: 
\be
\Delta S = \int d^4x \vnew[ h^R_C(x)] \sim \frac{1}{\Lambda^2 R^2} + \frac{1}{M^4 R^4}
\ee
By dimensional analysis, this correction will be suppressed by factors of $MR$ and $\Lambda R$ as shown. 
Even for Planck scale operators, this may not be much suppression, since in the Standard Model $R^{-1}\sim 10^{17}~\GeV$ which is not that far from $\mpl$. 

 In the limit that $M$ and $\Lambda$ are taken very large, one expects that the new physics should decouple and
the Standard Model bounce should be unaffected, which indeed is what happens. However, in this limit a new tunneling direction can open up; although the Standard Model bounce is unaffected, the decay rate (determined by the integration over all bounces) \emph{can} be affected if a new direction in field space has lower Euclidean action than the Standard Model bounce. Thus, the second effect which can happen when new
operators are added is that tunneling proceeds through an entirely new direction in field space.

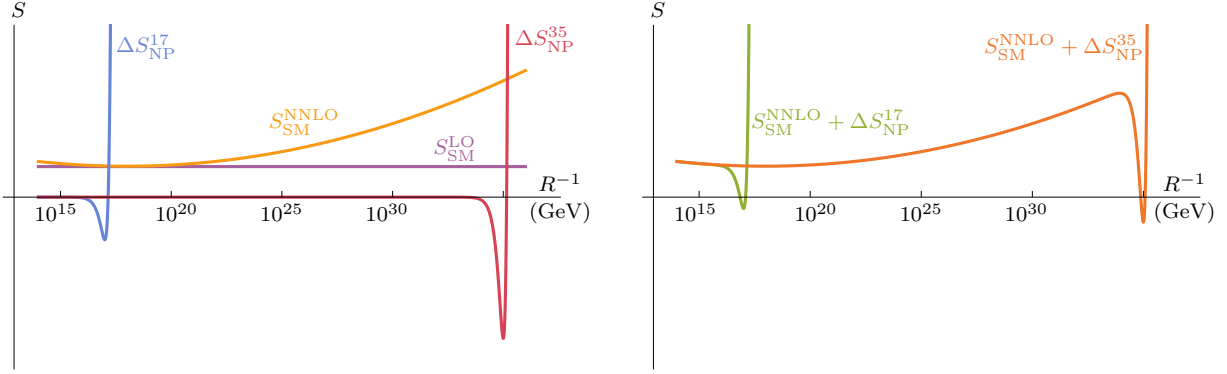
\begin{figure}
\begin{center}
\begin{tikzpicture}
\node at (-3,0) {\includegraphics[width=0.45\columnwidth]{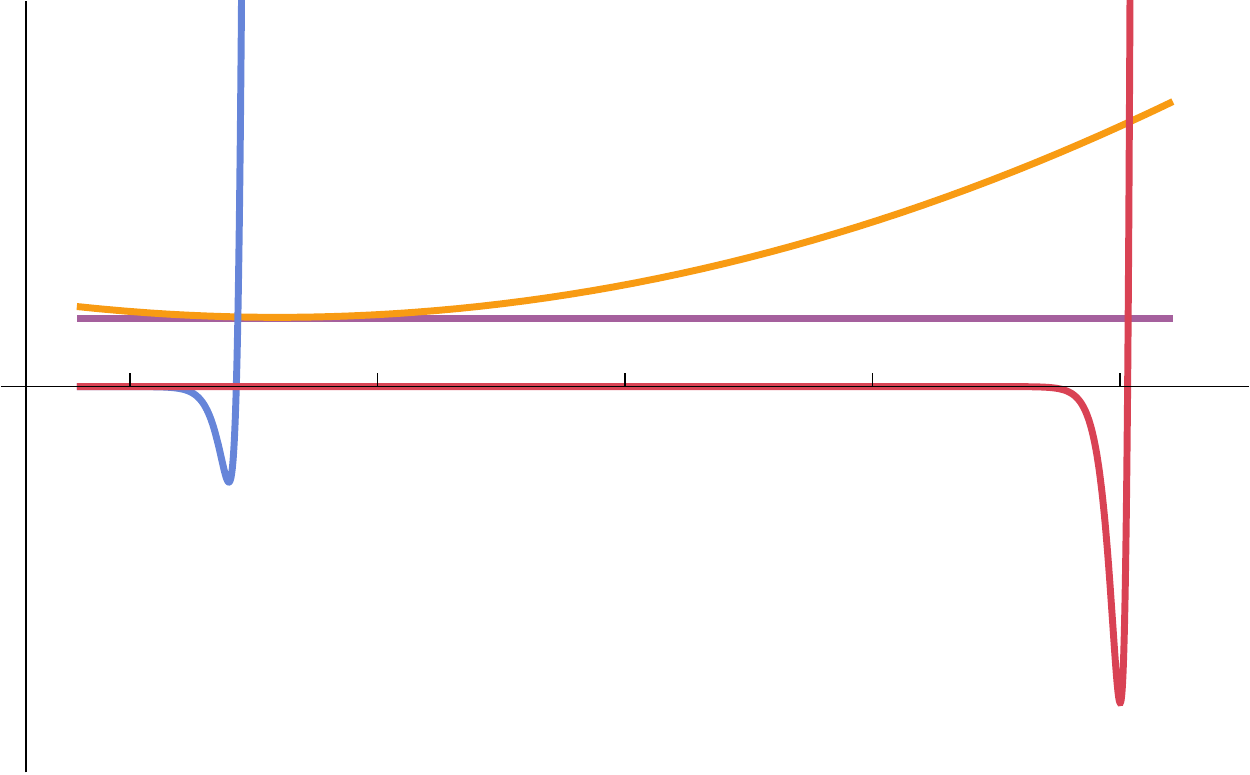}};
\node at (5.5,0) {\includegraphics[width=0.45\columnwidth]{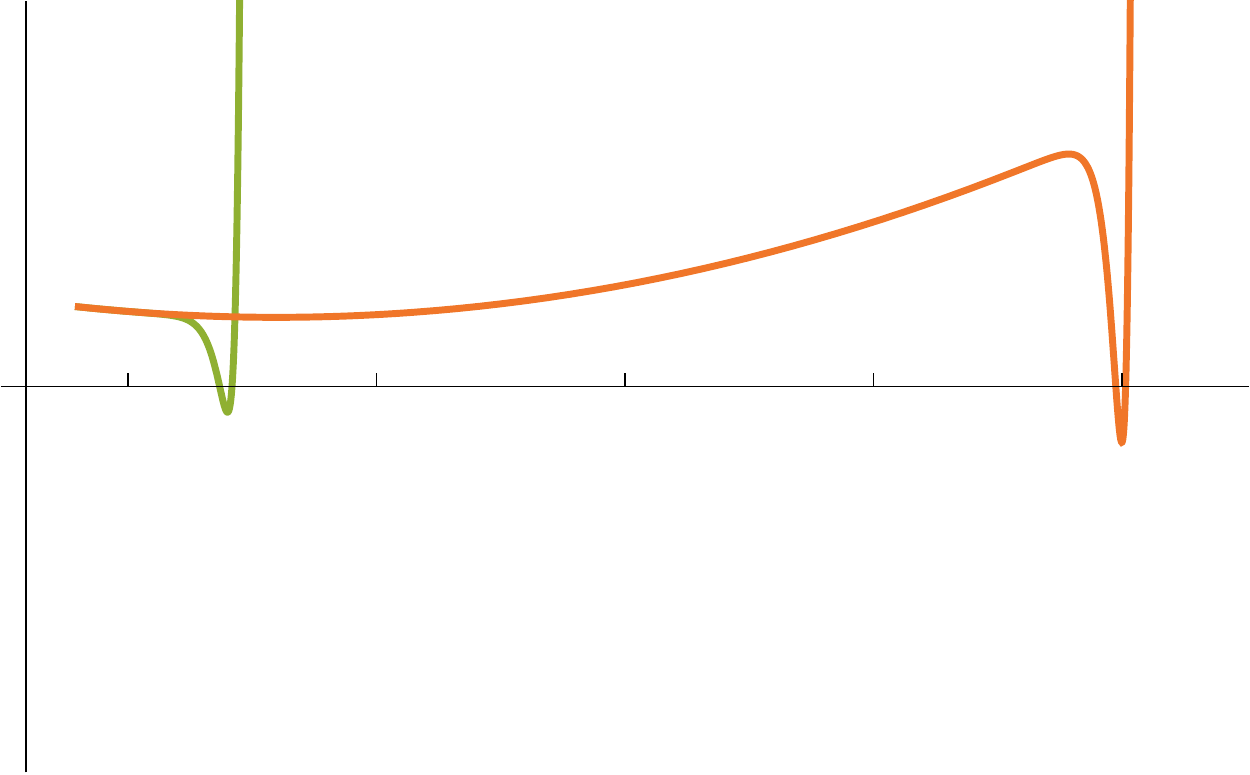}};
\node at (-6.5,2.5) {\scriptsize $S$};
\node at (2,2.5) {\scriptsize $S$}; 
\node at (0.7,0.2) {\scriptsize $R^{-1}$};
\node at (0.7,-0.2) {\scriptsize (GeV)};
\node at (9,0.2) {\scriptsize $R^{-1}$};
\node at (9,-0.2) {\scriptsize (GeV)};
\node at (-6,-0.2) {\scriptsize $10^{15}$};
\node at (2.5,-0.2) {\scriptsize $10^{15}$};
\node at (-4.4,-0.2) {\scriptsize $10^{20}$};
\node at (4.1,-0.2) {\scriptsize $10^{20}$};
\node at (-3,-0.2) {\scriptsize $10^{25}$};
\node at (5.5,-0.2) {\scriptsize $10^{25}$};
\node at (-1.6,-0.2) {\scriptsize $10^{30}$};
\node at (6.9,-0.2) {\scriptsize $10^{30}$};
\node at (-0.7,0.65) {\color{newpurple}\scriptsize $S_{\text{SM}}^{\text{LO}}$};
\node at (-2.7,1) {\color{newyellow}\scriptsize $S_{\text{SM}}^{\text{NNLO}}$}; 
\node at (0.5,2.1) {\color{newred}\scriptsize $\Delta S_{\text{NP}}^{35}$}; 
\node at (-4.8,2) {\color{newblue}\scriptsize $\Delta S_{\text{NP}}^{17}$}; 
\node at (7.4,2) {\color{neworange}\scriptsize $S_{\text{SM}}^{\text{NNLO}}+\Delta S_{\text{NP}}^{35}$}; 
\node at (4.3,1) {\color{SMgreen}\scriptsize $S_{\text{SM}}^{\text{NNLO}}+\Delta S_{\text{NP}}^{17}$};  
\end{tikzpicture}
\end{center}
\caption{
Left: Corrections to the standard model LO action (violet) for each $R$, using $\mu_0=10^{17}$ and $\lambda(\mu_0)=-0.015$, induced by the Standard Model logs at NNLO (orange), and also the higher-dimension potential terms from Eq~\eqref{eqn:Vpoly} with $\Lambda =5.09 \times 10^{18}~\text{GeV},~M=3.16 \times 10^{18}~\text{GeV}$ (blue), and $\Lambda =2.80 \times 10^{36}~\text{GeV},~M=2.34 \times 10^{36}~\text{GeV}$ (red). Right: Sum of standard model logs at NNLO and new physics. 
For $S_{\text{NP}}^{35}$ there are two distinct local minimum bubbles, the SM one with $R\sim 10^{17}$~GeV and the high-scale one with $R\sim 10^{35}$~GeV, while for $S_{\text{NP}}^{17}$ there is only one bubble. For these particular parameters for $S_{\text{NP}}^{35}$, the high-scale one has less total action and hence it dominates the rate.}
\label{fig:doublecorrectedaction}
\end{figure}

To see the new direction emerge, we need to keep track of two corrections to the tree-level quartic potential: the Standard Model logarithms and the high-scale operators. Each of these induces a slight change to the action, $\Delta S(R)$. The high-scale operators' correction can be calculated analytically; the standard model logarithms must be computed numerically. These corrections pick out a dominant value of $R$. Depending on the scale of new physics, we might get one or two local minima (see figure \ref{fig:doublecorrectedaction}).

Explicitly, taking the bounce to be the quartic bounce of size $R$ from Eq.~\eqref{confbubb} (with $\lambda\to \frac{\lambda}{4}$ and denoting this solution by $h_C^R$), the shift in the action from the high-scale operators is
\be
\Delta S_{\text{NP}}(R) = 
 \int \Omega_3 \rho^3d\rho \left(-\frac{1}{6 \Lambda^2} h^R_C(\rho)^6 + \frac{1}{8 M^4} h^R_C(\rho)^8 \right) = 
\frac{64 \pi ^2}{\lambda ^4} \left(\frac{4}{21 M^4 R^4}-\frac{\left| \lambda \right| }{15 \Lambda ^2 R^2}\right)
\ee
This action is minimized for:
\be
R_{\text{min}} = \sqrt{\frac{40}{7|\lambda|}} \frac{\Lambda}{M^2}
\ee
That is, as $M\to\infty$ and $\Lambda \to \infty$ holding $M/\Lambda$ fixed, tunneling is dominated by bubbles of smaller and smaller size.

Next we focus on the Standard Model corrections, which will pick out the minimum of $\leff = \frac{4}{h^4} V$. Of course, as we have argued, one cannot use the SM effective potential for tunneling calculations. However, there does exist a full effective action for the SM which one could in principle use, and we will assume the higher derivative terms are independent of $R$.
Evaluating this action on bubbles of size $R$, one finds a function $\Delta S_\text{SM}(R)$ which has its minimum at $R_\text{SM}$ where  $R_{\text{SM}}^{-1} = \mu_0 \approx 10^{17}~\GeV$.
Thus the full action is 
\be
S_\text{full} = S_{\text{SM}}^{\text{LO}}+\Delta S_{\text{SM}}(R) + \Delta S_{\text{NP}}(R)
\ee
The second term, $\Delta S_{\text{SM}}$, we know has a minimum around $R_\text{SM}$, and the third term, $\Delta S_{\text{NP}}$, has a minimum around $R\sim\frac{\Lambda}{M^2}$. So we generically expect the curve to look something like the right plot in Fig.~\ref{fig:doublecorrectedaction}.

When $\Lambda$ and $M$ are much larger than $\Mpl$, there are two local minima (see Fig.~\ref{fig:doublecorrectedaction}); the Standard Model one at $R_\text{SM}$, perturbed only slightly by the higher-dimension operators, and the higher-dimension one at $\sqrt{\frac{40}{7|\lambda|}} \frac{\Lambda}{M^2}$, perturbed only slightly by the standard model logs.

Since there are several bounces, the total tunneling rate is dominated by whichever has the smaller Euclidean action. The Standard Model bounce's action is given in the exponent of Eq.~\eqref{eqn:SMTunnelingRate}, and the higher-dimension operators' action is controlled by $\Lambda$ and $M$ according to Eq.~\eqref{eqn:num68}. Thus depending on the details of the higher-dimension operators, the Standard Model or the new physics bounce might dominate, \emph{no matter how high the scale $\Lambda$}.
This was the point illustrated by Fig.~\ref{fig:potentialtunneling}. 
Because of this possibility, the full tunneling rate is \emph{always} sensitive to physics at arbitrary high scales. The Standard Model calculations at low energy provides a lower bound on the tunneling rate, but arbitrarily high scale operators can always make the rate faster by adding new tunneling directions in field space.

\section{Summary and Conclusions\label{sec:conc}}
The first four sections of this paper attempt to provide a thorough exposition of the various ways tunneling rates are calculated in quantum mechanics and quantum field theory. While many of the methods discussed here are explained in the literature, 
we hoped that compiling them with explicit examples and additional commentary on subtle points rarely emphasized could be helpful. One path-integral method we call the {\it direct method} was introduced in~\cite{Andreassen:2016cff} and more details are given here for the first time. 
In the latter parts of this paper, we explored tunneling in quantum field theory and discussed the role that the effective potential plays. As an application to the Standard Model, we investigated the UV-sensitivity of the tunneling rate.

One critical observation about tunneling is that  two time scales must be well-separated for the tunneling-rate to even be well-defined. First, there is what we call the sloshing time $\sloshtime$, characterizing the time-scale for movement within the false vacuum well-region. One must average over times of order $\sloshtime$ in defining the rate $\Gamma$. The other time scale we call $\nonlintime$. It represents the timescale for the transmitted wavefunction to start propagating back into the false-vacuum. 

In many presentations, one takes $T \to \infty$ to find the decay rate. However, as we repeatedly emphasize, this has to be done carefully. In the strict $T\to \infty$ limit, the system ends up in the true vacuum, at rest. Much of the subtlety in calculating decay rates can be traced to enforcing the double limit $T\gg \sloshtime$ and $T \ll \nonlintime$.

In QM, one can enforce $T\ll \nonlintime$ by using radiative outgoing-wave only boundary conditions. These boundary conditions, by definition, prevent the transmitted wave from returning. Moreover, the boundary conditions are unphysical, and so the energy eigenstates of the Hermitian Hamiltonian can have complex energy eigenvalues. The imaginary part of the energy of a resonance localized in the false vacuum can be readily  identified with the decay rate. This was shown in Section~\ref{sec:QM}.

The outgoing-boundary condition way of enforcing $T\ll \nonlintime$ is not readily generalizable to multi-dimensional systems. In the path-integral derivation, originally due to Callan and Coleman~\cite{Callan:1977pt}, this time scale is left implicit. Nevertheless, we show that if one blindly takes $T\to\infty$, the energy $E_0$ picked out by this method is not the quasistable false-vacuum state of interest, but rather the true vacuum (in the path integral, the dominant saddle point is not the bounce but rather {\it the shot}, a solution that stays in the true vacuum for nearly all time). To isolate the false vacuum energy as $E_0$, one must change the contour on integration to be the steepest descent contour through the false vacuum. 

We spent some time in this paper discussing various subtleties in performing the analytic continuation/contour deviation and saddle point approximation properly. One must account for the locations of the all the relevant saddle points in the complex plane, how the integration contour deforms as the continuation is done, and how the imaginary part is to be extracted so that one is using an asymptotic expansion consistently. Importantly, the relevant contour is the contour of steepest descent through the false vacuum saddle point.

One hiccup of the potential-deformation method that we were not able to resolve is ``why is the imaginary number computed through this method the decay rate of the false vacuum?'' Unlike the outgoing-boundary conditions prescription, we found no intuitive physical narrative to connect this imaginary number to the decay rate of interest. It does seem that deforming the contour of integration to the steepest descent through the false vacuum has some connection to enforcing $T \ll \nonlintime$, which prevents the true vacuum from dominating. However, we do not know how to make this precise. Moreover, the imaginary part depends on how the analytic continuation is done.
We expect that there should be a  proof that this method, with some precise prescription for how the contour deformation is to be done, will always generate the decay rate of 
interest.
It would also be great to see a proof of the universality of the famous factor of $\frac{1}{2}$ (cf. Section~\ref{sec:potmethod}),
other than the indirect proof we provide here through the direct method. Such investigations could provide fruitful avenues for future research.

In the direct method, explored at length in Section~\ref{sec:potmethod},  the computation of tunneling rates is approached from the starting point of the casual propagator. Inspired by the WKB approximation, the direct method computes the rate to propagate from the false vacuum through a barrier. The condition $T \ll \nonlintime$ is imposed by assuming that once the particle exits the barrier, it will never pass back through. The end result of this method is a non-perturbative  formula for the tunneling rate.
It is exact, up to the exponentially small corrections that are an inherent ambiguity in the definition of the decay rate.
 The formula relates $\Gamma$ to a ratio of two path integrals. In the saddle point approximation to this formula, the numerator is dominated by the Euclidean bounce solution and the denominator by the static false vacuum solution. The final prescription is in agreement with previous results at NLO. One corollary of the direct derivation provided here is that it  
relates the bounce action to the actual tunneling rate thereby validating the universality of the factor of $\frac{1}{2}$. 

The remainder of this paper was devoted to reviewing and expounding some important aspects of tunneling calculations in quantum field theory. We discussed how the energy functional $U[\phi]$ can be productively visualized along a tunneling direction. We then showed how one can get good intuition for tunneling and bounces from approximate analytic solutions, rather than exact numerical ones. In particular, we derived that the field at the center of the bounce $\phi_0 = \phi_B(\rho=0)$ will generically be determined by the scale where $\partial_\phi [\frac{1}{\phi^4} V(\phi)] =0$. This is in accordance with renormalization-group based arguments predicated on minimizing $\Gamma \sim \exp[ - \frac{1}{\lambda(\mu)}]$, but is conceptually cleaner: it is just a shortcut to solving classical equations of motion. 

One point we thought worth clarifying is the use of the effective potential in tunneling calculations. It is {\it not} ok to use an effective potential for tunneling, e.g. using it to find bounces. We also showed that the terms in the effective action with derivatives are equally important as the NLO corrections to the potential, and the higher derivative terms are generally unknown. To verify this explicitly, we computed the 4-derivative terms at 1-loop in a scalar field theory and compared to what one gets using the full NLO bounce action. The potential terms,  the 4-derivatives, and even higher-order terms, all contribute to the decay rate at the same order. 

The last section was devoted to some comments on vacuum stability in the Standard Model. 
It would be good to check that the decay rate in the Standard Model is gauge-invariant by an explicit calculation.  Although the rate {\it must} be gauge invariant, and general non-perturbative proofs show that it is, there may be subtleties with the
power counting that require special care when working in perturbation theory. Such was the case for the absolute stability bound~\cite{Andreassen:2014gha}. As a small step in this direction, we argued that the appropriate power counting should
be the usual loop expansion in $\hbar$, not the $\lambda \sim \hbar$ counting of the Coleman-Weinberg model. Even then,
one cannot use a resummed potential or action, since these mix orders in perturbation theory. Conveniently, we showed
that for the effective potential at least resummation is not necessary: a fixed order expansion using $\mu\sim 10^{17} \GeV$
agrees qualitatively quite well with the full resummed potential.

Finally, we included some remarks on the UV sensitivity of the Standard Model decay rate. It has been argued that the rate is sensitive to Planck scale physics due to a coincidence between the critical bubble size in the gravity-free Standard Model, and the Planck scale. We showed that this is not true. No matter how high the Planck or UV scale is, tunneling rates will always be UV sensitive. Decoupling arguments simply do not apply to the lower bound on tunneling rates. This result is in agreement with other recent work \cite{DiLuzio:2015iua}. Without any gravitational or beyond-the-standard model physics, the lifetime of our universe appears to be around $10^{600}$ years. Although physics beyond $10^{18}~\GeV$ could make this lifetime shorter, it seems hard for new physics at this scale to make the lifetime longer. Thus if, for whatever reason (e.g.~\cite{Isidori:2001bm}), one can argue that our universe must be absolutely stable, then there must be sub-Planckian physics beyond the Standard Model. Otherwise, unless we get a better handle on quantum gravity, the  fate of our 
universe 
will remain uncertain.

\section*{Acknowledgements}
\label{sec:acknowledgements}
The authors would like to thank D. Harlow, M. Marino, A. Strumia, M. Unsal and E. Weinberg for helpful discussions.
This research was supported in part by the U.S. Department of Energy, under grant DE-SC0013607. AA is supported in part by the Stolt-Nielsen Fund for Education of the American-Scandinavian Foundation and the Norway-America Association.

\newpage
\appendix

\section{Complex energies and decay rates in a square well \label{app:Gammasimple}}
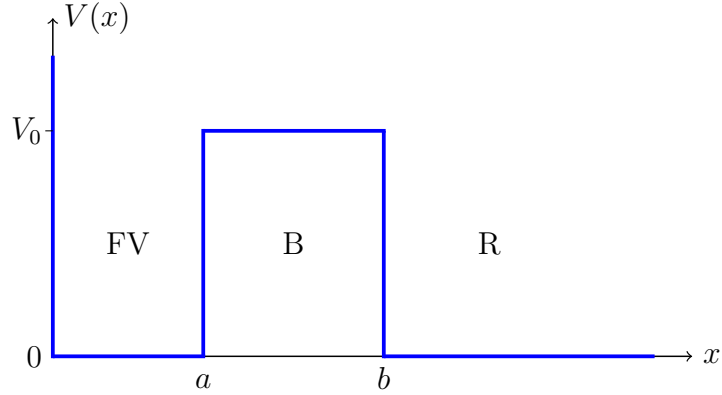
\begin{figure}[t]\begin{center}
\begin{tikzpicture}
\draw[line width = 0.2mm,->] (0,0)--(8.5,0);
\draw[line width = 0.2mm,->] (0,0)--(0,4.5);
\draw[blue,line width = 0.5mm,-] (0,-0.025)--(0,4); 
\draw[blue,line width = 0.5mm,-] (0,0)--(2.025,0);
\draw[blue,line width = 0.5mm,-] (2,0)--(2,3.025);
\draw[blue,line width = 0.5mm,-] (2,3)--(4.425,3);
\draw[blue,line width = 0.5mm,-] (4.4,-0.025)--(4.4,3);
\draw[blue,line width = 0.5mm,-] (4.4,0)--(8,0);
\node [right] at (0,4.5) {$V(x)$};
\node [below] at (2,-0.1) {$a$};
\node [below] at (4.4,0) {$b$};
\node [right] at (8.5,0) {$x$};
\draw[-] (0,3)--(-0.1,3);
\node [left] at (0,3) {$V_0$};
\node [left] at (0,0) {$0$};
\node at (1,1.5) {\FV};
\node at (3.2,1.5) {\B};
\node at (5.8,1.5) {\R};
\end{tikzpicture}
\caption{
Square potential well. The potential is divided into three regions: $\FV$ is the false vacuum, $\B$ is the barrier, and $\R$ is the destination region where to which a state initially
localized in $\FV$ will decay over time.
\label{fig:simp}
}
\end{center}
\end{figure}

In this appendix, we compute the decay rate for the simple square-well potential shown in Fig.~\ref{fig:simp}. The explicit results here may help
elucidate some of the general statements from Section~\ref{sec:def}. Related calculations can be found, for example, in~\cite{Razavy}.

Suppose we have an initial state $|\psi\rangle$ localized in the $\FV$ region at $t=0$.
 This can be a Gaussian, a delta function, or simply a constant in the $\FV$ region.
At a later time, the state is $|\psi(t)\rangle = e^{-i H t}|\psi\rangle$. 
The probability that we find the state in region $\FV$ at a later time is given by 
\be
\mc{P}_{\FV}(t) = \int_{\FV} dx~|\langle x|\psi(t)\rangle|^2 = \int^a_0 dx~|\psi(x,t)|^2\label{eq:prob}
\ee

To compute the probability, we start by decomposing the wavefunction into energy eigenstates.
Labeling these states $\phi_p$ by momentum $p$, we can write write $\psi(x,t)$
\be
\psi(x,t) = \int^\infty_0 \frac{dp}{2\pi}  \int^\infty_0dy~\psi(y)\phi_p^*(y)\phi_p(x) e^{-i \frac{p^2}{2m} t}\label{psiexpand}
\ee
where $\psi(y)\equiv \psi(y,t=0)$. 

Since region $\R$ is infinite in extent, there are energy eigenstates for any energy $E = \frac{p^2}{2m}$. Of these, some will be resonances. These resonances come in bands of width $\Gamma_i$ around
energies $E_i$. The resonant energies $E_i$ are close to what the bound state energies in the $\FV$ region would be if we disallowed tunneling, for example, with $b\rightarrow \infty$.
For finite $b$, the resonant energies broaden across a band and have support in the region $\R$. 

More precisely, the exact energy eigenstates are
\eqn{
\phi_p(x) =  \left\{
     \begin{array}{lcl}
       \phi_p^{\FV}(x) = \frac{2}{N_p}\sin(p x) &,& \qquad  0<x<a\\
       \phi_p^{\B}(x) = \frac{1}{N_p}\left[A_p e^{\kappa (x-a)} + B_p e^{-\kappa(x-a)}\right] &,& \qquad a<x<b\\
       \phi_p^\R(x) = \frac{1}{N_p}\left[C_p e^{ip(x-b)} + D_p e^{-ip(x-b)}\right]&,& \qquad b<x
      \end{array}
   \right.\label{eq:defphi}}
where $\kappa = \sqrt{2mV_0-p^2}$ and
\eqn{
A_p &= \sin(pa) +\frac{p}{\kappa}\cos(pa), \\
B_p &= \sin(pa) - \frac{p}{\kappa}\cos(pa) \\
C_p &= \frac{1}{2}\left(1 - i\frac{\kappa}{p}\right) A_p e^{W_p}+\frac{1}{2}\left(1 + i\frac{\kappa}{p}\right) B_p e^{-W_p},  \label{eq:Cform}\\
D_p &=  \frac{1}{2}\left(1 + i\frac{\kappa}{p}\right)A_p e^{W_p}+\frac{1}{2}\left(1 - i\frac{\kappa}{p}\right) B_p e^{-W_p}\label{eq:Dform}}
Here $W_p = \int_a^b dx \kappa  = (b-a)\kappa$ is the usual WKB exponent.

The factor $N_p$ can be computed by requiring that the states have the
usual normalization
\begin{equation}
  \int_0^{\infty} d x~\phi_p (x) \phi^*_{p'} (x) = \delta (p - p') \label{normdef}
\end{equation}
The only place such a $\delta$-function can come from is the integral over the
region $b < x < \infty$. To see this, write 
\begin{align}
  \int_0^{\infty} d x~\phi_p (x) \phi_{p'}^* (x) 
&= \int_b^{\infty} d x~
  \phi_p^{\R} (x) \phi^{\R *}_{p'} (x) + \int_0^b dx~ \phi_p
  (x) \phi_{p'}^* (x)
\\
  &= \pi \frac{C_p C_{p'}^* + D_p D_{p'}^*}{N_p N^*_{p'}} \delta (p - p') 
\end{align}
The second integral, from $0$ to $b$ has exactly vanished.
Comparing with Eq.~\eqref{normdef} and noting that $C_p=D_p^*$ for real $p$, we can write $|N_p|^2$ as an analytic function of $p$:
\begin{equation}
  | N_p |^2 =2\pi  C_p  D_p \label{Np}
\end{equation}

We now see, that up to an overall phase, $\phi_p^{\R} (x) = \cos(p x)$ for any $p$. That is, all the wavefunctions are order $1$  in the region $\R$. In the $\FV$ region, the wavefunctions vary in size as $\frac{1}{|N_p|}$.
Eqs.~\eqref{eq:Cform}, \eqref{eq:Dform} and \eqref{Np} imply that $N = \sqrt{2\pi} C \sim A e^W + B e^{- W}$. For $W \gg
1$, where the WKB approximation is supposed to work, $N \sim A e^W$. Thus,
generically, $\phi_p^\FV \sim e^{- W}$. So for most values of $p$, the
wavefunction has support almost entirely outside the well, and is
exponentially suppressed in the well. The only time it can have reasonable
support in the well is when $A\lesssim e^{-2W}$; indeed for $A\lesssim e^{-2W}$, we see  $N \lesssim e^{- W}$ and $\phi_p^\FV
\sim e^W$.

The momenta for which $|N_p|^2$ is minimized are the resonance momenta. They are exponentially close  (within $\mc{O}(e^{-2W})$) to the zeros of $A_p$. The zeros of $A_p$ correspond to the bound states in the limit $b\rightarrow \infty$ or $W\rightarrow \infty$. This can be seen by considering the condition for wavefunction normalization in this limit; it is precisely $A_p=0$ that stops the wavefunction's growth as $x$ increases, and discretizes the spectrum. For finite $b$, tunneling through the barrier shifts the bound state energies by $\mc{O}\left(e^{-2W}\right)$ and they become the resonant energies.

\subsection{Relating the probability to the pole}
To calculate the probability $P_\FV(t)$, let us assume for simplicity that the initial wavefunction $\psi(x)\equiv \psi(x,t=0)$ only has significant overlap
with modes whose energies are close to $E_0$, where $E_0$ is the smallest real energy for which $|N(E)|^2$ has a local minimum. 
Here, we are interchanging the momentum label $p$ with an energy label $E$, where $p^2 = 2mE$. So we will write $C(E),D(E)$ etc. rather than $C_p$, $D_p$, etc.

To compute the probability it is helpful to consider imaginary energies. To do so, we first we need to analytically continue $|N(E)|^2$ which can be done by writing it in the form of Eq.~\eqref{Np}: $|N(E)|^2 = 2\pi C(E) D(E)$. 
This analytic function has zeros in the complex plane, and the zeros come in pairs. Indeed, from looking at the form of $C$ and $D$ in Eq.~\eqref{eq:Cform}, we see that if $E = a + ib$ is a zero of $C$, then $E^*=a-ib$ will be a zero of $D$. From the form of $\phi^R$ in Eq.~\eqref{eq:defphi}, we see that $C=0$ corresponds to incoming boundary conditions and $D=0$ to outgoing boundary conditions. 
The first pair of zeros for $|N|^2 = 2\pi C D$ are at $E=E_0 \pm \frac{i}{2}\Gamma_0$.
As we will confirm, the $\frac{\Gamma_0}{E_0} \sim e^{-2W}\ll 1$ as in the WKB approximation. 

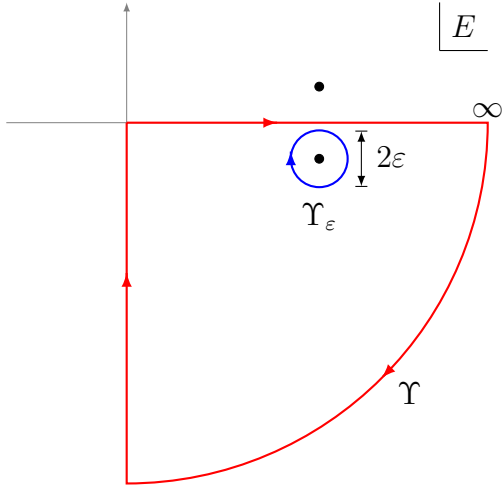
\begin{figure}[t]
\centering
\begin{tikzpicture}[scale=1.6]
\tikzset{>=latex}
\draw[help lines,-] (-1,0) -- (3,0) coordinate (xaxis);
\draw[help lines,->] (0,-3) -- (0,1) coordinate (yaxis);
\node at (2.8,0.8) {$E$};
\draw[-] (2.6,1)--(2.6,0.6);
\draw[-] (2.6,0.6)--(3,0.6);
\path[draw,line width=0.8pt,red] (0,0)  -- (3,0) arc (0:-90:3) -- (0,0);
\draw[->,thick,red] (1.25,0)--(1.25001,0);
\draw[->,thick,red] (0,-1.25001)--(0,-1.25);
\draw[->,thick,red] (2.13,-2.11)--(2.12,-2.12);
\draw[thick,blue] (1.6,-0.3) circle (1.3ex);
\draw[->,thick,blue] (1.364,-0.24)--(1.364,-0.239);
\draw[<->] (1.95,-0.53)--(1.95,-0.07);
\draw[-] (1.9,-0.53)--(2,-0.53);
\draw[-] (1.9,-0.07)--(2,-0.07);
\node at (2.2,-0.28) {$2\varepsilon$};
\draw[fill] (1.6,-0.3) circle (.2ex);
\draw[fill] (1.6,0.3) circle (.2ex);
\node at (2.35,-2.25) {$\Upsilon$};
\node at (1.6,-0.78) {$\Upsilon_\varepsilon$};
\node at (3,0.1) {$\infty$};
\end{tikzpicture}
\caption{The contour $\Upsilon$ used in Eq. (\ref{eq:Cintegral}) enclosing the pole $E = E_0 - \frac{i}{2}\Gamma_0$, and the smaller contour $\Upsilon_\varepsilon$ allowing the use the Laurent series. \label{fig:Econt}}
\end{figure}

Plugging Eq.~\eqref{eq:defphi}  into Eq.~\eqref{psiexpand} we find
\be
\psi(x,t) =\int^a_0 dy~\psi(y) \int^\infty_0 dE~\left(\frac{1}{2\pi}\sqrt{\frac{m}{2E}}\right)\frac{\sin(\sqrt{2mE}x)\sin(\sqrt{2mE}y)}{2\pi C(E) D(E)} e^{-iE t}
\ee
where the spatial integral is over $y \in (0,a)$ since $\psi(y,t=0)$ only has support in the \FV~region.  
This is convergent when $E$ has a negative imaginary part. So we can deform the contour to write
\be
\psi(x,t) =\int_0^a dy~\psi(y) \left[F(x,y,t)+G(x,y,t)\right]
\ee
where
\be
F(x,y,t) = \oint_\Upsilon dE~\left(\frac{1}{2\pi}\sqrt{\frac{m}{2E}}\right)\frac{\sin\left(x\sqrt{2mE}\right)\sin\left(y\sqrt{2mE}\right)}{2\pi C(E) D(E)} e^{-iE t}\label{eq:Cintegral}
\ee
and
\begin{align}
G(x,y,t) &= \int^{-i\infty}_0 dE~\left(\frac{1}{2\pi}\sqrt{\frac{m}{2E}}\right)\frac{\sin\left(x\sqrt{2mE}\right)\sin\left(y\sqrt{2mE}\right)}{2\pi C(E) D(E)} e^{-iE t}\nonumber\\
 &=  -i \int^\infty_0 d\cE ~\left(\frac{1}{2\pi}\sqrt{\frac{m}{-i2 \cE}}\right)\frac{\sin\left(x\sqrt{-i2m \cE}\right)\sin\left(y\sqrt{-2m\cE}\right)}{2\pi C(-i\cE) D(-i\cE)} e^{-\cE t}
\label{eq:Mintegral}
\end{align}
The contour $\Upsilon$ for the integral in $F(x,y,z)$ is shown  in Fig.~\ref{fig:Econt}.

The $F(x,y,t)$ integral can be calculated by replacing the contour $\Upsilon$ with $\Upsilon_\varepsilon$, a circle of radius $\varepsilon$ around the pole.
This lets us expand $C(E)D(E)$ around its zero and use its lowest order term. That is, we can write $C(E)D(E) =  C(E_0-\frac{i}{2}\Gamma_0)D^\prime(E_0-\frac{i}{2}\Gamma_0)(E-E_0+\frac{i}{2}\Gamma_0) +\cdots$ and use the residue theorem to give
\eqna{
F(x,y,t) &= 2\pi i\left(\frac{1}{2\pi}\sqrt{\frac{m}{2\left(E_0-\frac{i}{2}\Gamma_0\right)}}\right)\frac{\sin\left(x\sqrt{2m\left(E_0-\frac{i}{2}\Gamma_0\right)}\right)\sin\left(y\sqrt{2m\left(E_0-\frac{i}{2}\Gamma_0\right)}\right)}
{2\pi C(E_0-\frac{i}{2}\Gamma_0)D^\prime(E_0-\frac{i}{2}\Gamma_0)} e^{-i\left(E_0-\frac{i}{2}\Gamma_0\right) t}\\
&= \frac{i}{2\pi}\sqrt{\frac{m}{2E_0}}\frac{\sin\left(x\sqrt{2mE_0}\right)\sin\left(y\sqrt{2mE_0}\right)}{C(E_0)D^\prime(E_0)} e^{-iE_0 t}e^{-\frac{1}{2}\Gamma_0 t}\left[1 + \mc{O}\left(\frac{\Gamma_0}{E_0}\right)\right]
\label{eq:Capprox}}
where in the last line we have used $\frac{\Gamma_0}{E_0}\ll1$.

To evaluate $G(x,y,t)$, we begin by replacing $\cE \to E_0 \alpha$ in Eq.~\eqref{eq:Mintegral} where $\alpha$ is dimensionless. Then
\be
G(x,y,t) = -i E_0 \left(\frac{1}{2\pi}\sqrt{\frac{m}{-i2E_0}}\right) \int^\infty_0 d\alpha ~\frac{1}{\sqrt{\alpha}}\frac{\sin(\sqrt{-i2mE_0}x \sqrt{\alpha})\sin(\sqrt{-i2mE_0}y\sqrt{\alpha})}
{2\pi C(-iE_0 \alpha)D(-i E_0 \alpha)} e^{-\alpha(E_0t)}
\ee
Now let us assume that $t >> \frac{1}{E_0} \sim \sloshtime$. This is one of the required conditions for having a well-defined decay rate. Then 
$E_0 t \gg 1$ and, due to the exponential factor, only very small values of $\alpha$ contrite. Thus we can expand the prefactor for small $\alpha$, giving
\eqna{
G(x,y,t) &= -i E_0 \left(\frac{1}{2\pi}\sqrt{\frac{m}{-i2E_0}}\right) \int^\infty_0 d\alpha ~\frac{1}{\sqrt{\alpha}}\frac{-i2mE_0 x y \alpha}{|N(0)|^2} \left[1+ \mc{O}\left(\alpha\right)\right] e^{-\alpha(E_0t)}\\
&= -(1+i)(E_0 m)^\frac{3}{2}\frac{xy}{4\sqrt{\pi} |N(0)|^2} (E_0 t)^{-\frac{3}{2}}\left[1 + \mc{O}\left((E_0 t)^{-1}\right)\right]\label{eq:Mapprox}}

Now note that $F \sim \exp(-\frac{1}{2} \Gamma_0 t)$ while $G \sim (E_0 t)^{-3/2}$.  Thus if $t$ is not too large $t \lesssim \Gamma_0^{-1} \sim \nonlintime$, $F\gg G$ and
\eqn{
\psi(x,t) &\approx \left[\int_0^a dy~\psi(y)\sin(\sqrt{2mE_0}y)\right]\frac{1}{4\pi^2}\sqrt{\frac{m}{2E_0}}\frac{\sin(\sqrt{2mE_0}x)}{C(E_0)D^\prime(E_0)} e^{-iE_0 t}e^{-\frac{1}{2}\Gamma_0 t}
}
where $\approx$ means that terms $\mc{O}((E_0 t)^{-\frac{3}{2}})$ and higher order in $\frac{\Gamma_0}{E_0}$ have been dropped.
The probability in Eq.~\eqref{eq:prob} therefore takes the form
\eqn{\underset{E_0^{-1}\ll t\ll \Gamma_0^{-1}}{P_\FV(t)} \approx \text{const} \times e^{-\Gamma_0 t}}
and therefore $\Gamma_0$ is indeed the rate.

 In other words, we have established a direct connection between the complex zeros of $D(E)$ and the decay rate.
From Eq.~\eqref{eq:defphi}, the complex zeros of $D(E)$ correspond to outgoing-only plane waves in region $\R$. In this way, the connection between Gamow-Siegert outgoing boundary conditions and decay rates is made precise
(at least in this example).

Note that the assumption $t< \nonlintime \sim \Gamma_0^{-1}$ was essential. At very late times $t \gg \Gamma_0^{-1}$ then $G \gg F$. In this limit,
\eqn{
\underset{E_0^{-1}\ll \Gamma_0^{-1} \ll t}{P_\FV(t)} \approx \text{const} \times \frac{1}{(E_0 t)^3}
}
This is the non-linear behavior. In this regime there is not a well-defined decay rate.

In this calculation, we assumed we were dealing with an initial wavefunction dominated by energy eigenfunctions close to $E_0$. If this were not the case, we would have other resonances to worry about; each corresponding to poles of the form $E = E_n - \frac{i}{2}\Gamma_n$. The above analysis follows through in the same way, and the result is a sum of the form $\sum_n a_n e^{-\Gamma_n t}$ for some $a_n$. 
After enough time, only the dominant term, with $P \sim e^{-\Gamma_0 t}$ will remain, while the others decay away.

\subsection{Explicit computation of $\Gamma$}
We have shown that the decay rate $\Gamma$ is determined by the imaginary part of the energy for which $D(E)$ vanishes. Now let us calculate $\Gamma$ explicitly.

First, we observe that the complex zeros are exponentially close to the resonant energies. These resonant energies $E_R = \frac{p_R^2}{2m}$ are the zeros of $A(E)$ or equivalently of $A_p$.
Setting $A_p=0$ implies
\be
\sin(p_R a) = -\frac{p_R}{\sqrt{p_R^2+\kappa_R^2}},\qquad \cos(p_R a) = \frac{\kappa_R}{\sqrt{p_R^2+\kappa_R^2}},
\label{sincos}
\ee
where $\kappa_R \equiv \sqrt{2m V_0 - p_R^2}$.
We can then find the complex zeros of $D_p$ by expanding perturbatively in
\be
\delta \equiv e^{-W_p}=e^{-\kappa_R(b-a)}
\ee
Both the real and the imaginary part of the complex energies should differ from the resonant energies by amounts of order $\delta^2$. 

Next, we write $p = p_R + \delta^2 p_C + \cO(\delta^4)$ and expand 
\be
D_p =  \frac{1}{2}\left(1 + i\frac{\kappa}{p}\right)\left[ \sin(pa) +\frac{p}{\kappa}\cos(pa) \right] e^{W_p}+\frac{1}{2}\left(1 - i\frac{\kappa}{p}\right)\left[ \sin(pa) - \frac{p}{\kappa}\cos(pa) \right]e^{-W_p}
\ee
to order $\delta$. Setting $D_p=0$ and using Eq.~\eqref{sincos} to simplify the answer, we find
\be
p_C = \frac{2p_R \kappa_R^2}{(p_R + i \kappa_R)^2(1+ a \kappa_R)}
\ee
and therefore 
\be
\Gamma = -2 \Im\frac{p^2}{2m} = \frac{8 p_R^3 \kappa_R^3}{ m (1+a \kappa_R)(p_R^2+\kappa_R^2)^2 } e^{-2W}
\label{Gammadirect}
\ee
Note that unlike Eq.~\eqref{GofT} the prefactor does depend on the height of the barrier, $V_0$. 

It is perhaps informative to compare this calculation to the result of Eq.~\eqref{Gofpsi}. Note that Eq.~\eqref{Gofpsi} is independent of the normalization, so we can set $N_p=1$  in Eq.~\eqref{eq:defphi}. Then we find
\be
|\phi_E(b)|^2 = \frac{16 p_R^2 \kappa_R^2}{(p_R^2+\kappa_R^2)^2} \delta^2
\ee
\be
\int_0^a dx |\phi_E(x)|^2= 2a+ \frac{2\kappa_R}{p_R^2+\kappa_R^2} + \cO(\delta^2)
\ee
and
\be
\int_a^b dx |\phi_E(x)|^2= \frac{1}{p_R^2  + \kappa_R^2}\frac{2p_R^2}{\kappa_R} + \cO(\delta^2)
\ee
So that
\be
\Gamma = \frac{p_b}{m} \frac{|\phi_E(b)|^2}{\int_0^b d x|\phi_E(x)|^2 }
=  \frac{8 p_R^3 \kappa_R^3}{ m (1+a \kappa_R)(p_R^2+\kappa_R^2)^2 } e^{-2W}
\ee
in agreement with Eq.~\eqref{Gammadirect}.

On the other hand, if we use the real-energy eigenstate in Eq.~\eqref{Gofpsi}, we would get
\be
\Gamma_\text{real} = \frac{p_b}{m} \frac{|\phi_E(b)|^2}{\int_0^b d x|\phi_E(x)|^2 }
=  \frac{2 p_R^3 \kappa_R}{ m (1+a \kappa_R)(p_R^2+\kappa_R^2) } e^{-2W}
\ee
Taking the ratio we find
\be
\frac{\Gamma_\text{real}}{\Gamma} = \frac{1}{4} + \frac{p_R^2}{4\kappa_R^2}
\ee
thus using the real energy eigenstates gets the rate right to factor of order 1. The factor of 4 in the large $\kappa_R$ (large $V_0$) limit 
can be traced to the mode that grows inside the barrier from $a$ to $b$. This mode is exactly zero for the real energy state.

\section{Changing to collective coordinates}\label{app:jacobian}
The Gaussian integral $\int d\xi_0 \exp(- \lambda_0 \xi_0^2 )$ is divergent when $\lambda_0$ is zero. This happens for the zero mode corresponding
to translation invariance around a bounce. In order to regulate this divergence, the standard procedure is to
trade $\xi_0$ for a collective coordinate $\tau_0$ so that the integral over $\tau_0$ gives simply a factor of $\bigtau$ which can
then be divided out to get a decay rate~\cite{Callan:1977pt,Marino, ZinnJustin:2002ru}.
In trading $\xi_0$ for $\tau_0$ one also must adjust the other modes $\xi_i$ to a new orthonormal basis $\zeta_i$. The change
of variables from $\{\xi_i\}$ to $\{\tau_0, \zeta_i\}$ results in a  Jacobian factor $J(\tau_0,\zeta_i)$. 
This appendix addresses some subtleties in the change of variables that we have not seen in decay rate literature, but was understood in pioneering works on collective coordinates \cite{Gervais:1975yg}.

In the evaluation of the saddle point approximation around a bounce or instanton denoted $\bar{x}$, one naturally parametrizes paths by the eigenfunctions $x_n$ of $S_E''$ with eigenvalues $\lambda_n$:
\begin{equation}
\label{eqn:definexi}
x^{\xi_0, \xi_1,\dots}(\tau) = \overline{x}(\tau) + \sum_{n=0}^\infty\xi_n x_n(\tau) 
\end{equation}
One of these modes, $x_0 = \partial_\tau \bar{x}$ has an eigenvalue of exactly zero: $\lambda_0=0$. 
This zero mode corresponds to an infinitesimal shift. 
 Large shifts in $\tau$ are an approximate symmetry of the path integral using Dirichlet~\cite{Callan:1977pt} 
boundary conditions an an exact symmetry with periodic boundary conditions~\cite{Marino, ZinnJustin:2002ru}.
In order to make the symmetry manifest, it is helpful to have one of the coordinates
parametrizing paths be the shift by $\tau_0$ rather than the addition of an $\xi_0$ amount of the first derivative.

We might write
\begin{equation}
\label{eqn:definetxi}
x^{\tau_0, \xi_0, \xi_1,\dots}(\tau) = \bar x(\tau-\tau_0)+\sum_{i=1}^\infty\xi_i x_i(\tau)
\end{equation}
as in~\cite{Marino}. 
But unfortunately this parametrization is not complete. 
 To see that, integrate both sides of Eq.~\eqref{eqn:definetxi} against $x_0(t)$. Using Eq.~\eqref{eqn:definexi} for the left-hand side, we
 find $\xi_0(\tau_0)=\int d\tau x_0(\tau)\bar x(\tau-\tau_0)$, which is bounded. In contrast, the integration over $\xi_0$ should go from $-\infty$ to $\infty$ to parametrize all paths. In particular, the parametrization in 
 Eq.~\eqref{eqn:definetxi} does not cover exactly the large-$\xi_0$ fluctuations which caused the problem with the Gaussian integrations the first place.

A better parametrization is
\begin{equation}
\label{eqn:definetzeta}
x^{\tau_0,\zeta_1, \zeta_2,\dots}(\tau) = \overline{x}(\tau-\tau_0) + \sum_{n=1}^\infty\zeta_n x_n(\tau-\tau_0)
\end{equation}
Here we have used $\zeta_n$ instead of $\xi_n$ for $n>0$ since the coordinates for all the modes generically change when we change variables.  This parametrization is complete. 

Next, we want to calculate the Jacobian $J(\tau_0, \zeta)$ between the parametrizations Eq.~\eqref{eqn:definexi} and Eq.~\eqref{eqn:definetzeta}. There are two subtleties in this calculation that are often overlooked (e.g. in~\cite{Kleinert}):
\begin{enumerate}
\item Because the $\zeta_n$ are not the same as the $\xi_n$, this Jacobian is really the determinant of a nontrivial infinite-dimensional matrix; it is not simply equal to $d\xi_0/d\tau_0$. 
\item Because $\bar x(\tau)$ breaks the time-translation symmetry, one must show that  $J(\tau_0,\zeta)$ is independent of $\tau$. For example, in~\cite{Kleinert}, only $J(0,\zeta)$ is calculated and assumed
equal to  $J(\tau_0,\zeta)$ (cf. Eqs. (17.103) and (17.108)).\footnote{
Actually, in the direct path integral method, described in Section~\ref{sec:dirmethod}, only $J(0,0)$ is needed because
of the $\delta$-function in the path integral. In the conventional potential-deformation method, the full $J(\tau_0, \zeta)$ is needed.}
\end{enumerate}

To calculate the Jacobian, we write the $\xi_n$ as a function of $\zeta_m$ and $\tau_0$
\eqna{\xi_n &= \int \left[x(\tau)-\overline{x}(\tau)\right]x_n(\tau)~d\tau\\
&=  \int \left[\overline{x}(\tau-\tau_0)-\overline{x}(\tau) + \sum_{m=1}^\infty\zeta_m x_m(\tau-\tau_0)\right]x_n(\tau)~d\tau}
which means that
\eqna{\frac{\partial \xi_n}{\partial \zeta_m} &= \int x_n(\tau)x_m(\tau-\tau_0)~d\tau\\
\frac{\partial \xi_n}{\partial \tau_0} &=\int \left[-\dot{\overline{x}}(\tau-\tau_0)- \sum_{m=1}^\infty\zeta_m \dot{x}_m(\tau-\tau_0)\right]x_n(\tau)~d\tau\\
&= -\sqrt{\frac{S_E[\overline{x}]}{m}}\int x_n(\tau)x_0(\tau-\tau_0)~d\tau - \sum_{m=1}^\infty\zeta_m \int \dot{x}_m(\tau-\tau_0)x_n(\tau)~d\tau}
To proceed, it is useful to define the orthogonal matrix $U$:
\begin{equation}
U_{nm}(\tau_0) \equiv \int d\tau x_n(\tau)x_m(\tau-\tau_0)
\end{equation}
$U$ is orthogonal because both $\{x_i(\tau)\}$ and $\{x_i(\tau-\tau_0)\}$ are complete bases. The derivative matrix is:
\begin{equation}
\left(
\begin{array}{c|ccc}
\vdots & &\vdots &\\
-\sqrt{\frac{S_E}{m}}U_{n0} +\sum\zeta_m\dot U_{nm} & U_{n1} & U_{n2} & \cdots\\
\vdots & &\vdots &\\
\end{array}
\right)
\end{equation}
The determinant is linear in the first column, so the determinant is:
\begin{equation}
\label{eqn:vndet}
J(\tau_0, \zeta) = \abs{-\sqrt{\frac{S_E}{m}}\det U+\det\left(
\begin{array}{c|ccc}
\vdots & &\vdots &\\
v_n & U_{n1} & U_{n2} & \cdots\\
\vdots & &\vdots &\\
\end{array}
\right)}
\end{equation}
Where the vector $v$ is defined by $v_n\equiv\sum\zeta_m\dot U_{nm}$. Since $U$ is orthogonal, we can decompose $v$ in terms of the columns of $U$; $v_n=\sum_kc_k U_{nk}$. Then the second term in Eq.~\eqref{eqn:vndet} is a linear combination of determinants of $U$ with the zeroth column replaced by the $k^{\text{th}}$ column. This determinant is simply 0 if $k\neq0$ and $\det U$ if $k=0$. Since $U$ is orthogonal, $\det U=1$, so we have:
\begin{equation}
J(\tau_0, \zeta) = \abs{-\sqrt{\frac{S_E}{m}}+c_0}
\end{equation}

The coefficient $c_0$ is simply the $k=0$ component of the vector $v_n$ decomposed into the columns of $U$:
\begin{equation}
c_0 = \sum_{n=0}^\infty v_n U_{n0} = \sum_{n=0}^\infty \sum_{m=1}^\infty \zeta_m\dot U_{nm}U_{n0} = -\sum_{m=1}^\infty \zeta_mr_m
\end{equation}
where
\begin{equation}
r_m\equiv\int d\tau \dot x_m(\tau)x_0(\tau)
\end{equation}
So all together the Jacobian is exactly:
\begin{equation}
J(\tau_0,\zeta) = \sqrt{\frac{S_E}{m}}+\sum_{m=1}^\infty \zeta_mr_m
\end{equation}
And we see that it is indeed independent of $\tau_0$.


\section{Finding numerical bounce solutions \label{app:bounce} }
In this Appendix we discuss how to numerically find bounce solutions. We want to solve Eq. \eqref{bouncede}:
\be
\partial_\rho^2 \phi + \frac{3}{\rho} \partial_\rho- V'[\phi] =0 \label{bouncede2}
\ee
with boundary conditions $\phi'(0)=0$ and $\phi(\infty)=0$. Equivalently, we want to find an initial condition $\phi(0) = \phi_0$ for which the field rolls
down the potential $-V(\phi)$ ending at the origin $\phi=0$ at asymptotically late times. The usual shooting method suggests we try various values of $\phi_0$
until we find one initial condition $\phi_0^+$ for which the evolution overshoots (ends up with $\phi(\rho)<0$ for some $\rho$) and one initial
condition $\phi_0^-$ for which the evolution undershoots ($\phi(\rho)>0$ for all $\rho$).
Then we know the solution is between $\phi_0^+$ and $\phi_0^-$, so we simply have to refine this interval until the desired precision is reached.

One difficulty with the shooting method described above is that the $\frac{3}{\rho}$ coefficient in the differential equation makes the point $\rho=0$ singular. 
Thus when numerically solving the equation, one has to start at some small $\rho_0>0$, say $\rho_0=10^{-5}$. However, taking $\phi(\rho_0)=\phi_0$ and $\phi'(\rho_0)=0$ as boundary
conditions can be dangerous. These conditions imply that $\phi$ has rolled from some $\phi(0)$ to end up at $\phi_0$ {\it at rest} when $\rho=\rho_0$. But how is $\phi$ to have come
to rest at $\rho_0$? This is only possible if it rolls {\it up} the potential to get to $\phi_0$ and then turns around to roll back. Clearly, such a solution is not what
we were looking for and will depend on $\rho_0$. Often the effect of starting at $\rho_0$ is negligible, since the rolling starts off slow due to the friction term.
However, for improved convergence, or for situations like searching for multiple bounces in which high precision is necessary, it can be helpful to reduce the $\rho_0$ dependence.

This difficulty can be overcome by expanding the potential around $\phi_0$:
\be
V(\phi) \approx V_\text{lin} \equiv V(\phi_0) + (\phi-\phi_0) V'(\phi_0)
\ee
Using $V_{\text{lin}}(\phi)$ in Eq.~\eqref{bouncede2}, leads to an analytic solution
\be
\phi_{\text{lin}}(\rho)  = \phi_0 +\frac{1}{8}\rho^2 V'(\phi_0) 
\ee
So that if $\phi(0)= \phi_0$ with $\phi'(0)=0$, then
\be
\phi(\rho_0) = \phi_0 + \frac{1}{8}\rho_0^2 V'(\phi_0) \qquad \text{and} \qquad
\phi'(\rho_0) = \frac{1}{4}\rho_0 V'(\phi_0) 
\ee
Using these boundary conditions allows for an efficient numerical solution to the differential equation and a fast convergence towards the bounce. 
The solutions computed this way are very insensitive to $\rho_0$. 


\section{Higher Derivative Corrections \label{app:derivcorr}}
The effective action is constructed so that when used classically (at tree-level)
it reproduces the quantum physics (all loop-order) of a classical action.
Unfortunately, it is difficult, if not impossible, to calculate the effective action, exactly even at 1-loop. 
Diagrammatically the effective action can be computed by summing over 1PI graphs with any number of external legs with any momenta running through them. Even at 1-loop, there are an infinite number of relevant graphs, so computing the effective
action exactly is intractable.

Fortunately, a derivative expansion of the effective action is calculable. The 1PI effective action of a scalar field $\phi$ at up to 4 derivative order can be written as
\begin{multline}
\Seff[\phi] = \int d^4x \left[-V_{\text{eff}}(\phi) + \frac{1}{2}(\partial_\mu \phi)^2Z_2(\phi) \right.\\
\left.
+ \frac{1}{2}\left(\square \phi\right)^2 Z_4(\phi) + \frac{1}{2}(\partial_\mu \phi)^2\square\phi \rho(\phi)+\frac{1}{8}(\partial_\mu \phi)^2(\partial_\nu \phi)^2 \Omega(\phi) +\mc{O}(\partial^6) \right]
\label{GammaDerivExp}
\end{multline}
Using Lorentz invariance and integration by parts, we have reduced the action to depending on only 5 independent functions: $V_{\text{eff}}(\phi)$, $Z_2(\phi)$, $Z_4(\phi)$, $\rho(\phi)$, and $\Omega(\phi)$.

$V_{\text{eff}}$ is the well-known effective potential. 
To compute it, we expand the Lagrangian around a constant background field, $\mc{L}(\phi+\tilde{\phi})$, and calculate the vacuum diagrams where  $\phi$ propagates and $\tilde\phi$ is fixed.
(Throughout this appendix $\tilde{\phi}$ will represent a constant field.) 
To determine the other 4 functions, one might think to calculate the 1PI vertices using $\Seff$ and $\mc{L}$, and compare, but this is not so straightforward. The problem lies in the fact that $Z_2$, $Z_4$, $\rho$, and $\Omega$ are non-local functions of $\phi$, with terms like $\ln\frac{\phi}{\mu}$ or $\frac{1}{\phi^3}$ in them. One cannot derive
Feynman rules for such terms as one does for a local Lagrangian.

To proceed, we note that the effective potential is computed by expanding around a constant background field $\tilde\phi$, but with momentum dependence in $\phi$. In background-field calculations, the external lines, with or without momentum are always $\phi$. To find $\Seff$, we simply compute the same thing. We expand $\Seff[\phi + \tilde\phi]$ for constant $\tilde\phi$,
and compute diagrams with external $\phi$ legs, with or without momentum in them. The difference between the calculation
using $\Seff$ and using $\cL$ is that with the effective action, only tree-level graphs are ever evaluated.

Let's compute with the effective action first. We take $\phi \to \phi + \tilde\phi$ and series expand each function
around  $\tilde\phi=0$. Let us write 
$Z_i(\phi+\tilde{\phi}) = \sum_{i=0}^\infty \frac{1}{n!}Z_i^{(n)}(\tilde\phi) \phi^n$ and so on,
where for each function $f^{(n)}(\tilde\phi) \equiv \frac{d^n}{d (\tilde\phi)^n}f(\tilde\phi)$. 
Then,
\begin{multline}
\Seff[\phi+\tilde{\phi}] = \int d^4x \sum_{n=0}^\infty \frac{1}{n!} \phi^n \Big[-V_{\text{eff}}^{(n)}(\tilde{\phi}) + \frac{1}{2}(\partial_\mu \phi)^2Z_2^{(n)}(\tilde{\phi}) + \frac{1}{2}\left(\square \phi\right)^2 Z_4^{(n)}(\tilde{\phi}) 
\\
+ \frac{1}{2}(\partial_\mu \phi)^2\square\phi~\rho^{(n)}(\tilde{\phi})+\frac{1}{8}(\partial_\mu \phi)^2(\partial_\nu \phi)^2 \Omega^{(n)}(\tilde{\phi}) + \mc{O}(\partial^6) \Big]\label{eq:ExpandedGamma}
\end{multline}
The expanded Lagrangian is now local in $\phi$, so we can easily compute Feynman diagrams that have
 external $\phi$ lines with it. We get:
\eqn{\begin{tikzpicture}[baseline={([yshift=-.5ex]current bounding box.center)}]
    \node[anchor=west,inner sep=0] at (0,0) {\includegraphics[width=1.8cm]{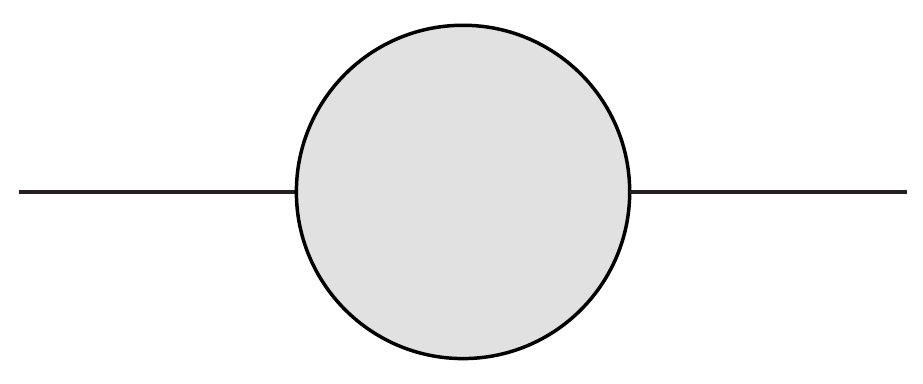}};
    \draw[->] (0.2,0.1)--(0.4,0.1);
    \node [below] at (0.3,0.05) {\scalebox{0.8}{$p_1$}};
        \draw[->] (1.6,0.1)--(1.4,0.1);
   \node [below] at (1.5,0.05) {\scalebox{0.8}{$p_2$}};
\end{tikzpicture}
 &= -i \frac{\partial^2}{\partial \tilde{\phi}^2}V_{\text{eff}}(\tilde{\phi}) - i p_1\cdot p_2Z_2(\tilde{\phi}) + i p_1^2 p_2^2 Z_4(\tilde{\phi}) + \mc{O}(p^6) \label{M2Gamma}\\
\begin{tikzpicture}[baseline={([yshift=-.5ex]current bounding box.center)}]
    \node[anchor=west,inner sep=0] at (0,0) {\includegraphics[width=1.5cm]{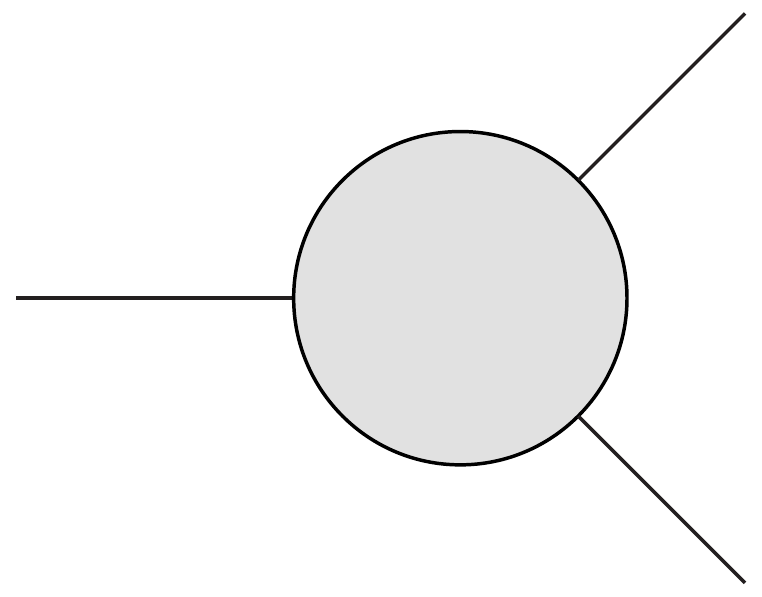}};
    \draw[->] (0.2,0.1)--(0.4,0.1);
    \node [below] at (0.3,0.05) {\scalebox{0.8}{$p_1$}};
        \draw[->] (1.3,0.55)--(1.15,0.4);
   \node [below] at (1.4,0.4) {\scalebox{0.8}{$p_2$}};
           \draw[->] (1.45,-0.4)--(1.29,-0.25);
   \node [left] at (1.4,-0.5) {\scalebox{0.8}{$p_3$}};
\end{tikzpicture}&= -i \frac{\partial^3}{\partial \tilde{\phi}^3}V_{\text{eff}}(\tilde{\phi}) - i \left(\sum_{i>j}p_i\cdot p_j\right)\frac{\partial}{\partial \tilde{\phi}}Z_2(\tilde{\phi}) + i \left(\sum_{i>j}p_i^2 p_j^2\right)\frac{\partial}{\partial \tilde{\phi}}Z_4(\tilde{\phi})\nonumber\\
&~~~~~~ + i\left(\sum_{\stackrel{i>j}{k\neq i,j}}p_i\cdot p_j p_k^2\right)\rho(\tilde{\phi}) + \mc{O}(p^6)\\
\begin{tikzpicture}[baseline={([yshift=-.5ex]current bounding box.center)}]
    \node[anchor=west,inner sep=0] at (0,0) {\includegraphics[width=1.2cm]{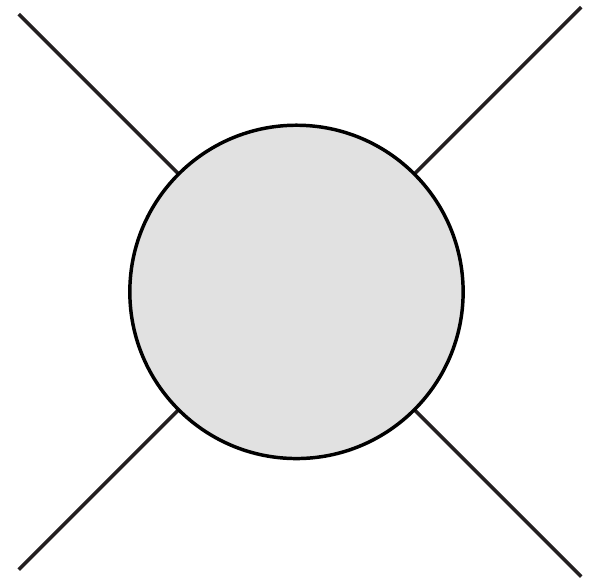}};
        \draw[->] (0.22,0.55)--(0.37,0.4);
   \node [below] at (0.125,0.4) {\scalebox{0.8}{$p_1$}};
        \draw[->] (1,0.55)--(0.85,0.4);
   \node [below] at (1.125,0.4) {\scalebox{0.8}{$p_2$}};
        \draw[->] (1.15,-0.4)--(1,-0.25);
   \node [below] at (0.85,-0.25) {\scalebox{0.8}{$p_3$}};
        \draw[->] (0.05,-0.4)--(0.2,-0.25);
   \node [below] at (0.4,-0.25) {\scalebox{0.8}{$p_4$}};
\end{tikzpicture}&= -i \frac{\partial^4}{\partial \tilde{\phi}^4}V_{\text{eff}}(\tilde{\phi}) - i \left(\sum_{i>j}p_i\cdot p_j\right)\frac{\partial^2}{\partial\tilde{\phi}^2}Z_2(\tilde{\phi})+ i\left(\sum_{i>j}p_i^2 p_j^2\right)\frac{\partial^2}{\partial\tilde{\phi}^2}Z_4(\tilde{\phi})\nonumber\\
&~~~~ + i\left(\sum_{\stackrel{i>j}{k\neq i,j}}p_i\cdot p_j p_k^2\right)\frac{\partial}{\partial\tilde{\phi}} \rho(\tilde{\phi})
+ i\left(\sum_{\stackrel{i>j,k>l}{\text{\tiny{all different}}}}p_i\cdot p_j p_k\cdot p_l\right)\Omega(\tilde{\phi}) + \mc{O}(p^6)\label{M4Gamma}}
This process can be continued to higher derivative terms if desired.

Now we compute the same $n$-point functions using the classical action, at 1-loop level, also with external $\phi$ field. 
For the example of massless $\phi^4$ theory, the Lagrangian density is
\eqn{\mc{L} = -\frac{1}{2}\phi \square \phi - \lambda\phi^4}
Following the process outlined above, we expand this with $\phi\rightarrow \phi+\tilde{\phi}$, resulting in
\eqn{\mc{L}_{\text{Expanded}} = - \lambda\tilde{\phi}^4 - 4\lambda\tilde{\phi}^3\phi -\frac{1}{2}\phi\left[\square + 12\lambda\tilde{\phi}^2\right]\phi  - 4\lambda\tilde{\phi} \phi^3 - \lambda \phi^4}
We can safely drop $- \lambda\tilde{\phi}^4 - 4\lambda\tilde{\phi}^3\phi$ because these terms cannot contribute to the diagrams we are calculating. Thus for our purposes, we have
\eqn{\mc{L}_{\text{Expanded}} = -\frac{1}{2}\phi\left[\square + m^2\right]\phi  - 4\lambda\tilde{\phi} \phi^3 - \lambda\phi^4}
with $m^2 = 12\lambda\tilde{\phi}^2$. The Feynman rules are
\eqn{D(k) = \fd{1.5cm}{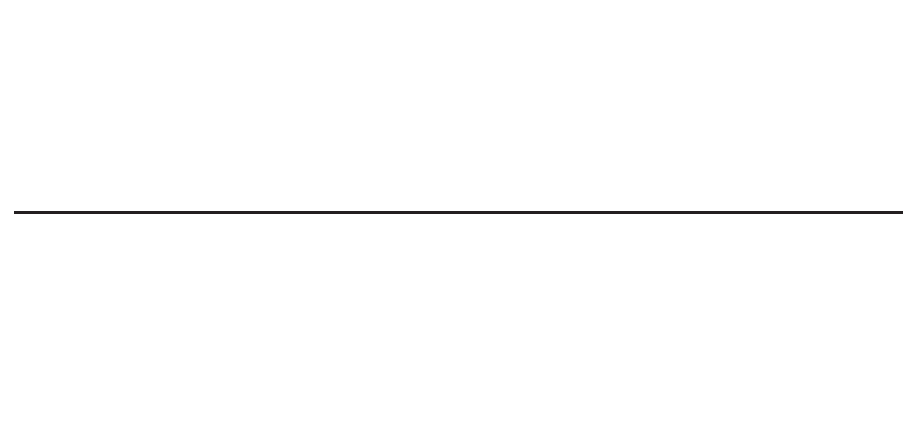} =  \frac{i}{k^2 - m^2}~, ~~\fd{2cm}{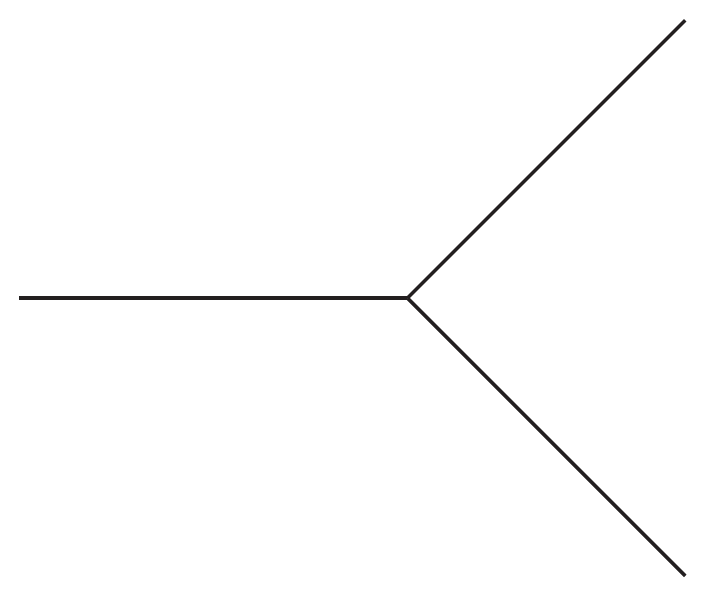} = -i24\lambda\tilde{\phi}~, ~~ \fd{2cm}{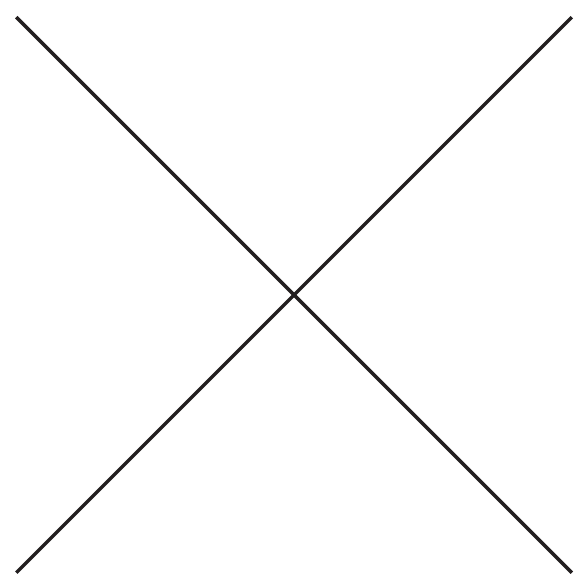}= -i24\lambda }

The 0-point diagrams  give  the well-known 1-loop effective potential\cite{Coleman:1973jx}:
\eqn{V_{\text{eff}} =\lambda\phi^4 -\frac{i}{2}\hbar\int \frac{d^4k}{(2\pi)^4} \ln\left(1-\frac{m^2}{k^2}\right) + \cdots = \lambda\phi^4 + \frac{9 \hbar}{4\pi^2}\lambda^2\phi^4\left(\ln\frac{12\lambda\phi^2}{\mu^2} - \frac{3}{2}\right) + \mc{O}(\hbar^2)}

There are two 1-loop diagrams for the 2-point function
\eqn{\fd{4cm}{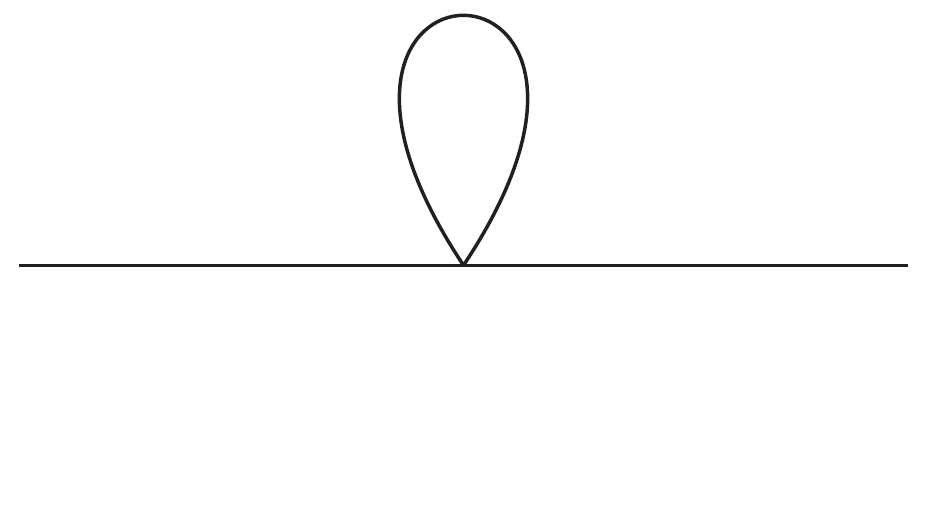}~~~~~~\fd{4cm}{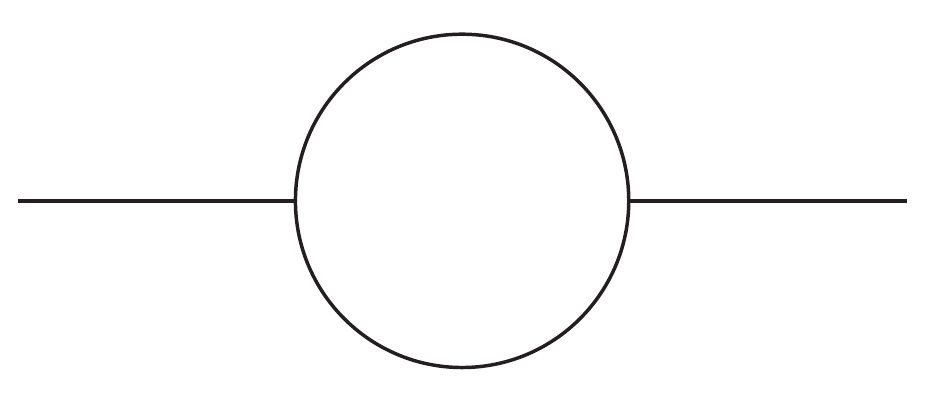}}
which along with the tree-level piece, is given by the amplitude
\eqn{\begin{tikzpicture}[baseline={([yshift=-.5ex]current bounding box.center)}]
    \node[anchor=west,inner sep=0] at (0,0) {\includegraphics[width=1.8cm]{2point1PI}};
    \draw[->] (0.2,0.1)--(0.4,0.1);
    \node [below] at (0.3,0.05) {\scalebox{0.8}{$p_1$}};
        \draw[->] (1.6,0.1)--(1.4,0.1);
   \node [below] at (1.5,0.05) {\scalebox{0.8}{$p_2$}};
\end{tikzpicture}&= -i (m^2-p_1^2) + \frac{(24 \lambda \phi)^2}{2}\hbar\int \frac{d^4k}{(2\pi)^4} \frac{1}{(k^2-m^2)((k-p_1)^2-m^2)} + 12\lambda\hbar\int\frac{d^4k}{(2\pi)^4}\frac{1}{k^2-m^2}\nonumber\\
&= -i(m^2-p_1^2) + i\frac{\hbar}{16\pi^2}\frac{(24\lambda\phi)^2}{2}\left[- \ln \left(m^2\right) + p_1^2\frac{1}{6 m^2} + (p_1^2)^2\frac{1}{60 (m^2)^2} + \cdots \right]\nonumber\\
&~~~~~~~~~~~~~~~~ + i\frac{\hbar}{16\pi^2}12\lambda m^2\left[1-\ln m^2\right]}
where in the last line, we have evaluated the integral expanded in momenta. We find that the momentum-free piece gives $-i\frac{\partial^2V_{\text{eff}}}{\partial \phi^2}$ as expected. Using $p_1=-p_2$, we can rewrite this in the form of 
Eq. (\ref{M2Gamma})
\eqn{\begin{tikzpicture}[baseline={([yshift=-.5ex]current bounding box.center)}]
    \node[anchor=west,inner sep=0] at (0,0) {\includegraphics[width=1.8cm]{2point1PI}};
    \draw[->] (0.2,0.1)--(0.4,0.1);
    \node [below] at (0.3,0.05) {\scalebox{0.8}{$p_1$}};
        \draw[->] (1.6,0.1)--(1.4,0.1);
   \node [below] at (1.5,0.05) {\scalebox{0.8}{$p_2$}};
\end{tikzpicture}&=-i(m^2+p_1\cdot p_2) + i\frac{\hbar}{16\pi^2}\frac{(24\lambda\phi)^2}{2}\left[- \ln \left(m^2\right) - p_1\cdot p_2\frac{1}{6 m^2} + p_1^2p_2^2\frac{1}{60 (m^2)^2} + \cdots \right]\nonumber\\
&~~~~~~~~~~~~ + i\frac{\hbar}{16\pi^2}12\lambda m^2\left[1-\ln m^2\right]}
from which we extract
\eqn{Z_2 &= 1 + \frac{\hbar}{16\pi^2}4\lambda \frac{12\lambda\phi^2}{m^2} = 1 + \frac{\hbar}{4\pi^2}\lambda \\
Z_4 &= \frac{\hbar}{4\pi^2}\frac{\lambda}{10}\frac{12\lambda\phi^2}{(m^2)^2} = \frac{\hbar}{480\pi^2\phi^2}}

For the 3-point function there are 2 types of 1-loop diagrams
\eqn{\fd{3cm}{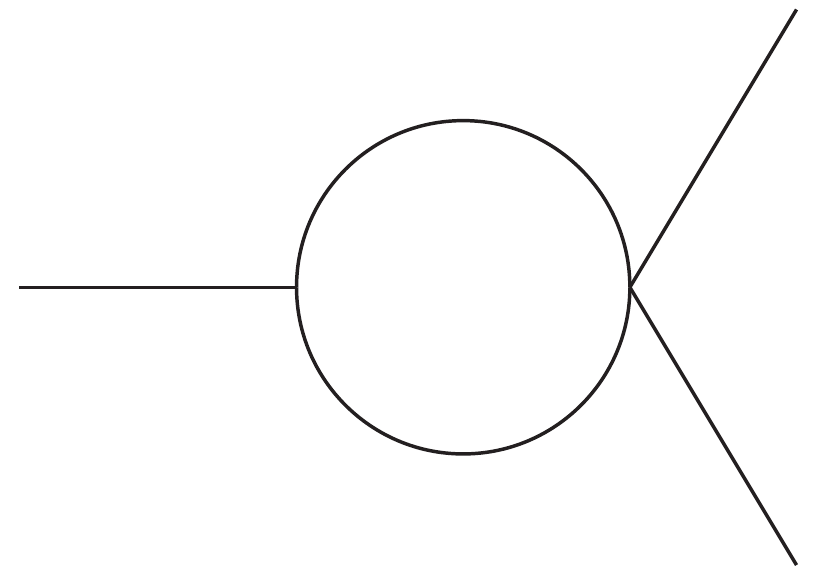}~~~~~~\fd{3cm}{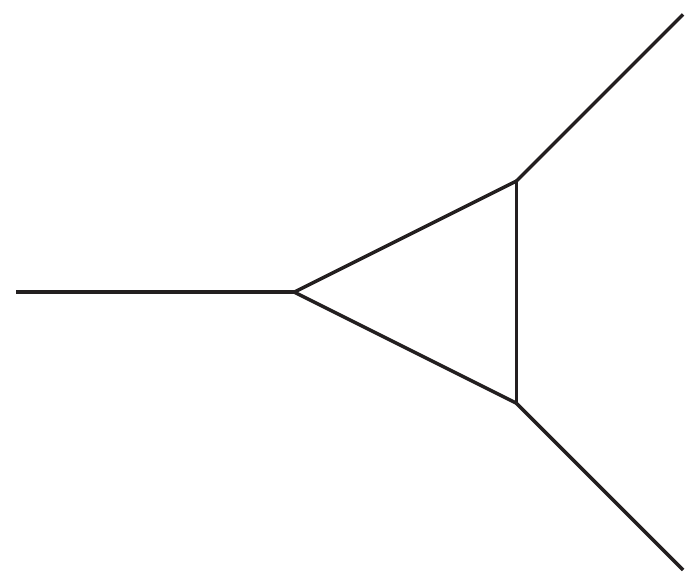}}
and for the 4-point function there are 3 types of 1-loop diagrams 
\eqn{\fd{3cm}{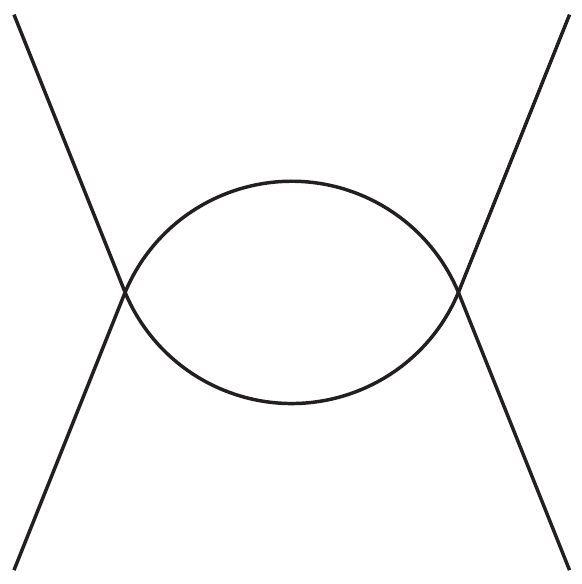}~~~~~~\fd{3cm}{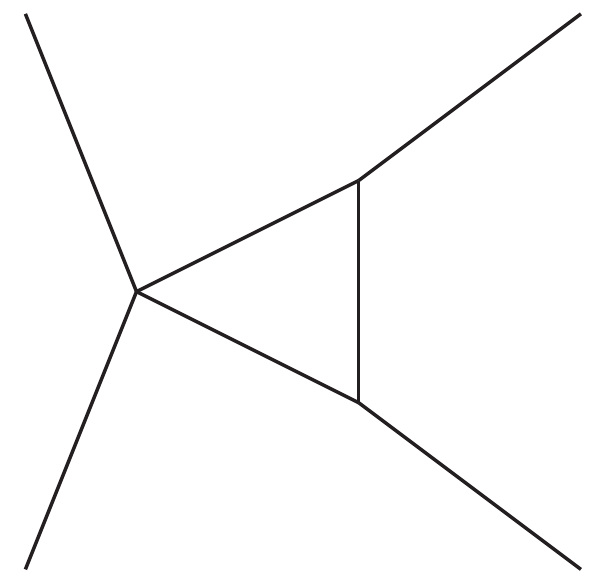}~~~~~~\fd{3cm}{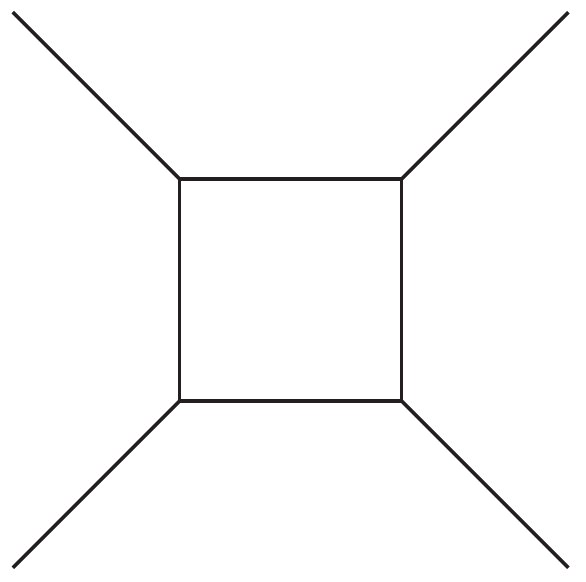}}
Following the same procedure as for the 2-point function we find the 3- and 4-point function amplitudes
\eqn{\begin{tikzpicture}[baseline={([yshift=-.5ex]current bounding box.center)}]
    \node[anchor=west,inner sep=0] at (0,0) {\includegraphics[width=1.5cm]{3point1pI}};
    \draw[->] (0.2,0.1)--(0.4,0.1);
    \node [below] at (0.3,0.05) {\scalebox{0.8}{$p_1$}};
        \draw[->] (1.3,0.55)--(1.15,0.4);
   \node [below] at (1.4,0.4) {\scalebox{0.8}{$p_2$}};
           \draw[->] (1.45,-0.4)--(1.29,-0.25);
   \node [left] at (1.4,-0.5) {\scalebox{0.8}{$p_3$}};
\end{tikzpicture} &= -i \left(24\lambda\phi + \frac{\hbar}{4\pi^2}144\lambda^2\phi\left(1+\frac{3}{2}\ln\frac{12\lambda\phi^2}{\mu^2}\right)\right)\nonumber\\
&~~~~~~ - i\left(\sum_{i>j}p_i^2 p_j^2\right)\frac{\hbar}{240\pi^2\phi^3} - i \left(\sum_{\stackrel{i>j}{k\neq i,j}}p_i\cdot p_j p_k^2\right)\frac{\hbar}{720\pi^2\phi^3}\\
\begin{tikzpicture}[baseline={([yshift=-.5ex]current bounding box.center)}]
    \node[anchor=west,inner sep=0] at (0,0) {\includegraphics[width=1.2cm]{4point1pI}};
        \draw[->] (0.22,0.55)--(0.37,0.4);
   \node [below] at (0.125,0.4) {\scalebox{0.8}{$p_1$}};
        \draw[->] (1,0.55)--(0.85,0.4);
   \node [below] at (1.125,0.4) {\scalebox{0.8}{$p_2$}};
        \draw[->] (1.15,-0.4)--(1,-0.25);
   \node [below] at (0.85,-0.25) {\scalebox{0.8}{$p_3$}};
        \draw[->] (0.05,-0.4)--(0.2,-0.25);
   \node [below] at (0.4,-0.25) {\scalebox{0.8}{$p_4$}};
\end{tikzpicture} &= -i\left(24\lambda + \frac{\hbar}{4\pi^2}(24\lambda)^2\left(1+\frac{3}{8}\ln\frac{12\lambda\phi^2}{\mu^2}\right)\right) + i\left(\sum_{i>j}p_i^2 p_j^2\right)\frac{\hbar}{80\pi^2\phi^4} \nonumber\\
&~~~~~~  +  i \left(\sum_{\stackrel{i>j}{k\neq i,j}}p_i\cdot p_j p_k^2\right)\frac{\hbar}{240\pi^2\phi^4} +i\left(\sum_{\stackrel{i>j,k>l}{\text{\tiny{all different}}}}p_i\cdot p_j p_k\cdot p_l\right)\frac{\hbar}{360\pi^2\phi^4}}
From these results we extract $\rho = \frac{\hbar}{720\pi^2\phi^3}$ and $\Omega=\frac{\hbar}{360\pi^2\phi^4}$.

To summarize our results, we found that up to 4 derivatives, the effective action for $\phi^4$ theory is given by
\eqn{\Seff[\phi] &= \int d^4x \Big[-\left(\lambda\phi^4 + \frac{9 \hbar}{4\pi^2}\lambda^2\phi^4\left(\ln\frac{12\lambda\phi^2}{\mu^2} - \frac{3}{2}\right)\right)\nonumber\\
&~~~~~~~~~~~~~~~~~~~~+ \frac{1}{2}(\partial_\mu \phi)^2\left(1+\frac{\hbar}{4\pi^2}\lambda\right)+ \frac{1}{2}\left(\square \phi\right)^2 \frac{\hbar}{480\pi^2}\frac{1}{\phi^2}\nonumber\\
&~~~~~~~~~~~~~~~~~~~~ - \frac{1}{2}(\partial_\mu \phi)^2\square\phi \frac{\hbar}{720\pi^2}\frac{1}{\phi^3}+\frac{1}{8}(\partial_\mu \phi)^2(\partial_\nu \phi)^2\frac{\hbar}{360\pi^2\phi^4}+\mc{O}(\partial^6) \Big]\label{EAphi4app}}
This is written in Minkowski space. Going to Euclidean space, we send $t\to -i\tau$, which changes $(\partial_\mu \phi)^2\to -(\partial_\mu \phi)^2$ and $\Box \phi \to -\Box \phi$. Pulling out an extra minus sign (since $i\Seff=-\Seff^E$), we find
\eqn{\Seff^E[\phi] &= \int d^4x \Big[\lambda\phi^4 + \frac{9 \hbar}{4\pi^2}\lambda^2\phi^4\left(\ln\frac{12\lambda\phi^2}{\mu^2} - \frac{3}{2}\right)\nonumber\\
&~~~~~~~~~~~~~~~~~~~~+ \frac{1}{2}(\partial_\mu \phi)^2\left(1+\frac{\hbar}{4\pi^2}\lambda\right)- \frac{1}{2}\left(\square \phi\right)^2 \frac{\hbar}{480\pi^2}\frac{1}{\phi^2}\nonumber\\
&~~~~~~~~~~~~~~~~~~~~ + \frac{1}{2}(\partial_\mu \phi)^2\square\phi \frac{\hbar}{720\pi^2}\frac{1}{\phi^3}
-\frac{1}{8}(\partial_\mu \phi)^2(\partial_\nu \phi)^2\frac{\hbar}{360\pi^2\phi^4}+\mc{O}(\partial^6) \Big]\label{EAphi4appEuclidean}}
where $d^4x=d\tau d^3\vec{x}$ and $\mu=0$ corresponds to $x^0=\tau$.

Our final result agrees with  \cite{Fraser:1984zb}, where the 4-derivative terms were computed using a different method.

\section{NLO Functional Determinants}\label{sec:nlofunctionaldeterminants}
In quantum field theory, the calculation of decay rates at NLO amounts to evaluating the ratio of functional determinants:
\begin{equation}
\frac{\Gamma^{\text{NLO}}}{V} = \frac{1}{V\bigtau}\frac{e^{-S_E[\bar{\phi}]}}{e^{-S_E[\phi_{\FV}]}}\abs{ \frac{\Det(-\partial^2+V''[\bar \phi(x]))}{\Det (-\partial^2+V''[\phi_{\FV}])}}^{-1/2}
\end{equation}
Here, $\bar\phi$ is the bounce solution to the Euclidean equations of motion, and $\phi_{\FV}$ is the field value at the false vacuum which we can always choose to be zero. Generally, $\bar\phi(x)$ is spherically symmetric, only depending on the Euclidean length $r$. So we define
\be
W(r) = V''[\bar\phi(r)]
\ee
One can evaluate the functional determinant in terms of Feynman diagrams
\begin{align}
\Delta S&=\frac{1}{2}\log\Det\left(-\partial^2+W(r)\right)-\frac{1}{2}\log\Det\left(-\partial^2\right)\\
&=\frac{1}{2}\Tr\log\left(-\partial^2+W(r)\right)-\frac{1}{2}\Tr\log\left(-\partial^2\right)\\
&=-\Tr\sum_{n=1}^\infty\frac{(-1)^{n}}{2n}\left[(-\partial^2)^{-1}W(r)\right]^n\\
&=
%
%
\begin{tikzpicture}[baseline={([yshift=-.5ex]current bounding box.center)}]
	\coordinate (c1) at (0,0);
	\newcommand*{\bigRadius}{0.8cm}
	\newcommand*{\smallRadius}{0.15cm}
	\draw[thick] ($(c1) + (0:\bigRadius)$) arc (0:360:\bigRadius);
	\draw[fill, white] ($(c1) + (180:\bigRadius)+(0:\smallRadius)$) arc (0:360:\smallRadius);
	\draw[thick] ($(c1) + (180:\bigRadius)+(0:\smallRadius)$) arc (0:360:\smallRadius);
	\draw[thick] ($(c1) + (180:\bigRadius)+(45:\smallRadius)$) -- ($(c1) + (180:\bigRadius)+(-135:\smallRadius)$);
	\draw[thick] ($(c1) + (180:\bigRadius)+(-45:\smallRadius)$) -- ($(c1) + (180:\bigRadius)+(135:\smallRadius)$);
	\node[left] at ($(c1) + (180:\bigRadius)$) {\footnotesize$\tilde{W}(0)$};
\end{tikzpicture} ~ + 
%
%
\begin{tikzpicture}[baseline={([yshift=-.5ex]current bounding box.center)}]
	\coordinate (c1) at (0,0);
	\newcommand*{\bigRadius}{0.8cm}
	\newcommand*{\smallRadius}{0.15cm}
	\draw[thick] ($(c1) + (0:\bigRadius)$) arc (0:360:\bigRadius);
	\draw[fill, white] ($(c1) + (180:\bigRadius)+(0:\smallRadius)$) arc (0:360:\smallRadius);
	\draw[thick] ($(c1) + (180:\bigRadius)+(0:\smallRadius)$) arc (0:360:\smallRadius);
	\draw[thick] ($(c1) + (180:\bigRadius)+(45:\smallRadius)$) -- ($(c1) + (180:\bigRadius)+(-135:\smallRadius)$);
	\draw[thick] ($(c1) + (180:\bigRadius)+(-45:\smallRadius)$) -- ($(c1) + (180:\bigRadius)+(135:\smallRadius)$);
	\node[left] at ($(c1) + (180:\bigRadius)$) {\footnotesize$\tilde{W}(q)$};
	\draw[fill, white] ($(c1) + (0:\bigRadius)+(0:\smallRadius)$) arc (0:360:\smallRadius);
	\draw[thick] ($(c1) + (0:\bigRadius)+(0:\smallRadius)$) arc (0:360:\smallRadius);
	\draw[thick] ($(c1) + (0:\bigRadius)+(45:\smallRadius)$) -- ($(c1) + (0:\bigRadius)+(-135:\smallRadius)$);
	\draw[thick] ($(c1) + (0:\bigRadius)+(-45:\smallRadius)$) -- ($(c1) + (0:\bigRadius)+(135:\smallRadius)$);
	\node[right] at ($(c1) + (0:\bigRadius)$) {\footnotesize$\tilde{W}(-q)$};
\end{tikzpicture}+
%
%
\begin{tikzpicture}[baseline={([yshift=-.5ex]current bounding box.center)}]
	\coordinate (c1) at (0,0);
	\newcommand*{\bigRadius}{0.8cm}
	\newcommand*{\smallRadius}{0.15cm}
	\draw[thick] ($(c1) + (0:\bigRadius)$) arc (0:360:\bigRadius);
	\draw[fill, white] ($(c1) + (180:\bigRadius)+(0:\smallRadius)$) arc (0:360:\smallRadius);
	\draw[thick] ($(c1) + (180:\bigRadius)+(0:\smallRadius)$) arc (0:360:\smallRadius);
	\draw[thick] ($(c1) + (180:\bigRadius)+(45:\smallRadius)$) -- ($(c1) + (180:\bigRadius)+(-135:\smallRadius)$);
	\draw[thick] ($(c1) + (180:\bigRadius)+(-45:\smallRadius)$) -- ($(c1) + (180:\bigRadius)+(135:\smallRadius)$);
	\node[left] at ($(c1) + (180:\bigRadius)$) {\footnotesize$\tilde{W}(p+q)$};
	\draw[fill, white] ($(c1) + (60:\bigRadius)+(0:\smallRadius)$) arc (0:360:\smallRadius);
	\draw[thick] ($(c1) + (60:\bigRadius)+(0:\smallRadius)$) arc (0:360:\smallRadius);
	\draw[thick] ($(c1) + (60:\bigRadius)+(45:\smallRadius)$) -- ($(c1) + (60:\bigRadius)+(-135:\smallRadius)$);
	\draw[thick] ($(c1) + (60:\bigRadius)+(-45:\smallRadius)$) -- ($(c1) + (60:\bigRadius)+(135:\smallRadius)$);
	\node[right] at ($(c1) + (60:\bigRadius)$) {\footnotesize$\tilde{W}(-q)$};
	\draw[fill, white] ($(c1) + (-60:\bigRadius)+(0:\smallRadius)$) arc (0:360:\smallRadius);
	\draw[thick] ($(c1) + (-60:\bigRadius)+(0:\smallRadius)$) arc (0:360:\smallRadius);
	\draw[thick] ($(c1) + (-60:\bigRadius)+(45:\smallRadius)$) -- ($(c1) + (-60:\bigRadius)+(-135:\smallRadius)$);
	\draw[thick] ($(c1) + (-60:\bigRadius)+(-45:\smallRadius)$) -- ($(c1) + (-60:\bigRadius)+(135:\smallRadius)$);
	\node[right] at ($(c1) + (-60:\bigRadius)$) {\footnotesize$\tilde{W}(-p)$};
\end{tikzpicture}
+\cdots\\
&=\frac{1}{2}\tilde{W}(0)\int\frac{d^dp}{(2\pi)^d}\frac{1}{p^2}-\frac{1}{4}\int\frac{d^dq}{(2\pi)^d}\frac{d^dk}{(2\pi)^d}\frac{\tilde{W}(q)\tilde{W}(-q)}{k^2(k+q)^2}+\cdots \label{eqn:deltaSdivergent}
\end{align}
We see that the first two terms are UV-divergent, and these divergences can easily be removed using $\msbar$ counterterms. All the other terms will be UV finite, but they are unfortunately very complicated to calculate. Note that we are not expanding in any small parameter, so, in general, all terms will be equally important. Hence, they are not only hard to calculate, we would also have to calculate infinitely many of them. 

There is an alternative way of calculating the functional determinants using the 
Gelfand-Yaglom theorem\cite{Dunne:2005rt, Isidori:2001bm} which makes it possible to calculate $\Delta S$ to all orders in $W$. 
Since $W(r)$ only depends on the Euclidean distance $r$, we can decompose $[-\partial^2+W(r)]$ into partial waves. We start by writing $\partial^2=\frac{d^2}{dr^2}+\frac{3}{r}\frac{d}{dr}-\frac{L^2}{r^2}\equiv \nabla^2_l$, where $L^2$ is the four-dimensional angular momentum operator with eigenvalue $l(l+2)$ and degeneracy $(l+1)^2$ for $l=0,1,2,\dots$. We can then write

\be
\Delta S \equiv\frac{1}{2}\ln\frac{\Det\left[-\partial^2+W(r)\right]}{\Det \left[-\partial^2\right]}=\frac{1}{2}\sum_l (l+1)^2 \ln\frac{\Det \left[-\nabla_l^2+W(r)\right]}{\Det\left[-\nabla_l^2\right]}
\ee
We can solve for the ratio of determinants for each $l$ by solving for the two functions $u_{W}^l(r)$  and $u_0^l(r)$, which are radial eigenfunctions of $-\partial^2+W$ and $-\partial^2$, respectively,
regular at $r=0$, with a given $l$ and zero eigenvalues. Then
\be
\rho_l \equiv \frac{\Det \left[-\nabla_l^2+W(r)\right]}{\Det\left[-\nabla_l^2\right]}  = \lim_{r\to \infty} \rho_l(r) = \lim_{r\to \infty} \frac{u_{W}^l(r)}{u_0^l(r)}
\label{rhoforms}
\ee
One can solve for $u_0^l(r)$, and express the differential equation for $u_W^l(r)$ in terms of $\rho_l(r)$ 
\be
\rho_l''(r)+\frac{2l+3}{r}\rho'_l(r)=W(r)\rho_l(r).\label{eq:rhol}
\ee
In summary, to calculate $\Delta S$ we need to find $\rho_l(r)$ from  Eq.~\eqref{eq:rhol} for each $l$ and then sum the asymptotic
values
\be
\Delta S=\frac{1}{2}\sum_{l=0}^\infty(l+1)^2 \ln \rho_l \,,
\label{eq:sumeigenvaluesrho}
\ee
where $\rho_l$ is related to $\rho_l(r)$ by Eq.~\eqref{rhoforms}

There are two complications with directly implementing this approach. First, for most cases of interest,  $\rho_l$ can be either negative or zero for $l=0,1$, so that $\ln \rho_l$ is infinite or complex. The negative eigenvalue is of course expected, as the imaginary part is supposed to give the decay rate. The zero eigenvalues are also expected, as they correspond to exact or approximate symmetries such as translation  or scale invariance. See \cite{Dunne:2005rt, Isidori:2001bm, Branchina:2013jra} for a discussion on how to evaluate the zero and negative eigenvalues.

Second, the sum over $l$ is divergent, as we already knew it had to be from Eq.~\eqref{eqn:deltaSdivergent}. The UV-divergence in Eq.~\eqref{eqn:deltaSdivergent} all came from the $\cO(W)$ and $\cO(W^2)$ terms which we easily can calculate analytically. So let us define $\Delta S^{[2]}\equiv\left[\Delta S\right]_{\cO(W^2)}$, i.e. formally truncated to second order in $W$. We then add and subtract $\Delta S^{[2]}$, as well as subtracting off the infinities using $\msbar$ counterterms
\be
\Delta S \to \left[\Delta S -\Delta S^{[2]}\right]+\left[\Delta S^{[2]}-\delta S_{\text{ct}}\right]
\ee 
The second bracket can be calculated analytically using Eq.~\eqref{eqn:deltaSdivergent} and the first bracket will be calculated numerically. 

To find $\rho_j$ truncated to second order in $W$, we write $\rho_l(r)=1+\rho_l^{(1)}(r)+\rho_l^{(2)}(r)+\cdots$, where $\rho_l^{(1)}(r)=\cO(W)$, $\rho_l^{(2)}(r)=\cO(W^2)$, etc. Solving Eq. \eqref{eq:rhol} order by order in $W$, we find the set of equations
\begin{align}
\rho^{(1)}_l{}''(r)+\frac{2l+3}{r}\rho^{(1)}_l{}'(r)&=W(r),\\
\rho^{(2)}_l{}''(r)+\frac{2l+3}{r}\rho^{(2)}_l{}'(r)&=W(r)\rho^{(1)}_l(r),
\end{align}
When we have obtained these solutions, we can calculate 
\begin{align}
\Delta S^{[2]}&=\frac{1}{2}\sum_{l=0}^\infty(l+1)^2 \left[\ln \rho_l\right]_{\cO(W^2)}\\
&=\frac{1}{2}\sum_{l=0}^\infty(l+1)^2 \left[\ln \left(1+\rho_l^{(1)}+\rho_l^{(2)}+\cdots\right)\right]_{\cO(W^2)}\\
&=\frac{1}{2}\sum_{l=0}^\infty(l+1)^2\left[\rho_l^{(1)}-\frac{1}{2}\left(\rho_l^{(1)}\right)^2+\rho_l^{(2)}\right].
\end{align}
Hence, we find that the numerical bracket is
\be
\left[\Delta S-\Delta S^{[2]}\right]=\frac{1}{2}\sum_{l=0}^\infty(l+1)^2\left[\ln\rho_l-\rho_l^{(1)}+\frac{1}{2}\left(\rho_l^{(1)}\right)^2-\rho_l^{(2)}\right]
\ee
which will be finite as $l\to \infty$. In practice, this sum rapidly converges, and one only has to sum a finite number of terms.  

We can simplify $\left[\Delta S^{[2]}-\delta S_{\text{ct}}\right]$ by noting that the $\cO(W)$ term in Eq.~\eqref{eqn:deltaSdivergent} has a scaleless momentum space integral, which is zero in dim.reg. using $\msbar$. The $\cO(W^2)$ integral can be simplified by first doing integral over $k$
\begin{align}
B_0(q^2)&=\mu^{4-d} \int \frac{d^d k}{(2\pi)^d}\frac{1}{k^2(k+q)^2}=\frac{1}{(4\pi)^2}\left[\frac{1}{\epsilon}+2+\ln\frac{\mu^2}{q^2}\right]
\end{align}
Removing the infinity using the counterterm, we find
\be
\left[\Delta S^{[2]}-\delta S_{\text{ct}}\right]
=-\frac{1}{4}\int \frac{d^4 q}{(2\pi)^4}[\tilde{W}(q)]^2 B_0(q^2)\label{eqn:analyticDeltaS2}
\ee
In the case of $V=-\frac{|\lambda|}{4!}\phi^4$ 
\be
W(r)=-\frac{|\lambda|}{2} \bar{\phi}(r)^2=-\frac{24R^2}{(r^2+R^2)^2}  \Rightarrow
\tilde{W}(p)=48 \pi^2 R^2 K_0(|p| R)
\ee
where $K_0$ is a BesselK function. 
Using Eq.~\eqref{eqn:analyticDeltaS2}, we then find
\be
\left[\Delta S^{[2]}-\delta S_{\text{ct}}\right]=-3L-\frac{5}{2}
\ee
where $L=\ln\frac{R\mu e^{\gamma_E}}{2}$.

\bibliography{stable}

\bibliographystyle{utphys}

\end{document}